\documentclass[graybox]{svmult}

\usepackage{mathptmx}       
\usepackage{helvet}         
\usepackage{courier}        
\usepackage{type1cm}        
\usepackage{makeidx}         
\usepackage{graphicx}        
\usepackage{multicol}        
\usepackage[bottom]{footmisc}

\usepackage{natbib}  
\usepackage{float}

\usepackage{url}

\makeindex             

\newcommand{\ha}{\hbox{H$\alpha$}}
\newcommand{\hb}{\hbox{H$\beta$}}
\def\oii{{\rm O{\sc ii}}}
\def\oiii{{\rm O{\sc iii}}}
\def\cii{{\rm C{\sc ii}}}
\def\ciii{{\rm C{\sc iii}}}
\def\civ{{\rm C{\sc iv}}}
\def\nii{{\rm N{\sc ii}}}
\def\nv{{\rm N{\sc v}}}
\def\ovi{{\rm O{\sc vi}}}

\def\hii{{\rm H{\sc ii}}}
\def\hi{{\rm H{\sc i}}}
\def\h2{{\rm H$_2$}}
\def\heii{{\rm He{\sc ii}}}

\newcommand\mnras{MNRAS}
\newcommand\apjl{ApJL}
\newcommand\apj{ApJ}
\newcommand\apjs{ApJS}
\newcommand\aj{AJ}
\newcommand\aap{A\&A}
\newcommand\pasp{PASP}
\newcommand\pasj{PASJ}
\newcommand\araa{ARAA}
\newcommand\nat{Nature}
\newcommand\physrep{PhR}
\newcommand\prd{PhRvD}
\newcommand\azh{AZh}
\newcommand \procspie{SPIE}

\def\ion#1#2{#1$\;${\sc\@roman{#2}}\relax}
\def\lesssim{\mathrel{\hbox{\rlap{\hbox{\lower4pt\hbox{$\sim$}}}\hbox{$<$}}}}
\def\gtrsim{\mathrel{\hbox{\rlap{\hbox{\lower4pt\hbox{$\sim$}}}\hbox{$>$}}}}

\begin{document}

\title*{Observations of Ly$\alpha$ Emitters at High Redshift}
\author{Masami Ouchi}
\institute{Masami Ouchi \at ICRR, The University of Tokyo, 5-1-5 Kashiwanoha, Kashiwa, 277-8582, Japan, \email{ouchims@icrr.u-tokyo.ac.jp}
}
%
%
\maketitle

\abstract{In this series of lectures, I review our observational understanding of high-$z$ Ly$\alpha$ emitters (LAEs) and relevant scientific
topics. Since the discovery of LAEs in the late 1990s, significant progresses in LAE studies have been made over the past two decades 
by deep multi-wavelength observations. More than ten (one) thousand(s) of LAEs 
have been identified photometrically (spectroscopically) in optical and near-infrared data, and the redshifts of these LAEs range from $z\sim 0$ to $z\sim 10$.
These large samples of LAEs are useful to address two major astrophysical issues, galaxy formation and cosmic reionization.
Statistical studies have revealed the general picture of LAEs' physical properties: young stellar populations, 
remarkable luminosity function evolutions, compact morphologies, highly ionized inter-stellar media (ISM) with low metal/dust contents, 
low masses of dark-matter halos. Typical LAEs represent low-mass high-$z$ galaxies, high-$z$ analogs of dwarf galaxies, some of which 
are thought to be candidates of population III galaxies. These observational studies have also pinpointed
rare bright Ly$\alpha$ sources extended over $\sim 10-100$ kpc, dubbed Ly$\alpha$ blobs,
whose physical origins are under debate.
%
LAEs are used as probes of cosmic reionization history 
through the Ly$\alpha$ damping wing absorption given by
the neutral hydrogen of the inter-galactic medium (IGM), which complement 
the cosmic microwave background radiation and 21cm observations targeting 
the epoch of reionization.
The low-mass and highly-ionized population of LAEs can be major sources of cosmic reionization,
and physical parameters including the ionizing photon escape fraction
have been extensively investigated. The budget of ionizing photons for cosmic reionization has been constrained,
although there remain large observational uncertainties in the parameters. Beyond these two established topics of LAEs,
galaxy formation and cosmic reionization, several new usages of LAEs for science frontiers have been suggested such as
the distribution of {\sc Hi} gas in the circum-galactic medium and filaments of large-scale structures.
On-going 10 m-class optical telescope programs and future telescope projects, such as JWST, ELTs, and SKA, 
will address the remaining open questions related to LAEs, and push the horizons of the science frontiers.
}

\section{Introduction}
\label{sec:introduction}

About two decades have passed since the observational discovery of Ly$\alpha$ emitters (LAEs) 
at high redshift. Before then, 
early theoretical studies focused on discussing young primordial galaxies 
with strong Ly$\alpha$ emission. However, after the discovery,
observations have revealed a number of exciting characteristics of LAEs, some of which 
are beyond the theoretical predictions. In this section, I show the growing importance of LAE studies, 
overviewing the LAE observation history through the early theoretical predictions, 
the discovery, and new problems in this observational field.
%
Throughout this lecture, magnitudes are in the AB system, if not otherwise specified.
All physical values are calculated with the concordance cosmology of
$H_0=100 h$ km s$^{-1}$ Mpc$^{-1}$ with $h\simeq 0.7$,
${\rm \Omega_m}\simeq 0.3$, ${\rm \Omega_\Lambda}\simeq 0.7$,
${\rm \Omega_b} h^2\simeq 0.02$, $\sigma_8\simeq 0.8$, and $n_s\simeq 1.0$
that are consistent with the latest Planck2016 cosmology \citep{planck2016}.

\subsection{Predawn of the LAE Observation History}
\label{sec:predawn}

\subsubsection{Theoretical Predictions}
\label{sec:theoretical_predictions}

\begin{figure}[H]
\centering
\includegraphics[scale=.40]{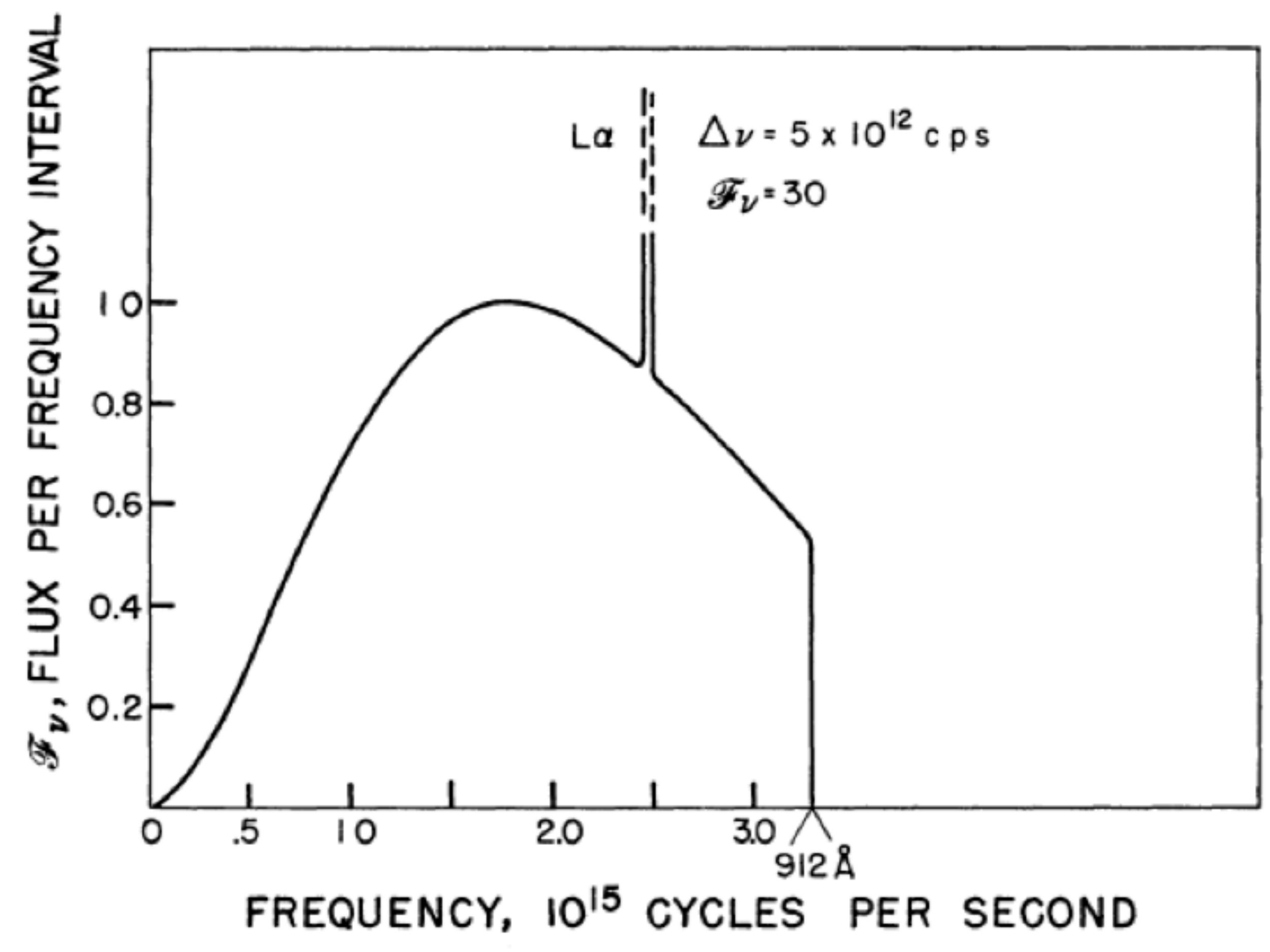}
\caption{
Expected spectrum of a young galaxy  
\citep{partridge1967a}.
Here, $\Delta \nu =0.002\nu$ is assumed for the line width.
This figure is reproduced by permission of the AAS.
}
\label{fig:partridge1976a_fig3}
\end{figure}

\citet{partridge1967a} is the first well-known study 
that discusses galaxies emitting strong Ly$\alpha$ at high redshift,
which are called LAEs today.
\citet{partridge1967a} predict that
an early galaxy emits a strong hydrogen Ly$\alpha$ line
through
the recombination process in the inter-stellar medium (ISM)
that is heated by young massive stars (Figure \ref{fig:partridge1976a_fig3}). 
%
As much as 6-7\% of the total galaxy luminosity can 
be converted to Ly$\alpha$ luminosity 
in a Milky Way mass halo.
Assuming a high star-formation rate that converts 
$2$\% of hydrogen to metal within $3\times 10^7$ yr in a Milky Way mass halo,
\citet{partridge1967a} suggest that such galaxies 
would have an extremely bright Ly$\alpha$ luminosity of $\sim 2\times 10^{45}$ erg s$^{-1}$
at $z\sim 10-30$.
Interestingly, \citet{partridge1967a} discuss the observability of those young Ly$\alpha$ emitting galaxies, 
taking the effects of cosmic reionization into account
(see also a companion paper, \citealt{partridge1967b}).
Here, \citet{partridge1967a} introduce a possible 
strong free-electron scattering in the ionized IGM that smears the radiation from the young galaxies,
which 
is an argument that differs from today's major discussion on the absorption of Ly$\alpha$ emission by the neutral IGM.
%
%
%
Although their discussion of young galaxies' Ly$\alpha$ luminosities and reionization effects on the Ly$\alpha$ observability
is very different from the present-day one, it is interesting to notice that the two major cosmological topics discussed today 
in relation with LAEs, namely galaxy formation and cosmic reionization, had already been studied theoretically in the 1960s.


\subsubsection{Early Searches for LAEs}
\label{sec:early_searches}

\begin{figure}[H]
\centering
\includegraphics[scale=.40]{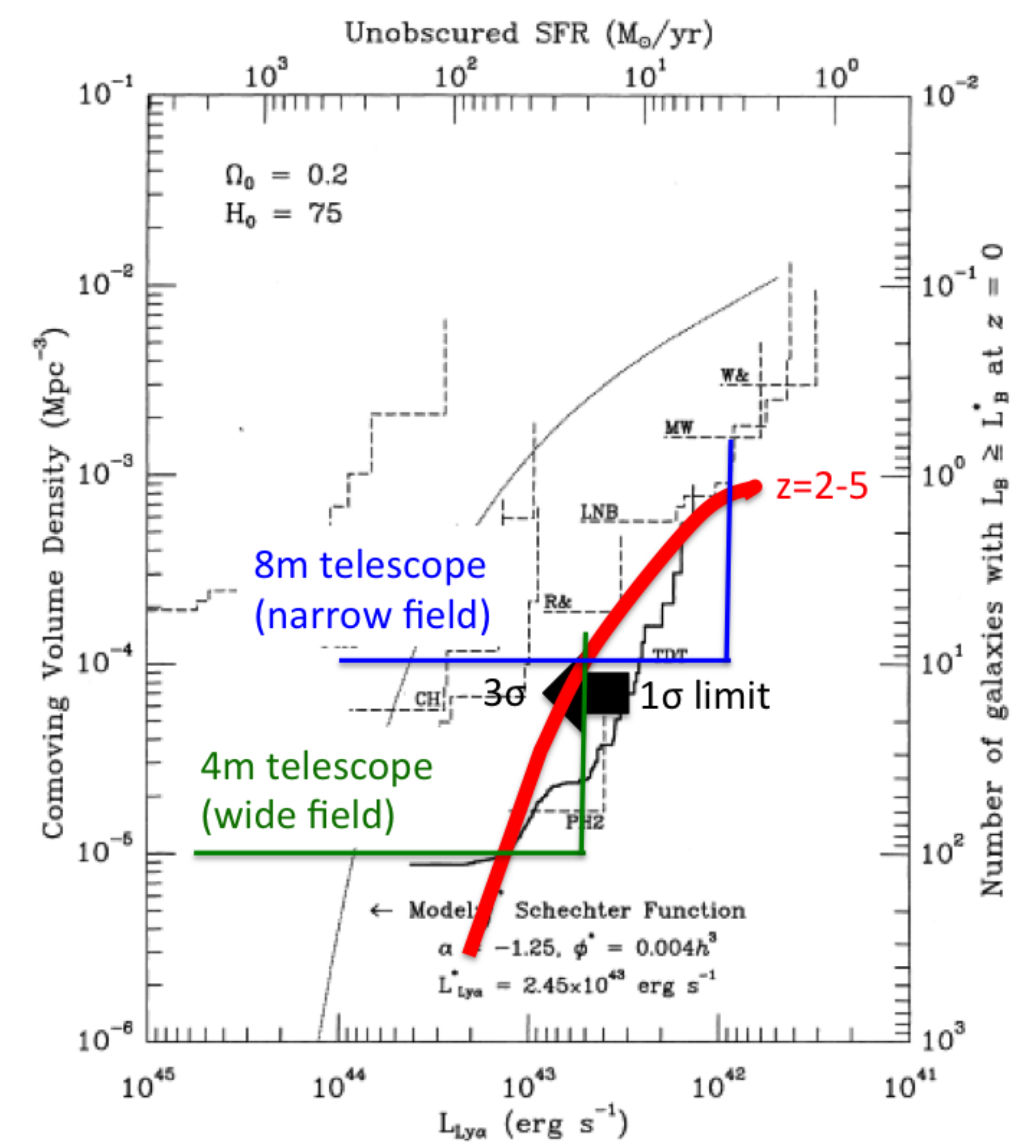}
\caption{
One sigma upper limits of Ly$\alpha$ luminosity functions at $z=2-9$ 
(black solid and dashed lines). The black arrow indicates the offset value
between one and three sigma levels that can allow us 
to convert the one to three sigma upper limits.
The red curve is the approximate Ly$\alpha$ luminosity function of LAEs at $z=2-5$ obtained thus far
(e.g. \citealt{gronwall2007,ouchi2008,cassata2011}).
The black curve is the model Ly$\alpha$ luminosity function.
The blue (green) line represents a typical observational limit of 
a narrow-field (wide-field) imager mounted on an 8m (4m) class telescope.
All numbers quoted in this figure hold for a cosmological model with 
$H_0=75$ km s$^{-1}$ Mpc$^{-1}$,
${\rm \Omega_m}=0.2$, and ${\rm \Omega_\Lambda}=0.0$.
This figure is adopted from Figure 13 of \citet{thompson1995},
and reproduced by permission of the AAS.
}
\label{fig:lya_lf_null_detection}
\end{figure}

Since the theoretical predictions of \citet{partridge1967a} were
published, a number of observational projects searched for
young LAEs at $z\sim 2-6$ (e.g. \citealt{koo1980,pritchet1987,pritchet1990,djorgovski1992,thompson1995}).
%
%
These observational searches were conducted in the 1980s and 1990s with 4m-class optical telescopes including 
the Palomar 200-inch Hale telescope that was the largest aperture telescope used for researches 
before the 10m Keck I telescope became
available. Although many candidates were pinpointed by narrowband imaging and slitless spectroscopy in these searches, 
no real young LAEs at $z\gtrsim 2$ were confirmed by spectroscopy.
However, these null-detection results placed meaningful upper limits on the luminosity function of 
LAEs (Figure \ref{fig:lya_lf_null_detection}).

\subsubsection{Discovery}
\label{sec:discoveries}

\begin{figure}[H]
\centering
\includegraphics[scale=.40,angle=-90]{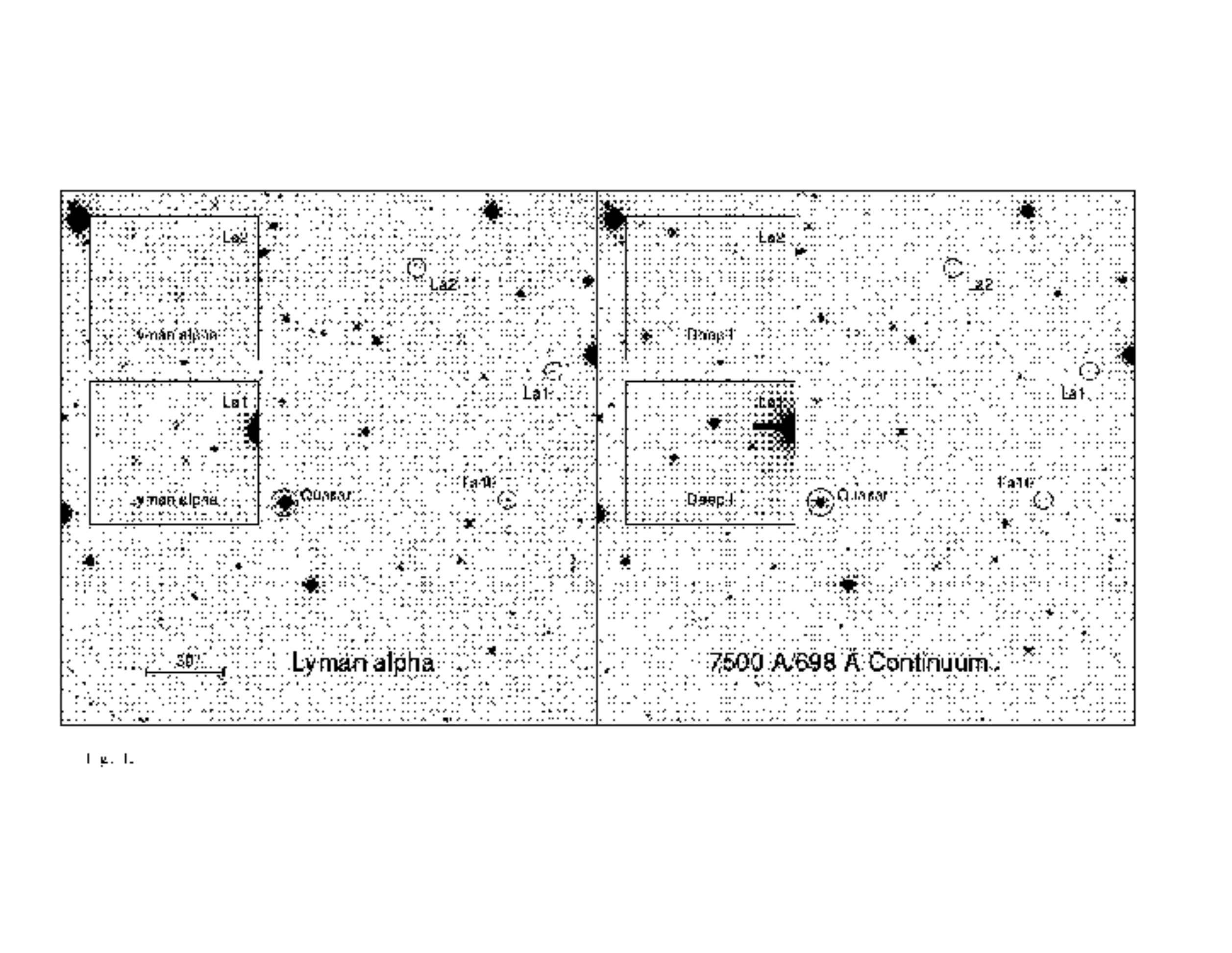}
\includegraphics[scale=.40,angle=-90]{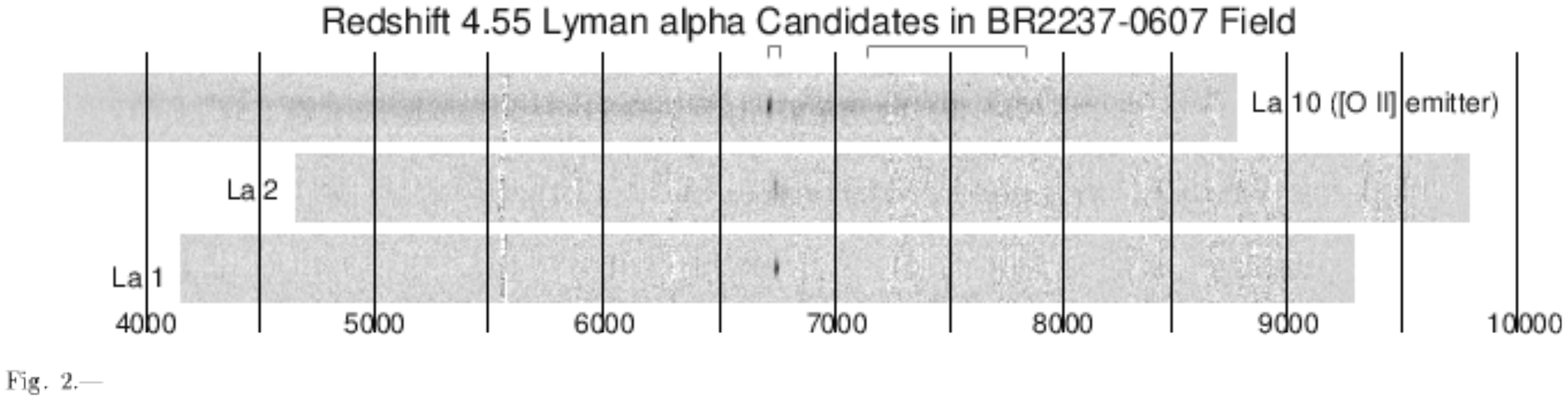}
%
%
\caption{
Top: Images of three LAE candidates in the QSO field of BR2237-0607 at $z=4.55$ 
marked with the small circles \citep{hu1996}. The QSO is indicated with the large circle.
The left and right panels present narrowband 
and broadband images, respectively, that cover the redshifted Ly$\alpha$ emission
and the rest-frame UV continua of the LAEs.
Bottom: Two-dimensional spectra of the three LAE candidates \citep{hu1996}. 
Clear Ly$\alpha$ signals of the LAE candidates 
are found in the wavelength between 6500 and 7000\AA. 
The two objects in the two lower panels
are LAEs, while the object in the top panel is classified as a low-$z$ [\oii] emitter.
This figure is reproduced by permission of the Nature Publishing Group.
}
\label{fig:hu1996_lalpha_fig1bm}       
\end{figure}

Since 1996, LAE observations have entered a new era.
\citet{hu1996} have identified two LAEs at $z=4.55$ 
around the QSO BR2237-0607 using the Keck LRIS 
spectrograph (Figure \ref{fig:hu1996_lalpha_fig1bm}), 
after having detected them using deep UH88 imaging 
through a narrow band filter with the central wavelength 
tuned to the wavelength 
of Ly$\alpha$ emission at $z=4.55$.
At the same time, \citet{pascarelle1996a} 
discovered 5 LAEs, including a weak AGN,
around the radio galaxy 53W002
at $z=2.39$ using
%
%
%
\noindent
the Hubble Space Telescope (HST) 
for broadband imaging, a ground based telescope 
for narrowband imaging, and the MMT for spectroscopy.
\citet{pascarelle1996b} claim that 
such LAEs form a galaxy group in the 53W002 region.
These discovered LAEs have Ly$\alpha$ luminosities of a few times $10^{42}$ 
erg\,s$^{-1}$
that is about $1/100$ of the young galaxies predicted by \citet{partridge1967a}.


These early studies found LAEs 
in the vicinity of AGNs that 
are thought to be signposts of 
high-$z$ galaxy overdensities. LAEs in blank fields were first identified
by \citet{cowie1998} and \citet{hu1998} who carried out
Keck LRIS narrowband imaging and spectroscopy in the SSA22 and 
Hubble Deep Field (HDF). Such deep field observations started to
identify LAEs at $z\gtrsim 2$ in blank fields routinely with the high sensitivities of 
8m class telescopes and the large-area survey capabilities of
4-8m class telescopes around the year 2000. These deep field observation
programs 
include the Hawaii Survey
\citep{cowie1998}, 
the Large Area Lyman Alpha Survey (LALA; \citealt{rhoads2000}),
the Subaru Surveys (e.g. \citealt{ouchi2003}), and
the Multiwavelength Survey by Yale-Chile (MUSYC; \citealt{gawiser2007}),
and recently an LAE search with a new technology has been demonstrated
by the HETDEX Pilot Survey (HPS: \citealt{adams2011})
\footnote{
In the early days of LAE observational studies, many names and abbreviations 
were used for LAEs, such as Ly$\alpha$ emitting galaxies,
Ly$\alpha$ galaxies, LEGOs etc. 
The present-day established name,
Ly$\alpha$ emitter, 
can be found in the early study of 
\citet{hu1996}, and the abbreviation, LAE, was first 
suggested in \citet{ouchi2003}.
%
}.
Recent space based observations even find LAEs at $z\sim 0$ 
whose Ly$\alpha$ emission lines fall in the far UV wavelength range.
There is an HST survey 
program, the so-called Lyman-Alpha Reference Sample
(LARS; \citealt{ostlin2014})
that investigates the Ly$\alpha$ properties of star-forming galaxies
originally selected as H$\alpha$ emitters at $z\sim 0$.
Moreover, far and near UV 
grism data from the Galaxy Evolution Explorer
(GALEX) are used to
detect LAEs at $z\sim 0-1$ 
and to build Ly$\alpha$ flux limited samples
for studies of Ly$\alpha$ luminosity functions (Section \ref{sec:luminosity_function}; 
\citealt{deharveng2008,cowie2010,cowie2011}). For more details about $z\sim 0$ LAE observations,
see M. Hayes' lectures in this course.


%
%
%
%
%
%
 


There is a question why no 
single LAE
at high-$z$ 
could be identified 
by the observations until 1996,
about two decades after the predictions of \citet{partridge1967a}.
As shown in Figure \ref{fig:lya_lf_null_detection}, the blank field
surveys conducted until 1995 found no LAEs at $z>2$ 
down to number densities of
$\sim 10^{-4}$ Mpc$^{-3}$ at $L({\rm Ly\alpha})=10^{43}$ erg s$^{-1}$
and
$\sim 10^{-3}$ Mpc$^{-3}$ at $L({\rm Ly\alpha})=10^{42}$ erg s$^{-1}$.
These number-density limits just touch the Ly$\alpha$ luminosity functions
at $z\sim 3-5$ 
that have been determined to date
(see Section \ref{sec:luminosity_function}).
In other words, 
one could have found LAEs 
in a blank field before 1996, 
if there was a one-more push of sensitivity or survey volume.
However, such a one-more push was not made until 1996.
In reality, there are two important approaches 
leading to these successful detections 
of LAEs.
The first approach is 
to focus on AGN regions.
The first LAE detections \citep{hu1996,pascarelle1996a} 
were accomplished in AGN regions, whose galaxy overdensities enhance the probability
of bright LAEs existing in the survey area, 
resulting in successful selections and spectroscopic confirmations even with 
2-4 m class telescopes.
The second approach is to exploit the great sensitivity of 8m-class telescopes newly available since the 1990s.
In fact, the Keck deep narrowband observations by \citet{cowie1998} successfully identified LAEs in blank fields.
Interestingly, these two approaches provided successful detections almost at the same time
in the late 1990s. 

Around the year 2000, wide-field optical imagers started operation
in 4m-class telescopes (e.g. KPNO/MOSAIC), 
allowing
the observers to detect LAEs 
in blank fields even with the moderately low sensitivity of 4m-class telescopes (\citealt{rhoads2000,gawiser2007}; Figure \ref{fig:lya_lf_null_detection}).
Moreover, after the first light of the wide-field optical imager Suprime-Cam on the 8m-Subaru telescope in 1999,
large LAE surveys cover wider sensitivity and volume ranges, in contrast with
the previous narrow-field 8m-class observations (Figure \ref{fig:lya_lf_null_detection}).
The deep spectroscopic capabilities of the Keck, VLT, and Subaru telescopes are also key for confirmation of 
faint LAE candidates that are found in blank fields.

%
%

\subsubsection{Definition of LAEs}
\label{sec:definitions}

Here I introduce the definition of LAEs, although it should be noted that some detailed 
%
definitions depend on the study considered.
Nowadays, the widely accepted definition of LAEs is:
{\it LAEs are galaxies with a Ly$\alpha$ rest-frame equivalent width ($EW_0$)
greater than $\simeq 20$ \AA.}
The criterion of Ly$\alpha$ $EW_0\gtrsim 20$ \AA\ is historically
determined by the realistic selection limit of narrowband imaging surveys for Ly$\alpha$ emitting
galaxies at $z\sim 3$. By this definition, 
the main contribution to the LAE population consists of star-forming galaxies,
some of which have AGN activity.

\subsubsection{LAE Search Techniques}
\label{sec:lae_search_methods}

\begin{figure}[H]
\centering
\includegraphics[scale=.40]{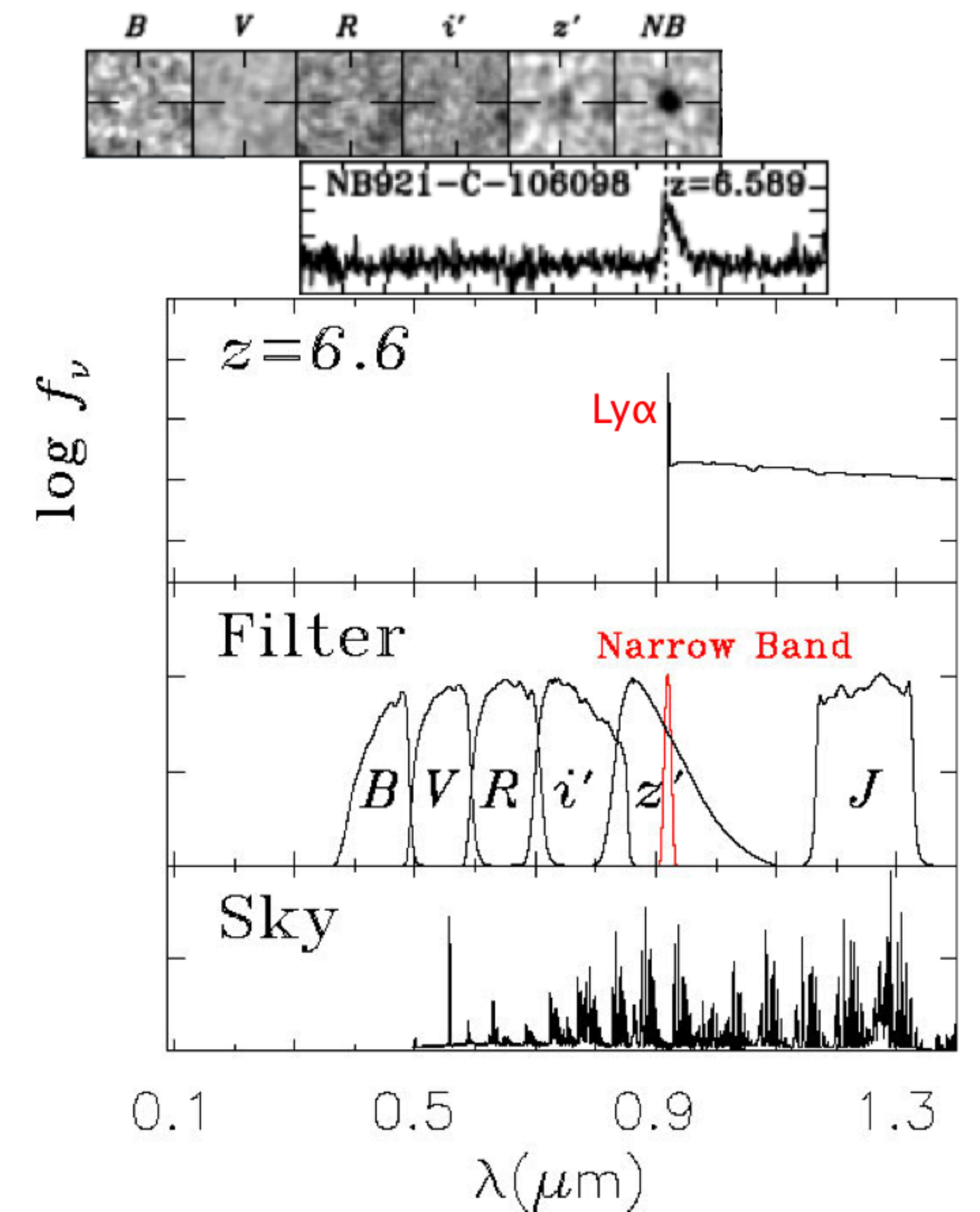}
\caption{
Illustration of the narrowband selection for an LAE at $z=6.6$ (NB921-C-106098).
The top panel presents images of this LAE observed with 
broadbands ($B$, $V$, $R$, $i'$, and $z'$) and a narrowband ($NB$) whose central wavelength is $\sim 9200$\AA.
The second top panel is a spectrum of this LAE in the wavelength range of $9050-9275$\AA.
The third top panel shows the model spectrum of a LAE redshifted to $z\sim 6.6$.
The second bottom panel exhibits the transmission curves of the broadbands and the narrowband.
The bottom panel presents the atmospheric OH lines.
Some images of this figure are taken from \citet{ouchi2010}.
This figure is reproduced by permission of the AAS.
}
\label{fig:nbselection}       
\end{figure}

There are two popular techniques to search for LAEs. One is 
narrowband imaging. Figure \ref{fig:nbselection} illustrates the idea.
The redshifted Ly$\alpha$ emission of LAEs is identified
by a flux excess 
in a narrowband image
over other wavelength images
(Figure \ref{fig:nbselection}). 
The central wavelength of the narrowband filter, $\lambda_{\rm c}$, determines the redshift of the target LAEs
that is roughly given by $\lambda_{\rm c}/1216-1$. The $\lambda_{\rm c}$ value of a narrowband filter
is chosen by a scientific requirement (i.e., redshift of target LAEs) and/or observational 
constraints (e.g. avoiding weak night-sky OH emission lines).
In most cases, 
$\lambda_{\rm c}$ is 
placed in an OH emission window (bottom panel of Figure \ref{fig:nbselection}) 
to realize a high sensitivity. 

\begin{figure}[H]
\centering
\includegraphics[scale=.40]{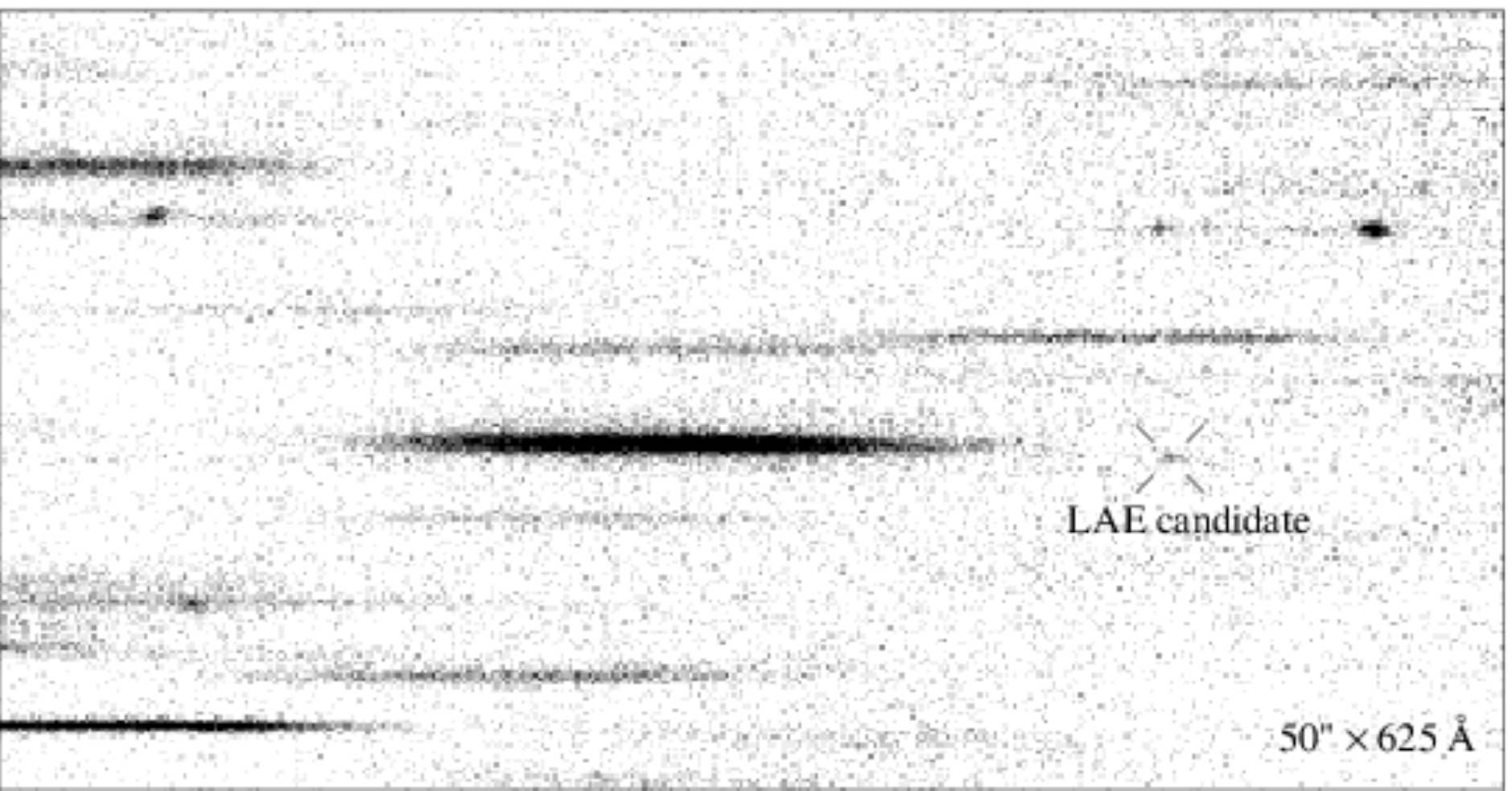}
\caption{
An LAE candidate found in VLT/FORS grism data in a blank field \citep{kurk2004}, 
labeled on the image. 
An [\oiii]5007 emitter is also
found at the upper right of this image.
This figure is reproduced by permission of A\&A.
}
\label{fig:kurk2004_fig1}
\end{figure}

The other technique is 
blind spectroscopy, including slitless spectroscopy.
Figure \ref{fig:kurk2004_fig1} shows one example
that uses a VLT/FORS grism 
targeting a blank sky field
with no prior positional 
information of a LAE candidate \citep{kurk2004}. The LAE candidate is found as a single-line emitter
in the grism image. Although this technique provides positions and spectra of LAEs
at the same time, the background sky level is high in the slitless data.
Thus, HST grism observations are popular to perform slitless spectroscopic
searches for LAEs, exploiting the low sky background in space \citep{pirzkal2004}.
The blind spectroscopy technique also includes slit spectroscopy such as
long-slit spectroscopy conducted on positions of critical lines of lensing clusters
searching for lensed LAEs \citep{santos2004}. Moreover, the recent advancement
of integral field spectrographs (IFSs) allows blind spectroscopic searches for LAEs 
in reasonably large areas, keeping the background sky sufficiently low \citep{vanbreukelen2005,bacon2015}.

Although these two techniques 
are major ones for identifying LAEs, recent deep spectroscopy has found
continuum-selected galaxies (e.g. dropouts or Lyman break galaxies; LBGs) 
with a spectroscopic measurement of Ly$\alpha$ $EW_0\gtrsim 20$\AA\ 
that are also classified as LAEs (e.g. \citealt{erb2014}). 
In this series of lectures, LAEs include continuum-selected galaxies
with Ly$\alpha$ $EW_0\gtrsim 20$\AA.



\subsection{Progresses in LAE Observational Studies After the Discovery}
\label{sec:progress}


Large survey programs have so far identified a total of more than $10^4$ LAEs up to $z\sim 8$ photometrically
(e.g. \citealt{yamada2012a,konno2016}), out of which about $10^3$ have been spectroscopically confirmed
(e.g. \citealt{hu2010,kashikawa2011}). Due to the high abundance (i.e. number density) 
of LAEs, $10^{-3}$ Mpc$^{-3}$ at $L_{\rm Ly\alpha}\sim 10^{42}-10^{43}$
erg s$^{-1}$, 
LAEs are thought to constitute one 
of the major populations of high-$z$ galaxies.
Below, I highlight progresses in LAE observations that are detailed in Sections \ref{sec:galaxy_formationII}-\ref{sec:cosmic_reionizationII}.

\begin{figure}[H]
\centering
\includegraphics[scale=.40]{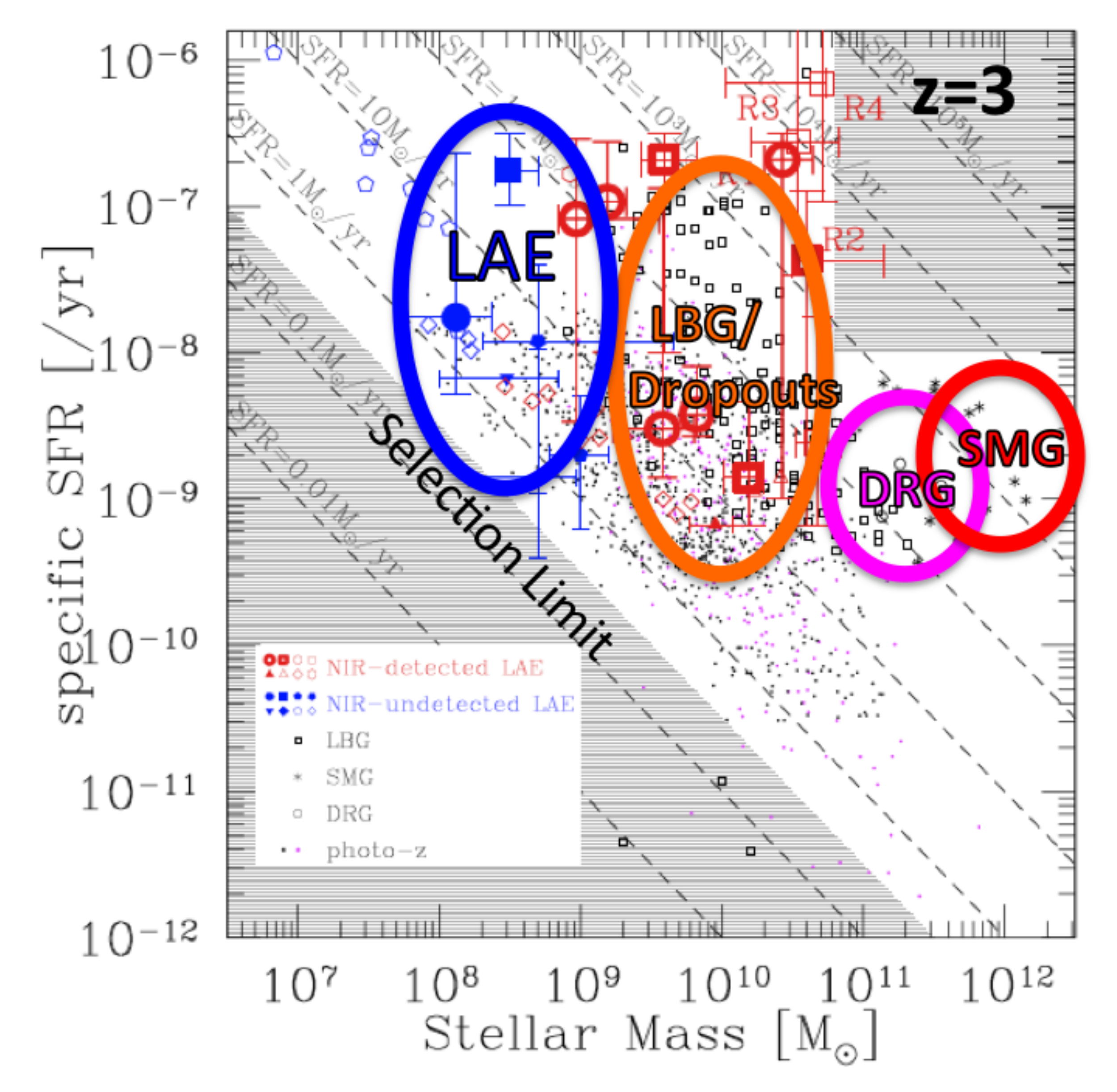}
\caption{
sSFR as a function of stellar mass for typical LAEs (blue symbols),
LBGs/dropouts (black open squares), DRGs (black open circles),
and SMGs (star marks) at $z=3$ \citep{ono2010a}. The thick ovals indicate the approximated distributions
of LAEs (blue), LBGs/dropouts (orange), DRGs (magenta), and SMGs (red).
The red squares and circles with error bars denote peculiar LAEs with bright NIR-band fluxes.
This figure is reproduced by permission of the Royal Astronomical Society.
}
\label{fig:ono2010a_fig12}       
\end{figure}

Deep photometric studies reveal the average spectral energy distribution (SED) of LAEs 
with deep optical and NIR photometric data. From comparisons with stellar population synthesis
models, stellar population, one of the basic properties of galaxies, is studied
(e.g. \citealt{gawiser2007,finkelstein2007,ono2010a,ono2010b,guaita2011,hagen2014,hagen2016}).
Figure \ref{fig:ono2010a_fig12} compares LAEs' average stellar masses ($M_{\rm s}$) and specific star-formation rates (sSFRs),
defined as the star-formation rate (SFR) divided by stellar mass,
with those of other galaxy populations: LBGs, distant-red galaxies (DRGs), and sub-millimeter galaxies (SMGs) at $z\sim 3$. 
The average $M_{\rm s}$ of LAEs is $10^8-10^9 M_\odot$,
which falls in the lowest mass range among the high-$z$ galaxy populations (Section \ref{sec:stellar_population}).
The low stellar masses of the LAEs suggest that LAEs are high-$z$ analogs of local star-forming dwarf galaxies.
The sSFR values of LAEs are comparable to or slightly higher than those of 
the other high-$z$ galaxies, although the distribution of LAEs at the low-mass limit of Figure \ref{fig:ono2010a_fig12} is biased by
the observational selection limits.


\begin{figure}[H]
\centering
\includegraphics[scale=.40]{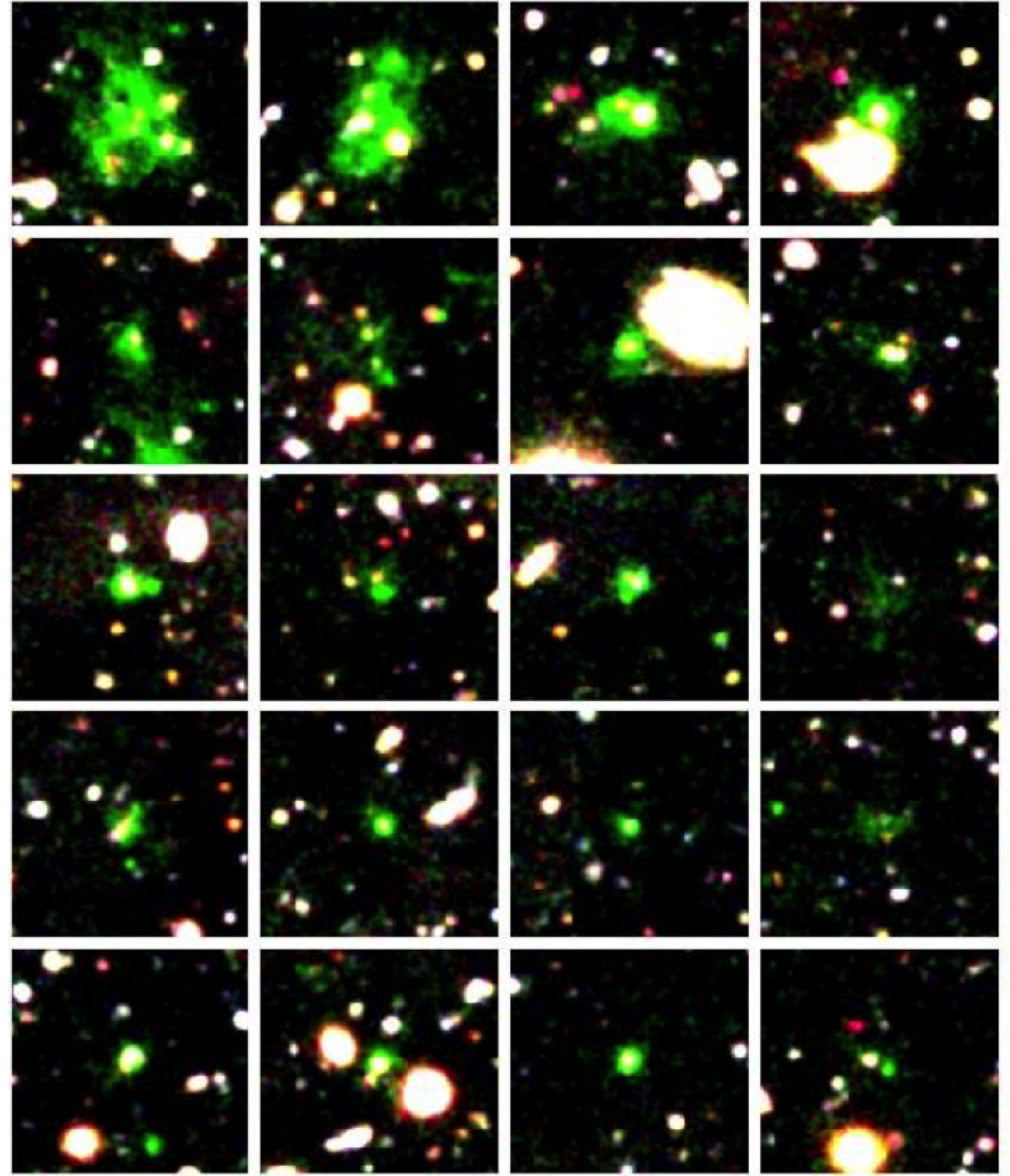}
\caption{
Color composite images of LABs at $z=3.1$ in the SSA22 field \citep{matsuda2004}.
The green color indicates Ly$\alpha$ emission, while the blue and red colors are
the continua bluer and redder than Ly$\alpha$ for $z\sim 3$ objects, respectively.
The top left (second-left) panel shows LAB1 (LAB2), the first LAB in star-forming galaxies
discovered by \citet{steidel2000}.
Note that the Ly$\alpha$ emission nebula of LAB1 extends over $\sim 200$ kpc.
The size of each image is about 200 kpc $\times$ 200 kpc.
This figure is adopted from
http://www.naoj.org/Pressrelease/2006/07/26/index.html
by permission of the National Astronomical Observatory of Japan.
}
\label{fig:matsuda2004_web}       
\end{figure}

A narrowband imaging search for LAEs has serendipitously identified
remarkable objects like Ly$\alpha$ blobs (LABs), many of which show no clear AGN signatures (Section \ref{sec:extended_lya_halo}),
that were first found in the LBG overdensity region SSA22 (Figure \ref{fig:matsuda2004_web}; \citealt{steidel2000}).
LABs consist of a large Ly$\alpha$ nebula with a spatial extent of $\sim 10-200$~kpc 
and a bright total Ly$\alpha$ luminosity
$L_{\rm Ly\alpha}\sim 10^{43}-10^{44}$ erg s$^{-1}$ \citep{matsuda2004}.
So far, a few tens of LABs are identified at $z\sim 2-7$ \citep{yang2009,scarlata2009,ouchi2009a}.
There are various models of LABs including \hi\ scattering clouds and cooling radiation.
However, the physical origins of the large Ly$\alpha$ nebulae are under debate.


\begin{figure}[H]
\centering
\includegraphics[scale=.48]{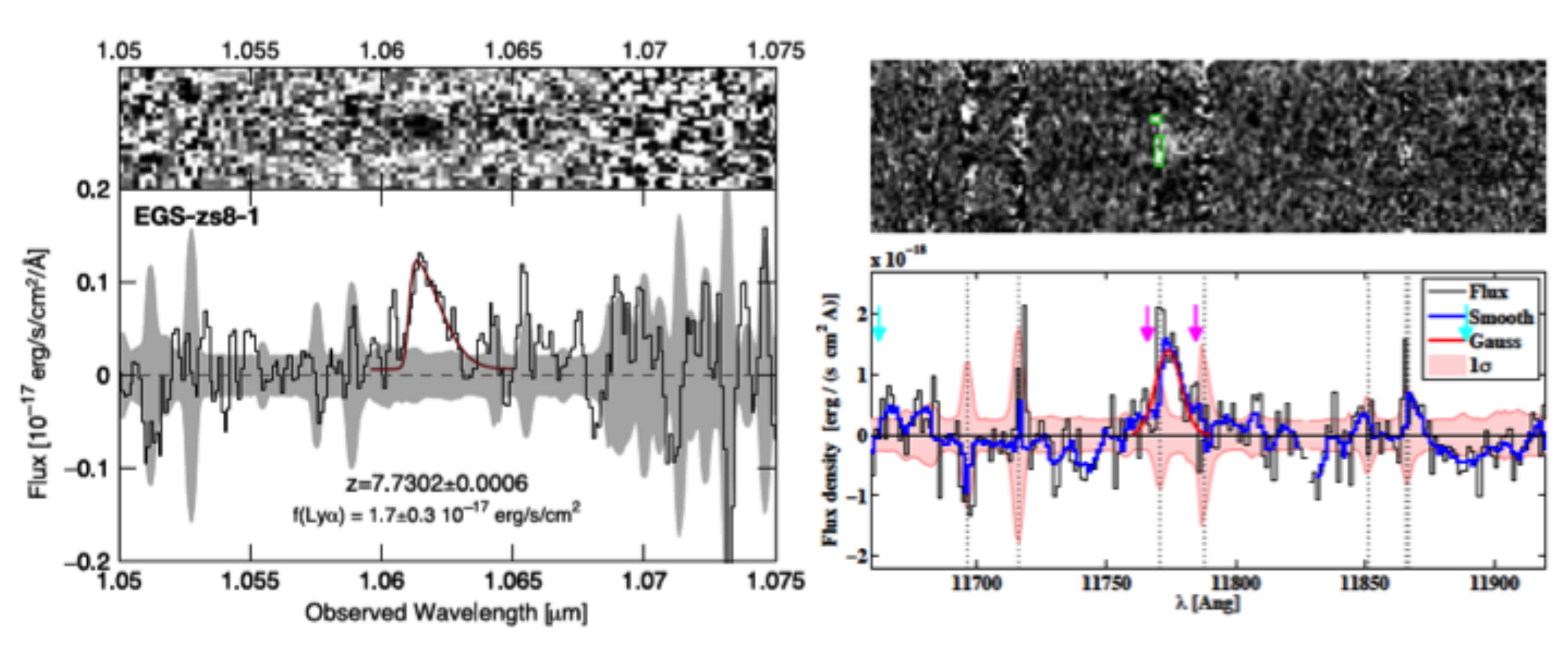}
\caption{
Spectroscopically identified LAEs at $z=7.73$ (left; \citealt{oesch2015}) and $z=8.68$ (right; \citealt{zitrin2015}).
The top and bottom panels present the two- and one-dimensional spectra, respectively.
A clear asymmetric line typical for high-$z$ LAEs is identified in the $z=7.73$ LAE (left). 
This figure is reproduced by permission of the AAS.
}
\label{fig:oesch2015_fig3_zitrin2015_fig1}       
\end{figure}

Since the late 1990s when the early observations detected LAEs, 
LAEs have remained the most distant galaxies known to date
(Figure \ref{fig:oesch2015_fig3_zitrin2015_fig1}; \citealt{hu1999,hu2002,kodaira2003,iye2006,vanzella2011,ono2012,shibuya2012,
finkelstein2013,oesch2015,zitrin2015}),
except for some examples of high-$z$ dropouts whose
redshifts are estimated with the Ly$\alpha$ continuum break
with an accuracy $\Delta z = 0.1-0.2$ \citep{watson2015,oesch2016}.
Most of the highest redshift galaxies confirmed by spectroscopy are
LAEs, because strong Ly$\alpha$ emission can be efficiently detected
in a very faint source at high redshift.
Some of the high-$z$ galaxies show intrinsically large Ly$\alpha$ $EW_0$ values,
suggestive of very young, population III (popIII)-like starbursts
such as those predicted by \citet{partridge1967a} (Section \ref{sec:theoretical_predictions}).


\begin{figure}[H]
\centering
\includegraphics[scale=.50]{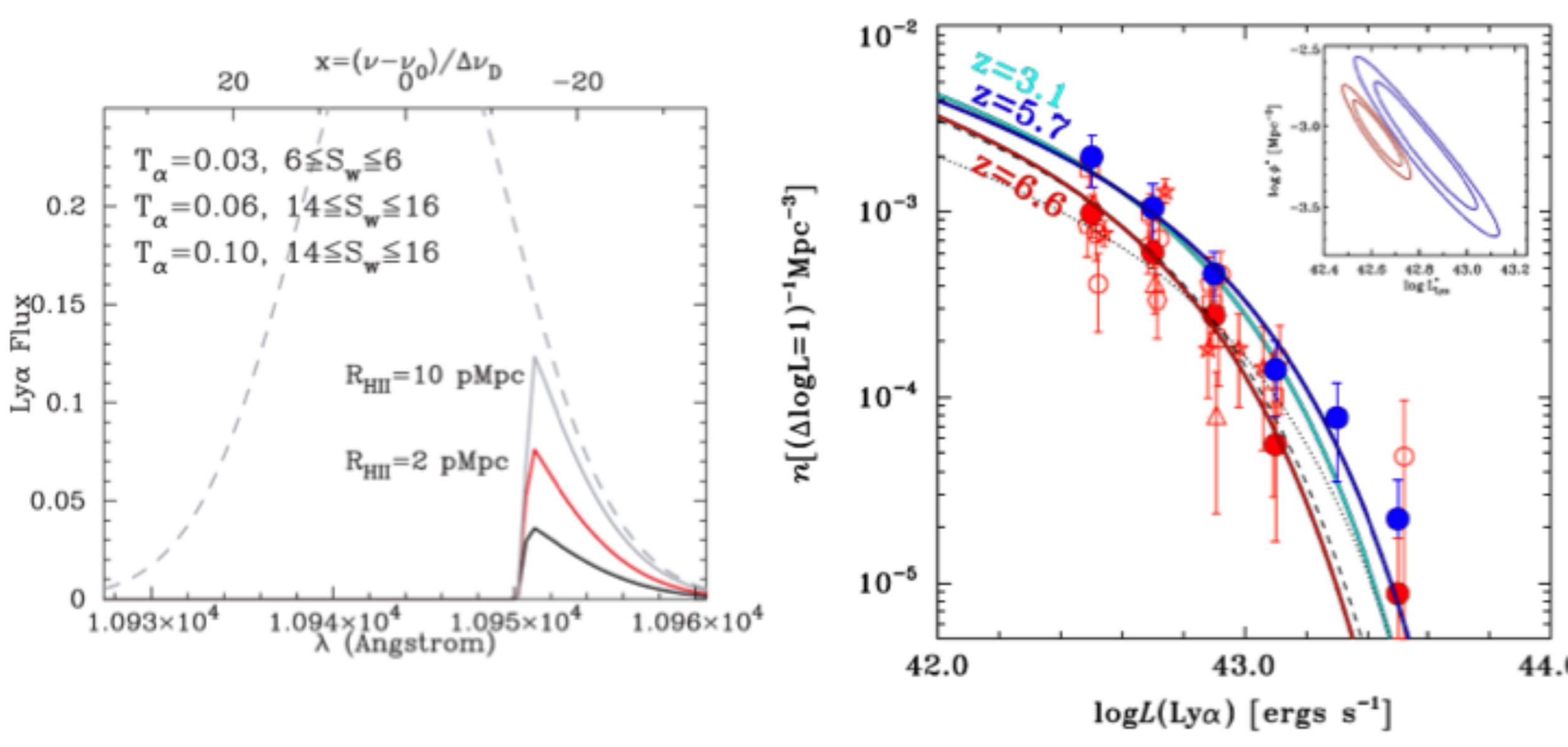}
\caption{
Left: Theoretical predictions of Ly$\alpha$ line profiles of a Ly$\alpha$ emitter at $z=8$ \citep{dijkstra2007}.
The black solid line represents the Ly$\alpha$ line from an LAE in the fully neutral IGM, while the 
gray dashed line denotes the intrinsic Ly$\alpha$ line. The red (gray) solid line shows the Ly$\alpha$ line
of an LAE escaping from the center of an ionized bubble with a radius of 2 (10) physical Mpc
in the neutral IGM.
Right: Ly$\alpha$ luminosity functions at $z=6.6$ (red), $5.7$ (blue), and $3.1$ (cyan) derived by
observations \citep{ouchi2010}. The red and blue filled circles are the best-estimate luminosity functions
that are the weighted average results of the luminosity functions estimated from LAEs in survey subfields,
while all of the red open symbols are luminosity functions of the subfields.
The inset panel presents the 68 and 90 percentile error contours 
for the Schechter function fitting of the best-estimate luminosity functions 
at $z=6.6$ (red contours) and $z=5.7$ (blue contours).
This figure is reproduced by permission of the the Royal Astronomical Society and the AAS.
}
\label{fig:dijkstra2007_fig7_ouchi2010_fig7_fig8}       
\end{figure}

A number of LAEs have been spectroscopically identified 
at the epoch of reionization (EoR) at $z\gtrsim 6$ (Section \ref{sec:cosmic_reionizationI}).
Because Ly$\alpha$ photons from LAEs are scattered by neutral hydrogen 
\hi\ that exists in the IGM at EoR, the detectability of Ly$\alpha$ from LAEs 
depends on the fraction of \hi\ in the IGM. 
In a statistical sense, weak Ly$\alpha$ emission of LAEs
suggests more Ly$\alpha$ scattering in the IGM 
or
lower Ly$\alpha$ production rates.
Exploiting this dependence,
LAEs are used as probes of cosmic reionization as well as galaxy formation
(Figure \ref{fig:dijkstra2007_fig7_ouchi2010_fig7_fig8}).

Cosmic reionization has been extensively investigated 
using LAEs, after isolating the effects of galaxy formation, 
namely the evolution of the Ly$\alpha$ luminosity, 
in conjunction with
complementary observational constraints
(Sections \ref{sec:cosmic_reionizationI}-\ref{sec:cosmic_reionizationII}; 
\citealt{malhotra2004,kashikawa2006,kashikawa2011,ouchi2010,
pentericci2011,ono2012,schenker2012,treu2013,schenker2014}).

\subsection{Goals of This Lecture Series}
\label{sec:goals}

The goal of this lecture series is to make the readers understand not only 
the established picture of LAEs, but also the cutting-edge results obtained 
from observations spanning the redshift range $z\sim 0-10$
covered to date.
As shown in Section \ref{sec:progress},
today's major LAE studies 
address questions about the physical properties of high-$z$ low mass galaxies, 
including popIII-like galaxies, sources of reionization, and 
the cosmic reionization history.
In other words, 
most LAE observational studies
discuss either galaxy formation or cosmic reionization.
%
This lecture series thus covers
\begin{itemize}
\item[1)]  Galaxy formation (Sections \ref{sec:galaxy_formationI}-\ref{sec:galaxy_formationIII}) and
\item[2)]  Cosmic reionization (Sections \ref{sec:cosmic_reionizationI}-\ref{sec:cosmic_reionizationII}).
\end{itemize}

Note that there are several promising studies of LAEs
that are growing in this field. One is the Ly$\alpha$ emission distribution
that traces the circum-galactic medium (CGM) extending 
along filaments
of large-scale structures \citep{cantalupo2014}.
Because Ly$\alpha$ is a resonance line, it is used as a probe of
the \hi\ distribution of the underlying cosmological structures.
The extended Ly$\alpha$ emission studies are detailed
in Section \ref{sec:galaxy_formationIII}, together with topics of Ly$\alpha$ blobs, diffuse Ly$\alpha$ halos,
Ly$\alpha$ fluorescence, proto-clusters, and large-scale structures (LSSs), all of which are closely related to galaxy formation.
Another important use of LAEs consists in probing properties
of dark energy with accurate measurements
of cosmic expansion history on the basis of baryon acoustic oscillations (BAO). Because no LAE studies have, so far,
successfully detected BAO, an on-going LAE BAO cosmology study project is
briefly touched in the section of future studies (Section \ref{sec:ongoing_and_future_projects}).

%





%

\section{Galaxy Formation I: Basic Theoretical Framework}
\label{sec:galaxy_formationI}

One of the major scientific drivers of LAE studies
is galaxy formation.
%
%
%
%
In this section, I show the basic theoretical framework of
galaxy formation and associated Ly$\alpha$ emission, 
and identify both established ideas and unresolved difficult issues.
This section mainly targets first-year graduate students
working on observations 
and those who know little 
about the modern picture of galaxy formation.
%

\subsection{Basic Picture of Galaxy Formation}
\label{sec:basic_picture}

\begin{figure}[H]
\centering
\includegraphics[scale=.40]{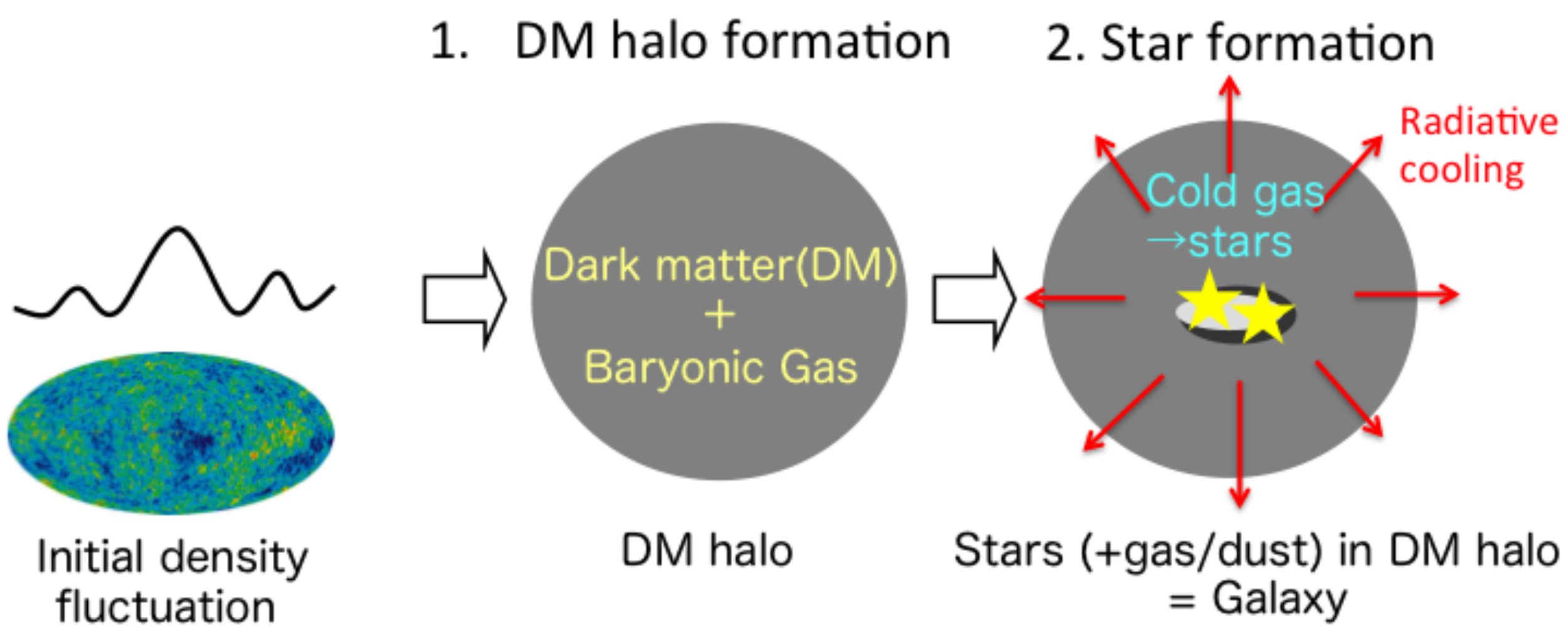}
\caption{
Conceptual diagram of the galaxy-formation processes.
}
\label{fig:DMH_SF}       
\end{figure}

Figure \ref{fig:DMH_SF} illustrates the basic picture of
galaxy formation that is believed in modern astronomy.
Generally, galaxy formation is made of two major processes,
dark-matter (DM) halo formation and star formation \citep{mo2010}.
First, DM halos are created from the initial density fluctuations, and then
star formation takes place in the cold dense gas clouds 
made by radiative cooling
in the DM halos.
These two processes are detailed in the following subsections.



\subsubsection{DM Halo Formation}
\label{sec:DM_halo}

The standard cosmological model of $\Lambda$ cold dark matter ($\Lambda$CDM)
suggests that the initial density fluctuations in the early universe
grow by gravity and produce cosmic structures \citep{peebles1993}. 
DM halos, virialized systems of DM, with baryon gas are 
created by gravitational collapses.
Low-mass DM halos are first made, 
and subsequently these low-mass DM halos increase their masses 
by merger and accretion processes. Because DM dominates the cosmic matter density,
this sequence of the cosmological structure formation is governed by DM.
DM physically interacts only by gravity, and the formation of 
cosmic structures including DM halos can be basically predicted 
with no serious systematics.

\begin{figure}[H]
\centering
\includegraphics[scale=.40]{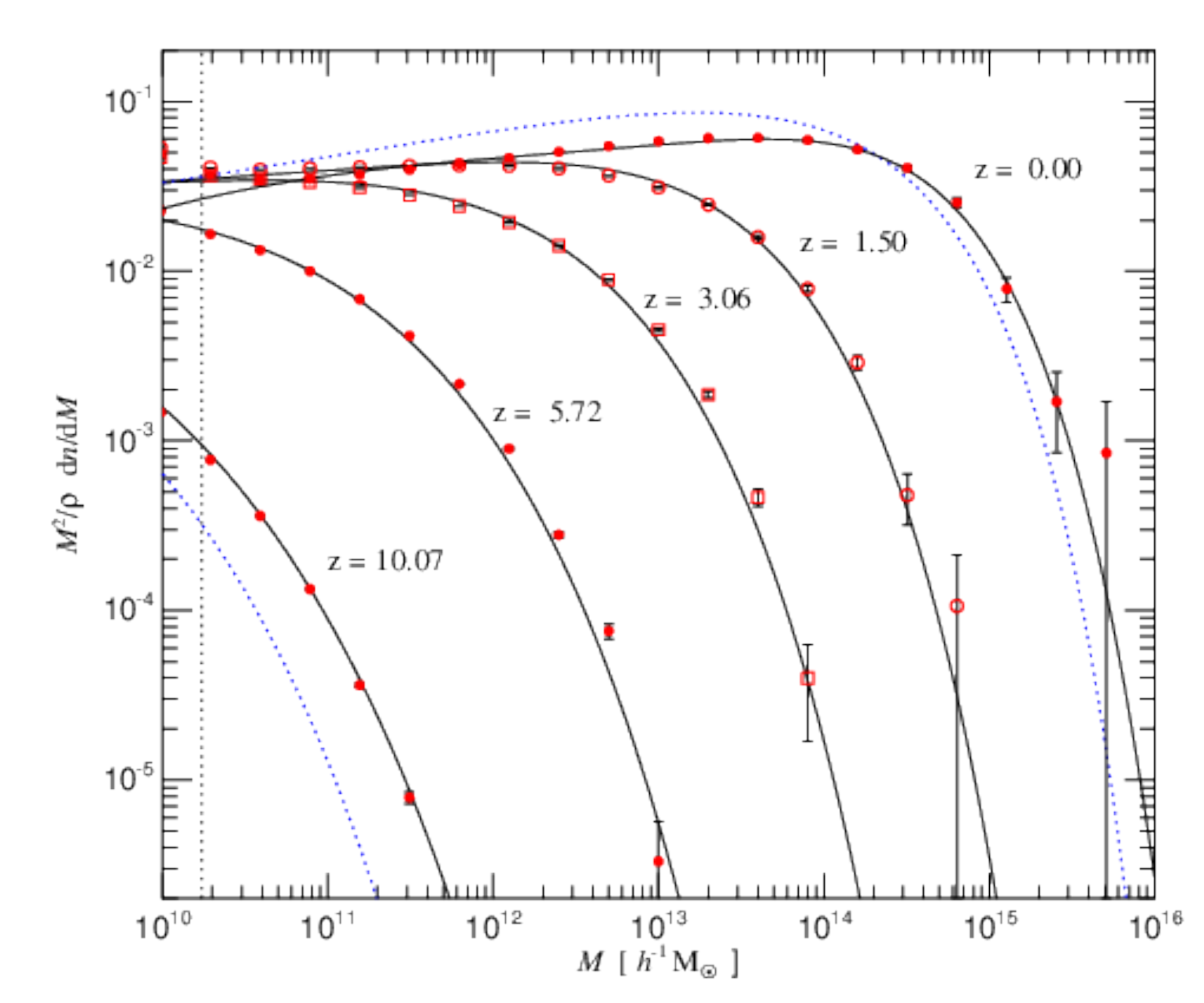}
\caption{
DM-halo mass function as a function of mass and redshift obtained
by numerical simulations \citep{springel2005}.
The ordinate is the differential number density ($dn/dM$) multiplied 
by $M^2 \rho^{-1}$ where $\rho$ is the mean density of the universe. 
The solid lines are the best-fit functions to the mass functions with the
analytical functional form of \citet{jenkins2001} (see also \citealt{sheth1999}).
The blue dashed lines are the Press-Schechter functions at $z=10.07$ and $0$ from left to right.
This figure is reproduced by permission of the Nature Publishing Group.
}
\label{fig:springel2005_fig2}       
\end{figure}

\begin{figure}[H]
\centering
\includegraphics[scale=.40]{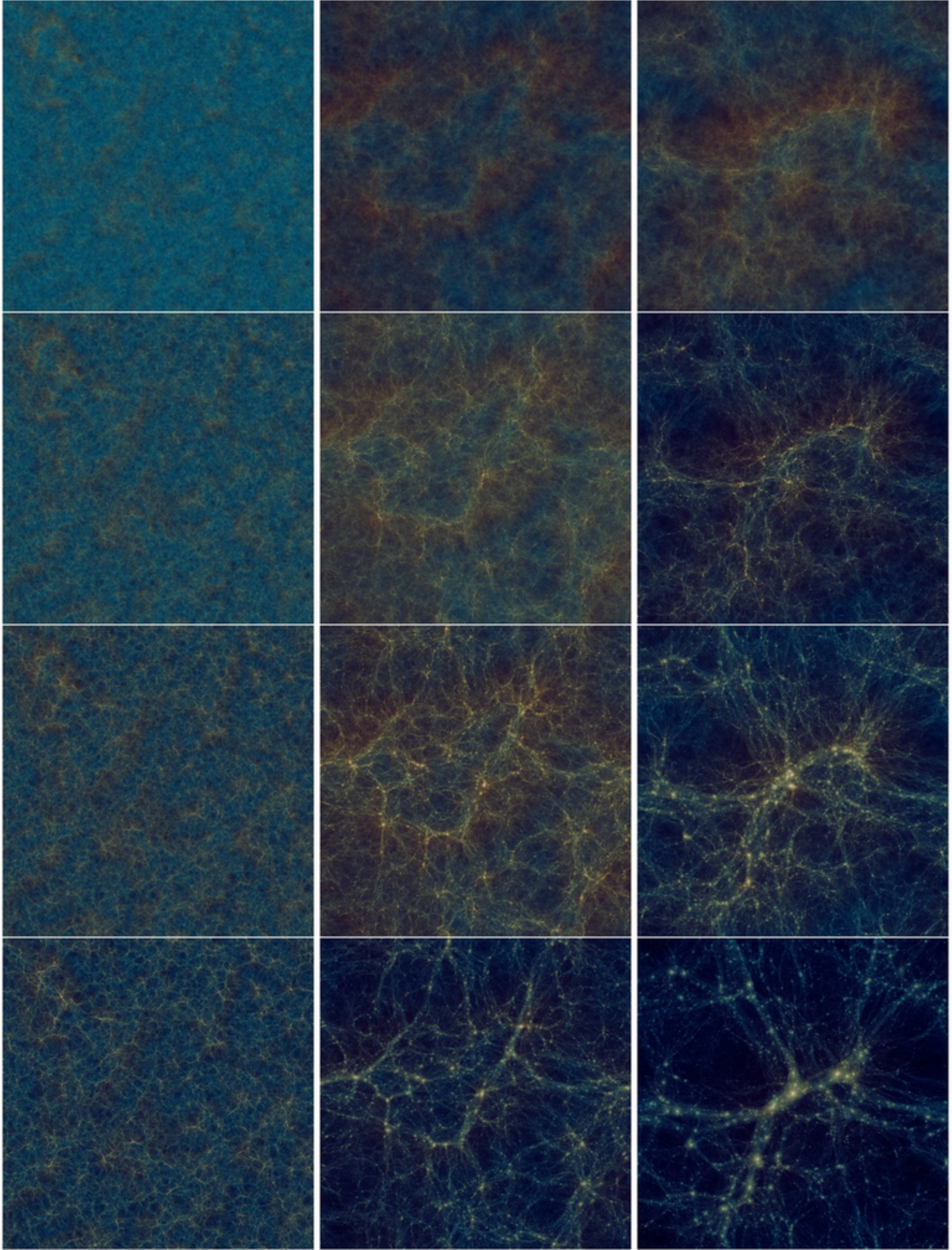}
\caption{
Cosmic structures including DM halos reproduced by the state-of-the-art
simulations \citep{ishiyama2015}. From top to bottom, cosmic structures at $z=7$, $3$, $1$, and $0$
are shown. The left, center, and right panels represent the simulation results with three different
DM-particle masses (box sizes),
$2.2\times 10^8 h^{-1} M_\odot$ ($1120 h^{-1}$ Mpc),
$2.8\times 10^7 h^{-1} M_\odot$ ($140 h^{-1}$ Mpc), and
$3.4\times 10^6 h^{-1} M_\odot$ ($70 h^{-1}$ Mpc), respectively.
This figure is reproduced by permission of the ASJ.
}
\label{fig:ishiyama2015_fig3}       
\end{figure}

Exploiting the great performance of computers today,
numerical simulations reproduce DM halos under the assumption that
DM is composed of collisionless particles that follow Newton's law of gravitation.
Figure \ref{fig:springel2005_fig2} presents the DM-halo mass functions 
calculated by large cosmological simulations \citep{springel2005}.
%
%
The state-of-the-art cosmological simulations 
(with a box size of a few-10 Mpc$^3$)
have a good mass resolution, and already make
DM halos with a mass of $\sim 10^7 M_\odot$ (Figure \ref{fig:ishiyama2015_fig3}; \citealt{ishiyama2013,ishiyama2015}) 
that is much smaller than those of most of the local dwarf galaxies 
and any high-$z$ galaxies observed, to date.
In other words, DM halos of galaxies are mostly recovered over cosmic time 
in numerical simulations.
%
%
It is true that the physical origin of DM is poorly understood.
However, under the collisionless DM particle assumption,
the DM-halo formation is established today.




The DM halo formation is understood not only by numerical simulations,
but also by analytic calculations. Starting from the Gaussian initial density fluctuations,
one can derive an approximation of the DM-halo mass function based on 
linear structure growth and spherical collapse. The analytic form is 
referred to as the Press-Schechter function \citep{press1974} that is,
\begin{equation}
n(M) dM= \frac{2}{\pi} \frac{a \rho_0}{{M^*}^2} \left( \frac{M}{M^*} \right)^{a-2} \exp \left[ - \left( \frac{M}{M^*} \right)^{2a} \right] dM
\label{eq:press_schechter} 
\end{equation}
where $M^*$ is the characteristic mass,  $a$ is a power-law index
of the mass fluctuations, and $\rho_0$ is the mean density of the universe.
Figure \ref{fig:springel2005_fig2} compares Press-Schechter functions (blue dashed lines) with numerical results (red points).
The Press-Schechter functions reasonably approximate the numerical results,
while there exist small departures. Note that theorists modify the analytic form of eq. (\ref{eq:press_schechter}),
with a few additional free parameters (e.g. \citealt{sheth1999}), and obtain 'modified' Press-Schechter functions
with the best-fit parameters determined by fitting the numerical results (solid lines in Figure \ref{fig:springel2005_fig2}).
A number of galaxy formation studies (including LAE modeling) exploit such modified Press-Schechter functions that are useful to reproduce 
DM-halo mass functions (mass vs. abundance) at any redshifts and cosmological parameter sets \citep{mao2007,samui2009}.
It should be also noted that these analytic formalisms can also provide reliable predictions in 
clustering of DM halos. In other words, once a redshift, mass, and cosmological parameter set 
are given, the abundance and clustering of DM halos are predicted by these formalisms
based on the $\Lambda$CDM structure formation scenario.


Because galaxies form in DM halos, galaxy luminosity functions and stellar-mass functions
should have a functional shape similar to DM halo mass functions.
Indeed, galaxy luminosity functions determined by observations can be fit well with the Schechter function \citep{schechter1976},
\begin{equation}
\phi (L) dL = \phi^* \left( \frac{L}{L^*} \right)^\alpha \exp \left( - \frac{L}{L^*} \right) d\left( \frac{L}{L^*} \right)
\label{eq:schechter}
\end{equation}
where $\phi (L)$ is the number density of galaxies at luminosity $L$
\footnote{
This is the Schechter function on the luminosity basis. The magnitude-based 
Schechter function is shown in, e.g., Equation 8 of \citet{ouchi2004}.
}.
The Schuchter function includes three free parameters, $\phi^*$, $L^*$, and $\alpha$,
that correspond to the characteristic number density, the characteristic luminosity, and 
the faint-end slope, respectively.
Similarly, stellar mass functions are expressed with stellar mass $M_{\rm s}$ and the characteristic 
stellar mass $M_{\rm s}^*$ that are in place of $L$ and $L^*$, respectively, in Equation (\ref{eq:schechter}) 
(Figure \ref{fig:read2005_fig4}).

It should be noted that Equation (\ref{eq:schechter}) has a functional form of the product
of an exponential cut-off and a power law for luminosity, which is the same 
as Equation (\ref{eq:press_schechter}) 
where halo mass is replaced with luminosity.

The three free parameters of the Schechter function 
reflect differences from the DM-halo mass function,
which depend on the baryonic processes of star-formation and feedback in galaxy formation (Section \ref{sec:star_formation}).
In LAE studies, the Schechter function is used to approximate the Ly$\alpha$ luminosity function and the continuum luminosity
function.






\subsubsection{Star Formation}
\label{sec:star_formation}

Star formation involves complicated physical processes of
gas cooling and feedback as detailed below. 
Moreover, star-formation is induced
by objects and matter outside of the galaxy by mergers and
gas accretion. I choose gas cooling, feedback, and cold accretion
that are key for understanding LAEs in the context of galaxy formation,
and introduce these physical processes below.\\

\noindent
{\bf Gas Cooling : }

Star formation requires a reservoir of cold dense gas
in a DM halo. However, such cold dense gas cannot
be easily produced in a DM halo. 
If an adiabatic gas contraction takes place
in the DM halo, gas temperature increases.
Then, the gas contraction stops by the thermal pressure,
and dense gas cannot be produced. In this way, 
adiabatic contractions do not make dense gas 
that is necessary for star formation. 
Star-formation thus requires a gas contraction 
associated with radiative cooling that 
reduces thermal energy in the gas cloud (Figure \ref{fig:DMH_SF}).
By the radiative cooling, gas temperature should decrease
from the virial temperature $T$ of the DM halo ($\gtrsim 10^4$ K)
to the temperature of molecular hydrogen \h2\ clouds
($\lesssim 10^2$ K). 

State-of-the-art simulations calculate the gas cooling 
processes numerically under realistic physical conditions, 
although these calculations cannot be described 
with simple analytical forms. 
Instead, I introduce a classic picture of gas cooling
with the free-fall time of the spherical model \citep{silk1993} that 
helps the readers understand the idea of gas cooling.

\begin{figure}[H]
\centering
\includegraphics[scale=.40]{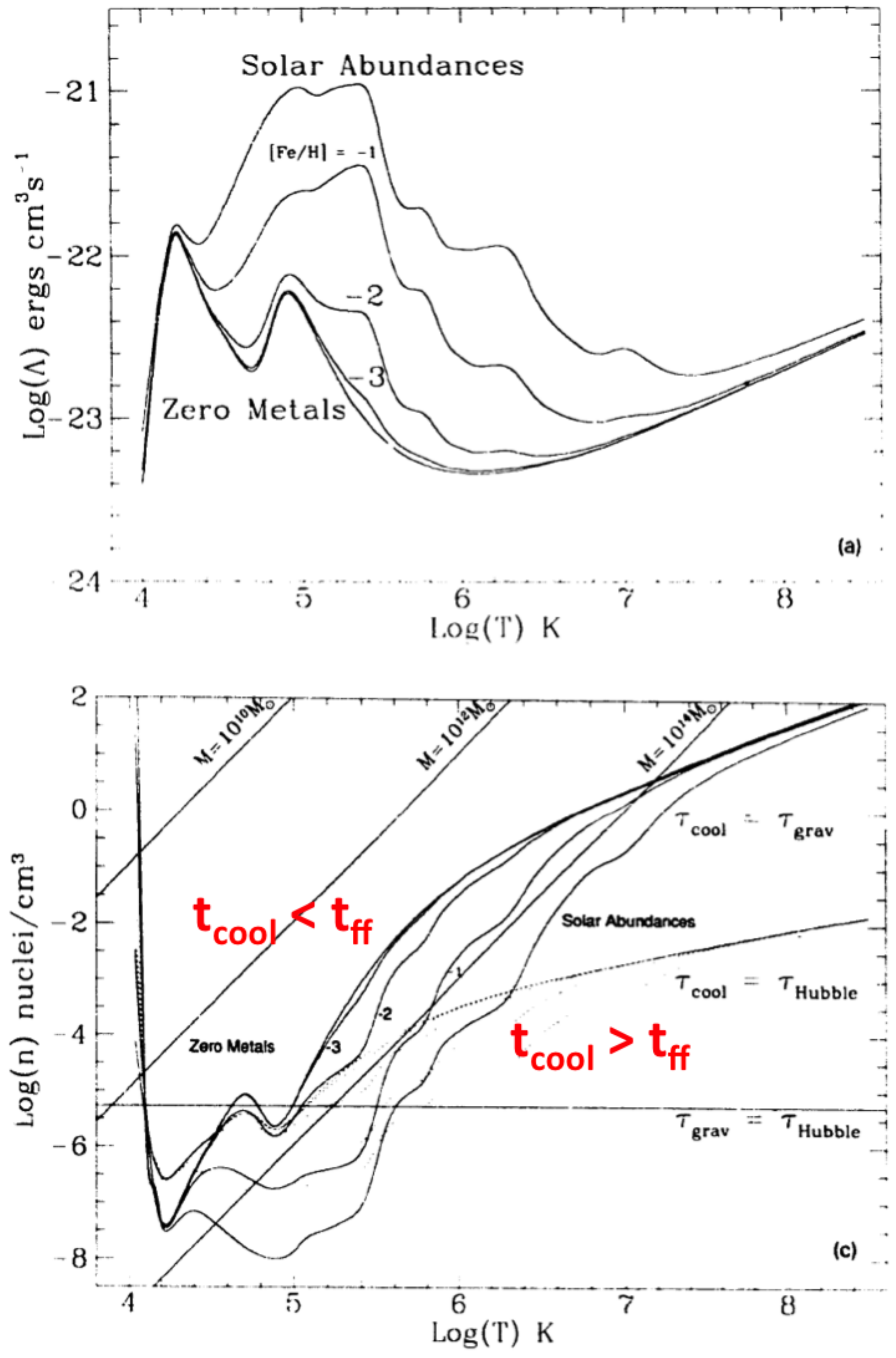}
\caption{
Cooling function (top) and gas density (bottom) as a function of virial temperature
in a spherical halo \citep{silk1993}. 
Top: The curves represent cooling functions with various metal abundances from
zero to one solar metallicities that are indicated by the labels.
Bottom: The solid curves denote $t_{\rm cool}=t_{\rm ff}$ for the metal abundances
from zero to one solar metallicities (same labels as in the top panel).
Beyond these curves,
the cooling time is shorter than the free-fall time, and gas collapses 
with a negligible thermal pressure. The dashed curve (the horizontal line) 
indicates that the cooling time (free-fall time) is equal to the present Hubble time.
Above the curve and the horizontal line, the gas contraction completes in the cosmic time.
The diagonal lines show the density and virial temperature that correspond
to the virial masses of the system, $10^{10}$, $10^{12}$, and $10^{14} M_\odot$. 
This figure is reproduced by permission of the Physics Reports.
}
\label{fig:silk1993_fig4}       
\end{figure}

The cooling function, $\Lambda(T)$, is defined by
\begin{equation}
|\dot{E}_{\rm cool}| = n^2 \Lambda(T),
\label{eq:cooling_function}
\end{equation}
where $\dot{E}_{\rm cool}$ and $n$ are the cooling rate (energy density divided by time) 
and the number density of particles, respectively. The cooling function is 
calculated based on quantum physics, and displayed 
in Figure \ref{fig:silk1993_fig4}. It is clear that
metal rich gas has a higher $\Lambda(T)$, because various atomic electron transitions
are allowed for heavy elements that enhance the efficiency of radiative cooling.
In zero-metal gas, there are two peaks in $\Lambda(T)$ near $10^4$ and $10^5$ K
that correspond to hydrogen and helium recombinations, respectively.
The upturn of $\Lambda(T)$ from $10^6$ to $10^8$ K is explained by
the cooling processes of Bremsstrahlung and Compton scattering.

In a virialized system, the kinetic energy $E_K$ of gas 
is given by $E_K=3nk_B T/2$, where
$k_B$ is the Boltzmann constant. 
Thus, the cooling time $t_{\rm cool}$ is given by
\begin{equation}
t_{\rm cool}= \frac{E_K}{|\dot{E}_{\rm cool}|}  = \frac{3}{2}\frac{k_B T}{n \Lambda(T)}.
\label{eq:cooling_time}
\end{equation}

In the spherical model, the free-fall time $t_{\rm ff}$, a simple dynamical time, is
\begin{equation}
t_{\rm ff} = \sqrt{\frac{3\pi}{32 G \rho_{\rm m}}},
\label{eq:free_fall_time}
\end{equation}
where $\rho_{\rm m}$ is the mass density that is proportional to $n$.
If the radiative cooling is very efficient,
the cooling time is shorter than the free-fall time, $t_{\rm cool}<t_{\rm ff}$.
In this case, gas collapse takes place with a negligible thermal pressure,
and cold dense gas necessary for star formation is produced.
This condition of $t_{\rm cool}<t_{\rm ff}$ for gas collapse is 
presented in the $T$ and $n$ plot of Figure \ref{fig:silk1993_fig4} (bottom).
Halo masses of $\sim 10^{12} M_\odot$ 
allow gas collapse down to the low gas densities, 
indicating an efficient gas cooling. It should be noted that
the halo mass of $10^{12} M_\odot$ coincides with
the mass of the Milky Way as well as the mass where the stellar-to-halo mass ratio is highest 
\citep{behroozi2013}.
Metals ease the conditions of gas collapse
in a massive halo with $>10^{12} M_\odot$.
In Figure \ref{fig:silk1993_fig4}, gas halos with $t_{\rm cool}>t_{\rm ff}$
cannot collapse but cause a quasi-static contraction
due to inefficient cooling, which can take time longer than the Hubble time.
As shown in Section \ref{sec:clustering}, typical LAEs have halo masses
of $10^{10}-10^{12} M_\odot$ and sub-solar metallicities. Figure \ref{fig:silk1993_fig4} indicates that
LAEs have physical parameters reasonably good for gas collapse, which enables
subsequent star formation.\\





%
%



\noindent
{\bf Feedback : }

\begin{figure}[H]
\centering
\includegraphics[scale=.40]{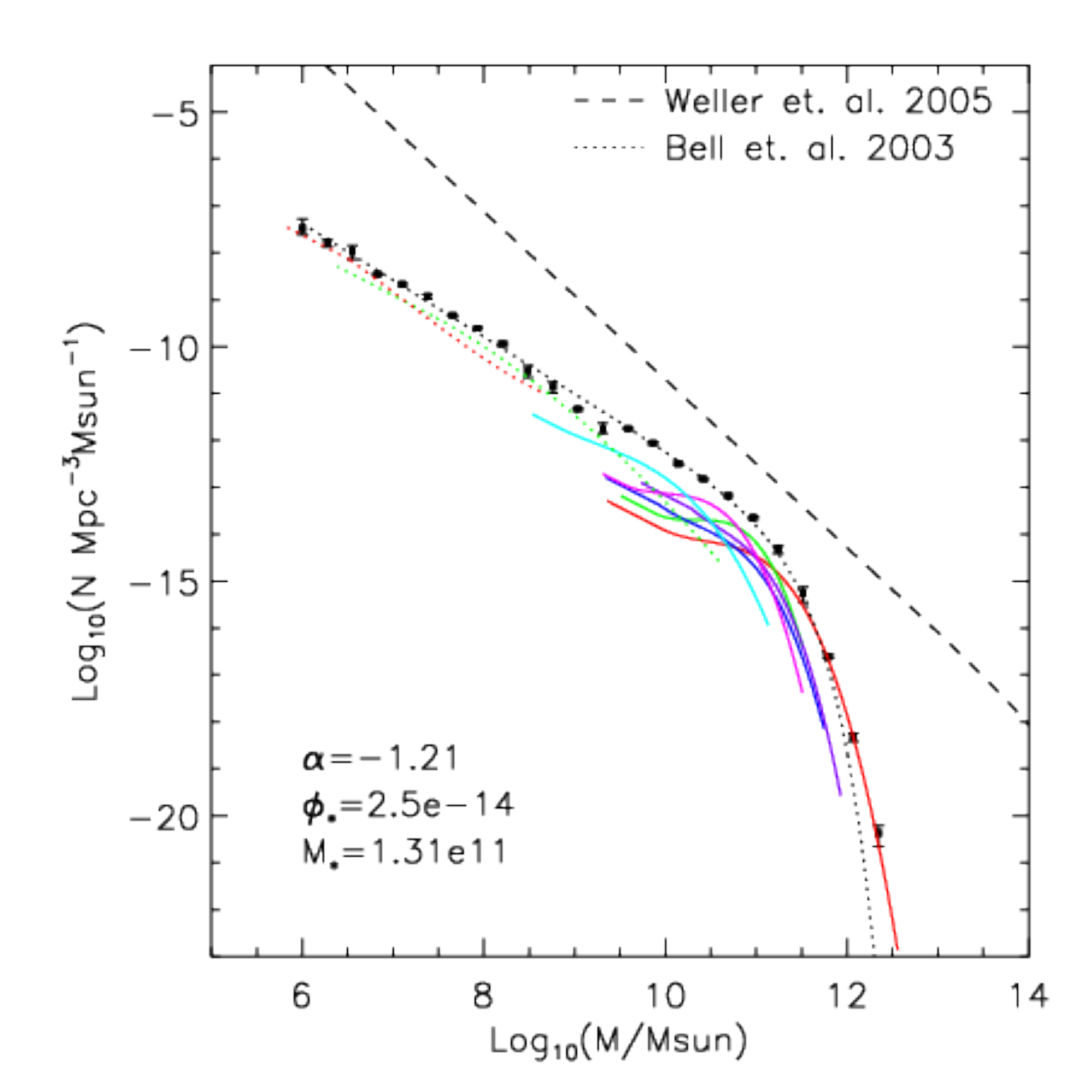}
\caption{
An observed baryonic (stars+cold gas) mass function of local galaxies (squares and the black dotted line; \citealt{bell2003})
and the DM-halo mass function (dashed line; \citealt{weller2005}). The colored solid and dotted lines
indicate baryonic mass functions of various Hubble type galaxies: ellipticals, spirals, irregulars, and
dwarf ellipticals. This figure is adopted from \citet{read2005}, and
reproduced by permission of the Royal Society.
}
\label{fig:read2005_fig4}       
\end{figure}

Feedback is known as one of the most important physical processes
involved in star formation. Figure \ref{fig:read2005_fig4} compares 
an observed galaxy stellar-mass function (filled squares)
with a DM halo mass function from numerical simulations (dashed line).
Because the cosmic baryon fraction $f_{\rm b}$ is $f_b \equiv \Omega_{\rm b}/\Omega_{\rm m} \simeq 0.16$
\citep{planck2015}, the DM-halo mass function should be at least about an order of magnitude
higher than the stellar-mass function, which can be clearly seen at $\sim 10^{11} M_\odot$ 
in Figure \ref{fig:read2005_fig4} (see the mass values at a constant number density of $\log N/{\rm [Mpc^3 M_\odot]} \sim -14$).
However, in Figure \ref{fig:read2005_fig4}, the stellar-mass function is 
flatter at the low-mass end and steeper at the massive end than 
the DM-halo mass function.
These shape differences are thought to be made by feedback effects that
suppress star-formation by gas heating and outflow
associated with star-formation and AGN activities 
in a galaxy \citep{bower2006}.

\begin{figure}[H]
\centering
\includegraphics[scale=.45]{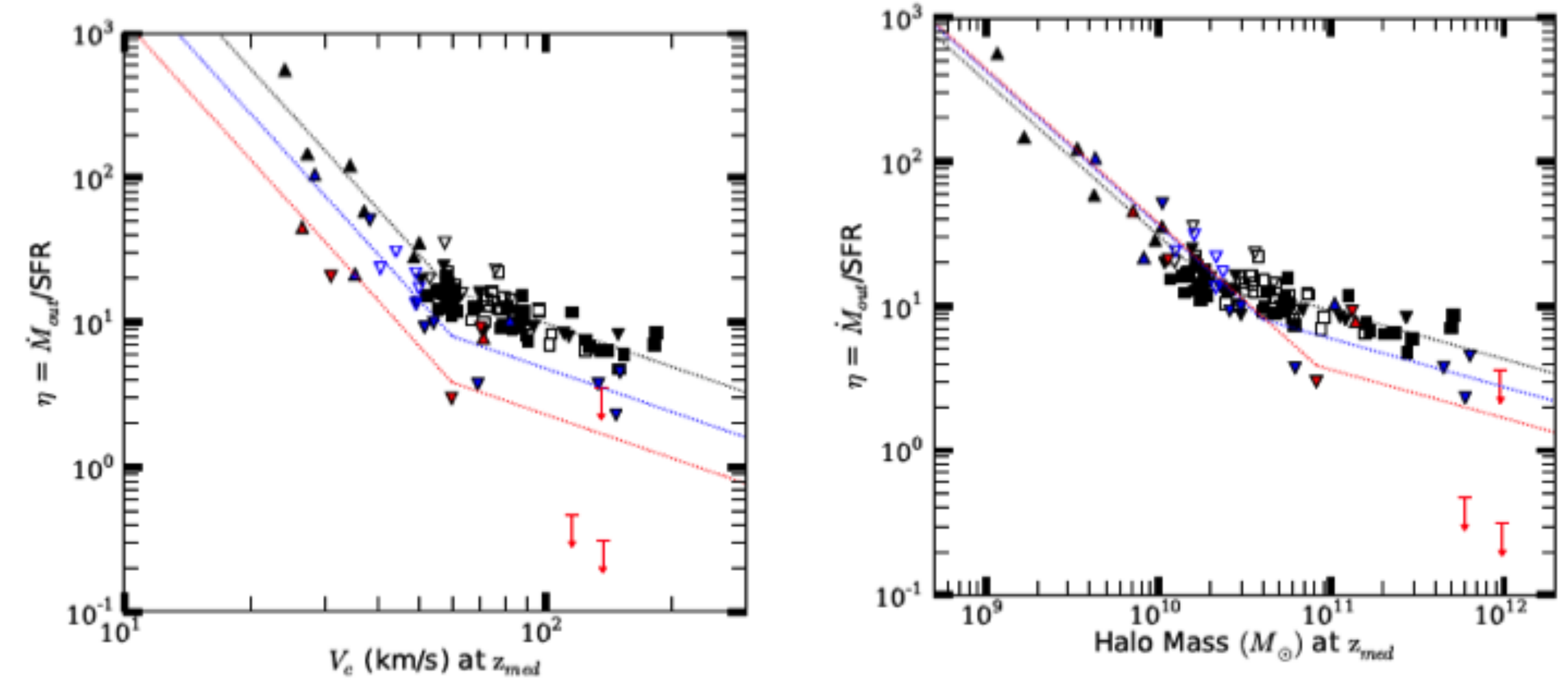}
\caption{
Mass loading factor as a function of $V_{\rm c}$ (left) and DM-halo mass (right)
predicted by numerical simulations \citep{muratov2015}.
The red, blue, and black symbols represent galaxies at $z=0-0.5$ $(z_{\rm med}=0.25)$,
$0.5-2$ $(1.25)$, and $2-4$ $(3)$, respectively, produced in the simulations.
The red, blue, and black dotted lines indicate a broken-power law fit to the simulated galaxies
at corresponding redshifts that roughly corresponds to the energy- and momentum-driven
feedbacks (Equations \ref{eq:energy_driven} and \ref{eq:momentum_driven}),
although the slope of the energy-driven feedback is slightly steeper than Equation (\ref{eq:energy_driven}).
This figure is reproduced by permission of the the Royal Astronomical Society.
}
\label{fig:muratov2015_fig5}       
\end{figure}

Theoretical studies 
assume
two feedback mechanisms in the low and high mass regimes, energy- and momentum-driven feedback
effects, respectively (Figure \ref{fig:muratov2015_fig5}; \citealt{muratov2015}).
The energy-driven feedback for low-mass galaxies is caused by 
thermal energy inputs from supernova (SN) explosions and stellar radiation.
The momentum-driven feedback for high-mass galaxies is activated by
kinetic energy inputs from stellar winds, radiative pressure, and AGN jets.
Defining the mass-loading factor $\eta \equiv \dot{M}_{\rm out}/SFR$
where $\dot{M}_{\rm out}$ is the outflow rate, numerical simulations show
\begin{eqnarray}
\label{eq:energy_driven}
\eta \propto V_{\rm circ}^{-2}\\
\label{eq:momentum_driven}
\eta \propto V_{\rm circ}^{-1}
\end{eqnarray}
for the energy and momentum driven feedbacks, respectively.
Here, $V_{\rm circ}$ is the circular velocity given by
\begin{equation}
V_{\rm circ}=\sqrt{\frac{G M_{\rm h}}{r_{\rm vir}}},
\label{eq:v_circ}
\end{equation}
where $M_{\rm h}$ and $r_{\rm vir}$ are the DM-halo mass and the virial radius,
respectively. In Figure  \ref{fig:muratov2015_fig5}, the energy and momentum driven feedbacks
are seen at $M_{\rm h} \lesssim 10^{10} M_\odot$ and $\gtrsim M_{\rm h} \sim 10^{10} M_\odot$,
respectively. From observations of local galaxies, the relation of Equation (\ref{eq:momentum_driven})
is confirmed \citep{heckman2015}, while no observations reach $\lesssim M_{\rm h} \sim 10^{10} M_\odot$
to test the relation of the energy-driven feedback (Equation \ref{eq:energy_driven}).
Because the average DM-halo mass of LAEs is estimated to be $M_{\rm h} \sim 10^{11} M_\odot$
by clustering analysis (Section \ref{sec:clustering}), feedbacks in typical LAEs are probably dominated by 
the momentum-driven feedback.

Note that some theoretical studies 
claim the existence of positive feedback effects 
that induce star-forming activities by, e.g., the shock cooling of AGN jets, 
radiation pressure, etc. \citep{silk2013,vitale2015}.\\

%


\noindent
{\bf Cold Accretion : }

\begin{figure}[H]
\centering
\includegraphics[scale=.48]{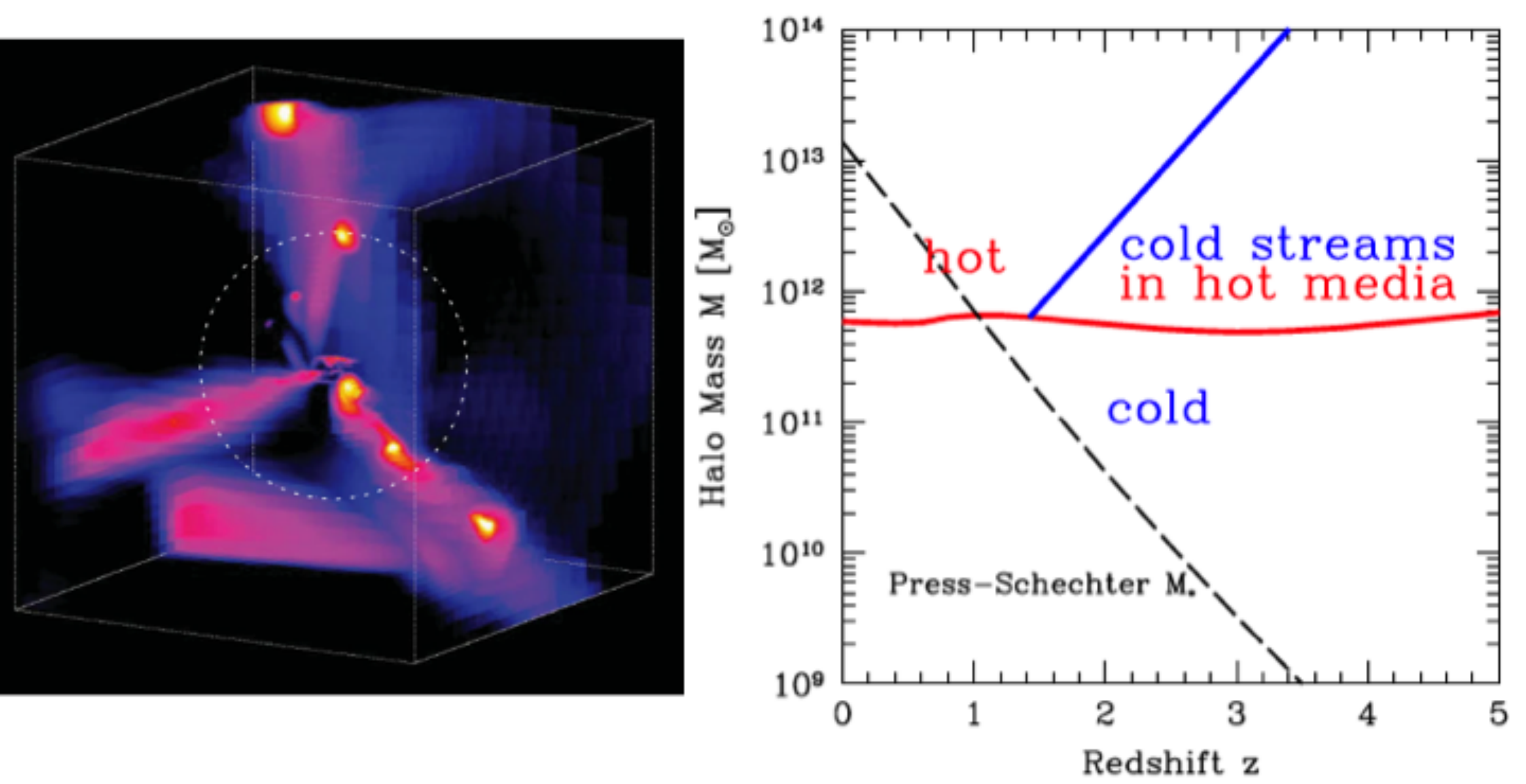}
\caption{
Left: Radial flux of cold gas accretion into the center of a DM-halo
predicted by numerical simulations \citep{dekel2009}.
The color scale indicates the inflow rate per solid angle.
The box size is 320 kpc. The dotted line represents 
the virial radius of the DM halo. There are three cold accretion streams 
clearly found in this figure, two of which include gas clumps with a  
mass ten times lower than that of the central galaxy.
Right: DM-halo mass as a function of redshift \citep{dekel2009}. The red horizontal curve
indicates the threshold mass above which infalling gas is shock-heated 
around the DM-halo virial radius. The blue line represents the limit of
cold accretion of gas whose density (temperature) is high (low) enough
to penetrate into the DM halo center through the diffuse shock-heated medium.
The dashed line denotes the characteristic DM-halo mass of the Press-Schechter mass function 
(eq. \ref{eq:press_schechter}) at a given redshift.
This figure is reproduced by permission of the Nature Publishing Group.
}
\label{fig:dekel2009_fig2_fig5}       
\end{figure}

Another important mechanism for star-formation is cold accretion.
In the standard picture of galaxy growth, gas infalling in a DM halo 
is heated to a virial temperature by shocks at around the DM-halo virial radius, 
and then 
reaches a quasi-hydrostatic equilibrium with $T\sim 10^6 (V_{\rm circ} / 167 $km s$^{-1})^2$ K.
The hot virialized gas cools by cooling radiation, and 
forms a cold gas disk that produces stars at the DM-halo center \citep{rees1977,white1978,fall1980}.
%
%
Recent theoretical studies suggest that, in this galaxy growth process,
the infalling gas can penetrate into the DM halo center through the diffuse
shock-heated medium, if the infalling gas is dense ($\sim 1$ cm$^{-3}$) 
and cold (a few $10^4$ K; \citealt{fardal2001,kravtsov2003,keres2005}),
which is referred to as cold accretion, cold mode accretion, or cold stream.
The theoretical studies predict that, beyond the virial radius of DM halos, 
there exist multiple cold dense gas streams to the DM-halo center through 
filaments of LSSs (Figure \ref{fig:dekel2009_fig2_fig5}).
These cold dense gas streams collide at the DM-halo center,
and cool very efficiently, which triggers intense star formation \citep{dekel2009}.
Such cold accretion is important, because about two-thirds of gas accretion in mass
has a form of smooth gas flows, in contrast with the rest of the gas accretion taking place
in a form of mergers with a $>1/10$-mass ratio
\citep{katz2003,keres2009,dekel2009}).
This theoretical picture would explain high SFR galaxies with no merger signatures,
such as bright LBGs and SMGs with an SFR of $\gtrsim 100M_\odot$ yr$^{-1}$,
and could be an answer to the question why 
the number density of high SFR galaxies at $z\sim 2$ is 
significantly larger than those expected from merger events.
Note that cold gas accretion is allowed in a massive halo 
only at $z\gtrsim 2$ (Figure \ref{fig:dekel2009_fig2_fig5}),
when the accretion gas is sufficiently cold.

%


Because the cold accretion is a theoretical picture, in the past decade
observers have searched for a signature of cold accretion in their observational data.
In LBGs at $z\sim 2$, velocities of low ionization metal absorption lines are 
mostly blueshifted from the galaxy systemic velocities, 
indicating gas outflow associated with star-forming activities \citep{steidel2010}.
Although there are several reports of cold accretion object candidates
in deep observational studies (e.g. \citealt{nilsson2006,rauch2011}), 
no definitive observational evidence for cold accretion has been found so far.
Because the cold accretion gas infalls along with filaments of LSSs, 
the covering fraction of cold accretion gas is very small, $\sim 1-2\%$ \citep{faucher-giguere2011}.
A large number of sightlines (i.e. a large sample of galaxies) would be needed to 
prove or disprove the existence of cold accretion.




\subsubsection{Role of Observations}
\label{sec:roles_of_observations}
\begin{figure}[H]
\centering
\includegraphics[scale=.40]{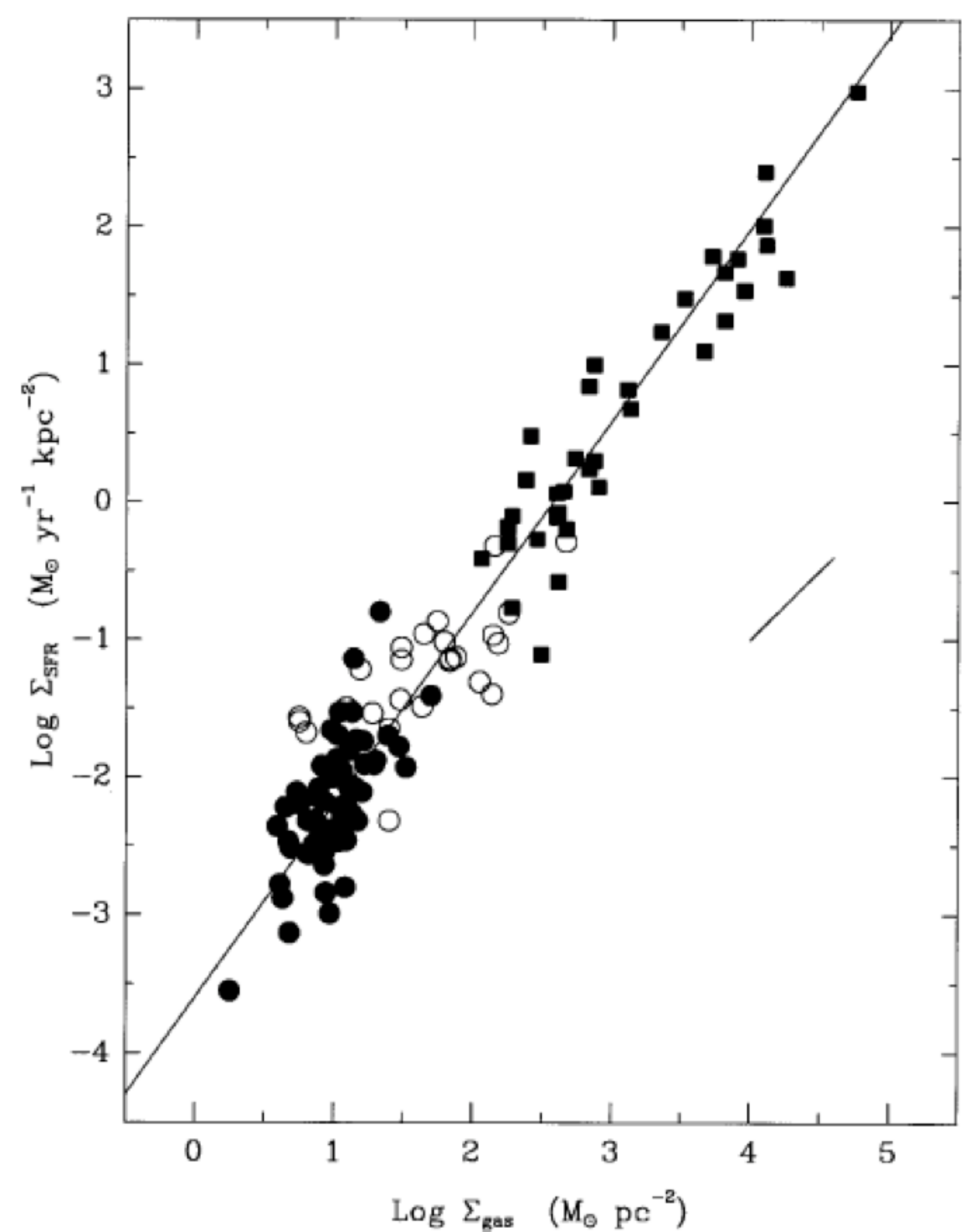}
\caption{
Star-formation surface density as a function of gas surface density \citep{kennicutt1998b}.
The gas surface density is defined by the average total surface density of atomic and molecular gas.
The filled circles and squares represent normal-disk and starburst galaxies, respectively, in the local universe.
The open circles denote the center of the normal-disk galaxies. The line is the least-squares fit
to the data points, $\Sigma_{\rm SFR} \propto \Sigma_{\rm gas}^{1.4}$, which is known as the Kennicutt-Schmidt law.
This figure is reproduced by permission of the AAS.
}
\label{fig:kennicutt1998b_fig6}       
\end{figure}

As introduced in Section \ref{sec:basic_picture},
the basic process of galaxy formation is
DM-halo formation and star formation (Figure \ref{fig:DMH_SF}).
DM-halo formation is well understood with no large systematics
by simple numerical simulations and analytic approximations 
(Section \ref{sec:DM_halo}), while star-formation is
poorly understood due to the complicated 
baryonic processes: gas cooling, feedback, and cold accretion as well as
merger induced star-formation.
The star-formation process involves a number of unknown
parameters, such as gas metallicity, density, temperature, outflow, and inflow.
Observations can obtain these key parameters tightly connected to star formation,
and constrain free parameters of galaxy formation models.
On the other hand,
many cosmological simulations including those for LAEs 
assume a simple relation between halo-mass and galaxy luminosity (e.g. \citealt{mcquinn2007}) as well as
an empirical relation between gas and star-formation density such as the Kennicutt-Schmidt (KS) law
(Figure \ref{fig:kennicutt1998b_fig6}). Such models with empirical relations can
derive the star-formation surface density $\Sigma_{\rm SFR}$ from the gas surface density $\Sigma_{\rm gas}$ that
is predicted by numerical simulations and semi-analytic models,
and aim to explain other various observational quantities of galaxies (e.g. \citealt{garel2012}). 
In this way, key observational parameters and empirical relations
are important to understand star-formation in galaxies.
Thus, the goal of observations is to determine
star-formation key parameters and empirical relations to develop 
a self-consistent physical picture of galaxy formation 




%


\subsection{Origins of Ly$\alpha$ Emission from LAEs}
\label{sec:origins_of_lya}

\begin{figure}[H]
\centering
\includegraphics[scale=.45]{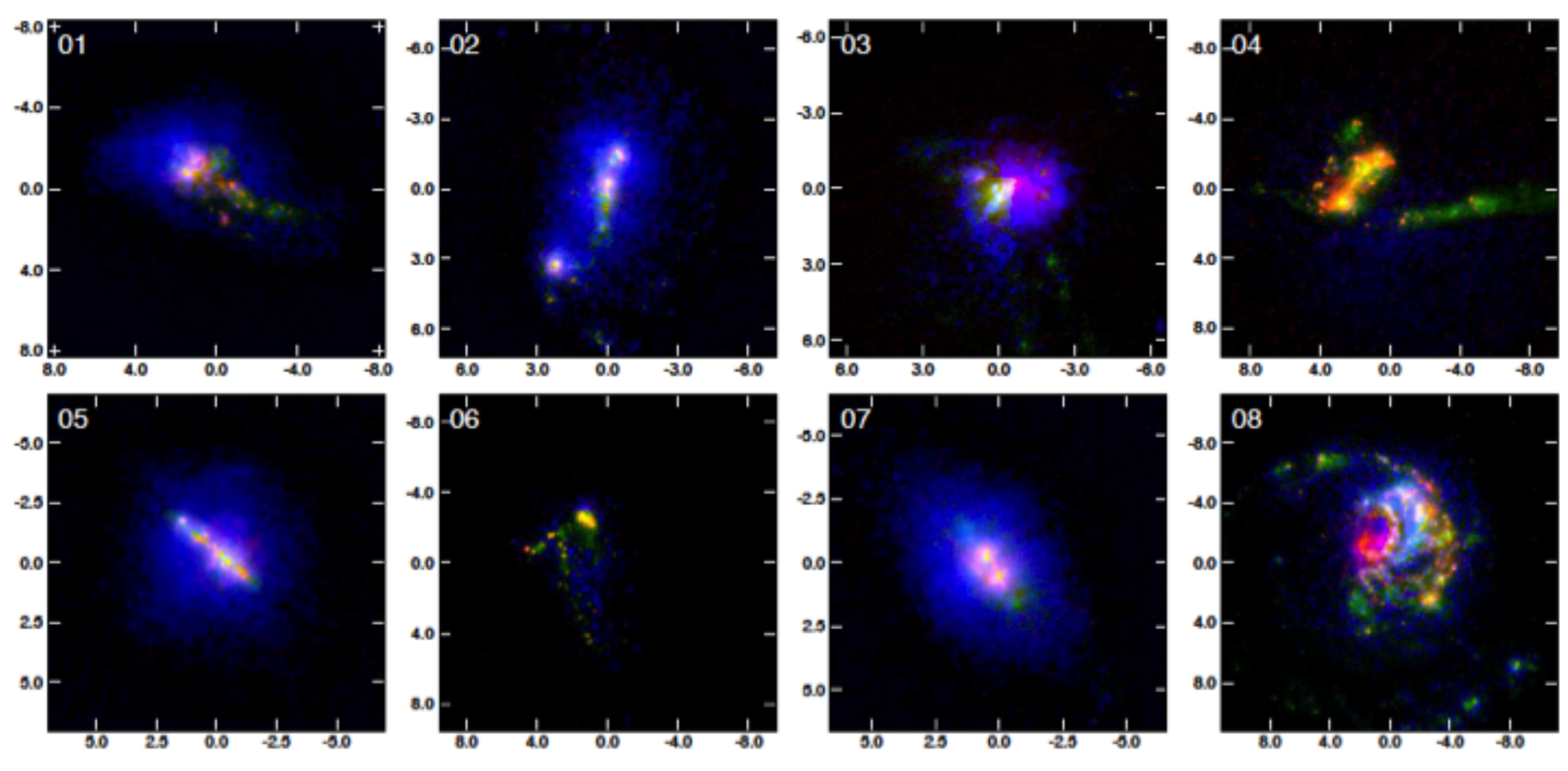}
\caption{
False-color images of local Ly$\alpha$ emitters obtained by the LARS survey \citep{hayes2013}.
The blue, red, and green colors indicate Ly$\alpha$ emission, H$\alpha$ emission,
and far-UV continuum, respectively. The scales in units of kpc are
shown in the vertical and horizontal axes.
This figure is reproduced by permission of the AAS.
}
\label{fig:hayes2013_fig1}       
\end{figure}

Once a galaxy formation model is developed, Ly$\alpha$ emission
of galaxies can be modeled. 
%
Theoretical studies suggest that in galaxies, Ly$\alpha$ emission can have five major origins, 
which probably explain the
diversity of the spatial distribution of Ly$\alpha$ emission revealed by observations
(Figure \ref{fig:hayes2013_fig1}).

Ly$\alpha$ emission following hydrogen recombination in the ISM near the center of a galaxy can 
result from two origins of ionizing sources: 
i) star formation that makes {\sc Hii} regions
and 
ii) nuclear activities (i.e. AGN), if any, producing highly ionized 
broad and narrow-line regions in the galaxy center.

The remaining three origins are dominated by
Ly$\alpha$ emission from the CGM to the outer halo:
iii) outflowing gas that collisionally excites hydrogen whose
Ly$\alpha$ to H$\alpha$ flux ratio is higher than the one of the optically thick
case B recombination $f_{\rm Ly \alpha}/f_{\rm Ly \alpha} > 8.7$ \citep{nakajima2013}.
iv) cooling radiation in the hot halo gas (Section \ref{sec:star_formation}), and
v) fluorescence emission produced by the halo and IGM neutral hydrogen gas photo-ionized 
by UV background radiation supplied, e.g., by QSOs \citep{kollmeier2010}.

Although Ly$\alpha$ emission can be made by these five photo-ionization and collisional excitation processes i)-v),
Ly$\alpha$ photons experience resonance scattering in the \hi\ gas of the ISM, the CGM, and the IGM,
due to the large Ly$\alpha$ cross section of \hi. Ly$\alpha$ photons are re-distributed 
in space and wavelength by the resonance scattering. For this reason, scattered Ly$\alpha$ emission
would dominate in the CGM, where the Ly$\alpha$ intensity of photo-ionization is relatively weak.
It should be noted that observations identify Ly$\alpha$ photons last scattered by \hi\ gas,
and largely miss the original Ly$\alpha$ source position and gas dynamics information.
However, this resonance nature of Ly$\alpha$ is also useful to probe the distribution and the kinematics 
of \hi\ gas by observations via theoretical modeling (Section \ref{sec:outflow_lya_profile}).

%
%


\subsection{Summary of Galaxy Formation I}
\label{sec:summary_galaxy_formationI}

This section overviews the basic theoretical framework of
galaxy formation and Ly$\alpha$ production, targeting young
observers with a limited theoretical background. This section
explains that galaxy formation is made of two physical processes
of DM-halo formation and star formation. DM-halo formation
is well understood with a simple robust model of cosmic structure formation
consistent with observations. However, star formation involves
complicated baryonic processes of gas cooling, feedback, and cold
accretion as well as mergers that include many physical parameters
difficult to determine.
It is concluded that observations should constrain important physical parameters
and empirical relations that are key for filling in the missing piece of
the picture of galaxy formation. To understand LAEs in the context of galaxy formation,
one also needs physical models of Ly$\alpha$ emission. Five Ly$\alpha$ emission origins
in theoretical models are introduced. Two origins are in ISM regions:
i) star formation (HII regions) and ii) AGN (highly-ionized gas in a galaxy center). 
The other three dominate in the CGM and outer halo regions: 
iii) outflowing gas, iv) cooling radiation, and v) fluorescence of UV background radiation.
LAE observations should also reveal the origins of Ly$\alpha$ photons 
in parallel with the efforts to address the general galaxy formation issues.




\section{Galaxy Formation II: LAEs Uncovered by Deep Observations}
\label{sec:galaxy_formationII}

Since the discovery of LAEs in the late 1990s,
various physical properties of LAEs have been revealed 
by exploiting deep optical to mid-infrared (MIR) imaging and spectroscopic capabilities
of 8m-class ground based telescopes, HST, and Spitzer Space Telescope (Spitzer)
in conjunction with observations at other wavelengths using the Chandra X-ray observatory (Chandra),
GALEX, Herschel Space Observatory (Herschel), 
Atacama Large Millimeter / submillimeter Array (ALMA), and
Very Large Array (VLA).
In this section, I review key physical properties of LAEs uncovered by those observations:
stellar population, luminosity function, morphology, ISM properties (metallicity, ionization parameter, dust),
AGN activity, and clustering.

\subsection{Stellar Population}
\label{sec:stellar_population}

\begin{figure}[H]
\centering
\includegraphics[scale=.55]{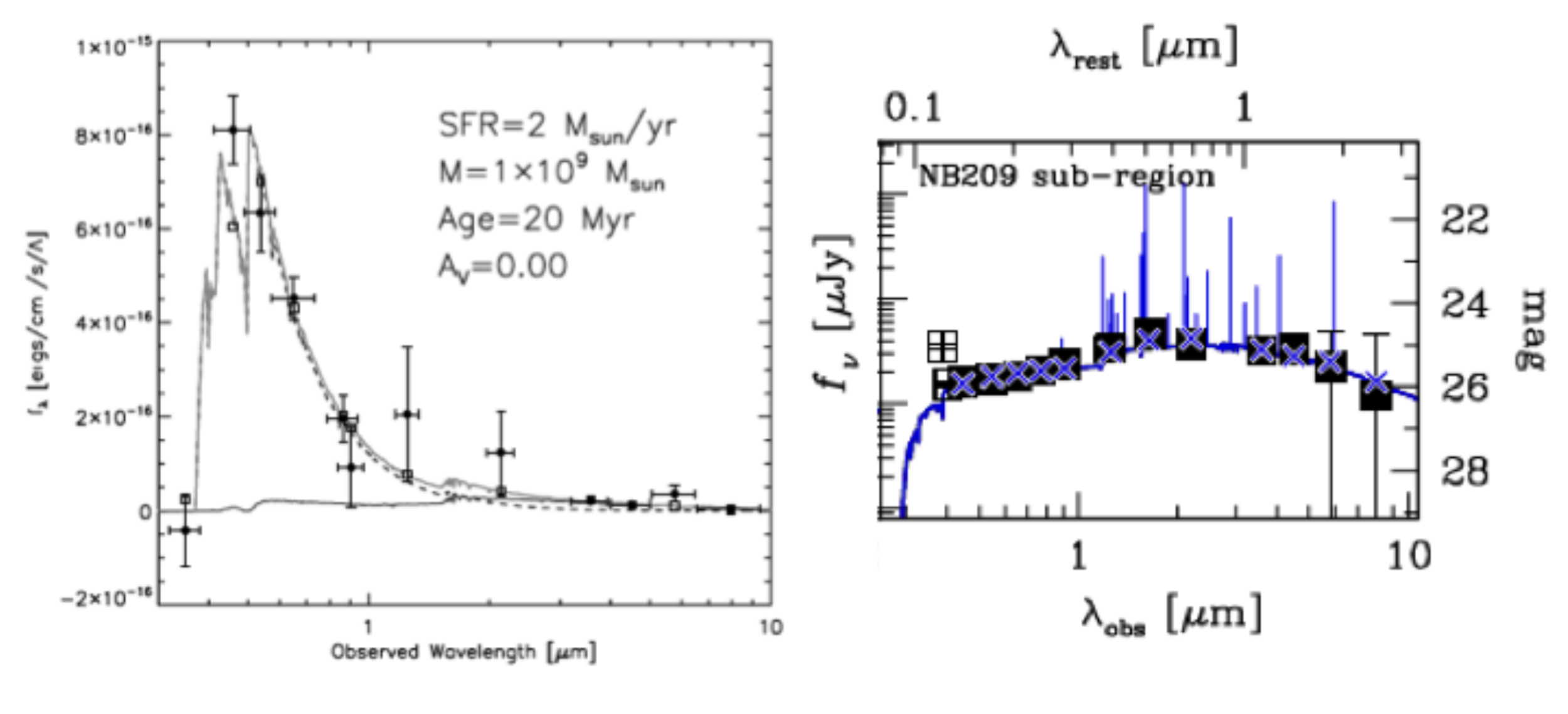}
\caption{
Left: A composite SED of LAEs at $z=3.1$ expressed in $f_\lambda$ \citep{gawiser2007}.
The thick solid line represents the best-fit SED that is made of an old (thin solid line) and young (dashed line)
stellar components.
Nebular emission is not included in modeling.
Right:  Same as the left panel, but for $z=2.2$ in $f_\nu$ \citep{nakajima2012}.
The blue line is the best-fit SED model that includes nebular emission. The blue crosses indicate the
expected broadband photometry from the best-fit model.
The two data points shown by open symbols are not used in the fitting 
in order to avoid a contamination from Ly$\alpha$ emission.
This figure is reproduced by permission of the AAS.
}
\label{fig:gawiser2007_fig4_nakajima2012_fig9}       
\end{figure}

The stellar population of galaxies is described by stellar mass, age, dust extinction, 
and some other parameters, and these parameters can be estimated by 
fitting broadband spectral energy distributions (SEDs) 
with stellar population synthesis models such as \citealt{bruzual2003}.
It is, however, difficult to investigate stellar populations of LAEs because 
most LAEs do not have detectable continuum emission even in deep images, 
although there do exist remarkably bright LAEs \citep{lai2008}.
Making a composite (average) SED of a number of continuum-faint LAEs by image stacking,
early studies have revealed that they have faint and blue SEDs on average.
An example SED is shown in the left panel of 
Figure \ref{fig:gawiser2007_fig4_nakajima2012_fig9} \citep{gawiser2007}, 
which is explained by a model with 
a low stellar mass of $\sim 10^9 M_\odot$, a young stellar age of $\sim 20$ Myr, and a negligibly
small dust extinction.

Because LAEs are young dust-poor star-forming galaxies, 
they often have strong nebular lines, such as \ha, \hb, [\oiii]5007, and [\oii]3727,
which contaminate continuum fluxes estimated from broadband photometry.
Because the observed-frame equivalent width $EW_{\rm obs}$ of nebular lines
increases with redshift as $EW_{\rm obs} \propto (1+z)$, nebular lines of high-$z$ LAEs
can cause serious systematic errors in broadband SEDs and hence in the calculation of 
stellar population parameters. Strong [\oiii], \hb, and [\oii] lines near 4000\AA\ 
mimic a Balmer break that is an indicator of stellar age, 
and an over/underestimated age leads to an over/underestimated stellar mass.
\citet{schaerer2009} introduce self-consistent 
population synthesis models with nebular lines where line ratios 
(as a function of metallicity) are fixed to the values of Galactic \hii\ regions.
Using a sample of $z\sim 6$ LBGs as an example, 
they claim that models without nebular lines overestimate stellar ages and 
masses by a factor of $3$.
Thus, considering nebular lines is critical to obtain
stellar population parameters of high-$z$ young star-forming galaxies including LAEs.
It should also be noted that there is another important source of contamination,
nebular {\it continuum}, that is the free-free/bound-free emission of hydrogen and helium and 
two photon continuum emission of hydrogen.
Because nebular continuum emission significantly changes UV-continuum colors for very young stellar populations with
a stellar age of $\lesssim 10$ Myr for instantaneous starbursts
(see Figures 3 and 4 of \citealt{bouwens2010}), it is usually
included in nebular emission modeling.

%

%



The right panel of Figure \ref{fig:gawiser2007_fig4_nakajima2012_fig9} presents an average SED of $z=2$ LAEs
and its best-fit stellar population synthesis model with nebular emission.
Table \ref{tab:stellar_population} summarizes the typical ranges of stellar population parameters of $z=2-7$ LAEs
that are obtained under the assumptions of constant star-formation history, 
a Salpeter IMF \citep{salpeter1955},
and Calzetti extinction law (\citealt{calzetti2000}; see \citealt{gawiser2007,ono2010a,ono2010b,guaita2011,hagen2014,hagen2016}).
Although different samples give different parameter values, 
Table \ref{tab:stellar_population} shows that LAEs are low-stellar mass galaxies 
with a low dust extinction, a medium-low SFR,
and a young stellar age.



\begin{figure}[H]
\centering
\includegraphics[scale=.55]{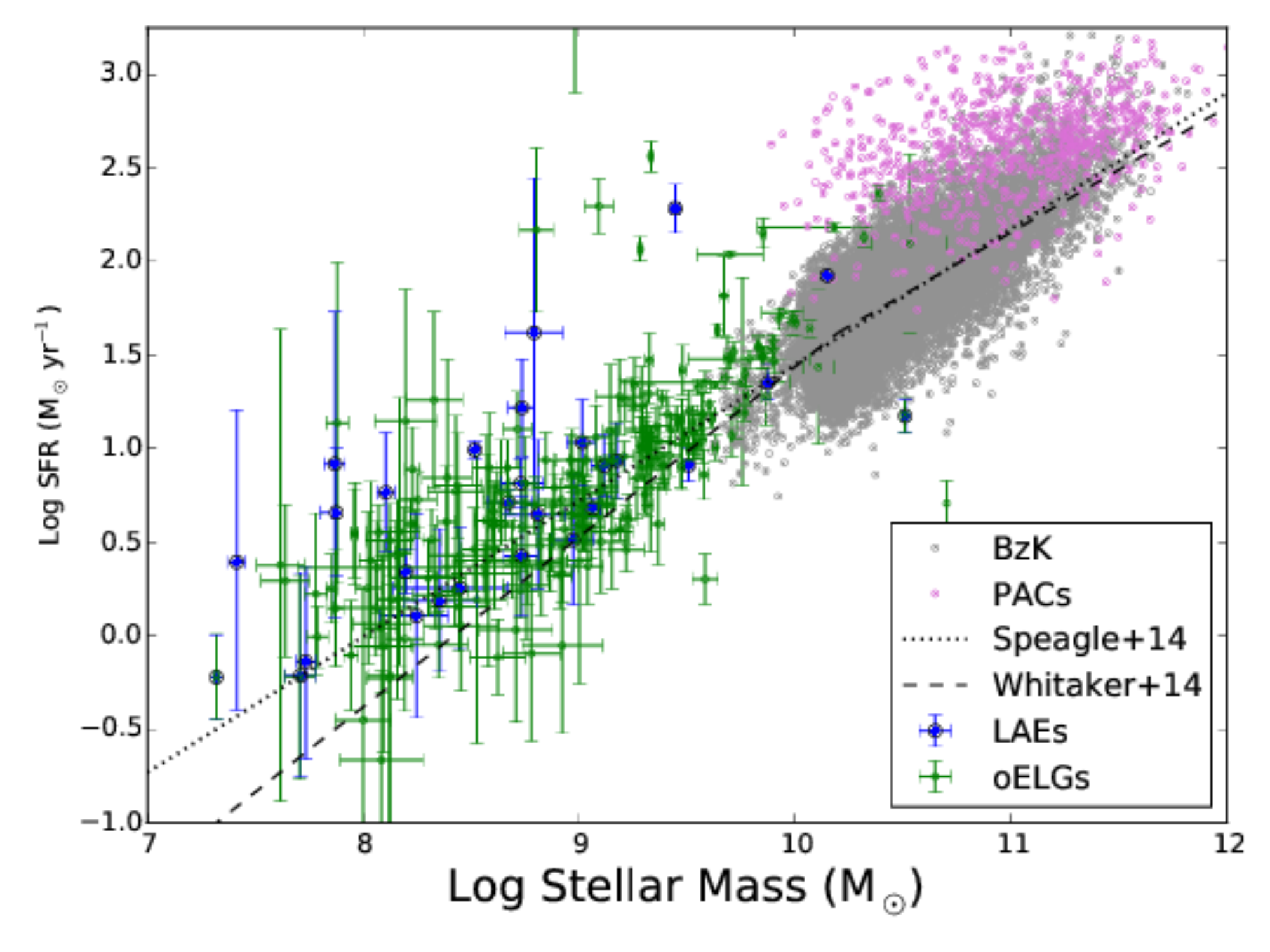}
\caption{
LAEs compared with the star-formation main sequence on the
SFR vs. stellar mass plot \citep{hagen2016}. The blue circles represent LAEs,
while the green circles denote optical emission galaxies (oELGs).
The gray and magenta circles indicate BzK and Herschel/PACS-detected
galaxies, respectively \citep{rodighiero2011}. The dotted and dashed
lines show the star-formation main sequences at $z=2$ obtained by
\citet{speagle2014} and \citet{whitaker2014}, respectively.
This figure is reproduced by permission of the AAS.
}
\label{fig:hagen2016_fig4}       
\end{figure}

Figure \ref{fig:hagen2016_fig4} compares LAEs (blue circles)
with other galaxies in the stellar mass vs. SFR plane.
At $M_{\rm s}\gtrsim 10^{10} M_\odot$ in Figure \ref{fig:hagen2016_fig4}, there is a star-formation (SF) main sequence,
a tight positive correlation between $M_{\rm s}$ and SFR \citep{daddi2007,elbaz2007}.
LAEs fall in the low mass regime of $M_{\rm s}\sim 10^7-10^{10} M_{\odot}$ 
slightly above an extrapolation of the SF main sequence found at $M\gtrsim 10^{10} M_\odot$
\citep{hagen2014,hagen2016}, suggesting that typical LAEs are
high-$z$ dwarf galaxies in a weak burst mode.
LAEs are located in a similar area in the stellar mass vs. SFR plane 
to other emission line galaxies, i.e., 
[\oii], \hb, and [\oiii] emitters, at $z\sim 2$  (green dots).


\begin{table}
\caption{General Properties of Typical LAEs$^\dagger$}
\label{tab:stellar_population}       
%
%
\begin{tabular}{p{2cm}p{2cm}p{2cm}p{2cm}p{2cm}}
\hline\noalign{\smallskip}
Stellar Mass  & $E(B-V)_{\rm s}$ $^a$ &  SFR & Stellar Age & Metallicity \\
$[M_\odot]$  &                 &  $[M_\odot$ yr$^{-1}]$ & $[$Myr$]$ & $Z_\odot$ \\
\noalign{\smallskip}\svhline\noalign{\smallskip}
$10^7-10^{10}$ & $0-0.2$ & $1-100$ & $1-100$ & $0.1-0.5$\\
\noalign{\smallskip}\hline\noalign{\smallskip}
\end{tabular}

$^a$ Color excess due to stellar extinction. 
The color excess due to nebular extinction, $E(B-V)_{\rm neb}$,
falls in the same range as $E(B-V)_{\rm s}$ (Section \ref{sec:dust_extinction}).
Calzetti's extinction law \citep{calzetti2000} is assumed.
$^\dagger$ LAEs at $z\sim 2-3$ with a Ly$\alpha$ luminosity near $L^*_{\rm Ly \alpha}$, $\simeq 10^{42}-10^{43}$ erg s$^{-1}$.
\end{table}

\subsection{Luminosity Function}
\label{sec:luminosity_function}

\begin{figure}[H]
\centering
\includegraphics[scale=.50]{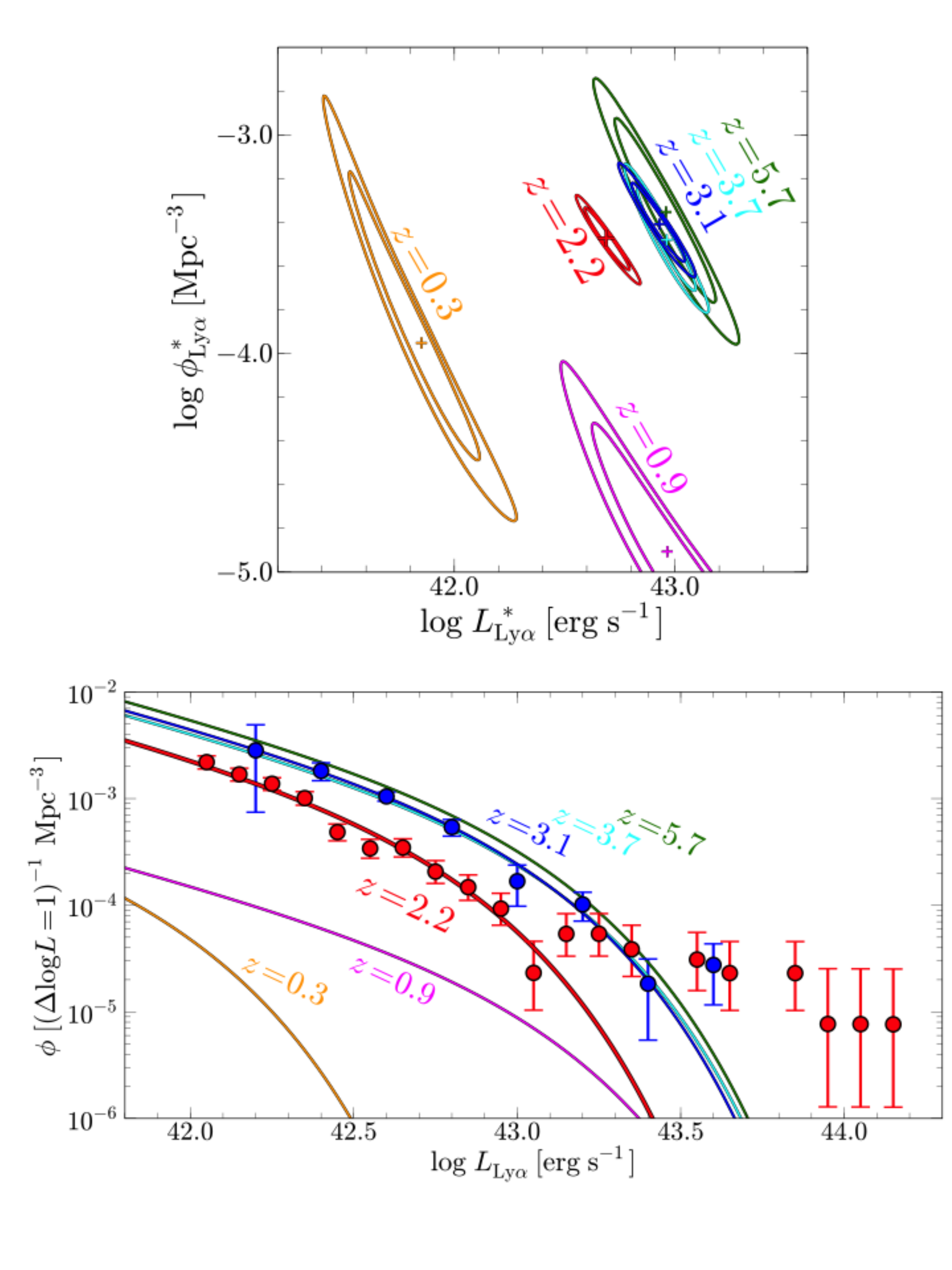}
\caption{
Top: Best-fit Schechter parameters and the error contours of Ly$\alpha$ luminosity functions at $z\sim 0-6$ \citep{konno2016}.
The pluses represent the best-fit values while the contours indicate their 
68\% and 90\% confidence levels. 
Redshift is coded by color: orange, $z=0.3$; magenta, 0.9; red, 2.2; blue, 3.1;
cyan, 3.7; green, 5.7.
Bottom: Ly$\alpha$ luminosity functions at $z\sim 0-6$ \citep{konno2016}. 
The curves indicate the best-fit Ly$\alpha$ luminosity functions, with the same color code
as the top panel.
The red and blue circles show the data points of the 
Ly$\alpha$ luminosity functions at $z=2.2$ and $3.1$, respectively.
This figure is reproduced by permission of the AAS.
}
\label{fig:konno2016_fig6}       
\end{figure}

The luminosity function and its evolution over time is one of the most fundamental 
properties for any galaxy population.
The Ly$\alpha$ luminosity function of LAEs has been derived at $z \sim 0-8$ 
by large survey programs (Section \ref{sec:discoveries}) 
since the discovery of LAEs in the late 1990s. The bottom and top panels of Figure \ref{fig:konno2016_fig6} 
present Ly$\alpha$ luminosity functions and their best-fit Schechter function parameters, respectively, 
from $z\sim 0$ to $6$,
where the Schechter function parameters are the characteristic Ly$\alpha$ luminosity $L_{\rm Ly\alpha}^*$ and 
the normalization $\phi_{\rm Ly\alpha}^*$ that determines the abundance
\footnote{
Ly$\alpha$ luminosity functions above $z\sim 6$ are discussed in 
the cosmic reionization section (Section \ref{sec:cosmic_reionizationI}).
}.
Two evolutionary trends are seen in Figure \ref{fig:konno2016_fig6}: 
a monotonic increase in the normalization from $z\sim 0$ to $3$ 
and no evolution in either the normalization or the shape over $z\sim 3-6$.
I explain details of these two trends in the following paragraphs.

\begin{figure}[h]
\centering
\includegraphics[scale=.40]{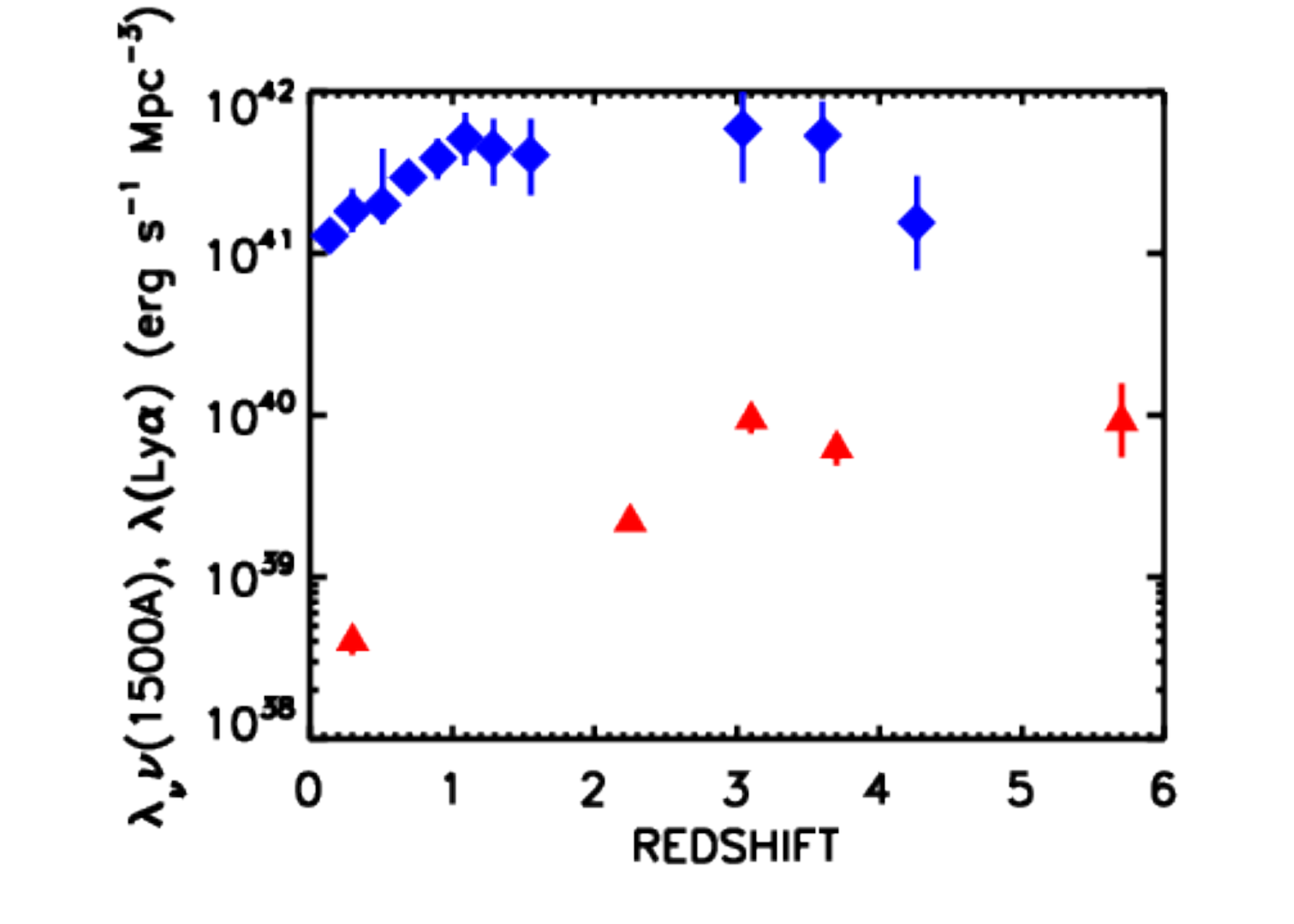}
\caption{
Ly$\alpha$-luminosity density (red triangles) and UV-continuum luminosity density (blue diamonds)
as a function of redshift \citep{cowie2010}.
This figure is reproduced by permission of the AAS.
}
\label{fig:cowie2010_fig30}       
\end{figure}

The first evolutionary trend is an increase found at $z\sim 0-3$ \citep{deharveng2008,cowie2010}. It is notable that 
the abundance of $z=0.3$ LAEs is very low with $\phi_{\rm Ly\alpha}^*=1\times 10^{-4}$ Mpc$^{-3}$, 
about $50$ times lower than that of $z\sim 0$ SDSS optical-continuum selected galaxies, 
$\phi^*=5\times 10^{-3}$ Mpc$^{-3}$ \citep{blanton2001},
meaning that LAEs are very rare in the local universe \citep{deharveng2008}.
The top panel of Figure \ref{fig:konno2016_fig6} suggests that over $z=0.3$ and $z=2.2$ 
the increase is
statistically more significant in Ly$\alpha$ luminosity ($L_{\rm Ly\alpha}^*$) 
than in the normalization ($\phi_{\rm Ly\alpha}^*$).
\footnote{
In this panel, the data for $z=0.9$ has a very low $\phi_{\rm Ly\alpha}^*$ value
that does not fall in the interpolation of the best-fit $\phi_{\rm Ly\alpha}^*$ values 
between $z=0.3$ and $z=2-3$. 
Although the Ly$\alpha$ luminosity function may have a truly very low $\phi_{\rm Ly\alpha}^*$ value 
at $z\sim 1$, there remains a possibility that the $z=0.9$ Ly$\alpha$ luminosity function could be 
biased toward a high $L_{\rm Ly\alpha}^*$, which gives
a low $\phi_{\rm Ly\alpha}^*$ value. In fact, this $z=0.9$ Ly$\alpha$ luminosity function is derived 
only with bright LAEs \citep{barger2012}. The result of the Ly$\alpha$ luminosity function at $z\sim 1$ 
is still under debate.
}
The evolution of the Ly$\alpha$ luminosity function is also quantified with the Ly$\alpha$ luminosity density,
\begin{equation}
\rho_{\rm Ly\alpha}= \int^{\infty}_{L_{\rm Ly\alpha}^{\rm lim}} L_{\rm Ly\alpha} \phi_{\rm Ly\alpha}(L_{\rm Ly\alpha}) dL_{\rm Ly\alpha},
\label{eq:lya_ld}
\end{equation}
where $L_{\rm Ly\alpha}$ and $L_{\rm Ly\alpha}^{\rm lim}$ are 
the Ly$\alpha$ luminosity and the limiting Ly$\alpha$ luminosity, respectively.
For reference, the UV continuum
\footnote{
The wavelength of the UV continuum is often chosen
at the far UV wavelength of $\sim 1500$\AA\ in the rest frame
that is longer than the Ly$\alpha$-line wavelength.
}
luminosity density $\rho_{\rm UV}$
is defined by
\begin{equation}
\rho_{\rm UV}= \int^{\infty}_{L_{\rm UV}^{\rm lim}} L_{\rm UV} \phi_{\rm UV}(L_{\rm UV}) dL_{\rm UV},
\label{eq:uv_ld}
\end{equation}
where $\phi_{\rm UV} (L_{\rm UV})$, $L_{\rm UV}$, and $L_{\rm UV}^{\rm lim}$
are the UV-continuum luminosity function, the UV-continuum luminosity, and the limiting UV-continuum luminosity, respectively.
Figure \ref{fig:cowie2010_fig30} compares the evolutions of Ly$\alpha$ and UV-continuum luminosity densities, 
where the latter is derived with UV-continuum selected galaxies \citep{tresse2007}.
Figure \ref{fig:cowie2010_fig30} clearly shows that, from $z\sim 0$ to $3$, 
the Ly$\alpha$ luminosity density increases by a factor of $\sim 20-30$, which is significantly faster
than the UV-continuum luminosity density evolution (a factor of $\sim 5-7$; \citealt{deharveng2008,cowie2010}). Similarly, in the
same redshift range, the Ly$\alpha$ luminosity density increases even faster than the cosmic SFR density 
(a factor of $\sim 10$) on the Madau-Lilly plot \citep{madau2014}, indicating that the Ly$\alpha$ luminosity density 
evolution cannot be explained by the cosmic SFR density evolution alone.




\begin{figure}[h]
\centering
\includegraphics[scale=.60]{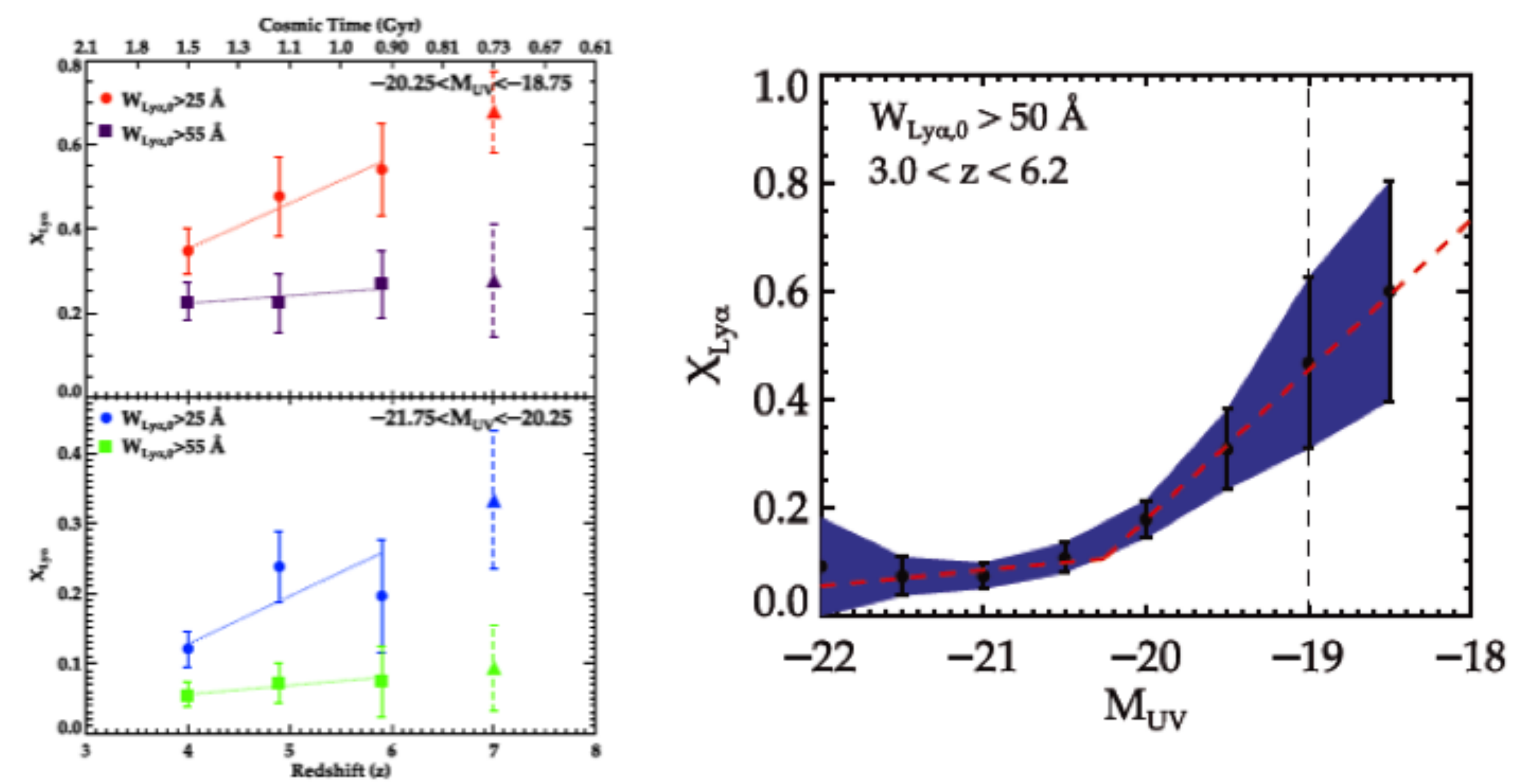}
\caption{
Left:  Ly$\alpha$ emitting galaxy fraction $x_{\rm Ly\alpha}$ as a function of redshift \citep{stark2011}.
UV-continuum faint ($-20.25<M_{\rm UV}<-18.75$) and bright ($-21.75<M_{\rm UV}<-20.25$)
galaxy results are shown in the top and the bottom panels. The red and blue circles (lines) 
represent the data points (the best-fit power law functions) for $EW_0 > 25$ \AA\ galaxies 
with faint and bright UV-continuum magnitudes, respectively, 
while the purple and green squares (lines) are 
for $EW_0 > 55$ \AA\ galaxies. The triangles with dashed error bars at $z \sim 7$ 
are predictions from the best-fit functions.
Right: $x_{\rm Ly\alpha}$ as a function of UV-continuum magnitude 
for $EW_0 > 50$ \AA\ galaxies at $z=3-6.2$ \citep{stark2010}.
The black circles represent measurements, while the red dashed line
corresponds to the best-fit first-order polynomial to the data points.
This figure is reproduced by permission of the AAS.
}
\label{fig:stark2011_fig2_stark2010_fig13}       
\end{figure}

\begin{figure}[h]
\centering
\includegraphics[scale=.50]{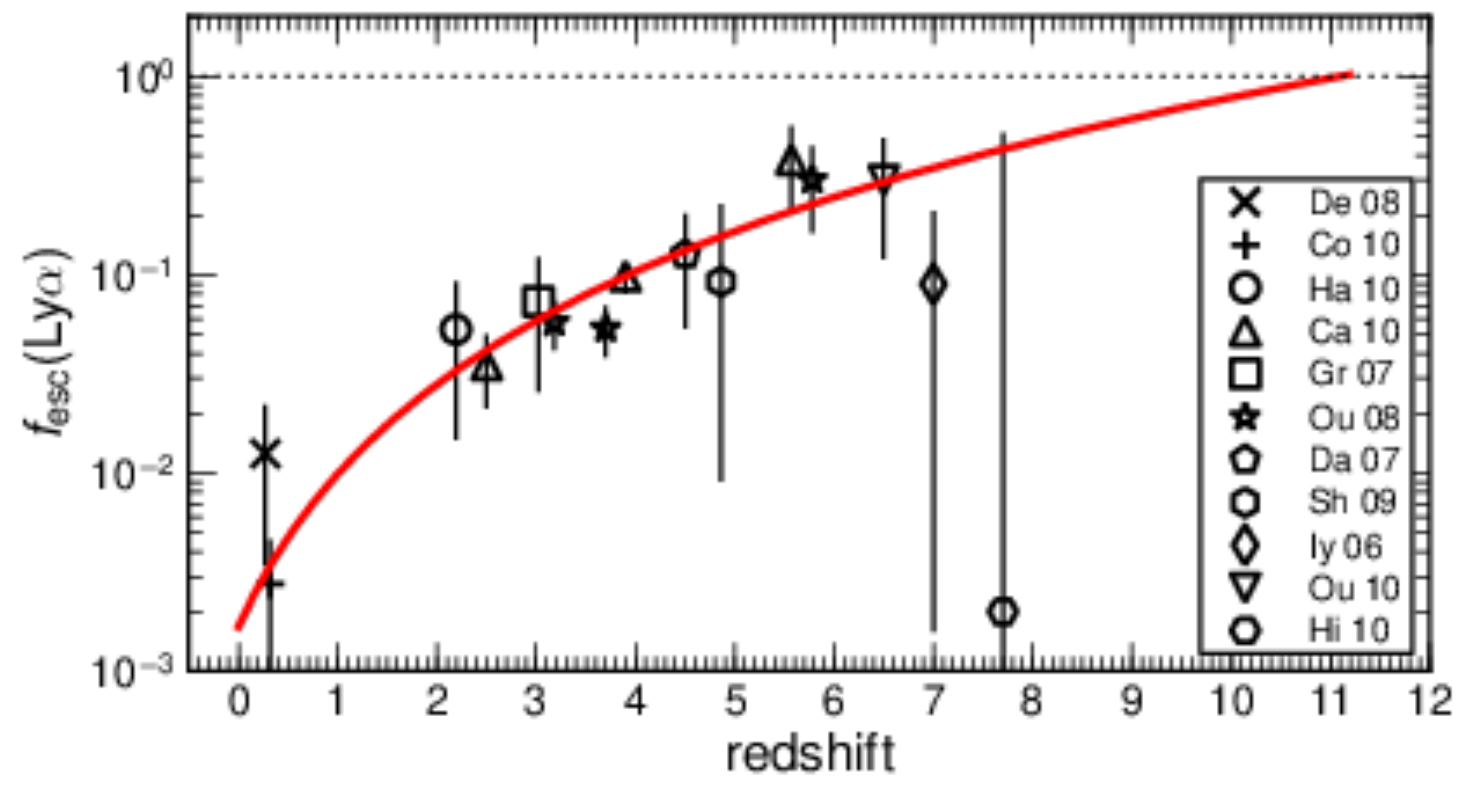}
\caption{
Average Ly$\alpha$ escape fraction as a function of redshift \citep{hayes2011a}.
The black symbols indicate average Ly$\alpha$ escape fraction estimates
based on various observational results while 
the red line is the best-fit power law function to them.
This figure is reproduced by permission of the AAS.
}
\label{fig:hayes2011a_fig1}       
\end{figure}

\begin{figure}[h]
\centering
\includegraphics[scale=.50]{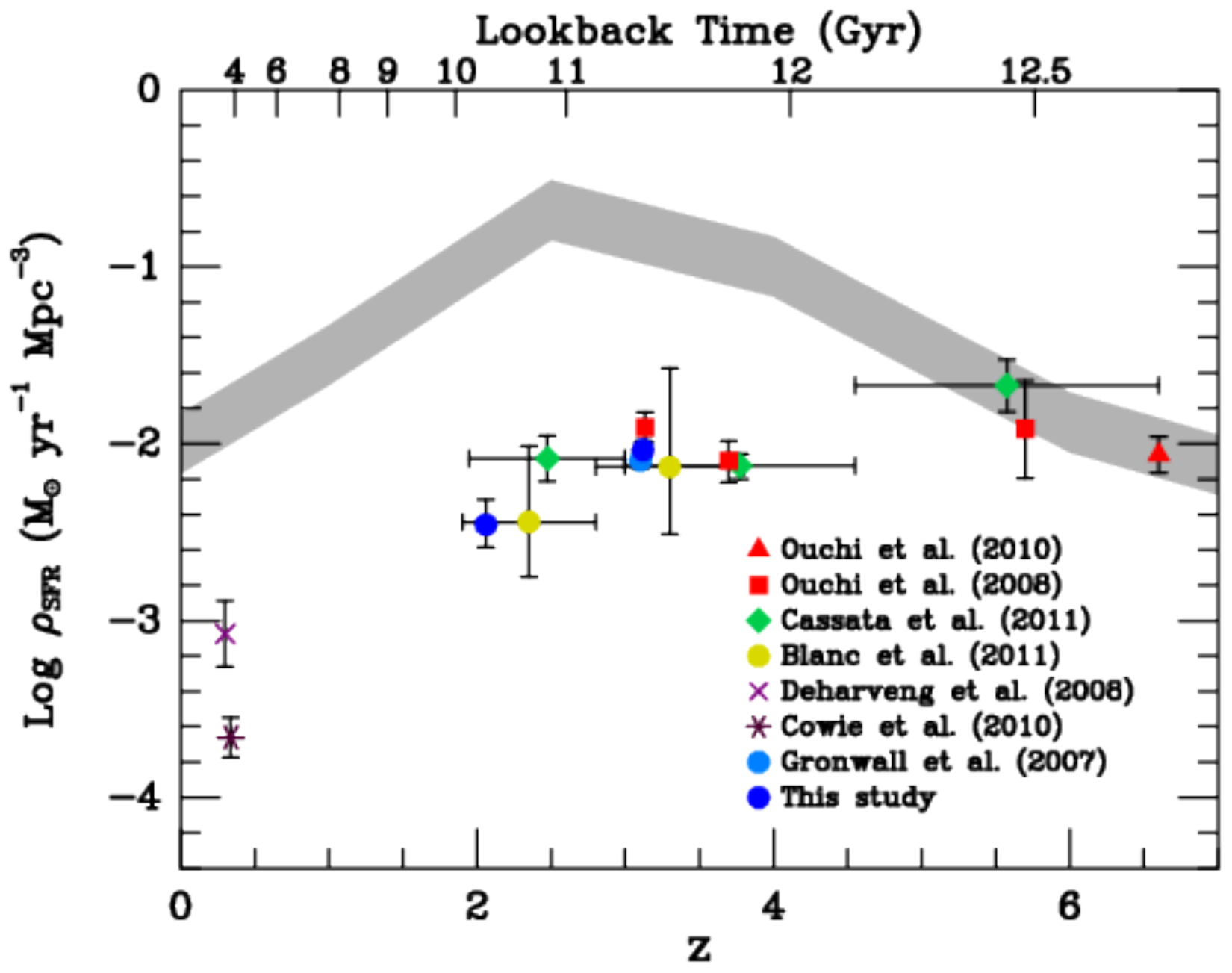}
\vspace{1.2cm}
\caption{
Cosmic SFR density as a function of redshift \citep{ciardullo2012}.
The data points with error bars represent the LAEs' contribution 
to the cosmic SFR density, while the gray region indicates the total
cosmic SFR density estimated with LBGs corrected for dust extinction.
This figure is reproduced by permission of the AAS.
}
\label{fig:ciardullo2012_fig12}       
\end{figure}

The second evolutionary trend is that the Ly$\alpha$ luminosity function is nearly
constant over $z\sim 3-6$ \citep{ouchi2008}. In contrast,
the UV-continuum luminosity function of UV continuum-selected LBGs decreases from $z\sim 3$ to $6$ and beyond, 
indicating that Ly$\alpha$ emitting galaxies dominate in number more at $z\sim 6$ than at $z\sim 3$ \citep{ouchi2008}. Indeed, 
deep spectroscopic surveys for UV-continuum selected LBGs suggest that the Ly$\alpha$ emitting (EW$_0 >25$\AA) 
galaxy fraction $x_{\rm Ly\alpha}$ increases from $x_{\rm Ly\alpha}\sim 0.3$ to $0.5$ over $z=4-6$ 
for galaxies with an absolute UV magnitude $M_{\rm UV}$ range of 
$-20.25<M_{\rm UV}<-18.75$ corresponding to $\lesssim L_{\rm UV}^*$ 
(the left panel of Figure \ref{fig:stark2011_fig2_stark2010_fig13}). In other words, about a half of $\lesssim L_{\rm UV}^*$-LBGs 
at $z\sim 6$ are LAEs with EW$_0>25$ \AA.
This $x_{\rm Ly\alpha}$ evolution result indicates an increase with redshift 
in either the fraction of Ly$\alpha$ emitting galaxies or the Ly$\alpha$ luminosity of galaxies or both.
Allowing both $\phi_{\rm Ly\alpha}^*$ and $L_{\rm Ly\alpha}^*$ to evolve,
one can derive the number- and luminosity-weighted average
Ly$\alpha$ escape fraction $\left < f_{\rm esc}^{\rm Ly\alpha} \right >$ from the
Ly$\alpha$ luminosity density (Equation \ref{eq:lya_ld}) as:
\begin{equation}
\left < f_{\rm esc}^{\rm Ly\alpha} \right > = \rho_{\rm Ly\alpha}/\rho_{\rm Ly\alpha}^{\rm int},
\label{eq:fesc_lya_avg}
\end{equation}
where $\rho_{\rm Ly\alpha}^{\rm int}$ is the intrinsic Ly$\alpha$ luminosity density
expected from the cosmic SFR density $\Psi_{\rm SFR}$ (e.g. \citealt{madau2014}).
The intrinsic Ly$\alpha$ luminosity density can be estimated by
$\rho_{\rm Ly\alpha}^{\rm int}$ [erg s$^{-1}$ Mpc$^{-3}$] $=1.1\times 10^{42} \Psi_{\rm SFR}$ [$M_\odot$ yr$^{-1}$ Mpc$^{-3}$]
under the assumption of the case B recombination ($L_{\rm Ly\alpha}/L_{\rm H \alpha}=8.7$; \citealt{brocklehurst1971}) 
and the \ha\ luminosity $L_{\rm H \alpha}$-SFR relation of \citet{kennicutt1998a}.
Figure \ref{fig:hayes2011a_fig1} presents $\left < f_{\rm esc}^{\rm Ly\alpha} \right >$ as a function of redshift \citep{hayes2011a},
and indicates a monotonic increase in $\left < f_{\rm esc}^{\rm Ly\alpha} \right >$ from $z\sim 0$ to $6$.

Here I address the issue whether all of these observational results at $z\sim 3-6$ are self-consistent. 
Observations of UV-continuum selected galaxies 
show that faint UV-continuum galaxies have a higher chance of
emitting strong Ly$\alpha$ in this redshift range (right panel of Figure \ref{fig:stark2011_fig2_stark2010_fig13}; 
\citealt{ando2006,stark2011}). In other words,
a majority of LAEs are faint UV-continuum galaxies. Although the abundance of 
bright ($>L^*$) UV-continuum galaxies drops significantly, the abundance of faint UV-continuum galaxies does not largely
decrease towards high-$z$, due to a steepening of the luminosity function slope 
$\alpha$ \citep{bouwens2015}. Because the abundance of LAEs is linked to the one of faint UV-continuum galaxies,
the Ly$\alpha$ luminosity function of LAEs does not evolve largely over $z\sim 3-6$.
In this way, all observational results suggest a self-consistent physical picture.


Because LAEs become a more dominant population at $z\sim 6$ than at $z\sim 3$,
LAEs contribute much to the cosmic SFR density at $z\sim 6$.
Figure \ref{fig:ciardullo2012_fig12} presents the evolution of the cosmic SFR density \citep{ciardullo2012}.
The contribution of LAEs is only $1/10$ of the total cosmic SFR density at $z\sim 3$ 
while it becomes the whole of it at $z\sim 6$.
%
If this trend continues at $z\gtrsim 6$ (i.e. the EoR),
LAEs may be a major population that emit ionizing photons for cosmic reionization.
Thus, it is probably important to study LAEs to understand the physical properties of ionizing sources for cosmic reionization,
although a large fraction of Ly$\alpha$ photons from LAEs (galaxies with {\it intrinsically} strong Ly$\alpha$ emission)
may not reach observers due to absorption
by neutral hydrogen in the IGM at the EoR  
(see Section \ref{sec:cosmic_reionizationII} for details of reionization sources).

An interesting approach to estimate the cosmic SFR density has been proposed 
by \citet{croft2016}. It is based on the so-called intensity mapping technique, 
and consists in determining the power spectrum of diffuse Ly$\alpha$ emission from star-forming galaxies 
that are too faint to be detected individually, but numerous enough to yield a significant signal 
in a statistical sense.
%
%
\citet{croft2016} have found that the cosmic SFR density at $z=2-3.5$ estimated from diffuse Ly$\alpha$ emission is 
about 30 times higher than those by \citet{ciardullo2012} and comparable to 
(or higher than) the dust-extinction corrected total cosmic SFR density. Because some amount of Ly$\alpha$ emission 
should be absorbed by dust, this result may be overestimating the true cosmic SFR density. 
Although being a powerful important technique,
intensity mapping requires a very careful evaluation of systematics.
Recently, \citet{croft2018} have updated the analysis with the systematics removals,
reducing the intensity measurement of the diffuse Ly$\alpha$ emission by a factor of 2.
\citet{croft2018} find that there is no correlation between the diffuse Ly$\alpha$ emission
and the Ly$\alpha$ forest, and
show that the diffuse Ly$\alpha$ emission 
is not explained by faint star-forming galaxies, 
but fluorescence Ly$\alpha$ emission  
around QSOs in a scale up to $15 h^{-1}$ Mpc.



\subsection{Morphology}
\label{sec:morphology}

\begin{figure}[H]
\centering
\includegraphics[scale=.60]{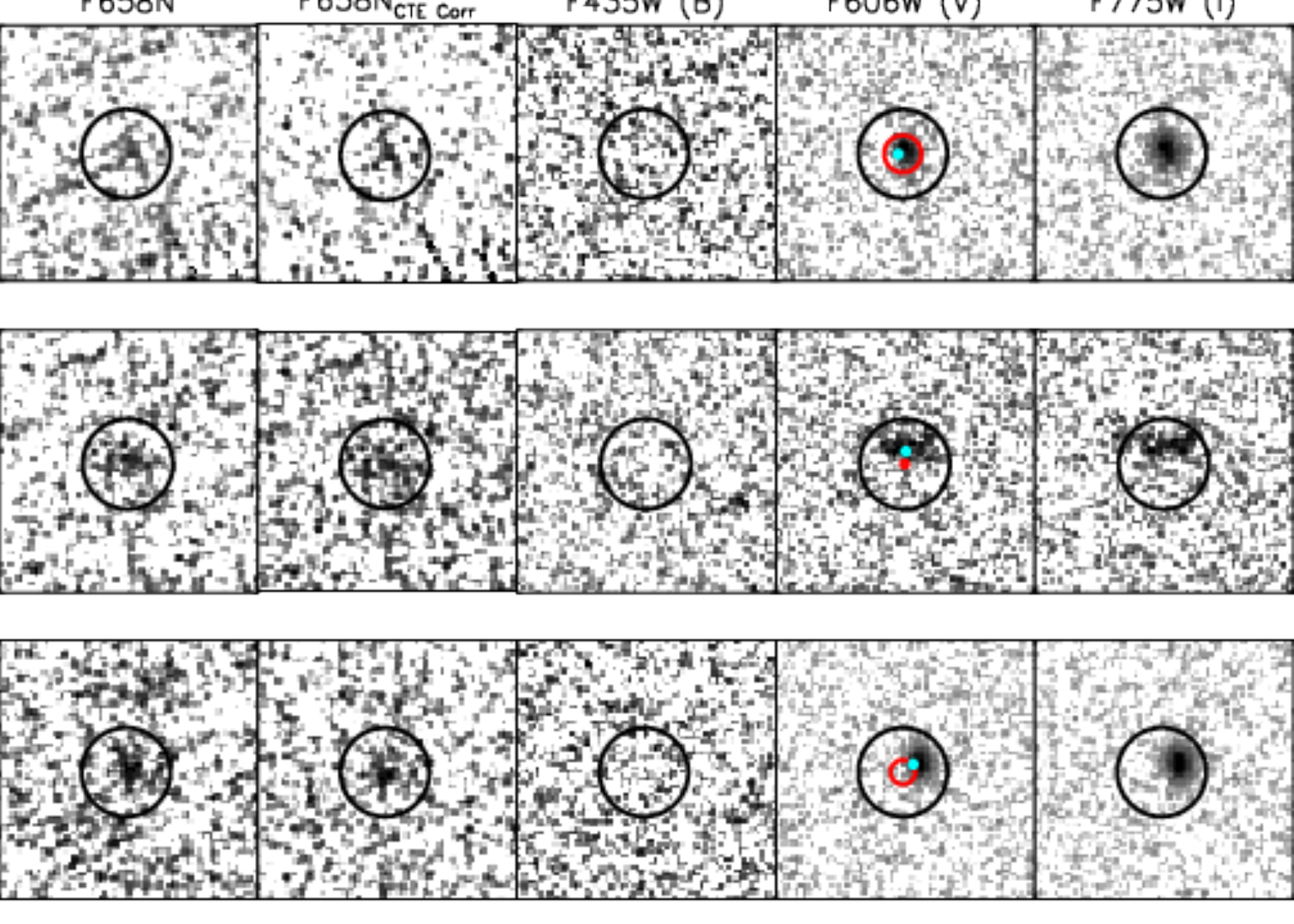}
\caption{
HST images of three LAEs dubbed CHa-2, CH8-1, and CH8-2 at $z=4.4$ \citep{finkelstein2011b}:
 from left to right,
narrowband $F658N$, narrowband $F658N$ corrected for the charge-transfer effect,
$B_{435}$, $V_{606}$, and $i_{775}$. The $F658N$ band includes the redshifted Ly$\alpha$ emission
of the LAEs. The centers of the black circles indicate the centroids of the LAEs in the $F658N$ band.
See \citet{finkelstein2011b} for more details.
This figure is reproduced by permission of the AAS.
}
\label{fig:finkelstein2011_fig4}       
\end{figure}

\begin{figure}[H]
\centering
\includegraphics[scale=.45]{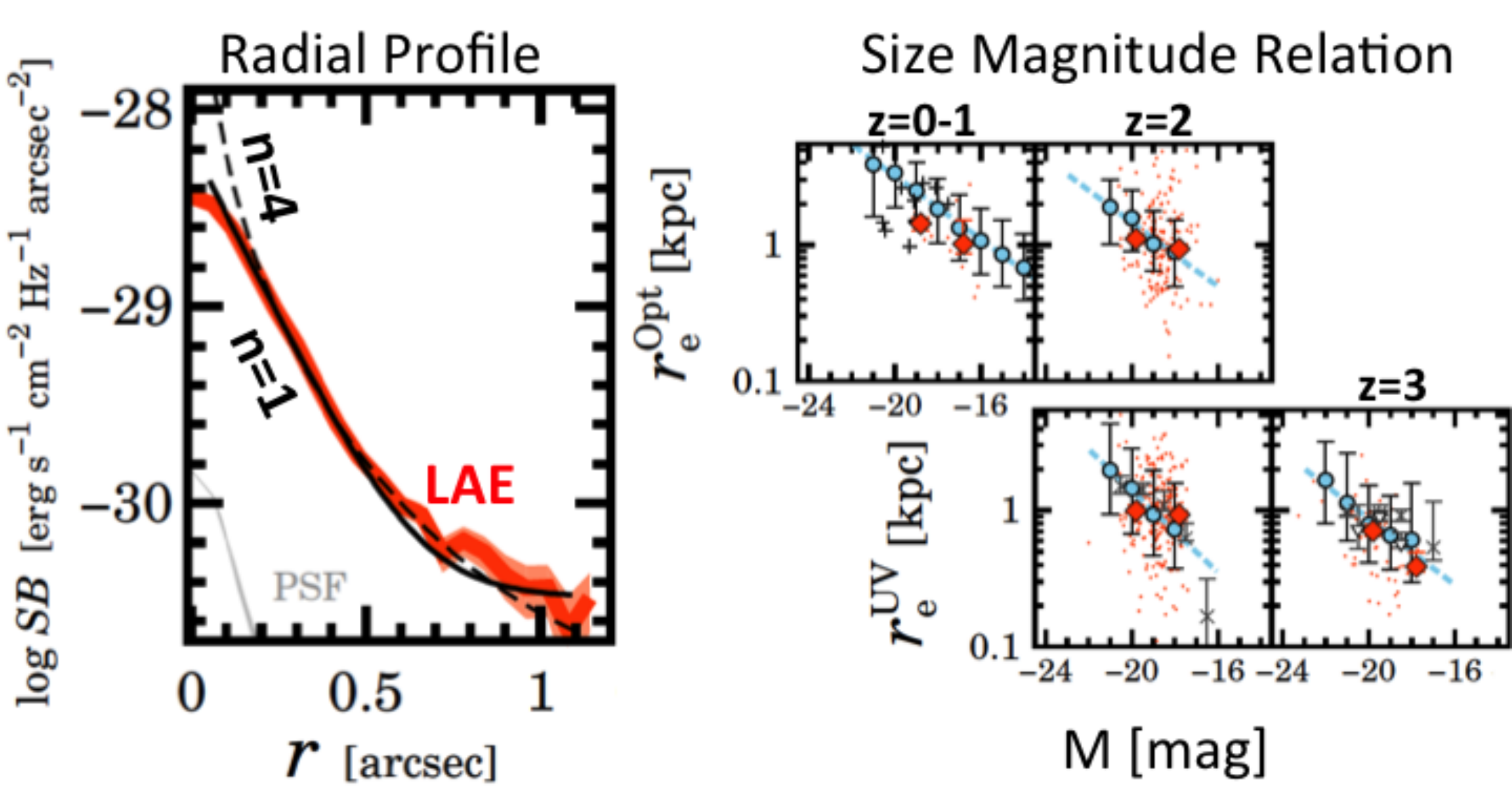}
\caption{ 
Left: Radial profile of UV-continuum surface brightness obtained
with the HST images \citep{shibuya2018c}.
The red shade represents the radial profile of the composite LAE
with the $1 \sigma$ uncertainty. The composite LAE consists of LAEs
with $0.12-1 L^*_{\rm z=3}$ at $z=0-7$, where $L^*_{\rm z=3}$ is 
the characteristic luminosity of the Schechter function
for the UV luminosity function at $z=3$.
The black solid and dashed lines indicate the S\'ersic profiles with $n=1$ and $4$,
respectively. The gray line denotes the radial profile of the point-spread function (PSF).
Right: Size-magnitude ($r_{\rm e}$-magnitude) relation \citep{shibuya2018c}.
The top two panels show the size-magnitude relations in the rest-frame optical wavelength,
while the bottom two panels present those in the rest-frame UV wavelength.
The red and cyan circles represent LAEs and continuum-selected star-forming galaxies, respectively,
where the error bars indicate the 16th- and 84th-percentiles of the $r_{\rm e}$ distribution.
The red dots indicate the size and magnitude measurements 
of individual LAEs.
The cyan dotted lines are the power law functions best fit to the data points of the cyan circles.
The gray symbols represent the measurements for LAEs obtained in the other studies.
This figure is reproduced by permission of the AAS.
}
\label{fig:shibuya2018c_fig5_fig7}
\end{figure}

It has been known that LAEs are generally very compact 
since first revealed by HST in the mid 90's (\citealt{pascarelle1996a,pascarelle1996b}; Figure \ref{fig:finkelstein2011_fig4}).
Deep HST images reveal small effective radii, $r_{\rm e}$, in rest-frame UV and optical continua, 
$\sim 1$ kpc
on average \citep{malhotra2012,paulino-afonso2018,shibuya2018c}.
The radial profiles of the rest-frame UV and optical continua typically show a disk morphology 
with a S\'ersic index $n$ of $n\simeq 1$, and follow the $r_{\rm e}$-magnitude relation similar to the one of LBGs, 
indicating that faint continuum LAEs have a small size in $r_{\rm e}$ (Figure \ref{fig:shibuya2018c_fig5_fig7}; \citealt{paulino-afonso2018,shibuya2018c}).
Because a majority of LAEs have a faint continuum, 
the $r_{\rm e}$-magnitude relation can explain the compact morphologies of LAEs.

\begin{figure}[H]
\centering
\includegraphics[scale=1.10]{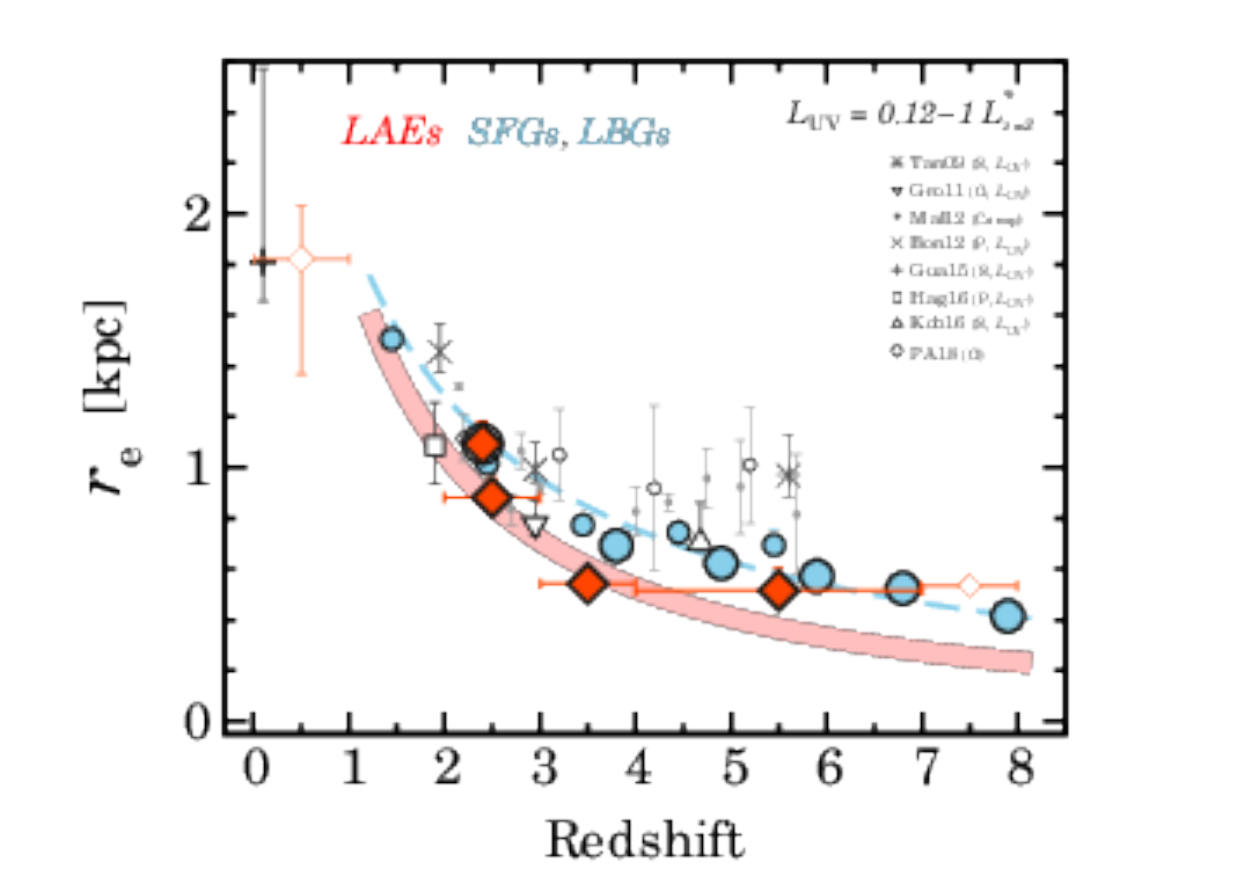}
\caption{
Effective radius as a function of redshift for LAEs and LBGs with
the same UV-continuum luminosity of $0.12-1\ L^*_{\rm z=3}$ \citep{shibuya2018c}.
Here, $L^*_{\rm z=3}$ is the characteristic luminosity of the Schechter function
for the UV luminosity function at $z=3$.
The red filled diamonds (with the black circle) indicate
LAEs in the rest-frame UV (optical) wavelength.
The red open diamond at $z\sim 0.5$ ($z\sim 7.5$) represents
the effective radius in the rest-frame optical (UV) wavelength
estimated from the extrapolation of the size-magnitude relation
(the small sample consisting of 3 LAEs).
The small and large cyan circles denote the effective radii of galaxies selected by
photo-$z$ and Lyman break techniques, respectively.
The red-broad and cyan-dashed curves show the power-law functions of $(1+z)^{\beta}$
for $\beta \simeq -1$ that are best-fit to the data points of the red filled diamonds and 
the cyan filled circles, respectively. The gray and black symbols present the effective radii
obtained by various studies whose magnitude limits are different.
%
This figure is reproduced by permission of the AAS.
}
\label{fig:shibuya2018c_fig8}
\end{figure}

Although previous HST studies claim no redshift evolution of $r_{\rm e}$ on average, a recent HST study \citep{shibuya2018c}
finds that the no-redshift evolution results may be produced by the sample selection bias. 
If a sample selection is not controlled, one can identify more LAEs with a faint continuum at low redshift. 
Because LAEs with a faint continuum have a $r_{\rm e}$ value smaller than LAEs with a bright continuum 
due to the $r_{\rm e}$-magnitude relation, 
$r_{\rm e}$ measurements of low-redshift LAEs are typically small, diminishing the trend of the redshift evolution.
Figure \ref{fig:shibuya2018c_fig8} shows the median $r_{\rm e}$ value as a function of redshift
that is obtained with the controlled samples whose LAEs fall in the same continuum luminosity range \citep{shibuya2018c}.
The median $r_{\rm e}$ value monotonically decreases as $\sim (1+z)^{-1}$
for a given continuum luminosity \citep{shibuya2018c}. 
This evolutionary trend of LAEs is similar to the one of LBGs
(e.g. \citealt{shibuya2015}).

%

The compact morphology of LAEs is not only found in continua, 
but also in Ly$\alpha$ emission by HST narrowband imaging studies \citep{bond2010,finkelstein2011b}.
It should be noted that deeper narrowband-imaging and spectroscopic observations identify very diffuse 
extended ($\gtrsim 10$ kpc) Ly$\alpha$ halos around LAEs that are detailed in Section \ref{sec:galaxy_formationIII}
\citep{hayashino2004,steidel2011,patricio2016,wisotzki2016}. 
A combination of these observational studies indicates that
the spatial structure of Ly$\alpha$ emission of LAEs 
is composed of a peaky Ly$\alpha$ core and a diffuse Ly$\alpha$ halo (see also \citealt{leclercq2017})



\subsection{ISM Properties}
\label{sec:ISM_state}

\begin{figure}[H]
\centering
\includegraphics[scale=.35]{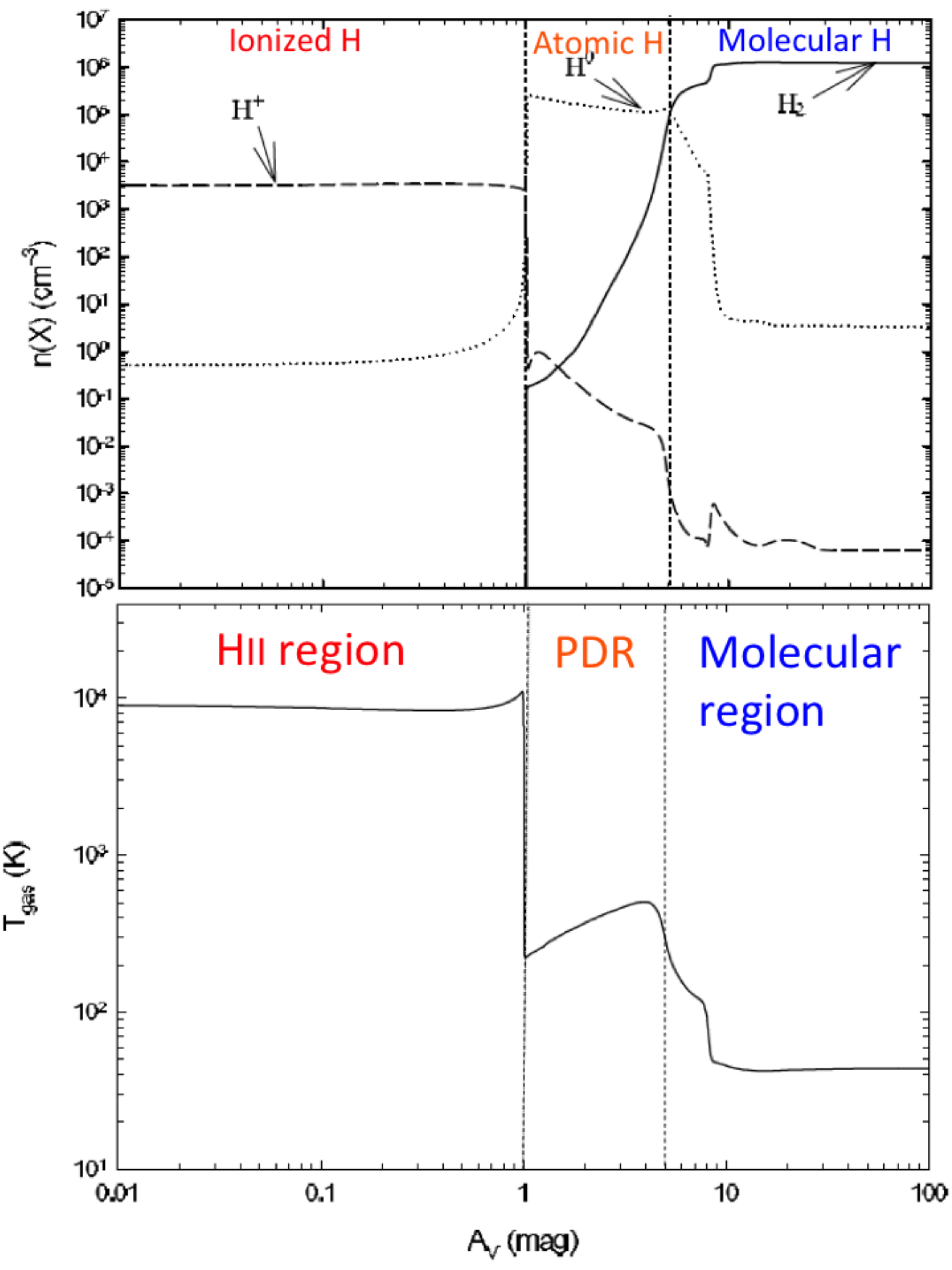}
\caption{
Top: Hydrogen density as a function of visual extinction $A_{\rm V}$
that corresponds to the illuminated face (small $A_{\rm V}$) to the shielded face (large $A_{\rm V}$; \citealt{abel2005}).
The dashed, dotted, and solid curves represent the model predictions of the ionized, atomic, and molecular
hydrogen densities, respectively. The two vertical lines indicate the hydrogen ionization front and 
the half-density molecular hydrogen point.
Bottom: Predicted gas temperature as a function of $A_{\rm V}$.
This figure is reproduced by permission of the AAS.
}
\label{fig:abel2005_fig3_fig8}
\end{figure}

Hydrogen in the ISM has three gas phases: 
H$^+$ ions, H atoms, and H$_2$ molecules.
Corresponding to these three phases,
the ISM is classified into three regions:
\hii\ regions (H$^+$), photodissociation regions (PDR; H), and molecular regions (H$_2$)
whose gas temperatures are $\sim 10^4$, $10^2-10^3$, and $10^1-10^2$ K,
respectively (Figure \ref{fig:abel2005_fig3_fig8}).

In local galaxies, most of the ISM is 
in PDRs, where UV radiation from stars photodissociates molecules. 
However, PDRs have large spatial variations of 
temperature and density (Figure \ref{fig:abel2005_fig3_fig8}), 
making it difficult to understand PDR properties by simple modeling. 
Moreover, there are many atomic and molecular transitions in
PDRs as well as in molecular regions. 
In contrast, \hii\ regions are moderately homogeneous media 
with a small number of ionization transitions, 
and thus can be modeled more simply than PDRs and molecular regions.
It should also be noted that \hii\ regions 
radiate emission lines falling in optical wavelengths 
where ground-based deep spectroscopy is possible.
These emission lines enable us to constrain physical parameters of \hii\ regions
such as gas-phase metallicity, electron temperature $T_{\rm e}$, ionization parameter $q_{\rm ion}$, 
and electron density $n_{\rm e}$. 

Although it is difficult to characterize ISM properties of LAEs that are generally faint,
recent LAE observations have constrained the gas-phase metallicity and ionization parameter 
in \hii\ regions. Moreover, there are some useful observations to constrain parameters of 
atomic gas and dust mainly found in PDRs and molecular regions.
Below I explain observational results as well as the methods used to probe the ISM properties.



\subsubsection{Gas-Phase Metallicity}
\label{sec:gas_phase_metallicity}

One of the most important ISM quantities that characterize galaxies 
is the gas-phase metallicity of \hii\ regions.
%
The metallicity of a galaxy is estimated 
from the ratio of appropriate lines using photoionization models.
Depending on the strength of the lines used,
there are two methods: the direct $T_{\rm e}$ method and the strong emission line method.
Note that all of the line ratios discussed below are corrected for dust extinction.\\

\noindent
{\bf Direct $T_{\rm e}$ Method} 

The direct $T_{\rm e}$ method mainly uses weak lines sensitive to electron temperature, 
\oiii]1661,1666, [\oiii]4363, [\nii]5755, and [\oii]7320,7330 etc. 
Because the most popular line among these 
is the auroral [\oiii]4363 line, below I explain the direct $T_{\rm e}$ method with this line.

\begin{figure}[h]
\centering
\includegraphics[scale=.50]{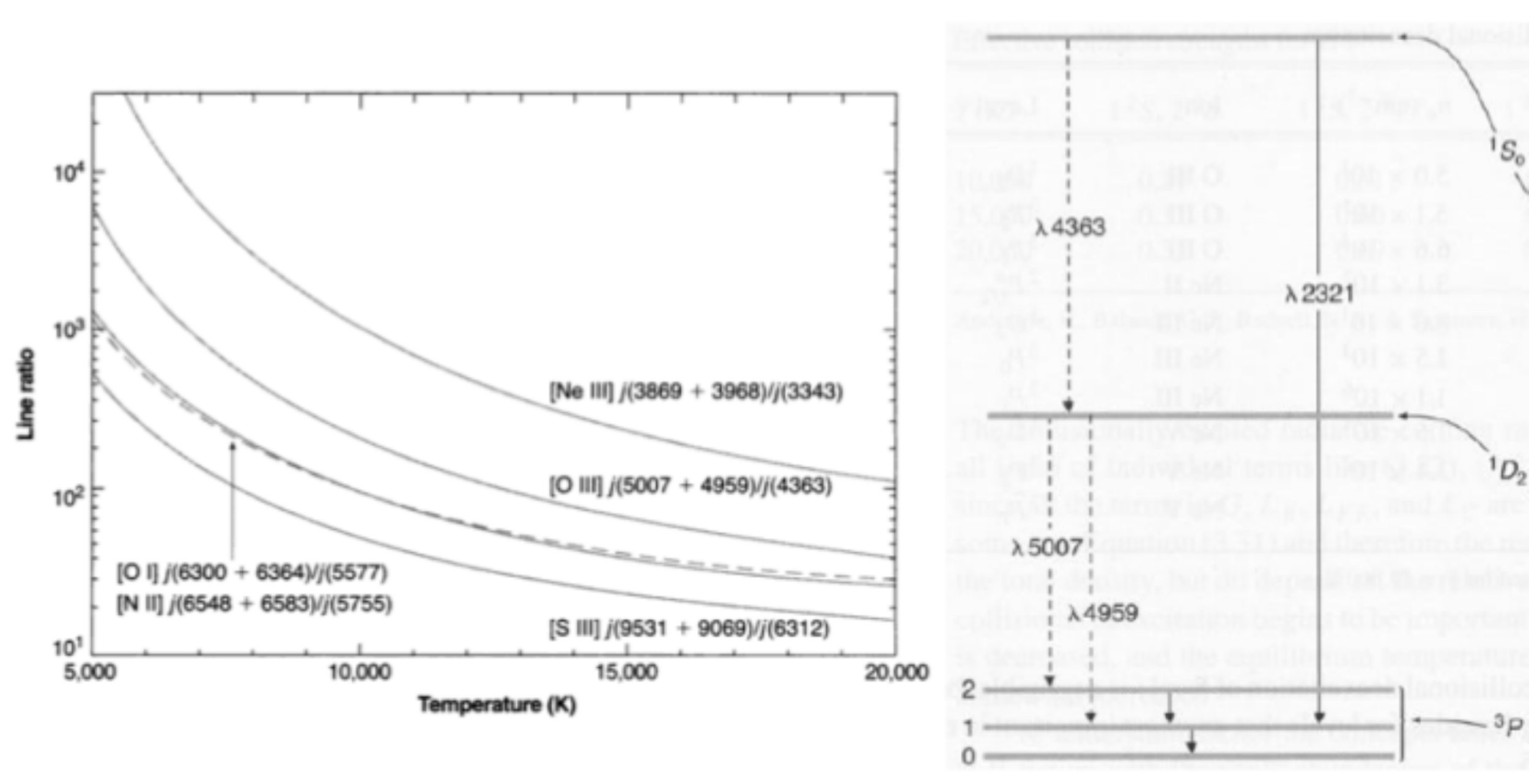}
\caption{
Left: Line ratios as a function of electron temperature \citep{osterbrock1989}.
The second top curve indicates the line ratio $(f_{[{\rm OIII}]4959}+f_{[{\rm OIII}]5007})/f_{[{\rm OIII}]4363}$,
while the other curves show line ratios that are not discussed in the text.
Right: Schematic diagram presenting the quantum states $^1$S$_0$, $^1$D$_2$, and $^3$P of an ${\rm O}^{2+}$ ion 
and emission lines produced by transitions between two states \citep{osterbrock1989}.
The solid and dashed-line arrows represent permitted and forbidden lines, respectively.
This figure is reproduced by permission of the University Science Books.
}
\label{fig:osterbrock1989_fig}
\end{figure}

One can estimate $T_{\rm e}$ from [\oiii]4363 and [\oiii]4959,5007 line fluxes using the following equation:
\begin{equation}
 (f_{[{\rm OIII}]4959}+f_{[{\rm OIII}]5007})/f_{[{\rm OIII}]4363} = \frac{7.90 \exp(3.29 \times 10^4/T_{\rm e})}{1+4.5\times 10^{-4} n_{\rm e}/T_{\rm e}^{1/2}},
\label{eq:electron_temperature}
\end{equation}
with a small uncertainty depending on $n_{\rm e}$ (left panel of Figure \ref{fig:osterbrock1989_fig}; \citealt{osterbrock1989}).
The $T_{\rm e}$ value is determined by the ratio $(f_{[{\rm OIII}]4959}+f_{[{\rm OIII}]5007})/f_{[{\rm OIII}]4363}$,
%
%
because the [\oiii]4363 ([\oiii]4959,5007) flux increases (decreases) 
when the rate of collisional excitation (de-excitation) 
from $^1$D$_2$ to $^1$S$_0$ (from $^1$D$_2$ to $^3$P) increases 
with $T_{\rm e}$ (right panel of Figure \ref{fig:osterbrock1989_fig})
%
%
%
%
\footnote{
Note that electrons stay at $^1$D$_2$
significantly longer than $^3$P.
}.
Because metals are major coolants of the gas in \hii\ regions, 
$T_{\rm e}$ is primarily determined by metallicity.
%
Therefore, if $T_{\rm e}$ is estimated, one can reliably derive the metallicity
based on photoionization models \citep{izotov2006},
\begin{eqnarray}
\label{eq:electron_temperature_main_oii}
12 + \log \frac{{\rm O}^+}{{\rm H}^+} & = & \log \frac{f_{[{\rm OII}]3727}}{f_{{\rm H}\beta}} + 5.961 + \frac{1.676}{t_2} \nonumber \\
 & & -0.40\log t_2 - 0.034t_2 + \log(1+1.35x), \\
\label{eq:electron_temperature_main_oiii}
12 + \log \frac{{\rm O}^{2+}}{{\rm H}^+} & = & \log \frac{f_{[{\rm OIII}]4959}+f_{[{\rm OIII}]5007}}{f_{{\rm H}\beta}} + 6.200 + \frac{1.251}{t_3} \nonumber \\
 &  & -0.55 \log t_3 - 0.014t_3, 
\end{eqnarray}
where 
\begin{eqnarray}
\label{eq:electron_temperature_t2}
t_2=10^{-4} T_{\rm e}{\rm [OII]}, \\
\label{eq:electron_temperature_t3}
t_3=10^{-4} T_{\rm e}{\rm [OIII]}, \\
\label{eq:electron_temperature_x}
x=10^{-4} n_{\rm e} t_2^{-0.5}, 
\end{eqnarray}
and ${\rm O}^+$, ${\rm O}^{2+}$, and ${\rm H}^+$ are
the abundances of singly-ionized oxygen, doubly-ionized oxygen, and
ionized hydrogen, respectively;
$T_{\rm e}$[\oii] and $T_{\rm e}$[\oiii] are the electron temperatures
in ${\rm O}^+$ and ${\rm O}^{2+}$ ion gas.
For simplicity, one can assume the relation \citep{campbell1986,garnett1992}
\begin{equation}
t_2 = 0.7 t_3 + 0.3
\label{eq:electron_temperature_t2t3}
\end{equation}
that generally does not change the metallicity estimate.
%
%
Because the last term of eq. (\ref{eq:electron_temperature_main_oii}),
$\log(1+1.35x)$,
is negligibly small,
this term can be practically omitted.

The oxygen abundance is calculated with
\begin{equation}
\frac{{\rm O}}{{\rm H}} =  \frac{{\rm O}^+}{{\rm H}^+}  + \frac{{\rm O}^{2+}}{{\rm H}^+}.
\label{eq:oxygen_abundance}
\end{equation}
It should be noted that the contribution of ${\rm O}^{3+}$ and higher-order ionized oxygen
is negligibly small, only $<1$\%, in \hii\ regions heated by stars.\\

%

\noindent
{\bf Strong Emission Line Method} 

\begin{figure}[h]
\centering
\includegraphics[scale=.32]{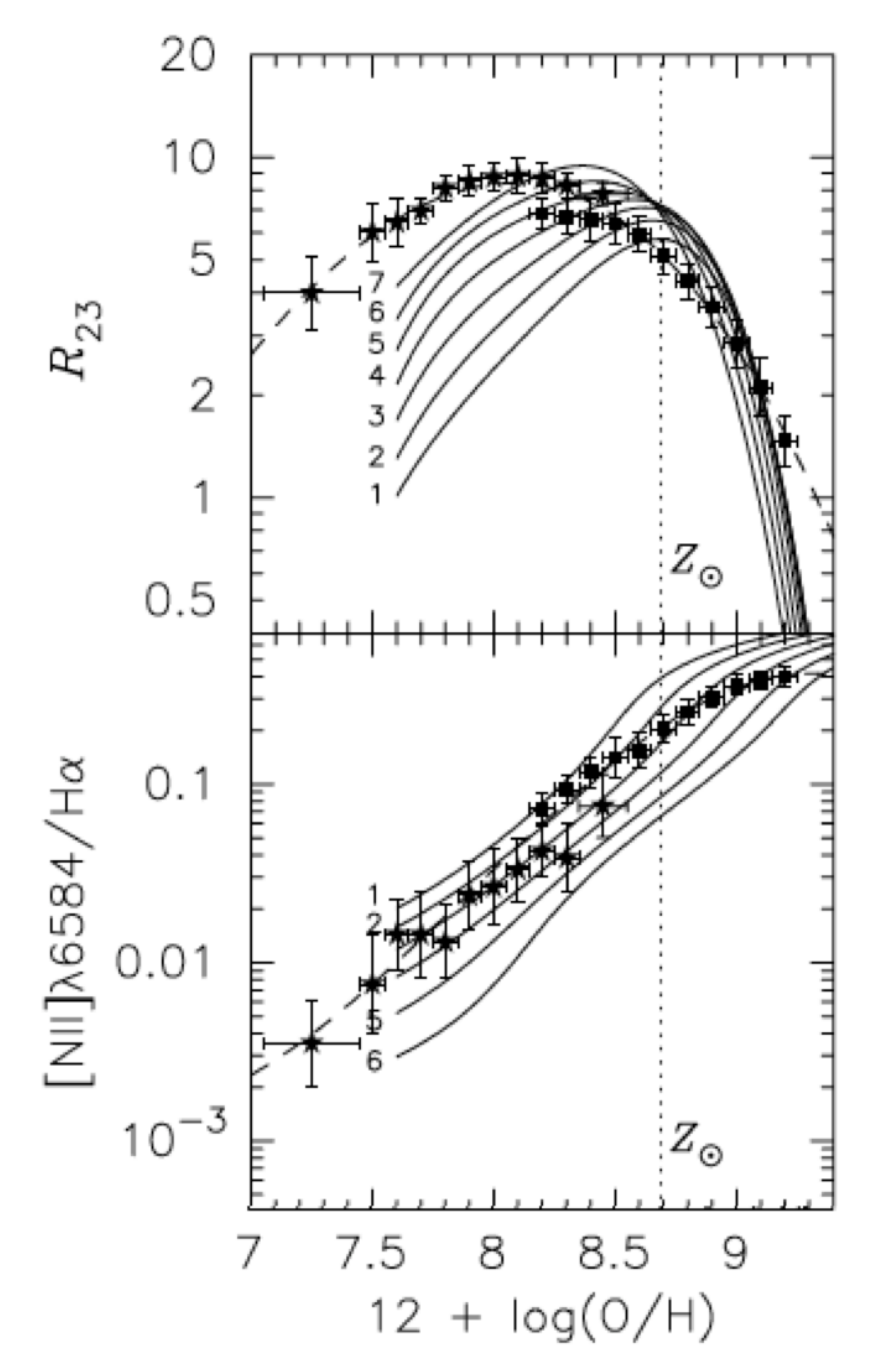}
\caption{
$R23$ ($N2$) index as a function of oxygen abundance
presented in the top (bottom) panel \citep{nagao2006}. The star marks denote
local galaxy averages obtained by the direct $T_{\rm e}$ method, 
and the dashed line represents an empirical fit to them.
The solid lines show photoionization model calculation results with
the normalized ionization parameter, $U\equiv q_{\rm ion}/c$, of $\log U =-3.8$, $-3.5$, $-3.2$, $-2.9$, $-2.6$, $-2.3$, and $-2.0$
from bottom to top (top to bottom) in the top (bottom) panel.
The dotted lines indicate solar metallicity.
This figure is reproduced by permission of the Astronomy \& Astrophysics journal.
}
\label{fig:nagao2006_fig13imp}
\end{figure}

The strong emission line method uses flux ratios of major emission lines of star-forming galaxies
that include [\oii]3727, H$\beta$4861, [\oiii]5007, H$\alpha$6563, and [\nii]6584. 
One of the most frequently used line ratios is the $R_{23}$ index 
defined by: 
\begin{equation}
R_{23} =  \frac{f_{\rm [OII]3727}+f_{\rm [OIII]4959}+f_{\rm [OIII]5007}}{f_{\rm H\beta}}.
\label{eq:r23}
\end{equation}
The top panel of Figure \ref{fig:nagao2006_fig13imp} presents $R_{23}$ as a function of oxygen abundance.
The solid lines in this panel 
show $R_{23}$-oxygen abundance
relations calculated by photoionization models with different ionization parameter ($q_{\rm ion}$) values.
Here, $q_{\rm ion}$ is defined by
\begin{equation}
q_{\rm ion} =  \frac{Q_{\rm H^0}}{4 \pi R_{\rm s}^2 n_{\rm H}},
\label{eq:q_ion}
\end{equation}
where $Q_{\rm H^0}$ is the number of hydrogen ionizing photons produced per unit time.
$R_{\rm s}$ is the Str$\ddot{\rm o}$mgren radius, and $n_H$ is the hydrogen density.
Because $R_{23}$ strongly depends on $q_{\rm ion}$ (see the top panel of Figure \ref{fig:nagao2006_fig13imp}), 
one needs to empirically calibrate the $R_{23}$-oxygen abundance relation with local galaxies 
that have $R_{23}$ and oxygen abundance measurements from the direct $T_{\rm e}$ method. 
In the top panel of Figure \ref{fig:nagao2006_fig13imp},
the star marks with error bars denote the average values of the local galaxies, 
and the dashed line is an empirical relation that fits the star marks.
In this way, a locally calibrated empirical relation 
is used to derive oxygen abundances from $R_{23}$ measurements.
However, it should be noted that the oxygen abundances of high-$z$ galaxies estimated in this manner have systematic errors 
because ISM properties of high-$z$ galaxies including $q_{\rm ion}$ 
are different from those of local galaxies \citep{nakajima2014}.

The top panel of Figure \ref{fig:nagao2006_fig13imp}
%
shows a degeneracy in 
all $R_{23}$-oxygen abundance 
relations because there are basically two possible oxygen abundances for a given $R_{23}$ measurement.
This is because $R_{23}$ increases with increasing metallicity up to $\sim 1Z_\odot$, 
while
at higher metallicities fine-structure cooling emission in far-infrared (FIR) wavelengths
dominates and reduces the collisionally excited line fluxes of [\oii] and [\oiii] that are the numerators of $R_{23}$.
To resolve this degeneracy,
one can use the $N2$ index defined by:
\begin{equation}
N2 =  \frac{f_{\rm [NII]6584}}{f_{\rm H\alpha}}.
\label{eq:n2}
\end{equation}
The bottom panel of Figure \ref{fig:nagao2006_fig13imp} displays photoionization model calculations (solid lines),
local galaxy averages (star marks), and an empirical $N2$-oxygen abundance relation that fits the local galaxy averages (dashed line).
Since the $N2$ index does not include oxygen line measurements i.e. only $f_{\rm [NII]6584}$ and $f_{\rm H\alpha}$,
oxygen abundances from the $N2$ index have a systematic uncertainty 
due to a possible variation of 
the nitrogen-to-oxygen abundance ratio,
implying that the $N2$ index alone is not a good estimator of oxygen abundance.
However, one can obtain a coarse oxygen abundance estimate from an $N2$ index (eq. \ref{eq:n2}) measurement
that is useful to resolve the degeneracy of the $R_{23}$-oxygen abundance relation discussed above.
Once the degeneracy is resolved by the $N2$ index, a single solution of oxygen abundance
can be obtained from the $R_{23}$ index. 
%

Because the strong emission line method does not use weak lines 
such as the [\oiii]4363 auroral line whose flux intensity is only $\sim 1/70$ of [\oiii]5007 \citep{izotov2006},
it can efficiently estimate the metallicities of faint galaxies including LAEs.
However, as described above, one should keep in mind that abundances based on the strong emission line method 
would be biased if the ISM properties of galaxies in question 
are different from those of local calibrators.\\










\noindent
{\bf Metallicity Estimates of LAEs} 

Although it is difficult to determine the metallicity of LAEs 
due to their faintness, some constraints have been obtained 
for a small number of LAEs 
\footnote{
Here, a solar metallicity of
$\log (Z/Z_\odot) = 12 + \log ({\rm O/H}) - 8.69$
is assumed
\citep{asplund2009}.
}.
\citet{finkelstein2011a} have obtained upper limits of
$Z<0.17Z_\odot$ and $<0.28Z_\odot$ for two LAEs
with the spectroscopic $N2$ index 
(see also \citealt{guaita2013}),
while \citet{nakajima2012} have placed a lower limit of $Z>0.09 Z_\odot$
for a stack of 105 LAE narrowband images that cover {[\oii]}3727, {[\nii]}6584, and  H$\alpha$ lines
based on a combination of two strong line methods,
the $N2$ index and the {[\oii]}/H$\beta$ index \citep{nagao2006}.
\citet{nakajima2014} have examined the metallicities of 6 LAEs at $z=2-3$ 
with the $R23$ index and the ratio of {[\oiii]}5007 to {[\oii]}3727 fluxes,
and found that they fall in the range of $12+\log ({\rm O/H}) = 7.98-8.81$,with an average of $Z\sim 0.5 Z_\odot$.
These metallicity constraints generally agree with the mass metallicity relation
\citep{finkelstein2011a} and an extrapolation of the SFR-mass metallicity relation to low mass of star-forming galaxies at similar redshifts \citep{nakajima2012}.

\begin{figure}[H]
\centering
\includegraphics[scale=.45]{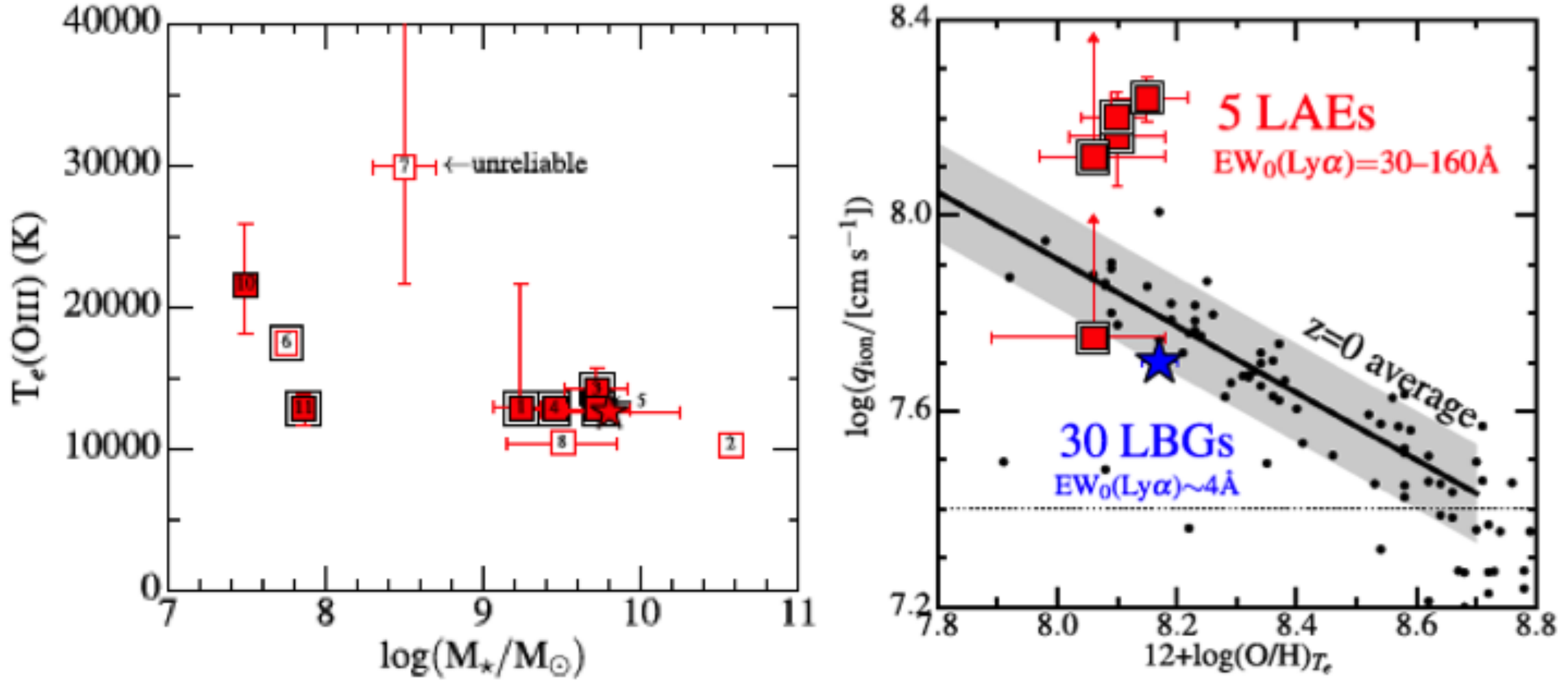}
\caption{
Left: Electron temperature as a function of stellar mass \citep{kojima2017}. The red filled/open squares marked with 
a black open square are one un-lensed LAE (ID 1) and five lensed LAEs (IDs 3, 4, 5, 6, and 11).
Note that ID 6 does not have reliable line flux estimates. The other squares and the star mark indicate 
individually-identified LBGs and a stacked LBG, respectively. 
Right: Oxygen abundance and ionization parameter of LAEs and LBGs that are estimated by the
reliable direct $T_{\rm e}$ method \citep{kojima2017}.
The red filled squares marked with a black square are LAEs (IDs 1, 3, 4, 5, and 11),
where ID 6 with an unreliable result ($Z=0.1 Z_\odot$) is removed. The blue star mark indicates 
the stacked LBG. The black circles denote the results of local galaxy stacks, while the black line (the gray region)
represents the local best-fit power law relation between oxygen abundance and ionization parameter
(the typical scatter around the best-fit power law). 
This figure is reproduced by permission of the PASJ.
}
\label{fig:kojima2017_fig5_fig12}       
\end{figure}

More recently, \citet{kojima2017} have measured the oxygen abundances of LAEs
by the direct $T_{\rm e}$ method (left panel of Figure \ref{fig:kojima2017_fig5_fig12}), 
to show that one LAE (and five lensed LAEs)
has (have) a metallicity (metallicity range) of $Z=0.26^{+0.08}_{-0.07} Z_\odot$
($Z=0.1-0.3 Z_\odot$). The right panel of Figure \ref{fig:kojima2017_fig5_fig12}
presents oxygen abundance and ionization parameter measurements for
the one LAE and four lensed LAEs, after carefully removing one lensed LAE (ID 6) whose value, $Z= 0.1 Z_\odot$, is based on unreliable flux estimates.
These results 
by the direct $T_{\rm e}$ method and the strong line method
suggest that the gas-phase metallicity of LAEs at $z\sim 2-3$
is typically $Z\simeq 0.1-0.5 Z_\odot$ (Table \ref{tab:stellar_population}).
So far, no extremely metal poor LAEs with $Z\lesssim 0.01 Z_\odot$ 
have been identified.
However, because most of the LAEs with metallicity estimates are 
moderately bright,
there remains a possibility that very metal poor LAEs are found
in future studies targeting faint objects.


\begin{figure}[h]
\centering
\includegraphics[scale=.45]{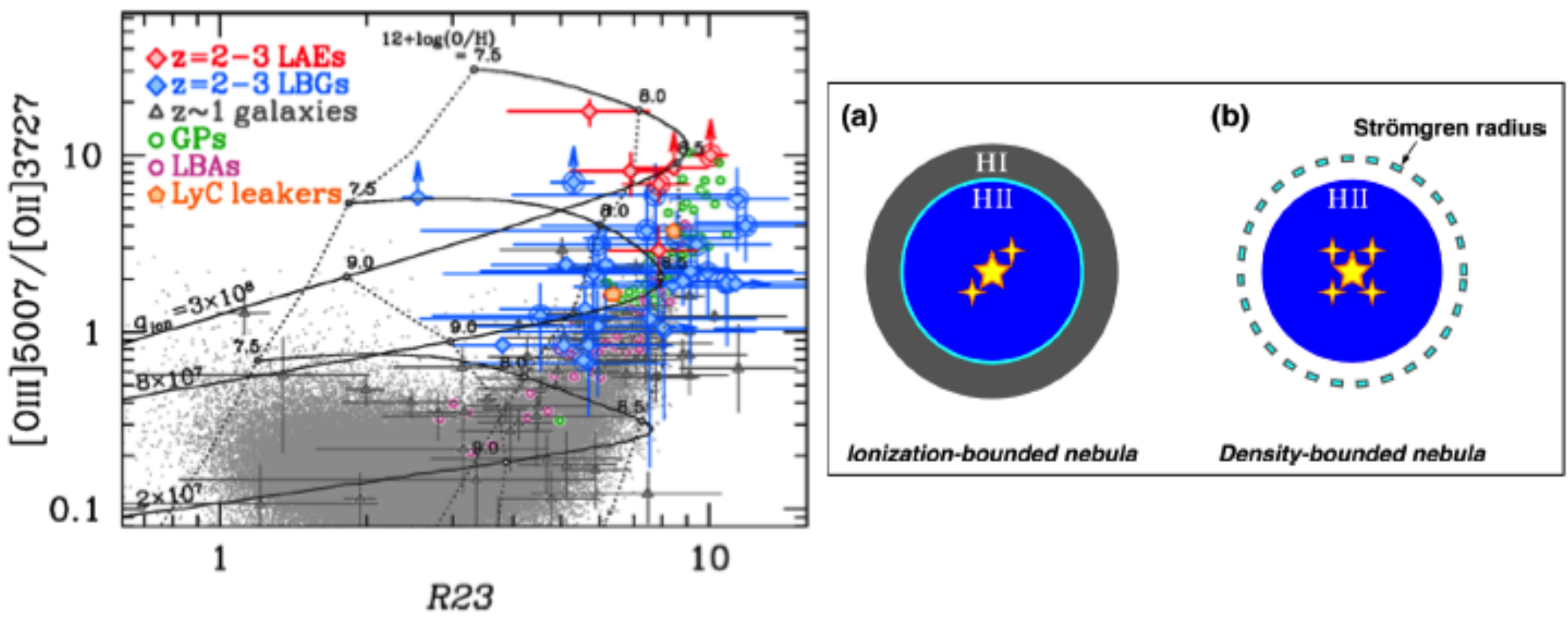}
\caption{
Left: $O_{32}$ $(=f_{\rm [OIII]5007}/f_{\rm [OII]3727} )$ as a function of $R_{23}$ \citep{nakajima2014}.
The Red and blue diamonds indicate LAEs and LBGs, respectively, at $z=2-3$.
The green circles, magenta circles, and orange pentagons denote, respectively, 
green-pea galaxies, Lyman-break analogs, and Lyman continuum leaking galaxies
in the local universe. 
The gray triangles and dots represent $z\sim 1$ galaxies and $z\sim 0$ SDSS galaxies, respectively.
The three solid lines show photoionization model tracks with the ionization parameter of $q_{\rm ion} = 3\times 10^8$,
$8\times 10^7$, and $2\times 10^7$ cm s$^{-1}$ from top to bottom. The model tracks cover the oxygen abundance,
from $12+\log({\rm O/H})=7.5$ to $9.0$ and beyond, 
with dotted lines connecting the same oxygen abundances.
Right: Schematic illustrations of two types of \hii\ regions: 
(a) an ionization bounded nebula and (b) a density-bounded nebula \citep{nakajima2014}.
The blue and gray regions denote \hii\ and \hi\ gas regions, respectively.
The star marks represent massive stars emitting ionizing photons.
The cyan solid (dotted) lines indicate the (expected) positions of the Str$\ddot{\rm o}$mgren radius.
This figure is reproduced by permission of the AAS.
}
\label{fig:nakajima2014_fig2_fig12}       
\end{figure}

\begin{figure}[h]
\centering
\includegraphics[scale=.45]{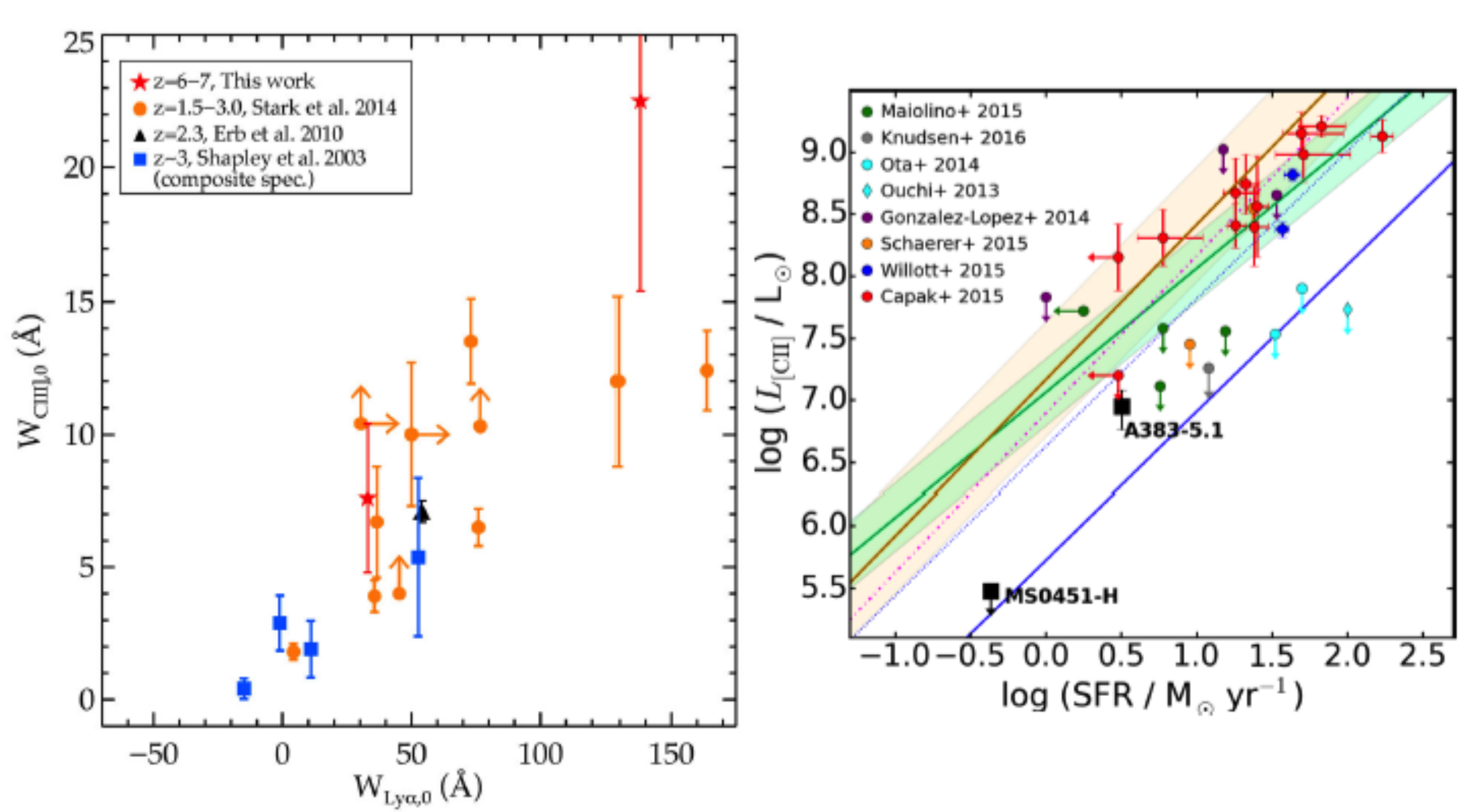}
\caption{
Left: \ciii]1909 rest-frame equivalent width as a function of Ly$\alpha$ rest-frame equivalent width \citep{stark2015a}.
The red stars, orange circles, black triangles, and blue squares
represent galaxies including LAEs at $z=6-7$, $1.5-3.0$, $2.3$, and $3$, respectively.
Right: [\cii]158$\mu$m luminosity as a function of SFR \citep{knudsen2016}. The cyan diamond/circles, purple circles,
orange circles, green circles, and black squares indicate LAEs. 
The blue and red circles are continuum-bright LBGs, some of which show Ly$\alpha$ emission.
This figure is reproduced by permission of MNRAS.
}
\label{fig:stark2015a_fig8_knudsen2016_fig3}       
\end{figure}

\begin{figure}[h]
\centering
\includegraphics[scale=.48]{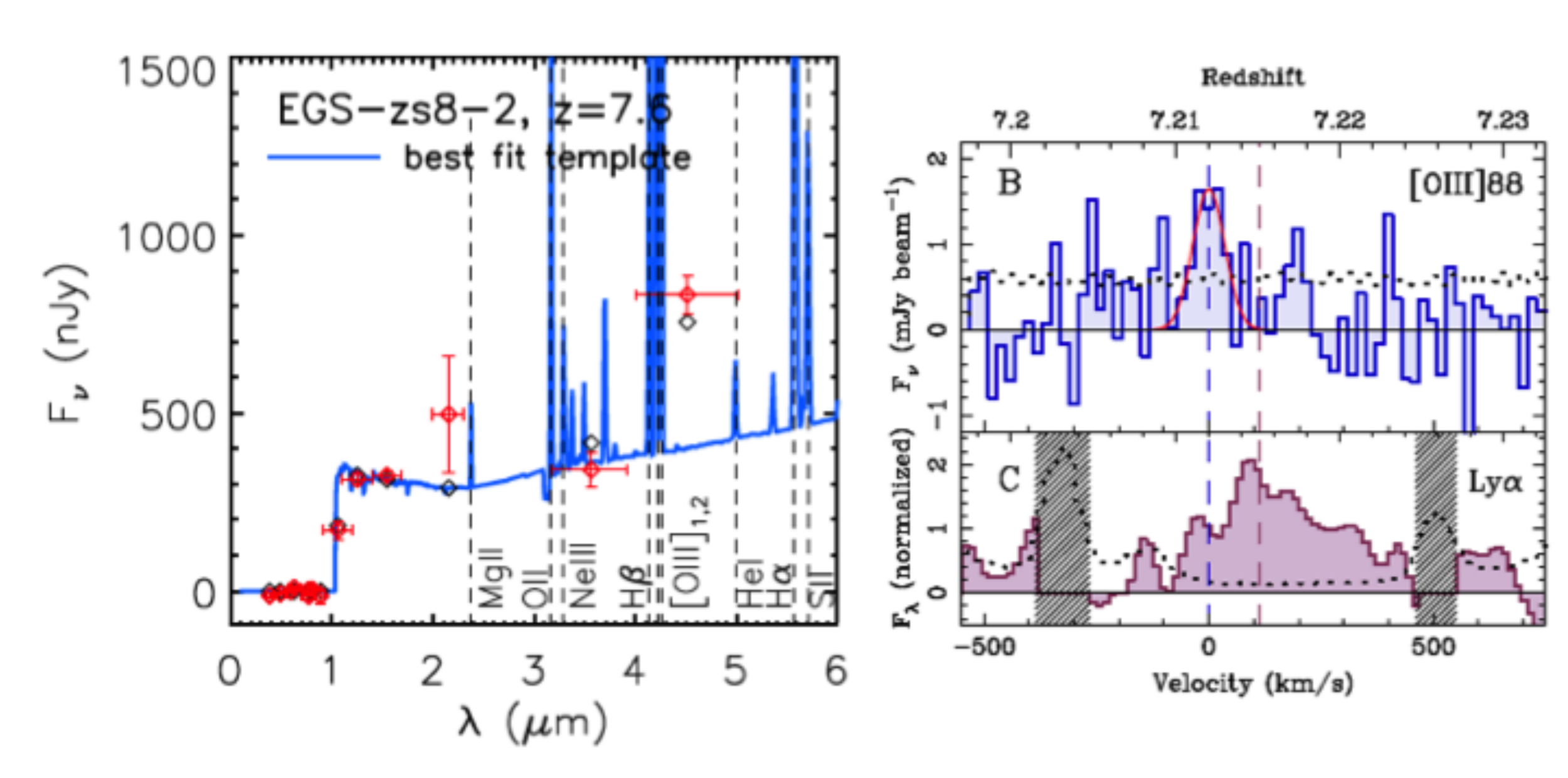}
\caption{
Left: Observed SED of EGS-zs8-2 at $z=7.48$ (red circles; \citealt{roberts-borsani2016}).
The blue line denotes the best-fit SED model with nebular emission lines.
The open diamonds are the expected photometry in the photometric bands
for the best-fit SED. The vertical dashed lines mark the wavelengths of 
major nebular emission lines. The observed flux density at $4.5\mu$m shows an
excess that is probably made by strong [\oiii]4959,5007 and H$\beta$ line emission.
Right: [\oiii]$88\mu$m (top) and Ly$\alpha$ lines of SXDF-NB1006-2 \citep{inoue2016}.
The dotted line denotes the $1\sigma$ noise level.
The blue and purple dashed lines represent the line-peak velocities of
[\oiii]$88\mu$m and Ly$\alpha$, respectively. The red curve in
the top panel indicates the best-fit Gaussian function to the [\oiii]$88\mu$m line.
The gray regions in the bottom panel show the wavelength range
with large sky subtraction systematics. 
This figure is reproduced by permission of the AAS and Science.
}
\label{fig:roberts-borsani2016_fig4_inoue2016_fig1}       
\end{figure}

\subsubsection{Ionization State}
\label{sec:ionization_state}

The ionization state of typical LAEs' \hii\ regions is 
clearly different from those of other types of high-$z$ galaxies.
Recent optical-NIR spectroscopy has revealed
that the $O_{32}$ ratios of $z\sim 2-3$ LBGs and LAEs are significantly
higher than those of local SDSS galaxies (left panel of Figure \ref{fig:nakajima2014_fig2_fig12}; \citealt{nakajima2013,nakajima2014}), where 
the $O_{32}$ ratio is defined by
\begin{equation}
O_{32} =  \frac{f_{\rm [OIII]5007}}{f_{\rm [OII]3727}}.
\label{eq:o32}
\end{equation}
Specifically, LAEs have extremely large values of $O_{32}\sim 10$, being about $\sim 10-100$ times higher 
than those of the local SDSS galaxies
and even higher than those of the LBGs on average.
In the left panel of Figure \ref{fig:nakajima2014_fig2_fig12}, photoionization models with various
metallicity and $q_{\rm ion}$ values are compared with LAEs.
Although the models predict that $O_{32}$ increases with decreasing metallicity,
the high $O_{32}$ values of LAEs cannot be explained by models 
that reproduce the local SDSS galaxies with $q_{\rm ion} = 0.1-1 \times 10^8$ cm s$^{-1}$.
The $z\sim 2-3$ LAEs are found to have 
$q_{\rm ion}=1-9 \times 10^8$ cm s$^{-1}$, about an order of magnitude larger than
those of the local SDSS galaxies \citep{nakajima2014}. The left panel of Figure \ref{fig:nakajima2014_fig2_fig12}
also shows that there exist local counterparts to LAEs, green pea galaxies (GPs;
\citealt{cardamone2009,jaskot2014}), whose $O_{32}$ and $R_{23}$ values are comparable with
those of LAEs.

The physical origin of the high $q_{\rm ion}$ values of LAEs is not
well understood. The ionization parameter defined by eq. (\ref{eq:q_ion})
is rewritten as
\begin{equation}
q_{\rm ion}^3 \propto  Q_{\rm H^0} n_{\rm H} \epsilon^2
\label{eq:q_ion_simple}
\end{equation}
by the substitution of the Str$\ddot{\rm o}$mgren radius.
Here, the Str$\ddot{\rm o}$mgren radius $R_{\rm s}$ is defined as
\begin{equation}
Q_{\rm H^0} = \frac{4}{3}\pi R_{\rm s}^3 n_{\rm H}^2 \alpha_{\rm B} \epsilon
\label{eq:stromgren_radius}
\end{equation}
with the coefficient of the total hydrogen recombination to the $n > 1$ levels 
$\alpha_{\rm B}$, where $\epsilon$ is the volume filling factor of the Str$\ddot{\rm o}$mgren sphere.
Eq. (\ref{eq:q_ion_simple}) indicates that 
either $Q_{\rm H^0}$,  $n_{\rm H}$,  or $\epsilon$
needs to increase by a factor of $10^3$, $10^3$, or $30$, respectively,
to explain the high $q_{\rm ion}$ values of LAEs.
With a moderately high SFR and metal-poor young stellar population, LAEs
produce ionizing photons more efficiently than the local SDSS galaxies.
However, 
it may not be possible that the $Q_{\rm H^0}$ of LAEs are $\sim 10^3$ times 
higher than those of the local SDSS galaxies. Some studies have reported an increase in the electron density from $\sim 25$ cm$^{-3}$ ($z\sim 0$) 
to $\sim 250$ cm$^{-3}$
($z\sim 2$; \citealt{steidel2014,shimakawa2015,sanders2016}), 
but these increase rates are not as high as $10^3$ times.
It is also unlikely that the average $\epsilon$ increases by a factor of 30
from $z\sim 0$ to $2$. I discuss the issue of high $q_{\rm ion}$ at the end of this subsection.


Some other observations also suggest that LAEs have high $q_{\rm ion}$ values.
LAEs with a large Ly$\alpha$ $EW_0$
tend to have high-ionization metal lines in rest-frame UV spectra.
\citet{stark2014} have identified moderately strong \ciii]1901,1909 lines in lensed LAEs at $z\sim 2$ by deep spectroscopy,
and revealed a positive correlation between Ly$\alpha$ and \ciii]1901,1909 $EW_0$.
The left panel of Figure \ref{fig:stark2015a_fig8_knudsen2016_fig3} suggests that high-ionization lines \ciii]1901,1909 
are strong for large-Ly$\alpha$ $EW_0$ galaxies such as LAEs. 
Highly ionized gas containing C$^{2+}$ is probably more abundant in LAEs compared to other types of galaxies.
Subsequently,  \citet{stark2015a} and \citet{stark2015b} have reported the detections of moderately strong lines of
\ciii]1907,1909 lines in two LAEs at $z=6-7$ and {\civ}1548 line in an LAE at $z=7$, respectively.
Although there still remains the possibility that the {\civ}1548 line is produced by a hidden AGN, not by young, massive stars 
(e.g. detection of {\nv}1239 for the definitive AGN identification; \citealt{laporte2017}),
these spectroscopic results suggest that ionization state of LAEs at $z=2-7$ is very high.
%
%

Most of the ALMA studies of LAEs have targeted the [\cii]158$\mu$m fine structure line
that originates from low-ionization C$^+$ gas. Because C has a lower ionization potential than H, it is thought that the majority of [\cii]158$\mu$m
photons are produced in PDRs that extend beyond \hii\ regions. 
These ALMA studies have found that LAEs have significantly fainter 
[\cii]158$\mu$m luminosities $L_{\rm [CII]}$ than local galaxies 
with similar SFRs (right panel of Figure \ref{fig:stark2015a_fig8_knudsen2016_fig3}; 
\citealt{ouchi2013,ota2014,knudsen2016,pentericci2016}; cf. \citealt{maiolino2015}).
In fact, \citet{harikane2018b} report a significant anti-correlation between 
a [\cii]-luminosity to SFR ratio ($L_{\rm [CII]}/SFR$) and Ly$\alpha$ $EW_0$.
The local galaxy relation in the right panel of Figure \ref{fig:stark2015a_fig8_knudsen2016_fig3} indicates that high SFR galaxies have
bright [\cii]158$\mu$m emission, 
because of high production rates of carbon ionizing photons 
from massive stars.
In this sense, 
LAEs' faint [\cii]158$\mu$m luminosities are puzzling, 
because they also have high SFRs. 
Although [\cii]158$\mu$m emission is relatively 
weak for galaxies with an AGN, there is no hint of AGN
in the LAEs observed by ALMA. Since [\cii]158$\mu$m is a forbidden line,
it can be weakened by collisional de-excitation in high density gas,
but no hint of a high gas density has been obtained for LAEs.
On the contrary, $z\sim 7$ LAEs show a hint of strong emission of 
the [{\oiii}]4959,5007 forbidden line (left panel of Figure \ref{fig:roberts-borsani2016_fig4_inoue2016_fig1}; \citealt{roberts-borsani2016}).
Moreover, recent ALMA studies have identified 
[{\oiii}]88$\mu$m fine-structure line emission in an LAE at $z=7$ (right panel of Figure \ref{fig:roberts-borsani2016_fig4_inoue2016_fig1}; \citealt{inoue2016}).


Regarding the ISM state of LAEs, 
the physical origins of two properties, high $O_{32}$ ratios (i.e. high $q_{\rm ion}$) 
and weak [\cii]158$\mu$m emission remain open questions 
as detailed above.
While there are no clear answers to them, 
it is suggested that these two properties can be consistently explained 
if the \hii\ regions of LAEs are density-bounded \citep{nakajima2014}.
Figure \ref{fig:nakajima2014_fig2_fig12} shows a conceptual diagram of
a density-bounded nebula,
compared with an ionization-bounded nebula
that is the standard picture of \hii\ regions.
In the standard picture,
the size of an ionized nebula is determined by the number of 
ionizing photons, which corresponds to the 
radius of the Str$\ddot{\rm o}$mgren sphere (eq. \ref{eq:stromgren_radius}).
On the other hand, the size of a density-bounded nebula
is determined by the amount of atomic gas around the 
ionizing source. In contrast with a ionization-bounded nebula,
a density-bounded nebula does not have an outer shell of
ionized hydrogen gas emitting low-ionization lines such as [\oii]3727,
but an inner shell of ionized hydrogen gas producing high-ionization lines
such as \ciii]1907,1909, [\oiii]5007, and [\oiii]88$\mu$m.
Moreover, PDRs, major sources of [\cii]158$\mu$m emission,
are not well developed. The density-bounded nebula scenario 
explains the high $O_{32}$ ratio (i.e. high $q_{\rm ion}$) 
and the weak [\cii]158$\mu$m emission. If this 
scenario applies to LAEs, ionizing photons escape easily from the ISM of LAEs.
Such ionizing photons can be major sources of cosmic reionization
(\citealt{nakajima2014,jaskot2014}; Section \ref{sec:escape_fraction_ionizing_photon}).
Although this scenario should be tested by theoretical models and
more observations, it is interesting that the ISM state of LAEs
may be important for the understanding of cosmic reionization.

\subsubsection{Dust and Extinction}
\label{sec:dust_extinction}

Stellar population analyses of LAEs suggest that 
the dust extinction of stellar continuum emission is as low as $E(B-V)_{\rm s}\simeq 0-0.2$ on average 
under the assumption of Calzetti's extinction law (\citealt{calzetti2000}; Section \ref{sec:stellar_population}).
One can also estimate the color excess of nebular lines, $E(B-V)_{\rm neb}$, with
the Balmer decrement. Note that $E(B-V)_{\rm neb}$ is not necessarily 
the same as $E(B-V)_{\rm s}$, because nebular lines originate 
from star-forming regions that are generally dustier 
than other regions in the galaxy. 
\citet{calzetti2000} claim
that local starbursts have $E(B-V)_{\rm neb}=E(B-V)_{\rm s}/0.44$, 
although the relation between $E(B-V)_{\rm neb}$ and 
$E(B-V)_{\rm s}$ for high-$z$ galaxies is poorly understood.

With a Balmer decrement measurement, H$\alpha$/H$\beta$,
the dust extinction of nebular lines is estimated with
\begin{equation}
E(B-V)_{\rm neb}=\frac{2.5}{k_{\rm H\beta} - k_{\rm H\alpha}} \log  \left ( \frac{ {\rm H\alpha} / {\rm H\beta} }{ 2.86 } \right ),
\label{eq:ebv_neb}
\end{equation}
where $k_{\rm H\alpha}$ and $k_{\rm H\beta}$ are coefficients depending on the dust extinction law.
The Calzetti extinction law gives $k_{\rm H\alpha}=3.325$ and $k_{\rm H\beta}=4.598$
\citep{momcheva2013,kashino2013}. 
Here, H$\alpha$/H$\beta = 2.86$ is an intrinsic (i.e., dust-free) line ratio
at $T_{\rm e}=10^4$ K and $n_{\rm e}=10^2$ cm$^{-3}$ 
for the case B recombination \citep{osterbrock2006}.
Since the H$\beta$ fluxes of LAEs are generally too faint to detect (e.g. \citealt{guaita2013}), 
only a small number of LAEs have $E(B-V)_{\rm neb}$ measurements;
they fall in the range of $E(B-V)_{\rm neb}=0-0.2$ \citep{kojima2017}.
So far, there are no studies of $E(B-V)_{\rm neb}$ statistics 
nor of the relation between $E(B-V)_{\rm neb}$ and $E(B-V)_{\rm s}$ for LAEs (cf. \citealt{erb2016}).
A statistical study addressing these issues of $E(B-V)_{\rm neb}$ should be
conducted for LAEs in the near future.



The dust extinction of high-$z$ galaxies is characterized with 
the UV-continuum slope, $\beta$, defined by
\begin{equation}
f_\lambda = \lambda^\beta,
\label{eq:uv_slope}
\end{equation}
where $f_\lambda$ is the UV-continuum spectrum of the galaxy
in the wavelength range $\simeq 1300-3000$\AA\ \citep{calzetti2001}.
The $\beta$ value is used as an indicator of the amount of dust extinction 
for moderately young star-forming galaxies such as LBGs and LAEs 
whose intrinsic UV-continuum slope is $\beta \simeq -2.2$.

\begin{figure}[H]
\centering
\includegraphics[scale=.60]{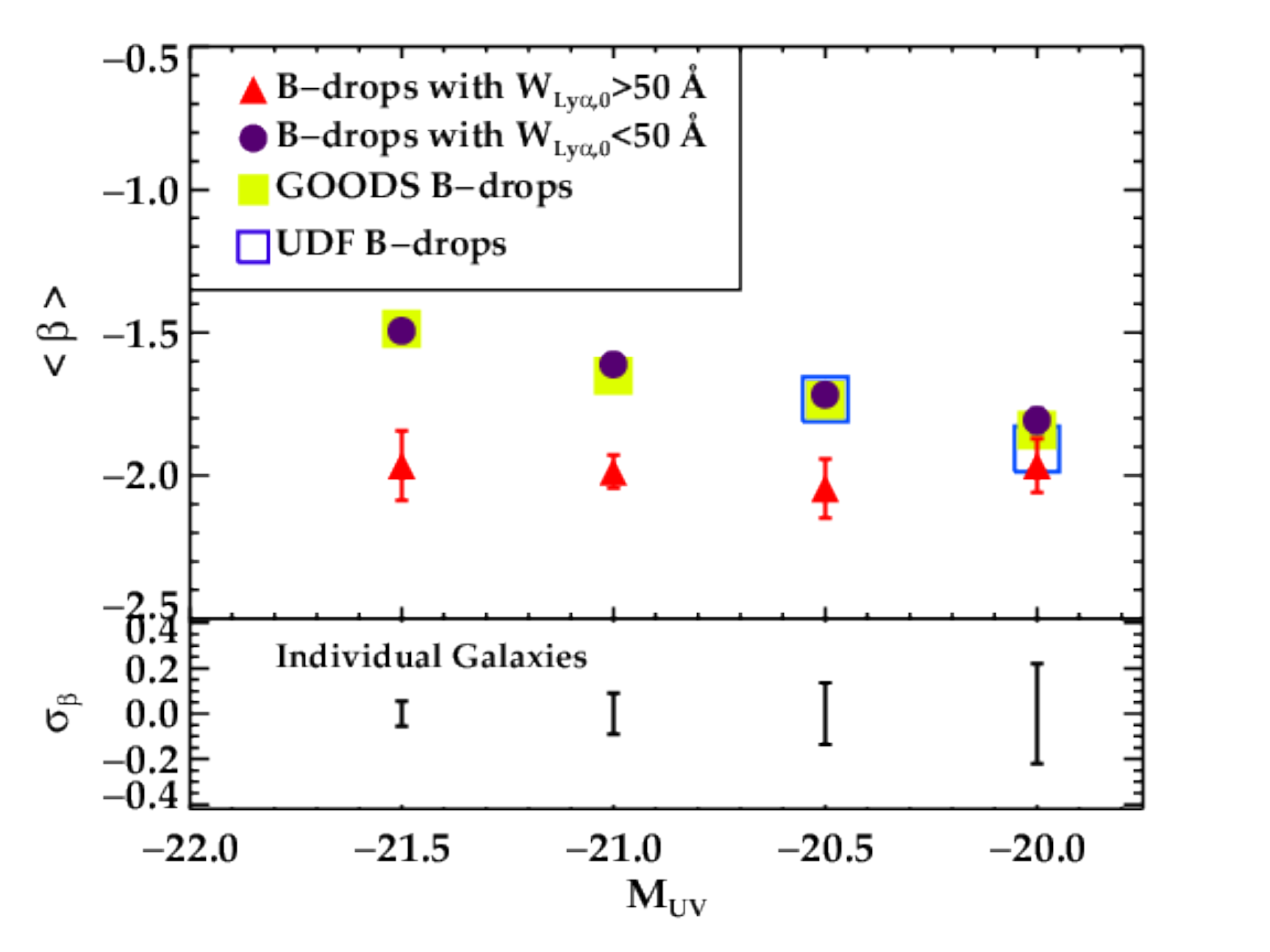}
\caption{
Top: $\beta$ as a function of UV magnitude \citep{stark2010}. 
The red triangles denote
LAEs with $EW_0>50$\AA, 
while the purple circles indicate dropout galaxies at $z\sim 4$
with $EW<50$\AA.
The green squares and blue squares represent 
dropout galaxies at $z\sim 4$.
Bottom: Typical uncertainties in $\beta$ measurements.
This figure is reproduced by permission of the AAS.
}
\label{fig:stark2010_fig14}       
\end{figure}

Figure \ref{fig:stark2010_fig14} presents the average $\beta$ values of LAEs with
high Ly$\alpha$ equivalent widths $EW_0>50$\AA, and compares them 
with those of LBGs. The LAEs have $\beta \simeq -2$ that is significantly smaller than those of the LBGs 
at the same UV luminosity, suggesting that LAEs are generally dust poor.
Figure \ref{fig:stark2010_fig14} also indicates that UV-continuum faint LAEs and LBGs with $M_{\rm UV}\sim -20$ 
have similar $\beta$ values, and probably 
similar extinction properties, supporting the idea 
that a high fraction of faint LBGs are LAEs (Section \ref{sec:luminosity_function}).

\begin{figure}[H]
\centering
\includegraphics[scale=.48]{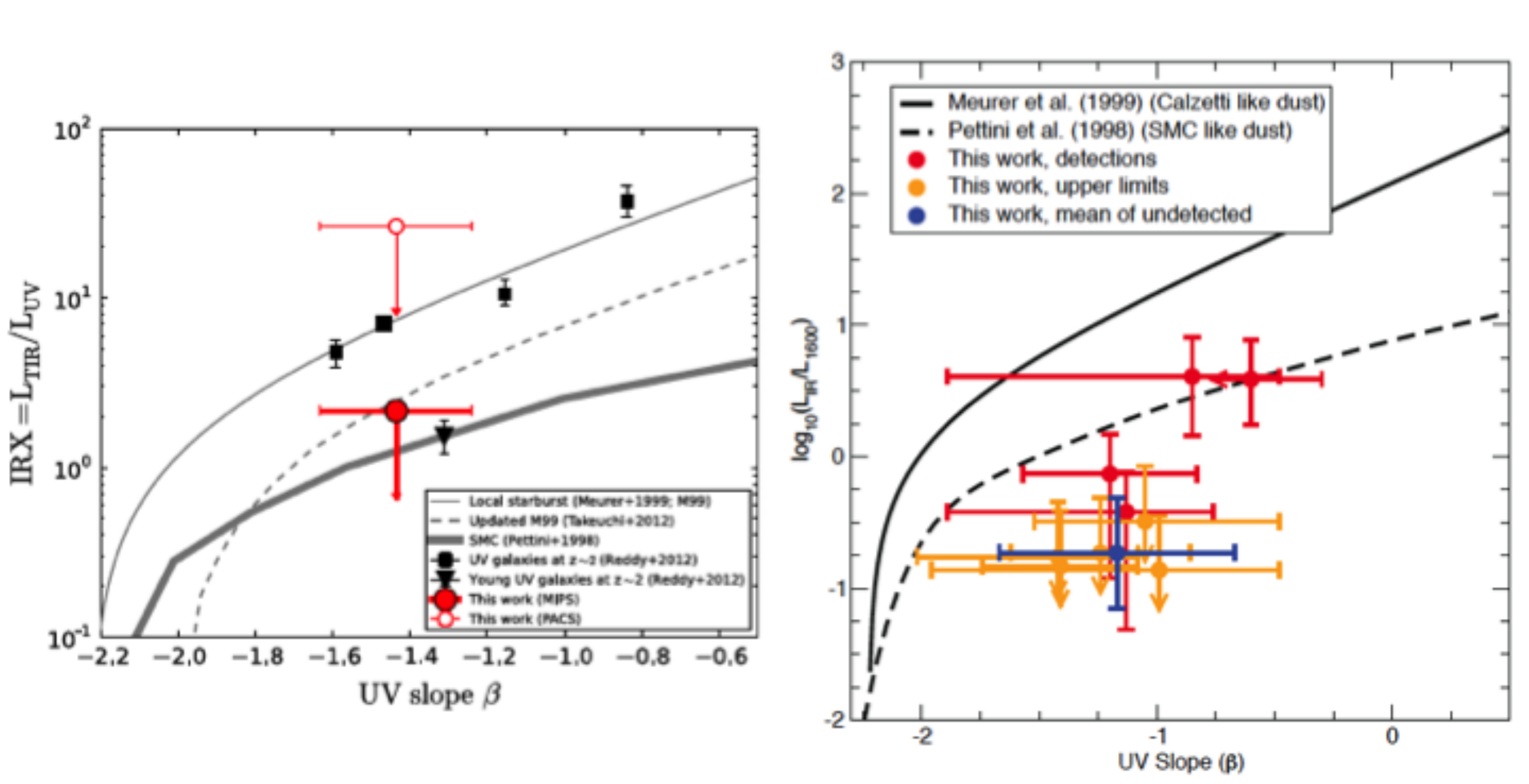}
\caption{
$IRX$ as a function of the UV slope $\beta$
for LAEs at $z\sim 2$ (left; \citealt{kusakabe2015}) and at $z\sim 5-6$ (right; \citealt{capak2015}).
Left: The red filled and open circles indicate, respectively, 
the upper limits of $IRX$ estimated
by the stack of Spitzer/MIPS and PACS images centered at the sky positions of 
213 LAEs at $z=2.2$. The gray thin-solid, dashed, and thick-solid lines 
denote, respectively, 
the dust extinction relations of Calzetti, Takeuchi \citep{takeuchi2012}, 
and the SMC. The squares (inverse-triangle) represent
UV-continuum bright (and young) galaxies at $z\sim 2$.
Right: The red, orange, and blue circles show the detections,
the upper limits, and the average of the no-detection data
for galaxies (including 3 LAEs) at $z\sim 5-6$.
The solid and dashed lines are the $IRX$-$\beta$ relations 
of Calzetti and the SMC.
This figure is reproduced by permission of the AAS and the Nature Publishing Group.
}
\label{fig:kusakabe2015_fig1_capak2015_fig2}       
\end{figure}

To evaluate the dust extinction law of a galaxy, one can use the $IRX$ ratio,
\begin{equation}
IRX = \frac{L_{\rm IR}}{L_{\rm UV}},
\label{eq:irx}
\end{equation}
where $L_{\rm IR}$ and $L_{\rm UV}$ are the total infrared (IR; $3-1000\mu$m) and UV ($\sim 1500$\AA) luminosities ,
respectively. The values of $L_{\rm IR}$ 
can be estimated from, e.g., Spitzer/MIPS, Herschel/SPIRE, APEX/LABOCA,
and ALMA photometry \citep{wardlow2014,kusakabe2015,capak2015}.
Figure \ref{fig:kusakabe2015_fig1_capak2015_fig2} presents the $IRX$-$\beta$ relation 
for LAEs and LBGs at $z\sim 2$ and $5-6$, together with the model curves of Calzetti and SMC dust extinction.
It is clear that LAEs have low $IRX$ values at a given $\beta$ on average.
The left panel of Figure \ref{fig:kusakabe2015_fig1_capak2015_fig2} indicates that
LAEs at $z\sim 2$ have an extinction curve similar to that of the SMC and different from those of Calzetti's local starbursts.
There are three LAEs at $z=5-6$ with $IRX$-$\beta$ measurements shown 
in the right panel of Figure \ref{fig:kusakabe2015_fig1_capak2015_fig2}.
This panel suggests that these three LAEs 
have $IRX$ values which fall close to or even below the SMC curve,
although these extremely low $IRX$ estimates are still under debate.
However, there is a consensus based on deep ALMA observations 
that LAEs at $z>5$ 
have faint $\sim 1$mm flux densities 
\citep{ouchi2013,ota2014,maiolino2015,capak2015,knudsen2016}.

\begin{figure}[H]
\centering
\includegraphics[scale=.50]{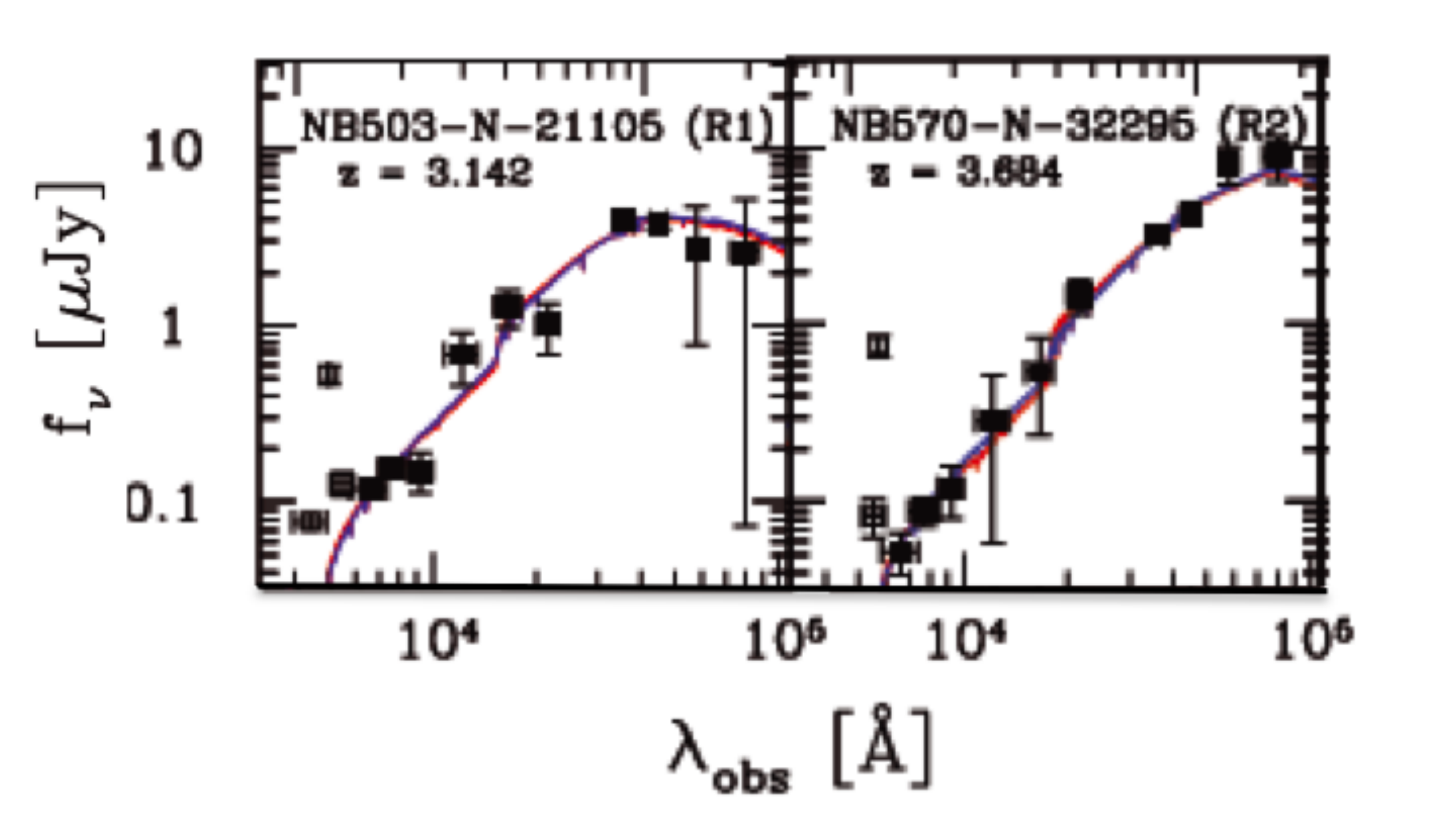}
\caption{
SEDs of two red LAEs, R1 and R2, at $z=3.142$ and $3.684$, respectively \citep{ono2010a}.
The squares denote photometric data obtained by deep optical and near-infrared
observations. 
The red and blue curves represent the best-fit SED models of exponentially-decaying 
and constant star-formation histories, respectively; 
data shown by open squares are not used for the fitting because they are contaminated
by either strong Ly$\alpha$ emission or the IGM Ly$\alpha$ forest absorption.
This figure is reproduced by permission of MNRAS.
}
\label{fig:ono2010a_fig4}       
\end{figure}

On average, LAEs have low extinction and low dust masses. However,
there exists a rare population of dusty LAEs 
with red stellar SEDs and bright submm luminosities.
Figure \ref{fig:ono2010a_fig4} shows the SEDs of two 
spectroscopically-confirmed LAEs (dubbed R1 and R2) at $z=3-4$ 
which have red SEDs and strong Ly$\alpha$ emission \citep{ono2010a}. 
It should be noted that some SMGs have strong Ly$\alpha$ emission that can 
be used for redshift determination \citep{chapman2005,capak2011}.
How Ly$\alpha$ photons can escape from those dusty starbursts
without significant extinction is an open question.
Dusty LAEs might have dust-poor star-forming regions that are 
spatially separated from usual dust-rich star-forming regions.

\subsection{Outflow and Ly$\alpha$ Profile}
\label{sec:outflow_lya_profile}

\begin{figure}[H]
\centering
\includegraphics[scale=.45]{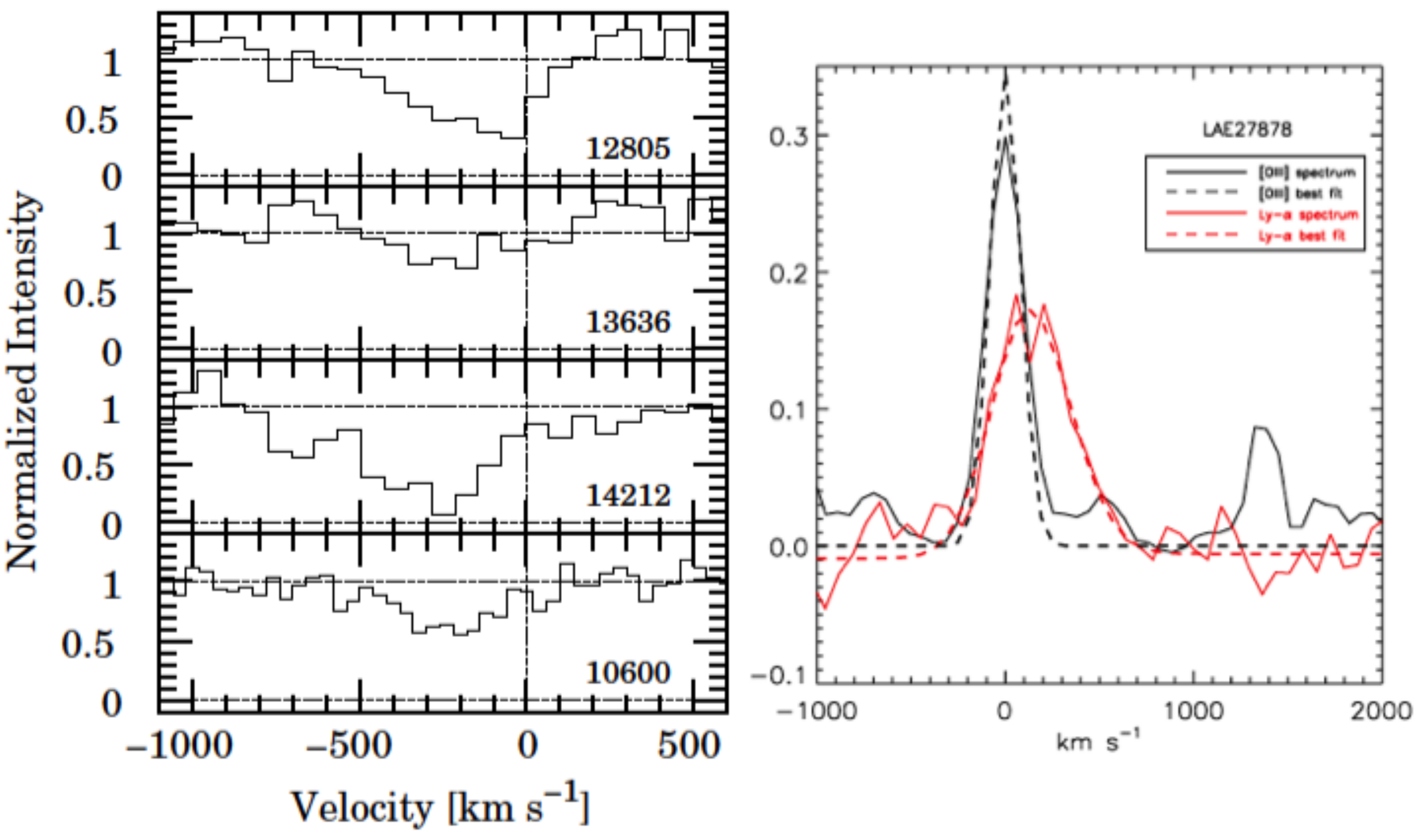}
\caption{
Left: Average UV low ionization absorption lines for four LAEs \citep{shibuya2014b} 
obtained by stacking the Si {\sc ii}$1260$, {\sc Cii}$1334$, and Si {\sc ii}$1526$ lines.
The systemic velocity is determined with nebular emission lines.
Right: Line profiles of Ly$\alpha$ (red solid line) and {\sc [Oiii]}5007 (black solid line)
observed by \citet{mclinden2011}. The red and black dotted lines represent the 
best-fit model profiles for the Ly$\alpha$ and {\sc [Oiii]} lines, respectively.
This figure is reproduced by permission of the AAS.
}
\label{fig:shibuya2014b_fig7_mclinden2011_fig2}       
\end{figure}

\footnote{In the Saas Fee lectures, 
the topics of this subsection were originally
included in Section \ref{sec:lya_escape_fraction}.
For the readers' convenience, I have moved these
topics here.} 
Using deep optical and near-infrared spectra, 
many researchers have investigated the velocities of 
the Ly$\alpha$ line, the low-ionization UV metal absorption lines, 
and the nebular emission lines in LAEs.
The average outflow velocity $V_{\rm out}$ of LAEs at $z\sim 2$ is
estimated to be 
$V_{\rm out}\simeq 200$ km s$^{-1}$
with low-ionization UV metal absorption lines blueshifted from
the systemic velocity (left panel of Figure \ref{fig:shibuya2014b_fig7_mclinden2011_fig2}; \citealt{hashimoto2013,shibuya2014b}).
Here, the systemic velocity is determined by strong nebular lines
such as H$\alpha$ 
that originate from \hii\ regions.
Blueshifted 
absorption lines are 
thought to form in the outflowing gas.
Another important feature in line velocities is that 
the Ly$\alpha$ line peak is generally redshifted
from the systemic velocity (right panel of Figure \ref{fig:shibuya2014b_fig7_mclinden2011_fig2}).
The Ly$\alpha$ line offset $\Delta V_{\rm Ly\alpha}$ is 
defined as the offset velocity of the Ly$\alpha$ line peak 
with respect to the systemic velocity.
The average Ly$\alpha$ line offset of $z\sim 2-3$ LAEs is 
$\Delta V_{\rm Ly\alpha} \simeq 200$ km s$^{-1}$
\citep{mclinden2011,hashimoto2013,shibuya2014b,erb2014}.
%
This average Ly$\alpha$ offset velocity
is comparable with the average outflow velocity, 
$\Delta V_{\rm Ly\alpha} \simeq V_{\rm out}$.

\begin{figure}[H]
\centering
\includegraphics[scale=.35]{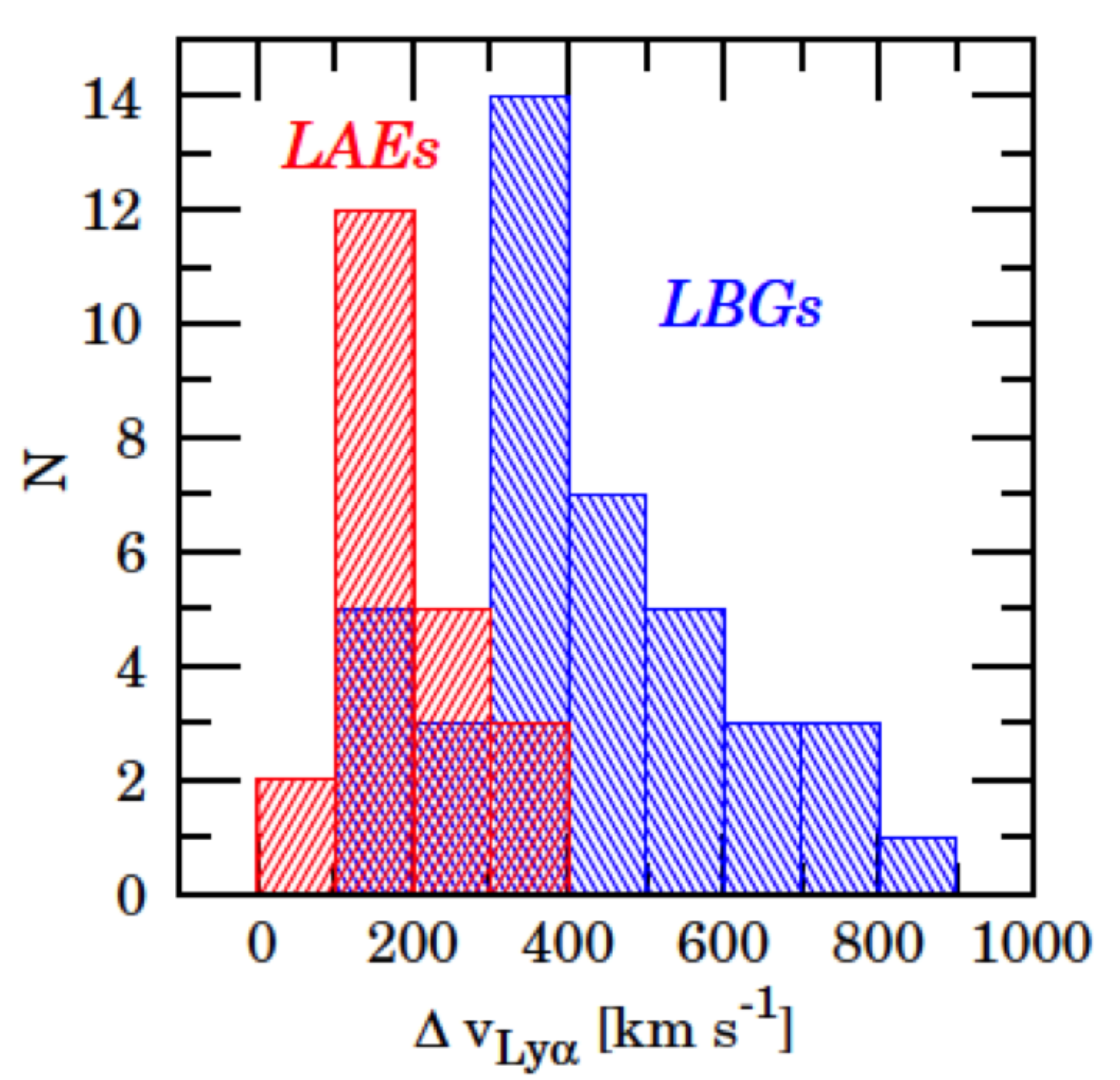}
\caption{
Ly$\alpha$ offset velocity $\Delta V_{\rm Ly\alpha}$ for
LAEs (red histograms) and LBGs (blue histograms) 
\citep{shibuya2014b}.
This figure is reproduced by permission of the AAS.
}
\label{fig:shibuya2014b_fig9}       
\end{figure}

Interestingly, typical LBGs ($L_{\rm UV}\gtrsim L*$) at $z\sim 2-3$ 
have 
$V_{\rm out}\simeq 200$ km s$^{-1}$ on average
\citep{pettini2001,steidel2010}, being comparable with that of LAEs.
However, the average Ly$\alpha$ offset velocity of LBGs
 is $\Delta V_{\rm Ly\alpha} \simeq 400$ km s$^{-1}$, 
about twice as large as LAEs' $\Delta V_{\rm Ly\alpha}$ 
(Figure \ref{fig:shibuya2014b_fig9}). Note that there is a negative correlation
between $\Delta V_{\rm Ly\alpha}$ and Ly$\alpha$ $EW_0$
(e.g. Figure 7 of \citealt{hashimoto2013}).
In contrast with LAEs, LBGs show
$\Delta V_{\rm Ly\alpha} \simeq 2 V_{\rm out}$.
The $\Delta V_{\rm Ly\alpha}$-$V_{\rm out}$ relation
is key to understanding the physical differences in
LAEs and LBGs via theoretical modeling as discussed below.

\begin{figure}[H]
\centering
\includegraphics[scale=.40]{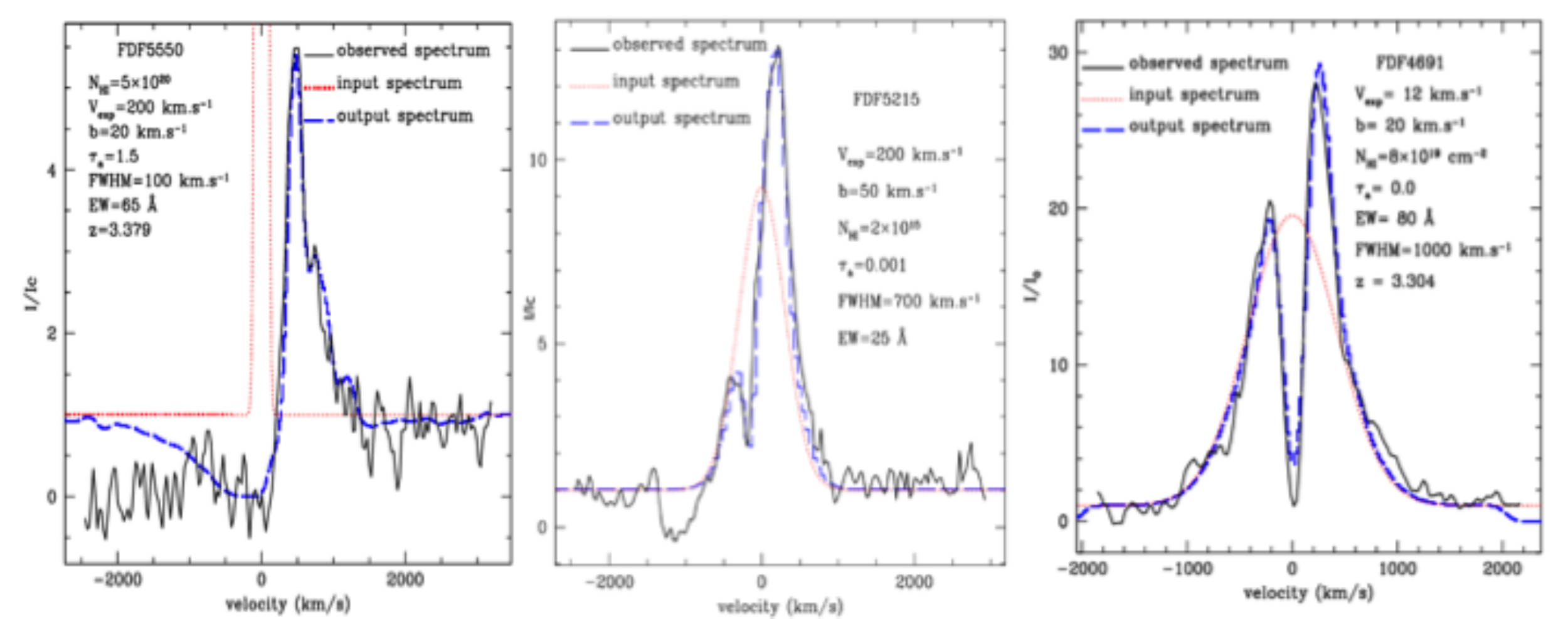}
\caption{
Ly$\alpha$ profiles obtained by observations (black lines)
and ES modeling (blue and red lines) for three LAEs \citep{verhamme2008}.
The blue and red lines represent the best-fit spectra
and the intrinsic spectra predicted by the ES models, respectively.
This figure is reproduced by permission of the A\&A.
}
\label{fig:verhamme2008_fig1_8_5}
\end{figure}

Detailed Ly$\alpha$ profiles of LAEs at $z\sim 2-3$
are investigated by medium-high resolution spectroscopy.
Such spectroscopic efforts have revealed that LAEs have a variety 
of Ly$\alpha$ profiles \citep{tapken2007,yamada2012b,hashimoto2015}.
Among those, three typical profiles are
a single asymmetric/symmetric line,
an asymmetric line with a weak blue peak, and
a double-peak line, as presented in
Figure \ref{fig:verhamme2008_fig1_8_5}.
%
%

Ly$\alpha$ profiles depend on physical parameters
relating to \hi\ resonance scattering of Ly$\alpha$, such as 
the \hi\ density, gas dynamics (including outflows), and 
dust extinction, and are quantitatively
investigated by modeling. One of the most popular models
for Ly$\alpha$ profiles is 
%
the expanding shell (ES) model \citep{ahn2004,verhamme2006}.
This model 
assumes a galaxy-scale spherical shell of outflowing gas around the Ly$\alpha$ source
that is described with four parameters: 
the \hi\ column density $N_{\rm HI}$,
the expansion (outflow) velocity corresponding to $V_{\rm out}$,
the doppler (thermal) velocity of gas in the shell $b$, and 
the optical depth of dust extinction $\tau_{\rm a}$.
The assumption that LAEs have an ES is supported by 
the fact that nearby starbursts 
have a galaxy-scale supershell made by 
multiple SNe in star-forming regions
\citep{marlowe1995,martin1998,kothes2002}.

\begin{figure}[H]
\centering
\includegraphics[scale=.40]{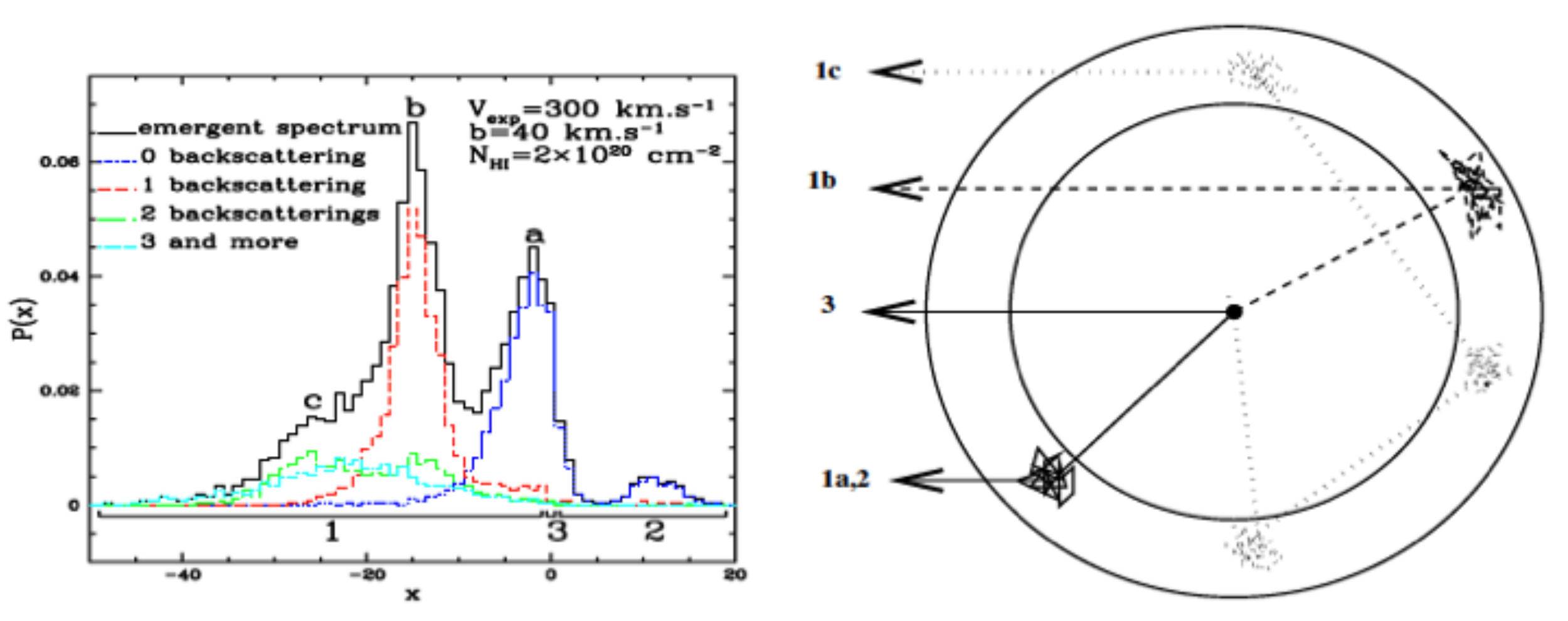}
\caption{
Left: Ly$\alpha$ profile predicted by the ES model \citep{verhamme2006}. 
Here, the abscissa axis is in units of the normalized velocity of the shell defined by $x=-2V_{\rm out}/b$ (see text).
A negative $x$ value indicates a wavelength longer than the systemic velocity ($x=0$).
The black line shows the total Ly$\alpha$ line profile that can be observed.
The blue, red, green, and cyan lines indicate the spectral components 
corresponding to the number of scattering events 
of none, once, twice, and 3 times.
The labels 1, 3, and 2 below the spectra represent the regimes of 
long, systemic, and short wavelengths.
Right: Illustration of the ES model \citep{verhamme2006}. The central dot indicates
the Ly$\alpha$ source position, while the annulus made of the two black circles represents the {\sc Hi} gas shell.
The solid, dashed, and dotted line arrows show examples of Ly$\alpha$ photon light paths (towards the observer)
that produce the spectral components whose labels correspond to those in the left panel.
This figure is reproduced by permission of the A\&A.
}
\label{fig:verhamme2006_fig12}       
\end{figure}

Figure \ref{fig:verhamme2006_fig12} illustrates the ES model and
predicted profiles of Ly$\alpha$ emission escaping to
the observer. The physical origins of the individual Ly$\alpha$ profiles
are explained below.
%
%
The light path "3" produces the profile "3",
where Ly$\alpha$ photons travel straight to the observer.
Note, however, that the profile is slightly redshifted 
because the blue side of the Ly$\alpha$ emission 
is efficiently scattered off by the \hi\ gas of the ES. 
%
The light path "1b" is back-scattered once by the ES,
providing a strong, redshifted peak in the predicted profile.
The velocity of the peak, $\sim 2V_{\rm out}$, is accomplished 
by two effects: 
(i) Ly$\alpha$ photons are scattered by the gas receding with 
$V_{\rm out}$ and hence their wavelengths are redshifted by 
$V_{\rm out}$ as seen from the gas, 
and (ii) the gas is receding from the observer by $V_{\rm out}$.
The light path "1c" indicates multiple scattering of Ly$\alpha$ photons that
gives the highly redshifted Ly$\alpha$ profile, but its contribution to the total flux 
is small in the reasonable range of \hi\ column density.

Back-scattered light dominates the total Ly$\alpha$ flux 
when the \hi\ column density is higher than 
$N_{\rm HII}\sim 10^{20}$ cm$^{-2}$. 
Therefore, the velocity offset of the total flux, 
$\Delta V_{\rm Ly\alpha}$, changes with $N_{\rm HII}$ 
from $\Delta V_{\rm Ly\alpha} \sim 0$ to $\sim 2V_{\rm out}$.
The value of $\Delta V_{\rm Ly\alpha} \sim 0$ is found
in low $N_{\rm HI}$ where the majority of Ly$\alpha$ photons 
take the path "3", 
while $\Delta V_{\rm Ly\alpha}$ has $\sim 2V_{\rm out}$ 
for $N_{\rm HI} \gtrsim 10^{20}$ cm$^{-2}$.
%
%

%
As demonstrated in Figure \ref{fig:verhamme2008_fig1_8_5},
the best-fit ES models reproduce the variety of Ly$\alpha$ profiles
with the only four physical parameters.


The ES models also explain the $\Delta V_{\rm Ly\alpha}$-$V_{\rm out}$ relations
of LAEs and LBGs. 
Because LAEs have the relation of $\Delta V_{\rm Ly\alpha} \simeq V_{\rm out}$,
the ES models suggest that their \hi\ column density is low, $N_{\rm HII}\lesssim 10^{20}$ cm$^{-2}$,
which produces weak back-scattered Ly$\alpha$ emission 
\citep{hashimoto2013,shibuya2014b,hashimoto2015}.
In contrast, LBGs have the relation of $\Delta V_{\rm Ly\alpha} \simeq 2 V_{\rm out}$.
The ES models indicate that back-scattered Ly$\alpha$ emission dominates
in LBGs, and that
their \hi\ column density 
is $N_{\rm HII}\gtrsim 10^{20}$ cm$^{-2}$ on average
that is higher than those of LAEs.

The low \hi\ column densities of LAEs may explain the 
large increase in the average Ly$\alpha$ escape fraction 
from $z\sim 0$ to $6$ 
shown in Figure \ref{fig:hayes2011a_fig1},
because the fraction of LAEs in the entire galaxy population increases
with redshift
(Figure \ref{fig:stark2011_fig2_stark2010_fig13}).
%
%
There are six possible mechanisms that control
the Ly$\alpha$ escape fraction:
1) IGM absorption,
2) stellar population,
3) outflow velocity,
4) gas-cloud clumpiness (Neufeld effect),
5) simple dust extinction, and
6) \hi\ gas resonance scattering in the ISM with dust.
Because the IGM absorption is stronger at higher $z$, 
mechanism 1) cannot explain
the increase in Ly$\alpha$ escape fraction towards high $z$.
The stellar population and outflow velocity of the mechanisms 2) and 3) 
evolve little for LAEs in the range $z\sim 3-6$ (Sections \ref{sec:stellar_population} and \ref{sec:outflow_lya_profile}),
which are not large enough to explain the evolution of two orders of magnitude
of the Ly$\alpha$ escape fraction in the range $z\sim 0-6$ (Figure \ref{fig:hayes2011a_fig1}). 
The mechanism 4) has been ruled out by
recent theoretical studies 
(Section \ref{sec:lya_escape_fraction}; \citealt{laursen2013,duval2014}).

\begin{figure}[H]
\centering
\includegraphics[scale=.45]{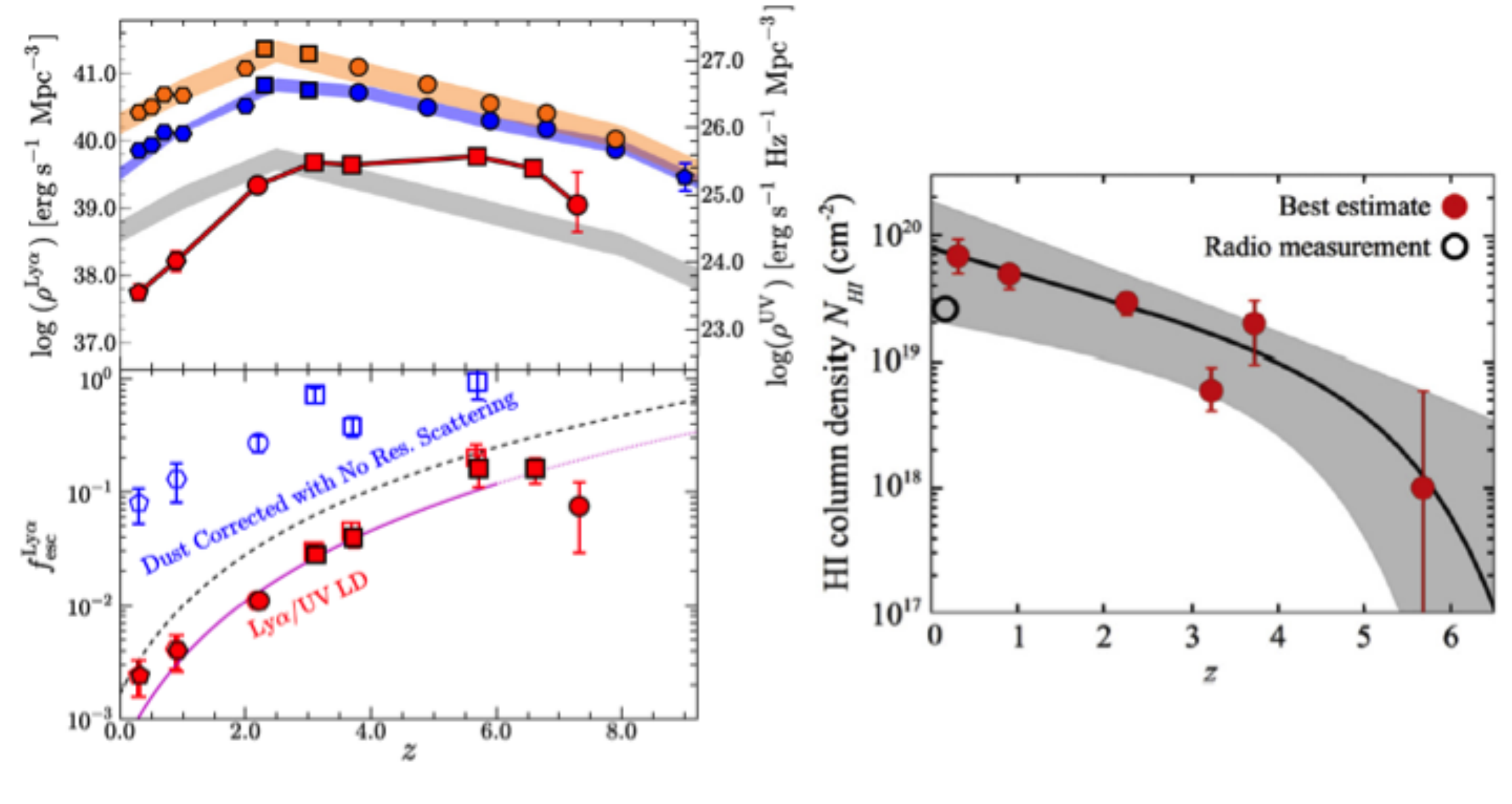}
\caption{
Top left: Evolution of the Ly$\alpha$ and UV luminosity densities whose
absolute values are presented in the labels in the left and right ordinate axes, respectively.
The red symbols and line indicate the Ly$\alpha$ luminosity density.
The blue and orange symbols/shades represent the observed and dust-corrected UV luminosity densities,
respectively. For comparison, the gray shade denotes the dust-corrected 
UV luminosity density scaled to the position of the Ly$\alpha$ luminosity density at $z\sim 3$.
Bottom left: Evolution of the Ly$\alpha$ escape fraction derived from the observed Ly$\alpha$ and dust-corrected
luminosity densities (red filled symbols). The best-fit function to the Ly$\alpha$ escape fraction evolution
is shown with the magenta and the black-dashed lines obtained by
\citet{konno2016} and \citet{hayes2011a}, respectively.
The blue symbols represent the Ly$\alpha$ escape fraction 
corrected for dust extinction under the simple assumption of no Ly$\alpha$ resonance scattering.
Right: Evolution of the average {\sc Hi} column density of galaxies inferred from
the Ly$\alpha$ escape fraction and the ES models (red circles). The black solid line
and gray shade are the best-fit function with the outflow velocity of 150 km s$^{-1}$ 
and the uncertainty raised by the different assumptions of the outflow velocity ranging from 50 to 200 km s$^{-1}$. 
The black open circle shows the average {\sc Hi} column density obtained by radio observations.
All of these plots are taken from \citet{konno2016}.
This figure is reproduced by permission of the AAS.
}
\label{fig:konno2016_fig7_fig10}       
\end{figure}

The top left panel of Figure \ref{fig:konno2016_fig7_fig10} presents 
the evolution of the Ly$\alpha$ escape fraction
corrected for dust extinction with a simple screen model (eq. \ref{eq:fesc_lya_indiv_screen}).
This simple dust extinction evolution of the mechanism 5) 
is found to predict only one order of magnitude evolution of 
the Ly$\alpha$ escape fraction.
Instead, the large evolution of the Ly$\alpha$ escape fraction can be probably 
explained by the mechanism 6). The mechanism 6) involves
\hi\ gas evolution with the effect of dust extinction in the ISM via Ly$\alpha$ resonance scattering. 
To reproduce the large Ly$\alpha$ escape fraction evolution based on
the ES models, it is suggested that the \hi\ column density should decrease by 1-2 orders of magnitude (Figure 
\ref{fig:konno2016_fig7_fig10})
from $z\sim 0$ to $6$ that reduces the effect of selective Ly$\alpha$ absorption for LAEs towards $z\sim 6$.
In this way, the \hi\ column density evolution is a major parameter of LAEs that determines 
not only $\Delta V_{\rm Ly\alpha}$, but also the Ly$\alpha$ escape fraction.

\subsection{AGN Activity}
\label{sec:AGN_activity}

\begin{figure}[H]
\centering
\includegraphics[scale=.40]{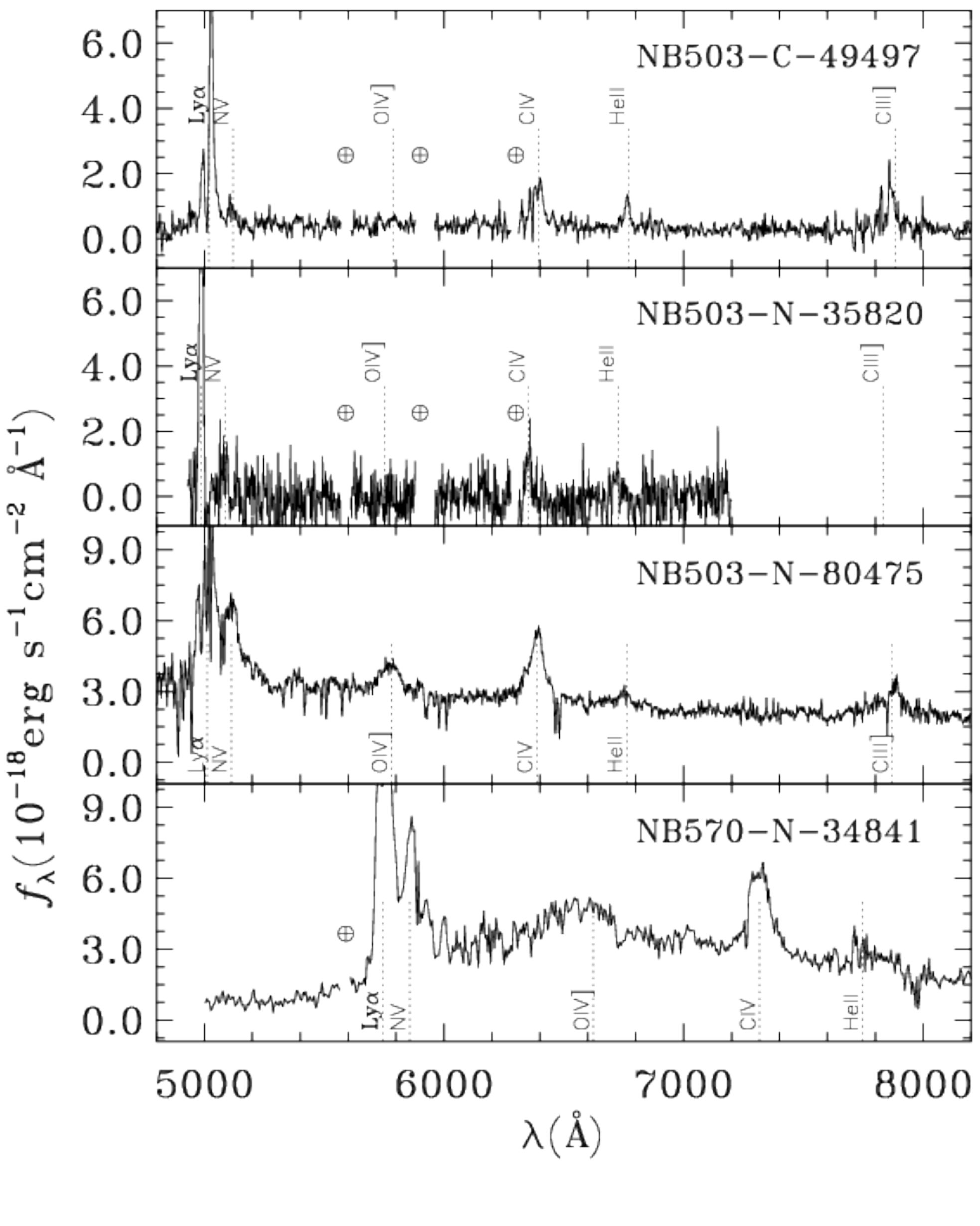}
\caption{
Example spectra of AGN-LAEs \citep{ouchi2008}.
This figure is reproduced by permission of the AAS.
}
\label{fig:ouchi2008_fig13}       
\end{figure}

LAE searches often find AGNs with strong Ly$\alpha$ emission
from the presence of 
broad ($\gtrsim 500$km s$^{-1}$) emission lines and
high ionization lines such as {\civ}1548 and {\nv}1240
as well as strong X-ray, far-UV, radio, and short-wavelength IR emission
(Figure \ref{fig:ouchi2008_fig13}). 
Diagnostics with nebular line ratios such as the BPT \citep{baldwin1981} 
diagram are also used 
\citep{finkelstein2011a,nakajima2013,guaita2013}.
AGNs with strong Ly$\alpha$ emission are referred to as AGN-LAEs. 

\begin{figure}[H]
\centering
\includegraphics[scale=.40]{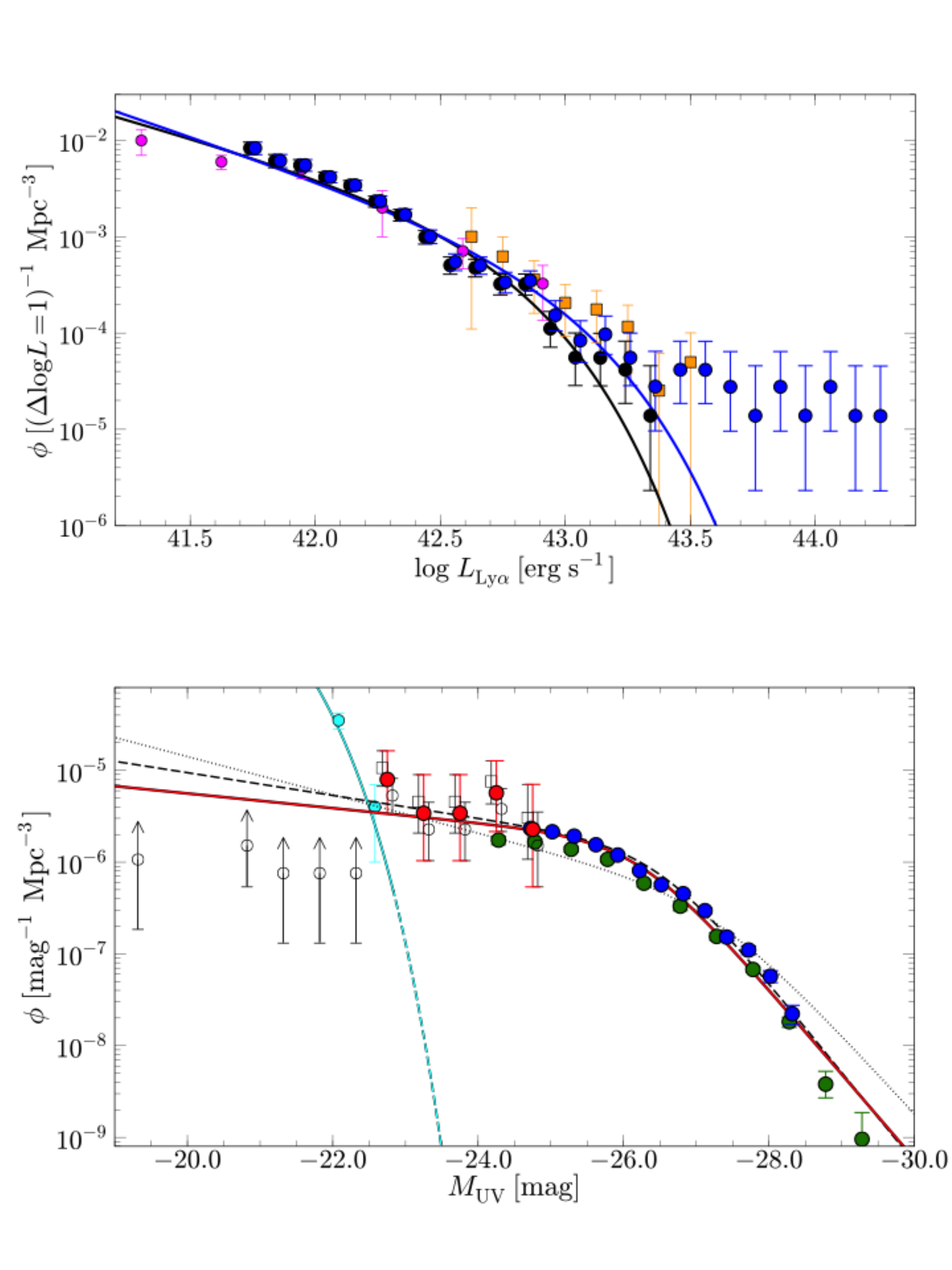}
\caption{
Top: Ly$\alpha$ luminosity function at $z=2$.
The blue and magenta (orange) circles (squares) represent
Ly$\alpha$ luminosity functions derived by three different studies.
The black circles are the same as the blue circles, but
based on LAEs with no AGN signature, i.e., 
with neither strong X-ray, far-UV, nor radio emission.
The blue and black curves are the best-fit Schechter functions
with the blue and black circle data points, respectively.
Bottom: UV luminosity function of AGN at $z=2$.
The red circles indicate the AGN UV luminosity function estimated
with the LAE-AGN whose errors include 
the systematic uncertainty 
raised by the incompleteness of AGN missed 
from the LAE selection because of weak or no Ly$\alpha$ emission.
The blue and green circles represent the AGN UV luminosity functions derived
with the SDSS and 2dF-SDSS LRG data, respectively.
The red curve shows the best-fit Schechter function.
The cyan circles and curve denote the UV luminosity function of LBGs.
These two plots are taken from \citet{konno2016}.
This figure is reproduced by permission of the AAS.
}
\label{fig:konno2016_fig3_fig8}       
\end{figure}

Early studies claim that about 1\% of narrowband-selected LAEs at $z\sim 2-3$
are AGN-LAEs, and that the AGN-LAE fraction increases 
with the Ly$\alpha$ luminosity
\citep{gawiser2007,ouchi2008}.
Recently, \citet{konno2016} 
have obtained statistical results on AGN-LAEs at $z=2$
based on a narrowband survey (see also \citealt{matthee2017a}).
The top panel of Figure \ref{fig:konno2016_fig3_fig8} presents the Ly$\alpha$ luminosity function of LAEs
that clearly shows an excess over the best-fit Schechter function 
at $\log L_{\rm Ly\alpha}\gtrsim 43.4$ erg s$^{-1}$.
Almost all objects in this luminosity range are bright 
in either X-ray, far-UV, or radio, thus being classified as AGN-LAEs.
The bottom panel of Figure \ref{fig:konno2016_fig3_fig8} is 
the AGN UV luminosity function derived with these AGN-LAEs, 
where the moderately large error bars 
include the systematic uncertainty 
raised by the incompleteness of AGN missed 
from the LAE selection because of weak or no Ly$\alpha$ emission.
The AGN UV luminosity function estimated from the AGN-LAE sample
agrees well with the faint-end UV luminosity function given by the SDSS study.
In summary, the bright end of the Ly$\alpha$ luminosity function ($\log L_{\rm Ly\alpha}\gtrsim 43.4$ erg s$^{-1}$) 
is dominated by AGN-LAEs, and AGN-LAEs may be used to estimate
the AGN UV luminosity function at the faint end.

\subsection{Overdensity and Large-Scale Structure}
\label{sec:overdensity}

\begin{figure}[H]
\centering
\includegraphics[scale=.40]{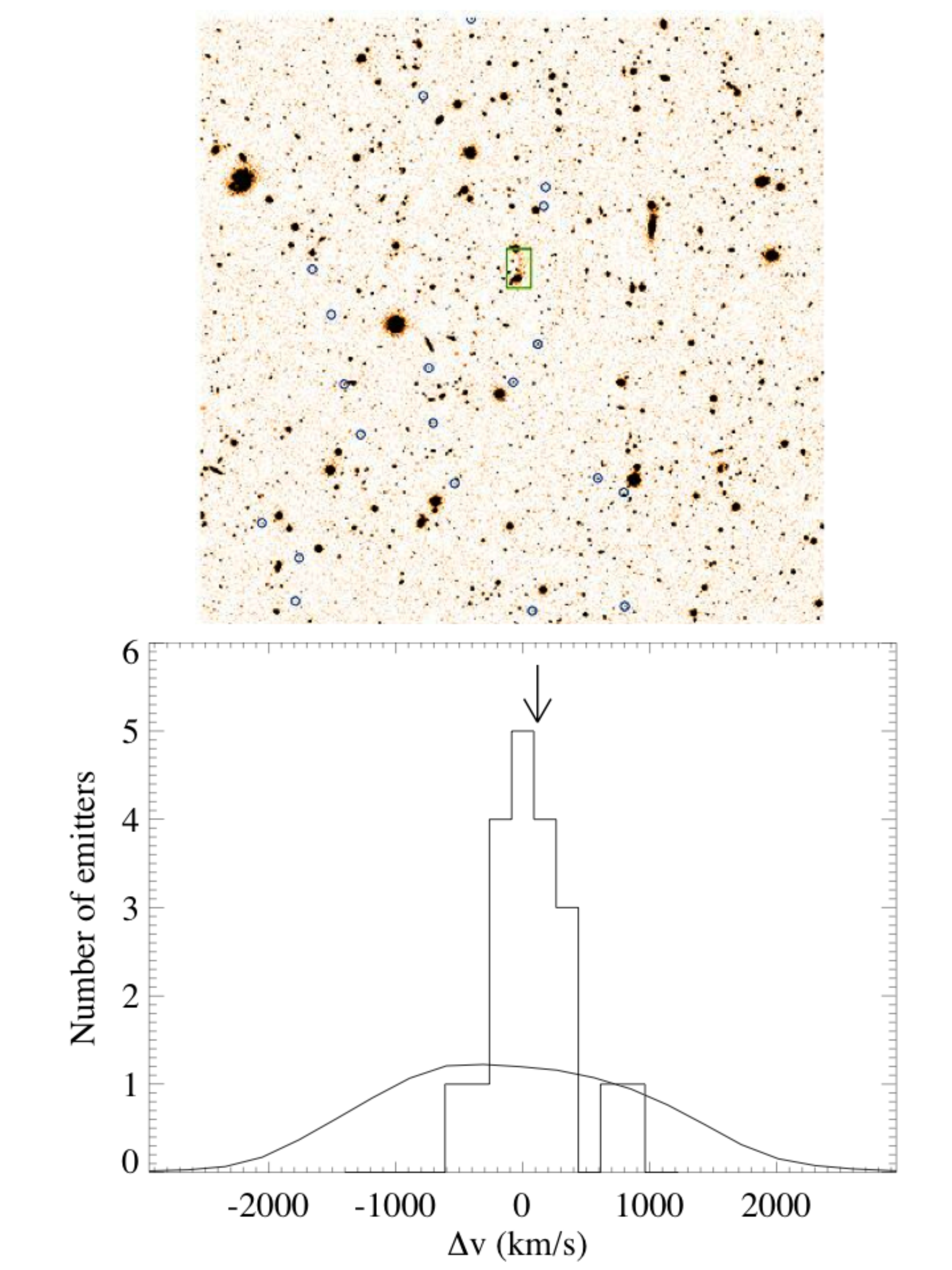}
\caption{
Overdensity of LAEs around the radio galaxy TN J1338 1942.
Top: VLT narrowband image centered at TN J1338 1942.
The green box and the blue circles indicate
the positions of TN J1338 1942 and LAEs, respectively.
The image is taken from https://www.eso.org/public/news/eso0212/ .
Bottom: Histogram of velocity differences $\Delta_v$ between LAEs
and TN J1338 1942 \citep{venemans2002}. 
This figure is reproduced by permission of the A\&A.
}
\label{fig:venemans2002_eso_fig2}      
\end{figure}

As explained in Section \ref{sec:discoveries},
early LAE searches targeted fields centered on
an AGN, especially a massive radio galaxy, that is 
thought to be a signpost of a high-$z$ galaxy overdensity.
I present one of the LAE overdensity examples in
Figure \ref{fig:venemans2002_eso_fig2} that 
\noindent
shows narrowband selected LAEs
around a luminous radio galaxy, TN J1338 1942, 
at $z\sim 4.1$ \citep{venemans2002}.
The number density of LAEs in this overdense region 
is about 15 times higher than the one in the average blank field.
The number-density peak is clearly 
found at the redshift of TN J1338 1942, while
the narrowband is capable to detect LAEs 
in a moderately broad redshift range (bottom panel of Figure \ref{fig:venemans2002_eso_fig2}).
Such an overdensity of high-$z$ galaxies is often refereed to as a 'proto-cluster' \citep{steidel2000}.
There are about 30 overdensities of high-$z$ galaxies in the range $z\sim 2-8$ reported to date 
(see Table 5 of \citealt{chiang2013}). About a half of them are LAE overdensities
like the one around TN J1338 1942.

\begin{figure}[H]
\centering
\includegraphics[scale=.60,angle=90]{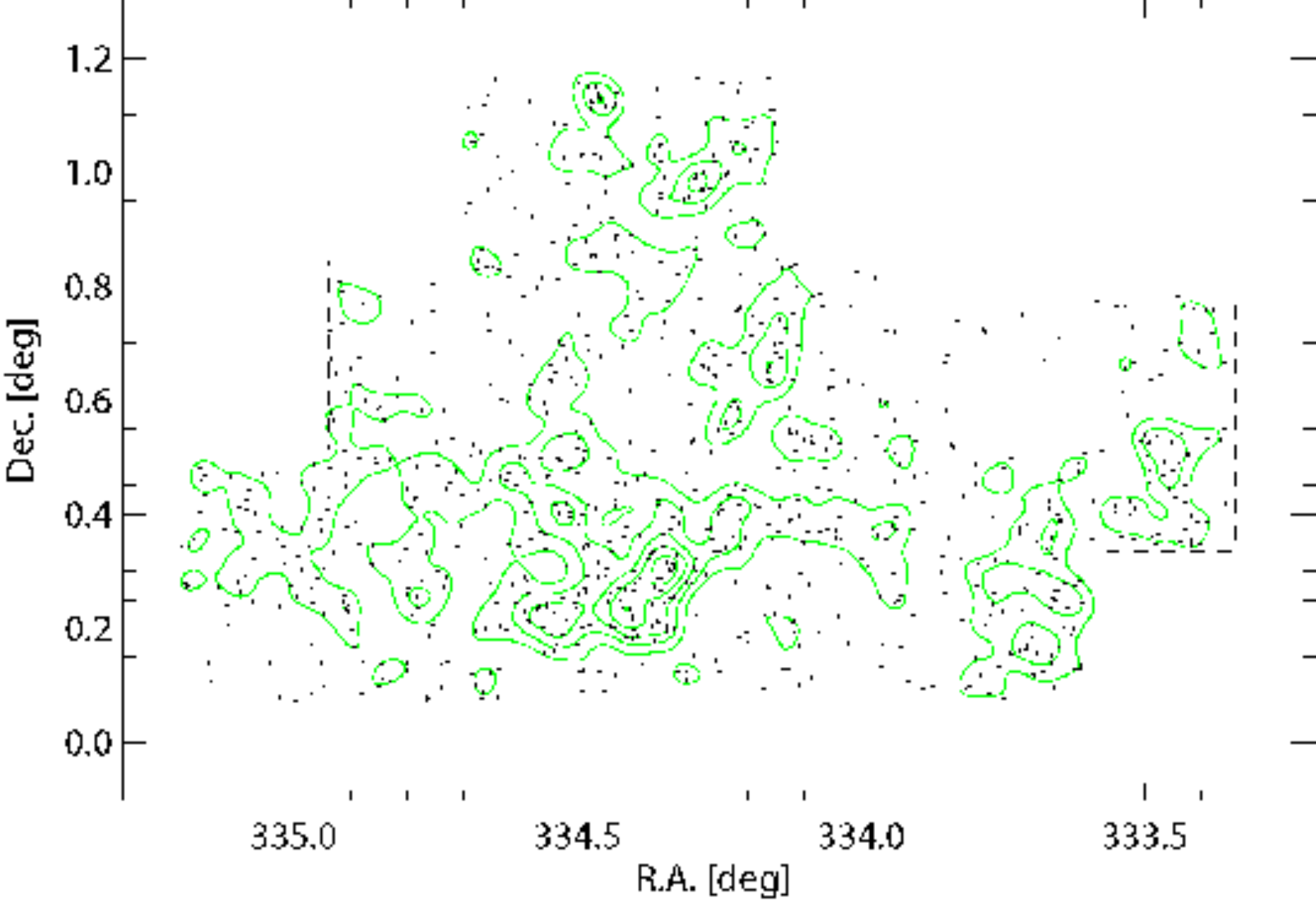}
\caption{
Large-scale ($>100$ Mpc) sky distribution of LAEs in and around the 
SSA22 overdensity of galaxies at $z=3.1$ \citep{yamada2012a}.
The dots indicate LAEs, while the green contours
represent the surface density of LAEs.
This figure is reproduced by permission of the AAS.
}
\label{fig:yamada2012a_fig11}
\end{figure}

Some systematic narrowband surveys of LAEs have covered a contiguous field 
with a size of $>100$ comoving Mpc, and discovered filamentary LSSs
in a flanking field of a 'proto-cluster' at $z\simeq 3$ (Figure \ref{fig:yamada2012a_fig11}; \citealt{yamada2012a}) 
and in a blank field at $z\simeq 6$ \citep{ouchi2005a}.
It is interesting that narrowband surveys can efficiently map out
a high-$z$ galaxy distribution in a large scale. Specifically, a mapping observation
in an unbiased blank field allows to obtain average features of LAE clustering
that can be used to constrain properties of the dark-matter halos hosting LAEs, with structure formation models (Section \ref{sec:clustering}).

\subsection{Clustering}
\label{sec:clustering}

\begin{figure}[H]
\centering
\includegraphics[scale=.60]{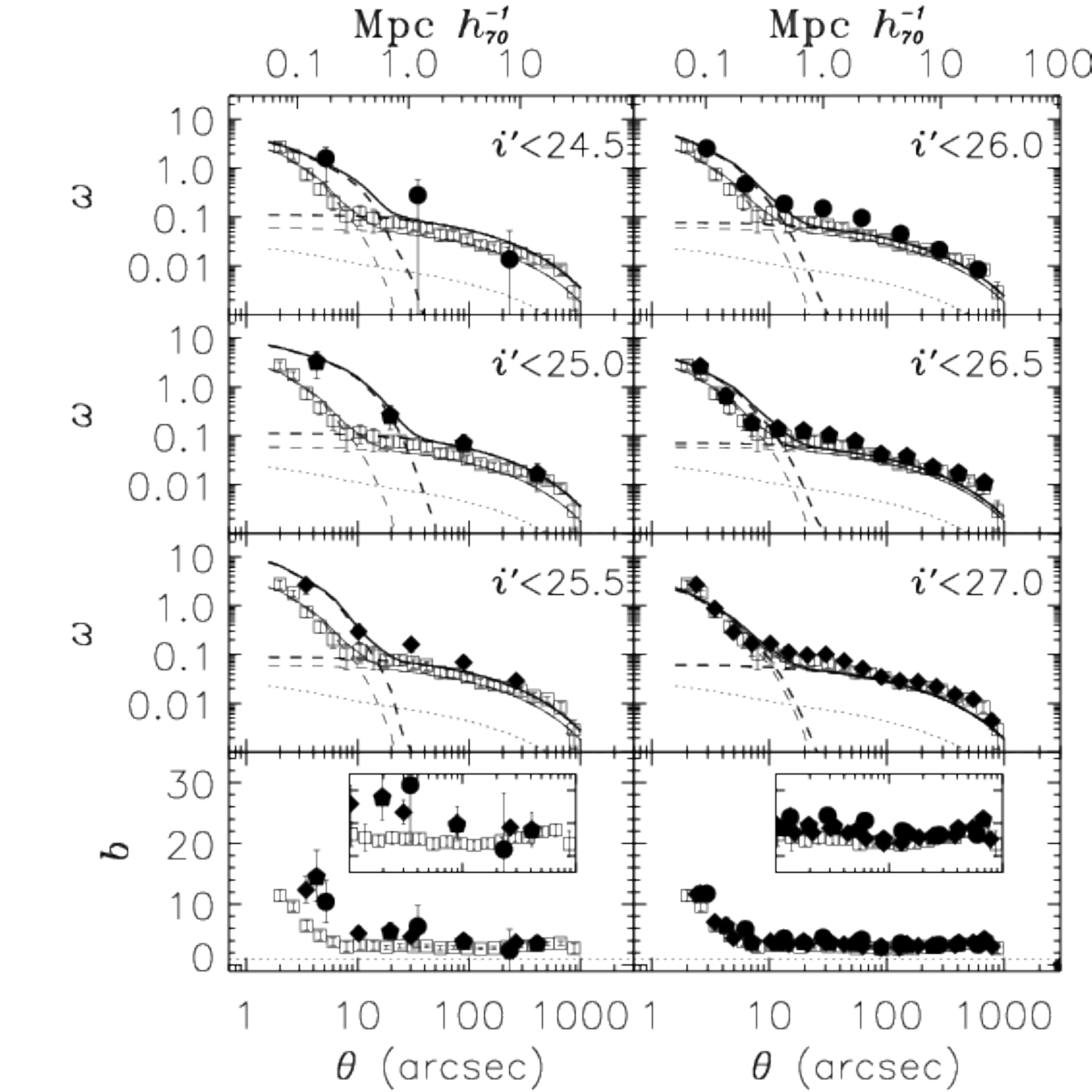}
\caption{
Angular correlation function (top three panels) and clustering bias (bottom panel)
as a function of angular distance for $z \sim 4$ LBGs \citep{ouchi2005b}. The open squares denote
the angular correlation function of objects with $i'<27.5$, 
while the filled circles are those for 
different magnitude limits indicated by labels.
The solid lines are the best-fit HOD models, while the dashed lines
indicate their breakdowns into one-halo and two-halo terms that are 
the correlations of galaxies in single halos and two different halos,
respectively. The dotted lines represent the DM angular correlation function (bias = 1).
This figure is reproduced by permission of the AAS.
}
\label{fig:ouchi2005b_fig3}      
\end{figure}

\begin{figure}[H]
\centering
\includegraphics[scale=.80]{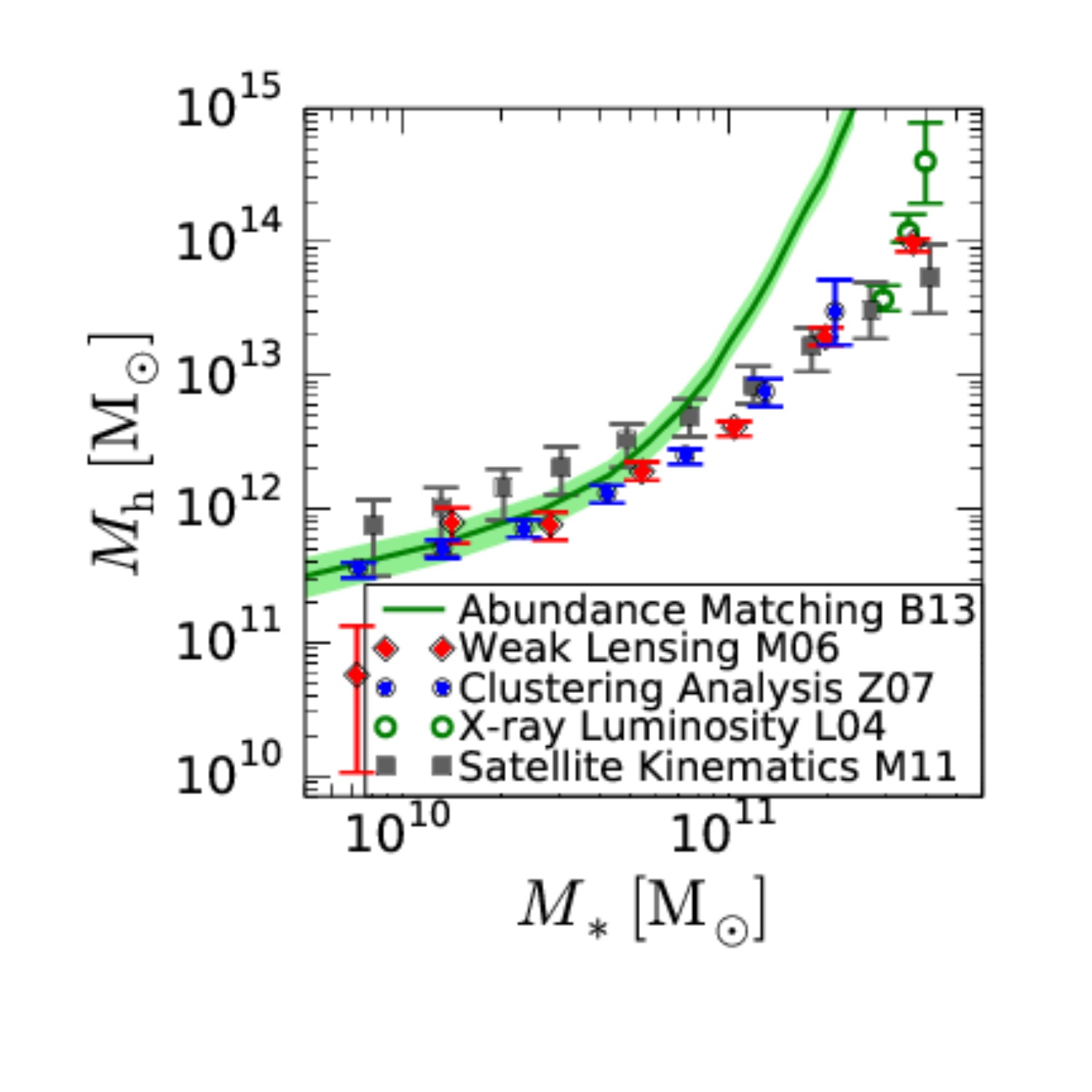}
\caption{
DM-halo masses estimated by five methods plotted as a function of stellar mass \citep{harikane2016a}.
The red diamonds represent estimates by the reliable
lensing technique. The blue circles, green open circles, and gray squares
denote DM-halo masses estimated from clustering, X-ray luminosity,
and satellite kinematics. The green line with a green shade indicate
the DM-halo mass estimated by the abundance matching technique.
This figure is reproduced by permission of the University of Tokyo.
}
\label{fig:harikane2016a_fig2imp}      
\end{figure}

Properties of galaxy-hosting DM halos are
key to understanding galaxy formation 
(Figure \ref{fig:DMH_SF}; Section \ref{sec:basic_picture}).
Invisible DM halos are not directly observed,
but there are various indirect methods to estimate their masses, $M_h$.
One promising approach is to use gravitational lensing
that provides reliable estimates of $M_h$. 
Other approaches include abundance matching and clustering
\footnote{
There are a number of methods to characterize hosting DM halos: e.g., 
satellite kinematics, X-ray luminosities, and 
Sunyaev-Zel'dovich effects. Here, I highlight
only lensing, abundance matching, and clustering
that require optical and NIR imaging data alone. 
}.
The abundance matching and clustering methods 
statistically estimate an average $M_h$ for a given galaxy population,
by comparing the abundance and the clustering amplitude, 
respectively, between galaxies and DM halos.
The abundance and the clustering amplitude are chosen 
as an indicator of $M_h$ because they are reliably calculated 
as a function of $M_h$ using the standard $\Lambda$CDM 
structure formation model (abundance: Figure \ref{fig:ishiyama2015_fig3}; 
clustering: Figure \ref{fig:ouchi2005b_fig3}) 
at any redshifts (Section \ref{sec:DM_halo}).
The basic idea of the abundance matching (clustering) method is 
to identify the ensemble of model dark-matter halos whose number density 
(clustering amplitude) is equal to that of observed galaxies.
Modern abundance matching methods take account of the contribution of sub-halos to
explain satellite galaxies as well as star-formation and merger histories,
and increase the reliability of $M_h$ estimates (e.g. \citealt{behroozi2013}).
Similarly, for the clustering method, it is popular to use clustering predictions
of the halo occupation distribution (HOD) of central and satellite galaxies
(e.g. \citealt{zheng2005}).
Figure \ref{fig:harikane2016a_fig2imp} compares $M_h$ of local galaxies
estimated by various techniques, and indicates that the results of the three techniques,
lensing, clustering, and abundance matching, agree very well 
in the galaxy DM halo mass scale up to $M_h \sim 10^{13} M_\odot$.
Table \ref{tab:DMhalo_technique} summarizes
the three techniques.

\begin{table}
\caption{DM-Halo Mass Estimate Techniques}
\label{tab:DMhalo_technique}       
%
%
\begin{tabular}{p{1.5cm}p{3cm}p{4cm}p{4cm}p{1.0cm}}
\hline\noalign{\smallskip}
Technique  & Key Quantity &  Advantage & Disadvantage (Requirement) & Redshift Range$^a$\\
\noalign{\smallskip}\svhline\noalign{\smallskip}
Lensing 				& Background object shear & Moderately simple gravity model & Large galaxy sample of high spatial resolution imaging data & $0-1$\\
Clustering 			& Correlation function 		& Virtually free from duty-cycle systematics & Large galaxy sample & $0-7$\\
Abundance Matching	& Luminosity function 		& Small galaxy sample & Many parameters constrained by star-formation/merger histories & $0-8$\\
\noalign{\smallskip}\hline\noalign{\smallskip}
\end{tabular}

$^a$ Redshift range that is covered by observations, to date.
\end{table}

\begin{figure}[H]
\centering
\includegraphics[scale=.50]{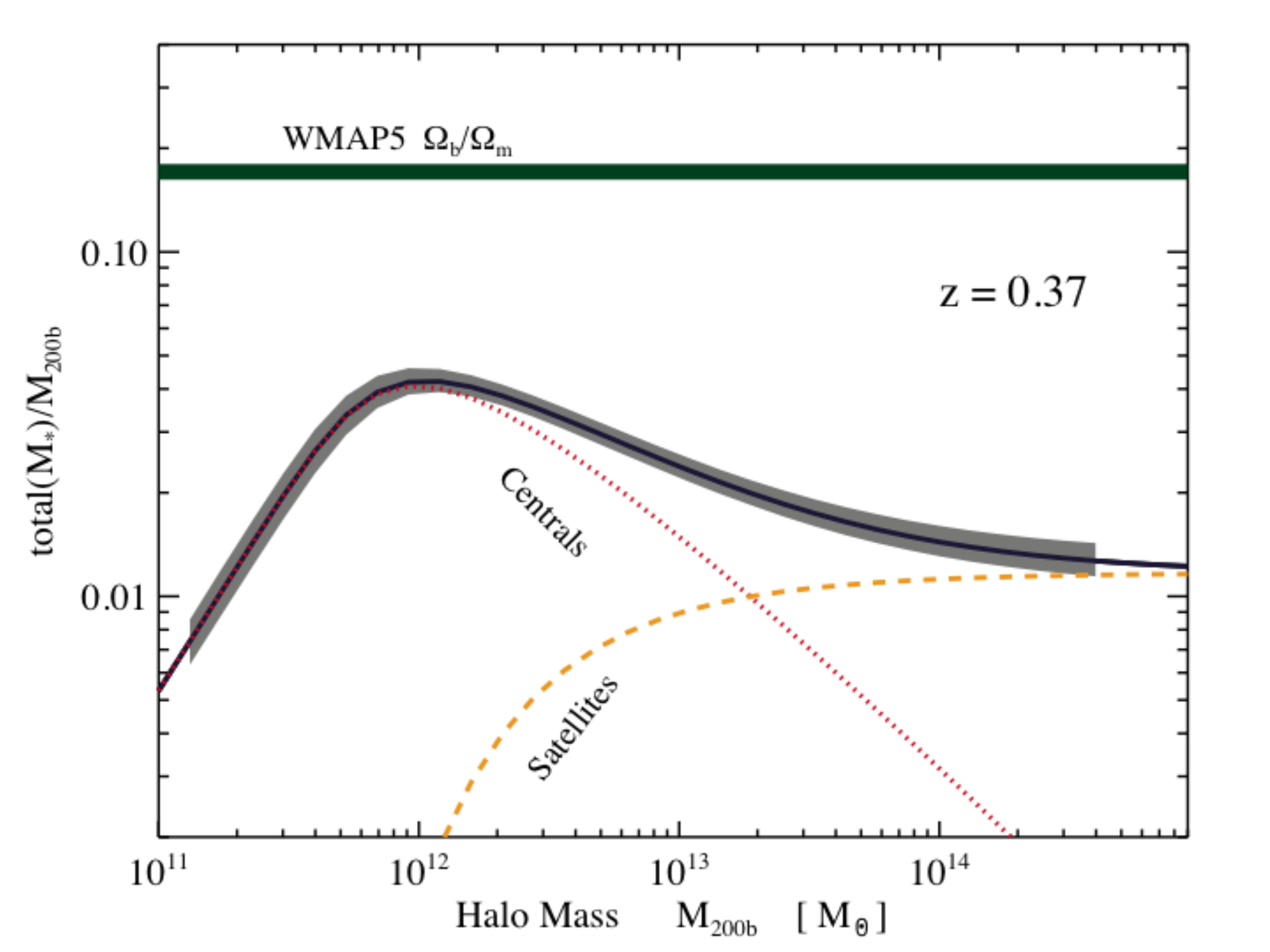}
\caption{
SHMR as a function of DM-halo mass at $z\sim 0$ \citep{leauthaud2012}.
The black line and the gray shade represent the total SHMR and its uncertainty,
while the red-dotted and orange-dashed lines indicate the SHMRs contributed by
central and satellite galaxies, respectively.
The green horizontal line denotes the cosmic baryon mass to total mass ratio. 
This figure is reproduced by permission of the AAS.
}
\label{fig:leauthaud2012_fig16}     
\end{figure}

\begin{figure}[H]
\centering
\includegraphics[scale=1.0]{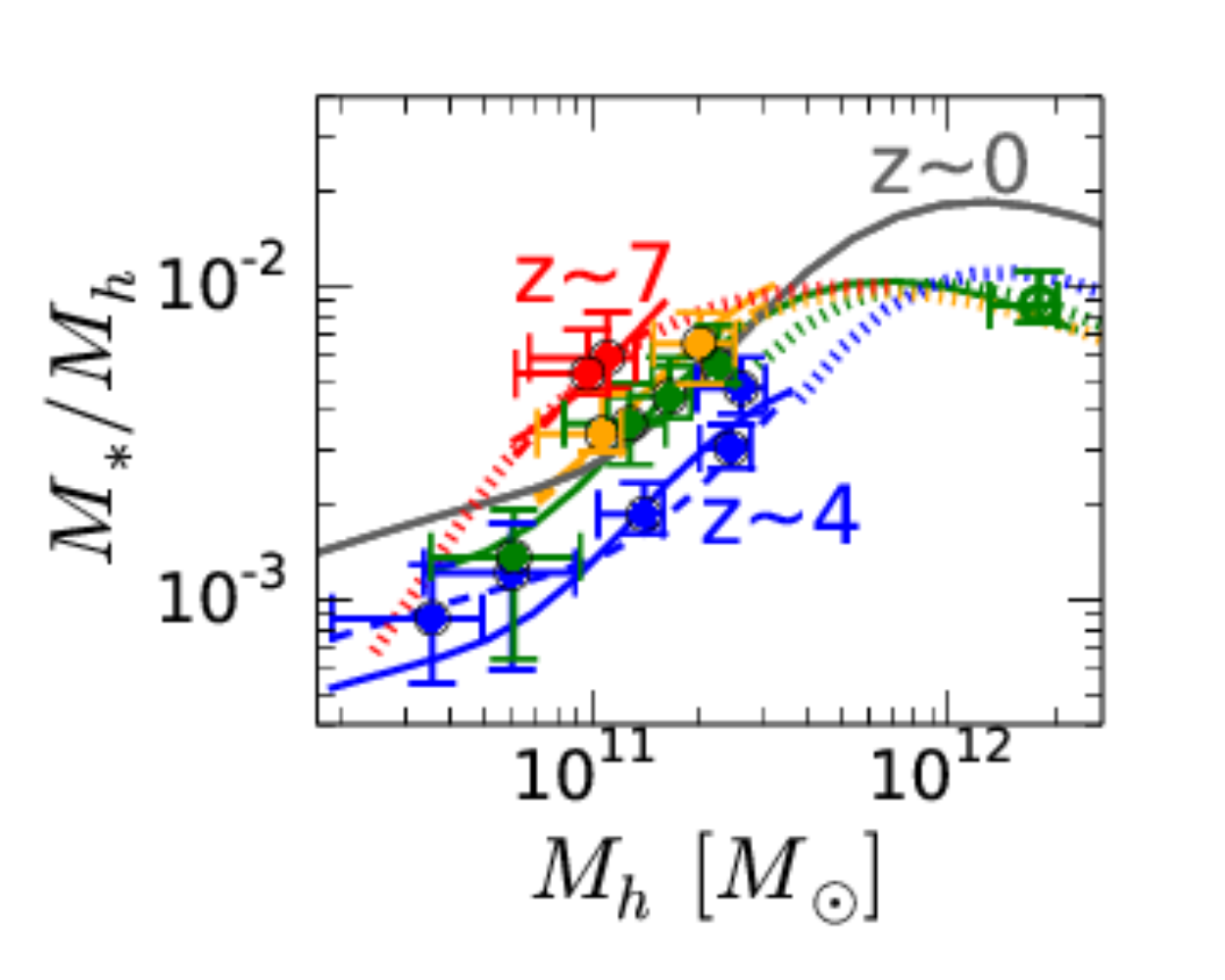}
\caption{
Redshift evolution of the SHMR over $z=0-7$ 
(\citealt{harikane2016b}; see also \citealt{harikane2018a}).
The red, orange, green, and blue circles and solid lines indicate
SHMRs at $z\sim 7$, $6$, $5$, and $4$, respectively,
derived from a combination of the clustering and 
abundance matching techniques. The dotted lines are the same as the solid lines,
but for SHMRs obtained by the abundance matching technique alone.
The gray solid line denotes the SHMR at $z\sim 0$.
This figure is reproduced by permission of the AAS.
}
\label{fig:harikane2016b_fig10imp}    
\end{figure}

Figure \ref{fig:leauthaud2012_fig16} presents the stellar to DM-halo mass ratio (SHMR)
as a function of $M_h$ for $z\sim 0$ galaxies. This result is obtained by
the combination of these three techniques that enhances the reliability
of $M_h$ estimates. In Figure \ref{fig:leauthaud2012_fig16}, 
the SHMR has a peak at $M_h\sim 10^{12} M_\odot$,
meaning that stars are most efficiently formed in DM halos  
whose present-day mass is $M_h\sim 10^{12} M_\odot$. The shape of the SHMR plot
reflects $M_h$-dependent gas cooling and feedback
and thus is essential to understand star-formation
processes in DM halos (Section \ref{sec:star_formation}).
Properties of galaxy-hosting DM halos are investigated by a combination of these three techniques
up to $z\sim 1$ in, e.g., the COSMOS field where wide and deep HST data are
available \citep{leauthaud2012}.
However, for more distant galaxies at $z\gtrsim 2$,
it is difficult to apply the lensing technique that requires 
shear measurements of a large number of background objects 
in high-sensitivity and high-spatial resolution images.
In contrast, the clustering and abundance matching techniques can still 
be used at $z\gtrsim 2$.
Figure \ref{fig:harikane2016b_fig10imp}
presents the evolution of the SHMR obtained with HST and Subaru data
by these two techniques. Although the
DM-halo mass range is limited, a clear evolution
of the SHMR is identified at $M_h\sim 10^{11}M_\odot$.
Beyond $z\sim 7$, there are no clustering measurements obtained to date,
because clustering analyses require a large number of galaxies.
Only requiring the number density of galaxies,
the abundance matching technique has been applied up to $z\sim 8$ to date
\citep{behroozi2013}.
Figure \ref{fig:harikane2016b_fig10imp} compares DM-halo masses estimated by
the abundance matching alone to those from the clustering+abundance matching, and
suggests that the abundance matching provides good estimates of $M_h$ at $z\gtrsim 4$
within an uncertainty of a factor of $\sim 3$ \citep{harikane2016b}.

DM-halo masses of LAEs have not been estimated by either lensing or 
abundance matching.
Lensing analyses cannot be performed for high-$z$ LAEs at $z\gtrsim 2$ due to the
limited quality and amount of imaging data. Moreover, abundance matching does not
work because LAEs have a very small duty cycle of strong Ly$\alpha$ emission
and hence have a significantly smaller abundance than DM halos 
(Section \ref{sec:duty_cycle}).
Thus, only the clustering method has been used for LAEs.

\begin{figure}[H]
\centering
\includegraphics[scale=.60]{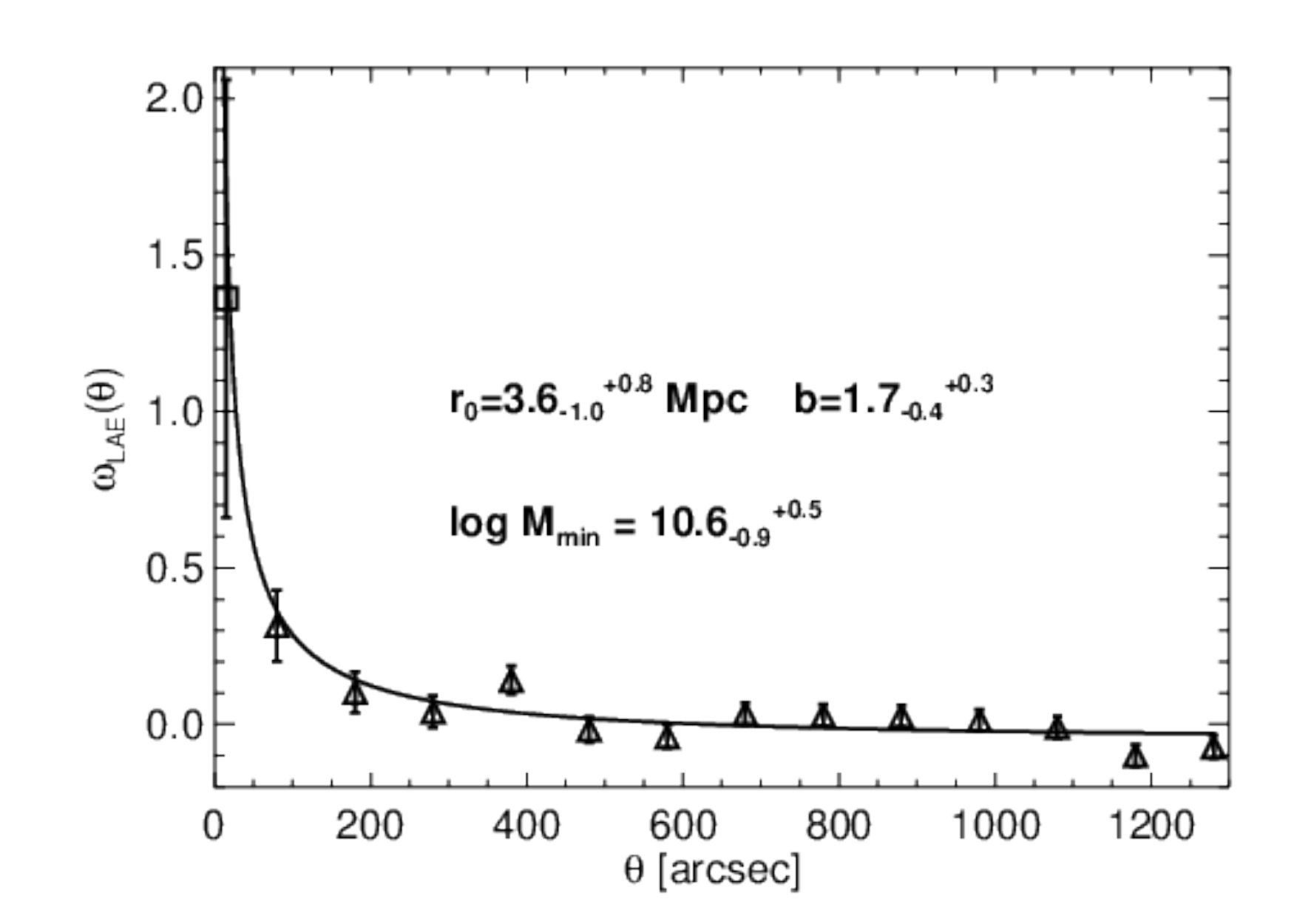}
\caption{
ACF of LAEs at $z=3.1$ \citep{gawiser2007}. The data points
and the best-fit power law function are shown with triangles
and a solid line, respectively. The square at $<30"$ is the data point that is not used for
the fitting.
This figure is reproduced by permission of the AAS.
}
\label{fig:gawiser2007_fig2}   
\end{figure}

\begin{figure}[H]
\centering
\includegraphics[scale=.50]{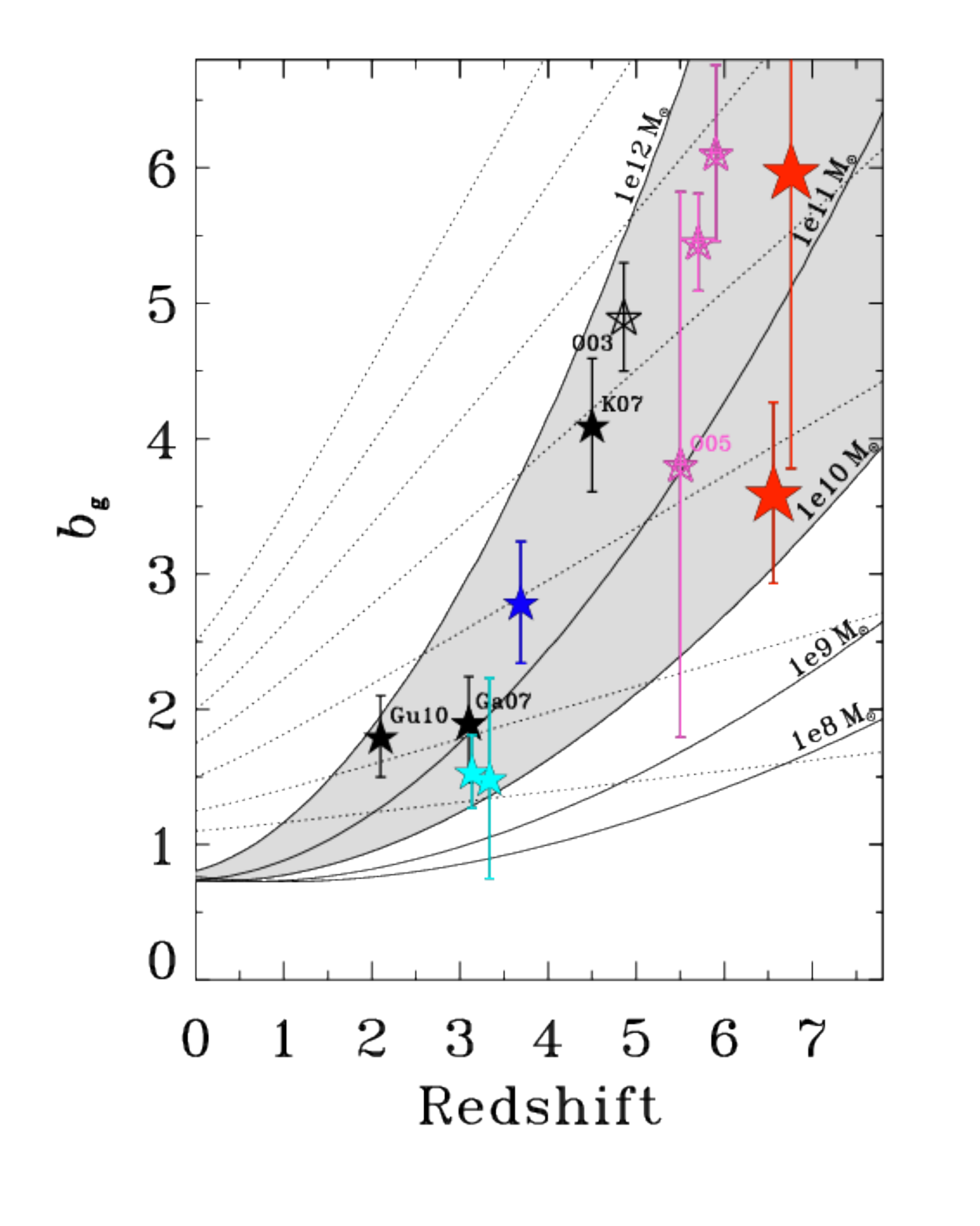}
\caption{
Bias of clustering $b_{\rm g}$ as a function of redshift \citep{ouchi2010}.
The star marks represent $b_{\rm g}$ measurements of LAEs with
$\gtrsim L_{\rm Ly\alpha}^*$ in the range $z=2-7$.
The solid lines represent $b_{\rm g}$ for DM halos with
a mass of $10^{8}$, $10^{9}$, $10^{10}$, $10^{11}$, and $10^{12} M_\odot$ 
predicted by the model of \citet{sheth1999}, 
while the gray area indicates the DM halo mass range of $10^{10}-10^{12} M_\odot$
corresponding to the typical DM halos of LAEs.
The dotted lines denote the evolutionary tracks of $b_{\rm g}$
for the galaxy-number conserving model.
This figure is reproduced by permission of the AAS.
}
\label{fig:ouchi2010_fig13}
\end{figure}

The clustering amplitude of a given galaxy sample can be evaluated 
with the angular correlation function (ACF), $\omega_{\rm obs} (\theta)$, defined 
as the excess probability of finding galaxies in two solid angles 
$d\Omega_1$ and $d\Omega_2$ separated by the angular distance $\theta$,
\begin{equation}
dP= N^2 [1+\omega_{\rm obs}(\theta)] d\Omega_1 d\Omega_2,
\label{eq:acf}
\end{equation}
where $dP$ is the probability finding galaxies and
$N$ the mean galaxy density per steradian \citep{groth1977}.
Large-area surveys have derived ACFs of LAEs with the \citet{landy1993}
estimator \citep{ouchi2003,gawiser2007,kovac2007}.
As an example, Figure \ref{fig:gawiser2007_fig2} presents ACF measurements 
for $z=3.1$ LAEs.
Although the statistics is not very good due to moderately
small sample sizes, it is well known that 
the clustering of LAEs is weak, $b_g\sim 2$, at $z\sim 2-3$. 
Here, $b_g$ is the large scale galaxy bias 
defined by
\begin{equation}
b_g^2= \omega_{\rm obs}/\omega_{\rm DM},
\label{eq:acf}
\end{equation}
where $\omega_{\rm DM}$ is the dark-matter ACF predicted by
the structure formation model (e.g. \citealt{peacock1996}). 
ACFs and $b_g$ have been derived for LAEs up to $z=6.6$.
Figure \ref{fig:ouchi2010_fig13} summarizes the bias of 
$z\simeq 2-7$ LAEs with $L_{\rm Ly\alpha} \gtrsim$ a few $\times 10^{42}$ erg s$^{-1}$.
Including the moderately large errors,
the estimated halo masses of LAEs
are typically $M_h=10^{11\pm 1} M_\odot$ over
$z\simeq 2-7$ \citep{ouchi2010}.
In Figure \ref{fig:ouchi2010_fig13}, $b_g$ increases with redshift, suggesting that
LAEs in earlier universes ($z\sim 5-7$) form from higher density fluctuation peaks 
and are progenitors of present-day massive 
\noindent
elliptical galaxies \citep{ouchi2010}.
On the other hand, the small $b_g$ of LAEs at $z\sim 2-3$ indicate that 
LAEs at $z\sim 2-3$ may be progenitors of today's Milky-Way like galaxies
based on the average evolution of $b_g$ (Figure \ref{fig:ouchi2010_fig13}; \citealt{gawiser2007,ouchi2010}).
Note that, due to the relatively small samples, the ACF measurements of LAEs still include large statistical errors
as found from a comparison of Figure \ref{fig:gawiser2007_fig2} and 
the bottom right panel of Figure \ref{fig:ouchi2005b_fig3}.
So far, no studies of LAE clustering 
have identified the one-halo term made by LAEs 
residing in single halos (see Figure \ref{fig:ouchi2005b_fig3} and the caption).
There remains an open question whether LAEs have a moderately strong one-halo term
similar to continuum-selected galaxies (Figure \ref{fig:ouchi2005b_fig3}).

\subsection{Ly$\alpha$ Duty Cycle}
\label{sec:duty_cycle}

%
%
Although stars can be formed in all DM halos (except in the least massive ones), 
not all halos with stars can be observed as LAEs for the following two reasons.
%
%
First, if galaxies tend to have an intermittent star-formation history 
as suggested by theoretical models, they can produce strong Ly$\alpha$ 
emission only over a limited fraction of cosmic time.
Second, it is not easy for Ly$\alpha$ photons produced in a galaxy 
to escape from it because of their resonant nature.
%
%
To quantify the first effect, let us introduce 
the duty cycle of strong Ly$\alpha$ emission, $DC_{\rm Ly\alpha}$: 
\begin{equation}
DC_{\rm Ly\alpha} (M_h) = \frac{n_{\rm Ly\alpha}^{\rm model}}{n_{\rm All}^{\rm model}},
\label{eq:dc_lya}
\end{equation}
where $n_{\rm Ly\alpha}^{\rm model}$ is the number density of DM halos with $M_h$
which are producing strong enough Ly$\alpha$ emission 
to be observed as LAEs if all Ly$\alpha$ photons escape, 
and $n_{\rm All}^{\rm model}$ is the number density of all DM halos with 
the same mass calculated from the DM halo mass function.
%
%

\begin{figure}[H]
\centering
\includegraphics[scale=.45]{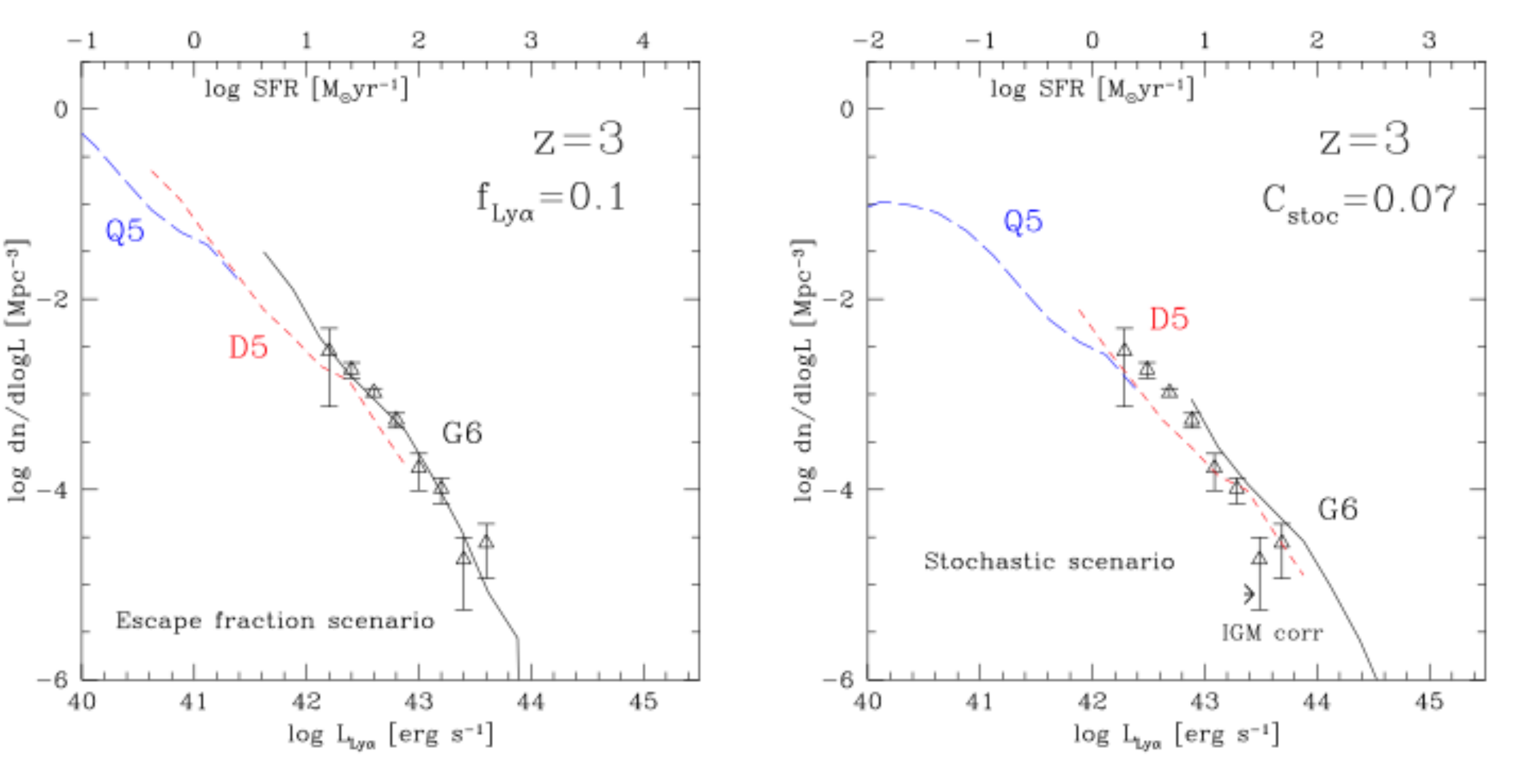}
\caption{
Ly$\alpha$ luminosity functions at $z=3$ obtained by observations (triangles) and cosmological
simulations (curves, color-coded by mass resolution and simulation box size: \citealt{nagamine2010}).
%
Left: The model Ly$\alpha$ luminosity functions are calculated by converting SFRs into Ly$\alpha$ luminosities with the Ly$\alpha$ escape fraction 
$\left < f_{\rm esc}^{\rm Ly\alpha} \right > =0.1$ and the duty cycle of strong Ly$\alpha$ emission 
$DC_{\rm Ly\alpha}=1$. 
Right: Same as the left panel, but with $\left < f_{\rm esc}^{\rm Ly\alpha} \right > =1$ 
and $DC_{\rm Ly\alpha}=0.07$. It should be noted that the simulation results 
shown in the two panels are indistinguishable 
despite very different $\left < f_{\rm esc}^{\rm Ly\alpha} \right >$ and $DC_{\rm Ly\alpha}$ values. 
In other words, the observed Ly$\alpha$ luminosity function (triangles) can be explained 
by adjusting only one of these two parameters.
This figure is reproduced by permission of the PASJ.
}
\label{fig:nagamine2010_fig4}
\end{figure}

\begin{figure}[H]
\centering
\includegraphics[scale=.40]{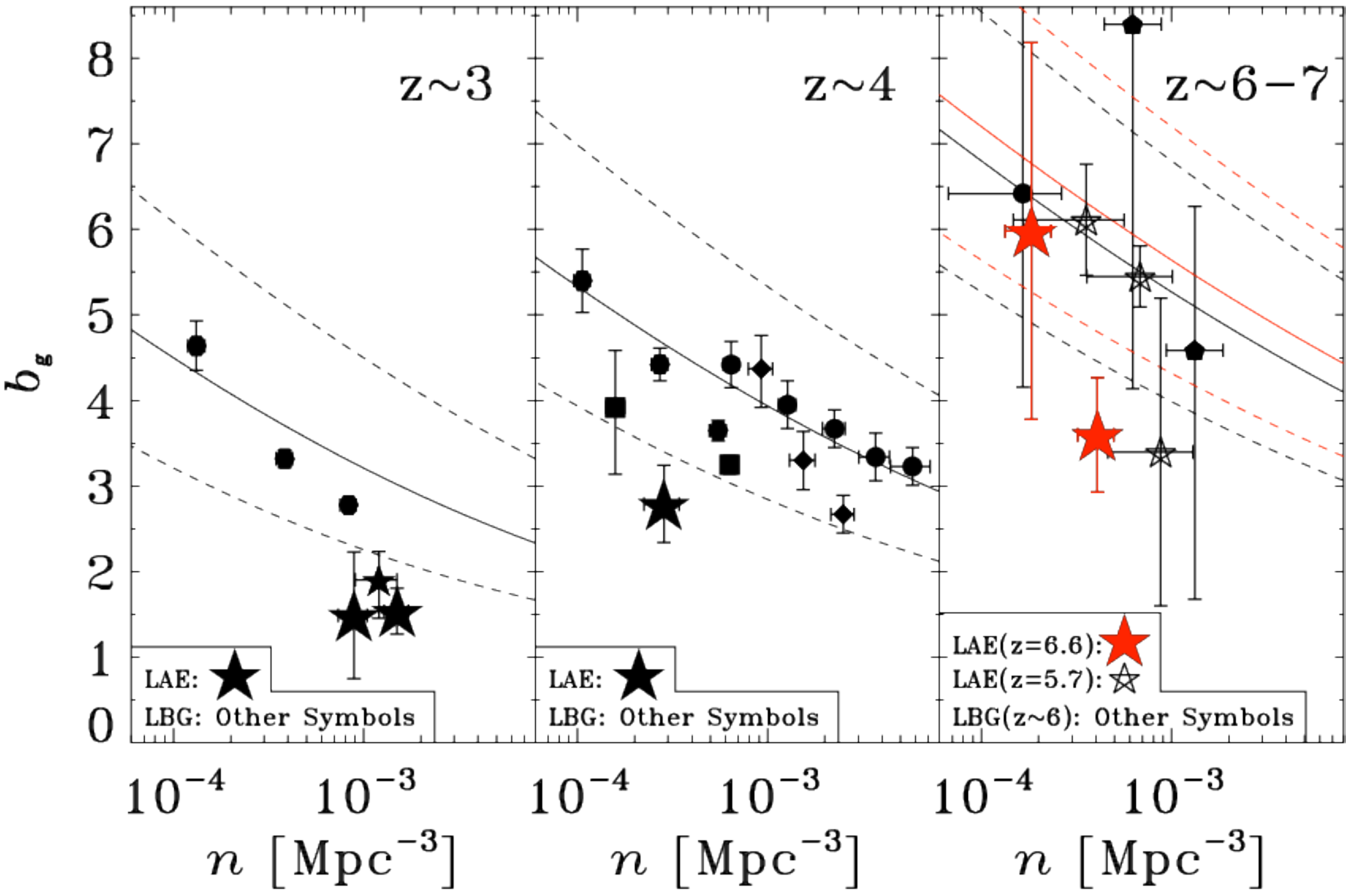}
\caption{
$b_g$ as a function of abundance \citep{ouchi2010}
at $z\sim 3$ (left), $z\sim 4$ (center), and $z\sim 6-7$ (right).
The star marks represent LAEs, while the other symbols (e.g. circles)
denote dropout galaxies. 
The solid lines are for DM halos. 
The dashed lines are the same as the solid lines, but the abundance is
multiplied by 1/10 and 10. 
In the right panel, black and red colors correspond to 
$z\sim 6$ and $7$, respectively. 
Note that 
the black star marks for $z\sim 6$ 
correspond to a highly clustered region of LAEs
whose $b_g$ values are probably higher than the average.
%
This figure is reproduced by permission of the AAS.
}
\label{fig:ouchi2010_fig14}
\end{figure}

\citet{nagamine2010} have used cosmological numerical simulations to study the effects of $DC_{\rm Ly\alpha}$ 
on the Ly$\alpha$ luminosity function (Figure \ref{fig:nagamine2010_fig4}) and the ACF. 
In the simulated luminosity functions, one can find two trends due to 
changing $DC_{\rm Ly\alpha}$ and 
$\left < f_{\rm esc}^{\rm Ly\alpha} \right >$ (Section \ref{sec:luminosity_function}).
Lowering $DC_{\rm Ly\alpha}$ decreases the number density of LAEs irrespective of their luminosity, thus uniformly lowering the luminosity function.
%
On the other hand, the number density of LAEs
also decreases by 
reducing $\left < f_{\rm esc}^{\rm Ly\alpha} \right >$ 
because of a uniform reduction of Ly$\alpha$ luminosities.
%
These results mean that the luminosity function alone cannot distinguish a change in
$DC_{\rm Ly\alpha}$ from a change in $\left < f_{\rm esc}^{\rm Ly\alpha} \right >$.
%
%
However, this degeneracy 
can be resolved with clustering measurements.
%
%
In a galaxy-formation model, LAEs are populated 
from the most-massive DM halos to low-mass DM halos
until the LAE number density becomes as large as the one
given by observations.
If a $DC_{\rm Ly\alpha}$ value is high (low), 
LAEs are hosted by high-mass (low-mass) DM halos on average 
for a given LAE number density, which show a strong (weak) LAE clustering signal.
Although various combinations of $\left < f_{\rm esc}^{\rm Ly\alpha} \right >$ and $DC_{\rm Ly\alpha}$
can explain the LAE number density (or Ly$\alpha$ luminosity function),
the choice of $DC_{\rm Ly\alpha}$ changes the LAE clustering signal that can be
tested with observational results.
\citet{nagamine2010} have made two competing LAE models:
a high $DC_{\rm Ly\alpha}$ ($=1$) and a low $\left < f_{\rm esc}^{\rm Ly\alpha} \right >=0.1$ 
(left panel of Figure \ref{fig:nagamine2010_fig4})
and a low $DC_{\rm Ly\alpha}$ ($=0.07$) and a high $\left < f_{\rm esc}^{\rm Ly\alpha} \right >=1$ 
(right panel of Figure \ref{fig:nagamine2010_fig4}),
both of which well reproduce the observed Ly$\alpha$ luminosity function
\footnote{
\citet{nagamine2010} refer to the Ly$\alpha$ duty cycle as the stochasticity of LAEs.
}.
They find that 
the LAEs in the former model are clustered much more strongly than observed ones (Section \ref{sec:clustering}),
while those in the latter model reproduce the observed weak clustering.
These models suggest that $DC_{\rm Ly\alpha}$ is an important parameter
less than unity.
With the same idea, the value of $DC_{\rm Ly\alpha}$ can be estimated 
by a simple comparison of number density and $b_g$ (i.e. the clustering strength).
Figure \ref{fig:ouchi2010_fig14}
shows that the number density of observed LAEs $n_{\rm Ly\alpha}^{\rm obs}$ 
is smaller than $n_{\rm All}^{\rm model}$ with the same 
$b_g$ by two orders of magnitude. This is a sharp contrast to LBGs at the same redshift (Figure \ref{fig:ouchi2010_fig14}).
Based on results of the kind shown in Figure \ref{fig:ouchi2010_fig14},
\citet{gawiser2007} and \citet{ouchi2010} estimate $DC_{\rm Ly\alpha}$ to be about $1$\%,
replacing $n_{\rm Ly\alpha}^{\rm model}$ with $n_{\rm Ly\alpha}^{\rm obs}$ in eq. (\ref{eq:dc_lya}).

\subsection{Summary of Galaxy Formation II}
\label{sec:summary_galaxy_formationII}

Section \ref{sec:galaxy_formationII}
has reviewed 
%
%
the basic physical properties of LAEs
characterized by observations. 
Observed SEDs 
indicate that
typical LAEs have a low stellar mass ($\sim 10^7-10^{10} M_\odot$)
and a moderately low SFR ($\sim 1-10 M_\odot$). 
They are thus distributed 
%
in the lowest mass regime 
%
of the star-formation main sequence
(Section \ref{sec:stellar_population}).
The Ly$\alpha$ luminosity function of LAEs 
%
%
rapidly increases from $z\sim 0$ to $3$, but 
shows no significant evolution over $z\sim 3-6$. On the other hand, the UV luminosity
function of UV-continuum selected galaxies (i.e. dropouts) shows a moderate increase from 
$z\sim 0$ to $3$, 
followed by a decrease to $z \sim 6$ and beyond.
These Ly$\alpha$ and UV luminosity function evolution results suggest 
a monotonic increase in the Ly$\alpha$ escape fraction $f_{\rm esc}^{\rm Ly\alpha}$ 
from $z\sim 0$ to $6$ (Section \ref{sec:luminosity_function}).
The morphology of LAEs is very compact on average, 
with $r_e\sim 1$ 
(Section \ref{sec:morphology}).
Showing strong high ionization lines such as [\oiii]5007 and \ciii]1907,1909,
typical LAEs are metal-poor ($\sim 0.3 Z_\odot$) and highly ionized ($\log q [{\rm s/cm}]\simeq 8-9$)
star-forming galaxies with negligibly small dust extinction ($A_V\sim 0$; Section \ref{sec:ISM_state}).
Deep spectra of LAEs show a signature of an outflow 
that is as strong as
that of LBGs. Through theoretical modeling, Ly$\alpha$ profiles and 
%
luminosities
are
useful to constrain the outflow velocity, hydrogen column density, dust extinction, 
and the spatial distribution of gas clouds (Section \ref{sec:outflow_lya_profile}).
Multi-wavelength data 
suggest that an AGN is found in about 1\% of LAEs in a given unbiased sample, 
while a majority of bright LAEs with $\log L_{\rm Ly\alpha} \gtrsim 43.5$ erg s$^{-1}$ 
host an AGN at $z\sim 2-3$ (Section \ref{sec:AGN_activity}).
Various studies use LAEs as low-mass galaxies associated with proto-clusters and LSSs
to probe the high-$z$ galaxy distribution. Clustering analyses of LAEs suggest that
LAEs are more weakly clustered than typical LBGs. 
The masses of LAE-hosting DM halos are estimated to be $M_{\rm h} \sim 10^{11\pm 1} M_\odot$, 
about an order of magnitude
smaller than for typical LBGs. 
Because LAEs are $\sim 10^2$ times less abundant than DM halos 
with the same bias value (or the same halo mass),
%
the duty cycle of the LAE phase (i.e. the phase when a galaxy is observed as a dust-poor star-forming galaxy with strong Ly$\alpha$)
is only $\sim 1$\% (Sections \ref{sec:overdensity}-\ref{sec:duty_cycle}).

%
 effects exist, but do not explain the strong Lya em. 
%

\section{Galaxy Formation III: Challenges of LAE Observations}
\label{sec:galaxy_formationIII}

There exist many open questions about the observational properties and the physical origins
of LAEs. In this section, I highlight three important questions about LAEs that are being actively
discussed:
%
extended Ly$\alpha$ halos, Ly$\alpha$ escape mechanisms, and the 
connection between LAEs and pop III star formation.
I explain major observational and theoretical 
progresses achieved to date about these issues.

\subsection{Extended Ly$\alpha$ Halos}
\label{sec:extended_lya_halo}

Deep observations in the 2000s (1980s-1990s) 
discovered 
$\sim 10-100$-kpc large Ly$\alpha$ nebulae 
associated with star-forming galaxies (and AGNs) at $z\gtrsim 2$ 
that spatially extend beyond the stellar components.
These nebulae are categorized into two classes, 
Ly$\alpha$ blobs (LABs) and diffuse Ly$\alpha$ halos (LAHs), 
%
according to their size and luminosity, with the former being 
larger and brighter.

\subsubsection{Ly$\alpha$ Blobs}
\label{sec:lya_blob}

\begin{figure}[H]
\centering
\includegraphics[scale=.47]{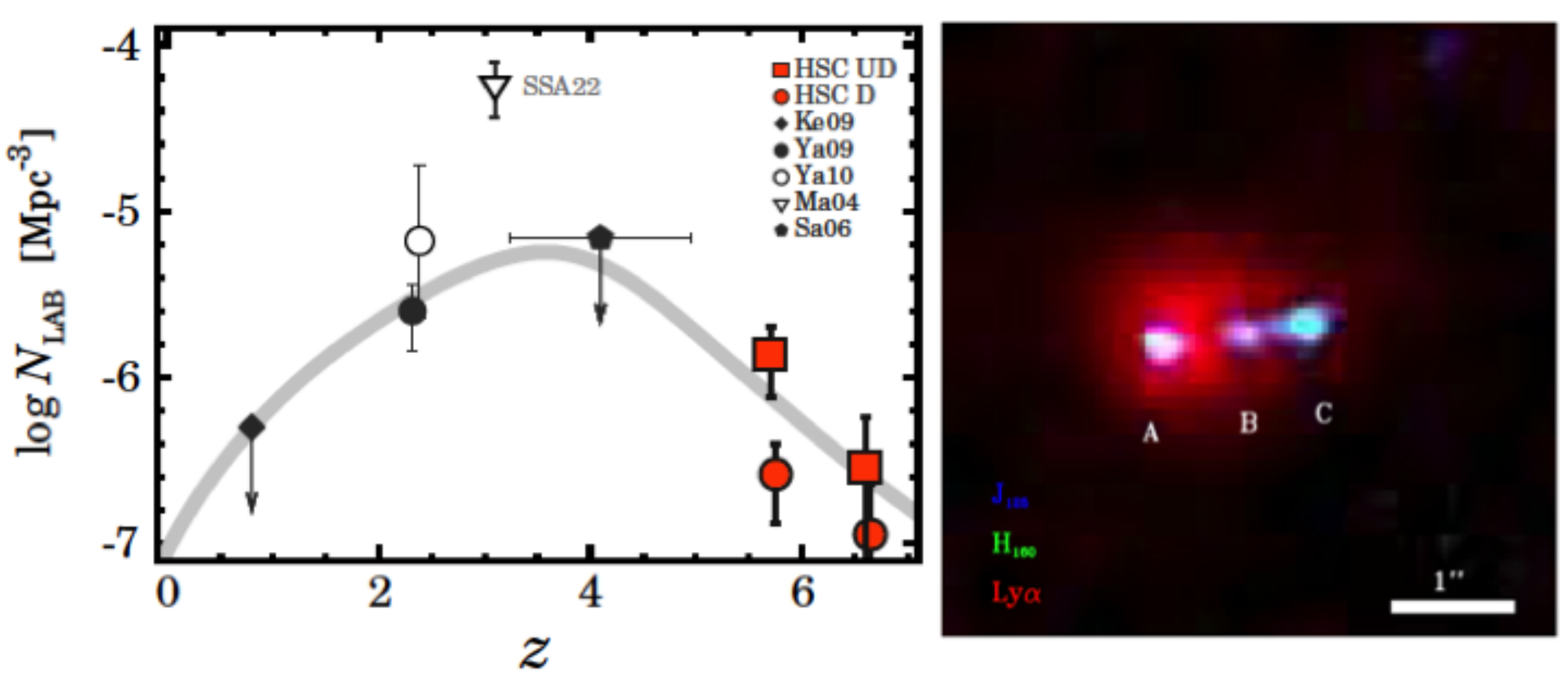}
\caption{
Left: Abundance of LABs identified at $z\sim 2-7$ \citep{shibuya2018a}. 
The inverse triangle is for LABs in the overdensity region SSA22, 
and the other symbols for those in field regions.
The grey curve illustrates a possible abundance evolution of
LABs. Because the selection criteria of LABs are heterogeneous, this evolutionary trend is
not conclusive.
Right: Color composite image of an LAB at $z=6.6$ dubbed Himiko \citep{ouchi2013}.
The red color shows a Ly$\alpha$ nebula extending over 17kpc,
while the blue and green colors represent rest-frame UV continua observed in the HST J and H bands, respectively.
The labels A, B, and C, indicate the three continuum clumps in Himiko.
This figure is reproduced by permission of the AAS.
}
\label{fig:shibuya2017a_fig10_ouchi2013_fig3}
\end{figure}

The first LABs discovered are
Blob1 and Blob2 (dubbed LAB1 and LAB2) 
in a LBG overdensity field at $z=3$ in the SSA22 field
(\citealt{steidel2000}; Section \ref{sec:progress}).
LAB1 and LAB2 
are each 
a huge ($>100$ kpc in physical length), 
bright ($10^{-15}$ erg s$^{-1}$) Ly$\alpha$ nebula 
belonging to the largest class of LABs.
Although similar extended Ly$\alpha$ emission 
has 
been found around
high-$z$ radio-loud galaxies since the 1980s (e.g. \citealt{mccarthy1987,
vanojik1997}), these two LABs are 
accompanied only by star-forming galaxies
with no clear AGN signature (see below for more details about the connection between LABs and AGNs).
The sizes and luminosities of these two LABs are, respectively, 
two and one order(s) of magnitude larger than 
those of $L^*$ LAEs at $z \sim 3$, 
$\sim 1$ kpc and $\sim 10^{43}$ erg s$^{-1}$ 
(Figures \ref{fig:shibuya2018c_fig8} and \ref{fig:konno2016_fig6}, respectively). 
%
%
Now a few tens of LABs are known 
(e.g. \citealt{matsuda2004}; Figure \ref{fig:matsuda2004_web}).
The 
definition
of LABs has not been quantitatively determined yet,
but galaxies with a spatially extended Ly$\alpha$ halo with a size of $>10$ kpc
are usually referred to as LABs in high-$z$ galaxy studies. 
The LABs found so far have large diversities in 
Ly$\alpha$ size and luminosity. Moreover, 
it should be noted that the Ly$\alpha$ sizes and luminosities 
follow a continuous distribution extending from regular LAEs to the largest LABs
%
such as LAB1 and LAB2
(Figure \ref{fig:matsuda2004_web}).
Although LABs are so far identified at $z\sim 2-7$
(Figure \ref{fig:shibuya2017a_fig10_ouchi2013_fig3}),
their physical origins 
are under debate. 
Three possible physical origins are suggested;
AGN photoionization (Section \ref{sec:AGN_activity}), 
cooling radiation (Section \ref{sec:star_formation}), 
and Ly$\alpha$ scattering HI clouds (Section \ref{sec:outflow_lya_profile}).

It is known that some AGNs 
are surrounded by
%
an extended Ly$\alpha$ nebula, but
the question is 
whether all LABs owe their luminosity to an AGN.
Deep X-ray follow-up observations find 
that a number of LABs host an AGN,
and that about $\sim 20$\% of LABs
show AGN activities \citep{basu-zych2004,geach2009}. 
%
Conversely,
a large fraction of LABs including LAB1 have no AGN signature.
Although bright AGNs would contribute to making large extended Ly$\alpha$ nebulae
in some cases,
there should exist other physical mechanisms to create 
LABs without an AGN.

Theoretical studies claim that cooling radiation can be the origin of
LABs \citep{fardal2001,dijkstra2009,goerdt2010}. Although the Ly$\alpha$ luminosity of cooling radiation 
around a galaxy 
is usually fainter
than the one of young stars
in it, 
these two luminosities  
are comparable in massive galaxies. Moreover, cooling radiation would 
dominate 
in the outer region of a galaxy, 
because 
it
is produced 
primarily in the outer DM halo 
where Ly$\alpha$ photons are not absorbed by dust \citep{fardal2001}.
\citet{goerdt2010} suggest that the luminosity and morphology of 
LABs
are reproduced by models that produce Ly$\alpha$ by collisional excitation in cold accretion gas.
Although theoretical studies reproduce the characteristics of LABs with cooling radiation,
so far there 
is
no observational evidence that clearly 
supports 
the cooling radiation scenario.
\citet{yang2006} claim that the {\heii}1640 line is useful to test 
if LABs at $z\sim 2-3$ originate from cooling radiation, 
because narrow-line ($<400$ km s$^{-1}$) 
{\heii}1640 emission can be produced neither by strong galactic outflow
nor by population-II star
photoionization, but only by cooling radiation.
%
However,  {\heii}1640 emission alone 
is not sufficient to distinguish cooling radiation from
a narrow-line (type II) AGN and population-III star formation (Section \ref{sec:popIII_in_lae}).

%



\begin{figure}[H]
\centering
\includegraphics[scale=.47]{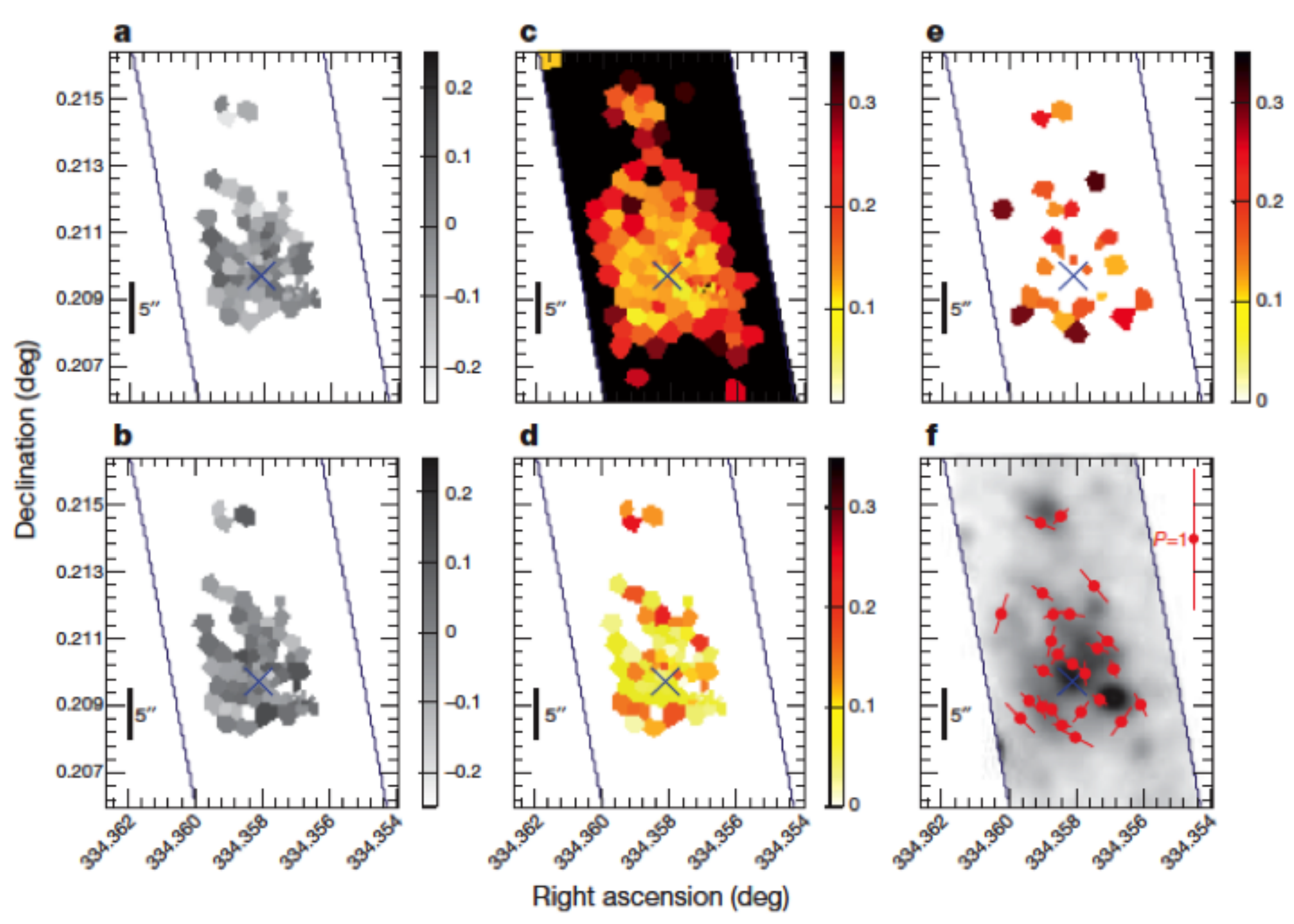}
\caption{
Polarization observation results of LAB1 \citep{hayes2011b}.
Panels (a) and (b) present the Stokes Q and U parameters, respectively.
Panel (c) shows a map of polarization fraction measurements beyond
the $2\sigma$ confidence level, 
with errors color-coded by values.
Panels (d) and (e) are 
maps of the polarization fraction $P$ for signals with an absolute ($2\sigma$) error smaller than 16\%, 
and with a relative error smaller than $50$\% respectively.
Panel (f) indicates an intensity image (gray scale) with polarization
vectors (red bars) whose lengths correspond to the amount of polarization.
The blue cross denotes the central position of LAB1.
This figure is reproduced by permission of the Nature Publishing Group.
}
\label{fig:hayes2011b_fig2}
\end{figure}

\begin{figure}[H]
\centering
\includegraphics[scale=.47]{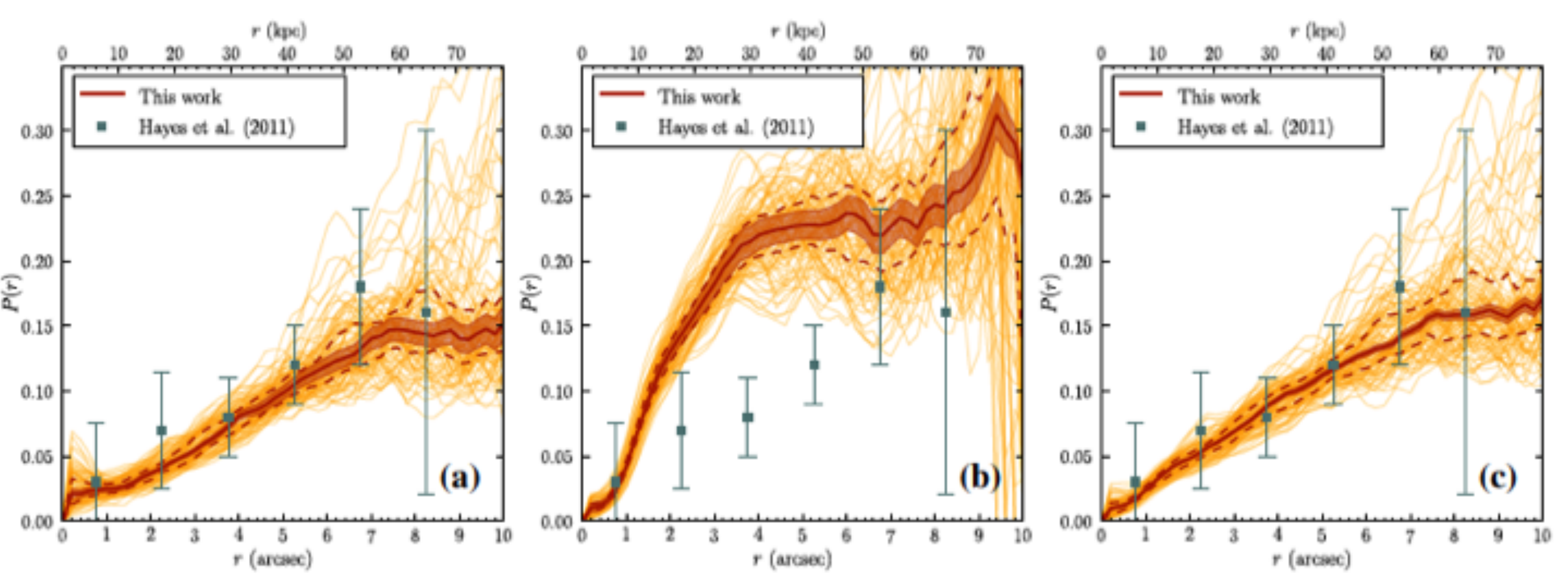}
\caption{
Ly$\alpha$ photon polarization as a function of 
the distance 
from the central galaxy \citep{trebitsch2016}.
The orange lines represent polarization values along various line of sights in the simulation box.
In each panel, 
the red line denotes the median of the orange lines, while the red region and the red-dashed lines 
indicate the $1\sigma$ dispersion and the first/third quartiles, respectively. The green data points 
are the observational measurements of \citet{hayes2011b}.
The left and central panels present polarization profiles for Ly$\alpha$ photons
produced in the extragalactic gas and the galaxy, respectively,
while the right panel shows the sum of these two components.
This figure is reproduced by permission of the A\&A.
%
%
}
\label{fig:trebitsch2016_fig5}
\end{figure}

Recent observations have advanced the understanding of LABs.
\citet{hayes2011b} have detected a tangential polarization signal of 0-20\%
in the Ly$\alpha$ emission 
of LAB1 (Figure \ref{fig:hayes2011b_fig2}). Because it is predicted that
resonance scattering of Ly$\alpha$ in \hi\ clouds makes a tangential 
polarization \citep{dijkstra2008}, the polarization signal in LAB1
suggests that extended Ly$\alpha$ nebulae are produced
by Ly$\alpha$ resonance scattering in \hi\ clouds around galaxies.
However, there remains a question about the source of Ly$\alpha$ photons.
The \hi\ cloud scattering scenario usually 
assumes that Ly$\alpha$ photons are produced in the central galaxy of the LABs 
\citep{hayes2011b}, but \hi\ cloud scattering also takes place
for Ly$\alpha$ photons 
produced {\it in situ} by 
gas cooling. \citet{trebitsch2016} have performed radiative hydrodynamics simulations
for Ly$\alpha$ photons from the central galaxy and gas cooling,
and calculated the polarization and surface brightness (SB) of Ly$\alpha$ emission
that are shown in Figure \ref{fig:trebitsch2016_fig5}. If LAB1 is made by
the \hi\ scattering of Ly$\alpha$ photons from the central galaxy,
the polarization signal is larger than $20$\% at $>40$ kpc,
which is significantly larger than the observational results.
Moreover, \citet{trebitsch2016} find that Ly$\alpha$ photons
from the cooling radiation are also scattered and polarized to a  
level of $10-15$\%. To explain the moderately small polarization 
and the large SB values, \citet{trebitsch2016} suggest 
that a significant contribution from cooling radiation is necessary.
In a way like this, 
the origins of LABs 
are still being actively discussed.
%

\begin{figure}[H]
\centering
\includegraphics[scale=.48]{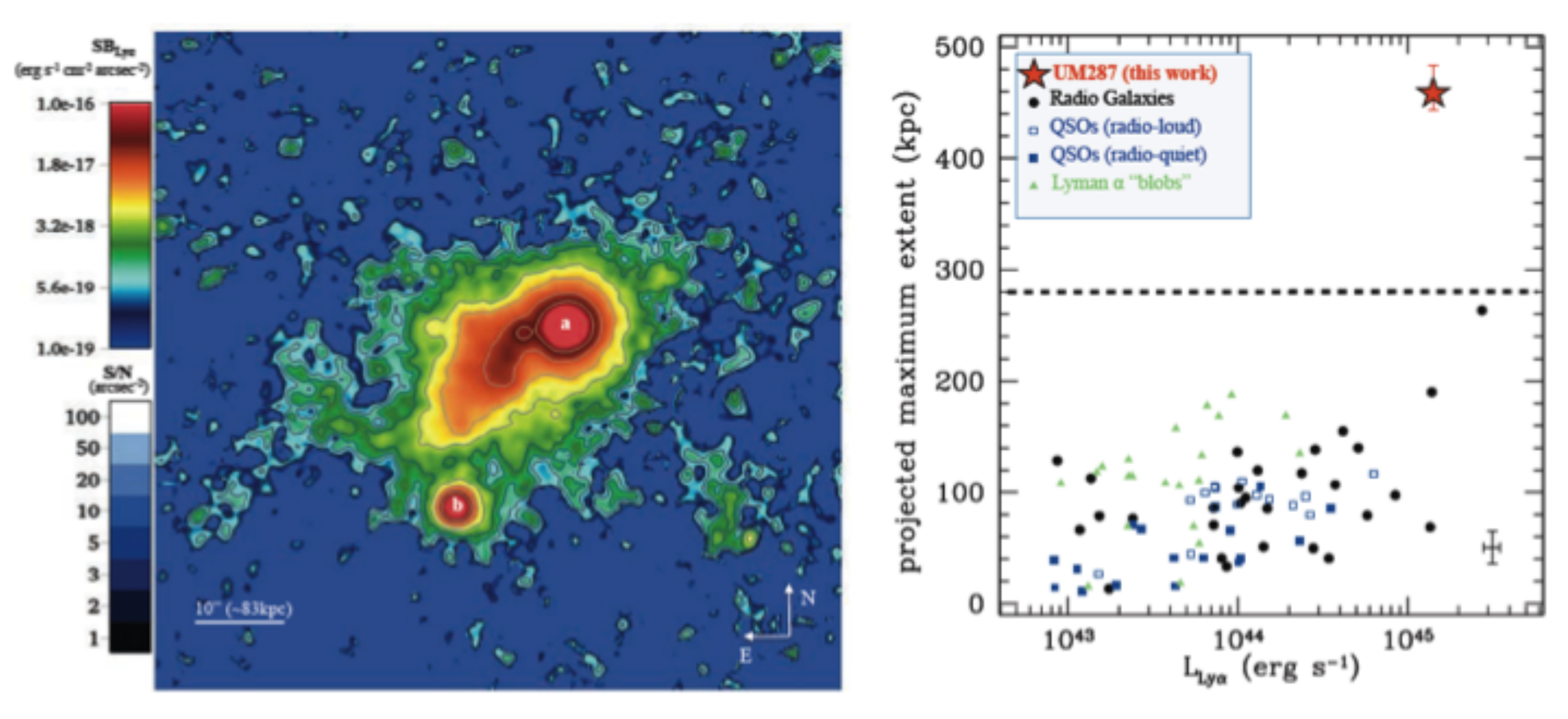}
\caption{
Left: Large Ly$\alpha$ nebula around the radio-quiet quasar UM287
that is dubbed Slug nebula \citep{cantalupo2014}. This is the continuum-subtracted Ly$\alpha$
image, presenting the Ly$\alpha$ surface brightness. The bottom-left scale bar represents $10"$ ($\sim 80$ kpc).
Right: Projected maximum sizes of the Ly$\alpha$ nebulae as a function of Ly$\alpha$ luminosity \citep{cantalupo2014}.
The size and luminosity of the Slug nebula is indicated with the red star mark. The black circles
and the blue squares are radio galaxies and QSOs, respectively. The green triangles denote
LABs with or without AGN mainly identified by narrowband imaging.
This figure is reproduced by permission of the Nature Publishing Group.
}
\label{fig:cantalupo2014_fig2_fig3}
\end{figure}

Moreover, there is an interesting problem not only about the origin of LABs without AGN signatures, 
but also of LABs harboring an AGN. \citet{cantalupo2014} report the discovery of a gigantic LAB around a radio-quiet QSO,
UM287, at $z=2.3$ (Figure \ref{fig:cantalupo2014_fig2_fig3}). 
This Ly$\alpha$ halo is 460 kpc in size 
that is larger than the virial diameter, $\sim 280$ kpc,
of the DM halo hosting this QSO, 
and may even extend to a filament of the LSS.
%
%
If this Ly$\alpha$ nebula is produced by Ly$\alpha$ photons from 
the recombination of a large cloud that was initially ionized by the QSO, 
and subsequent scattering in the now neutral cloud, 
then a very high gas mass or clumping factor is required to explain 
its high Ly$\alpha$ SB.
%
%
It is, however, not clear whether this object truly has such a very high gas mass or clumping factor.
Thus, the physical origin of this gigantic LAB is also under debate. 
There is also a report of the identification of large LABs 
with intermediate sizes of $100-300$ kpc around radio quiet QSOs \citep{borisova2016}.
%
These large LABs fill the gap between small and gigantic LABs and 
may facilitate our understanding of the whole LAB zoo.


\subsubsection{Diffuse Ly$\alpha$ Halos}
\label{sec:diffuse_lya_halo}

\begin{figure}[H]
\centering
\includegraphics[scale=.48]{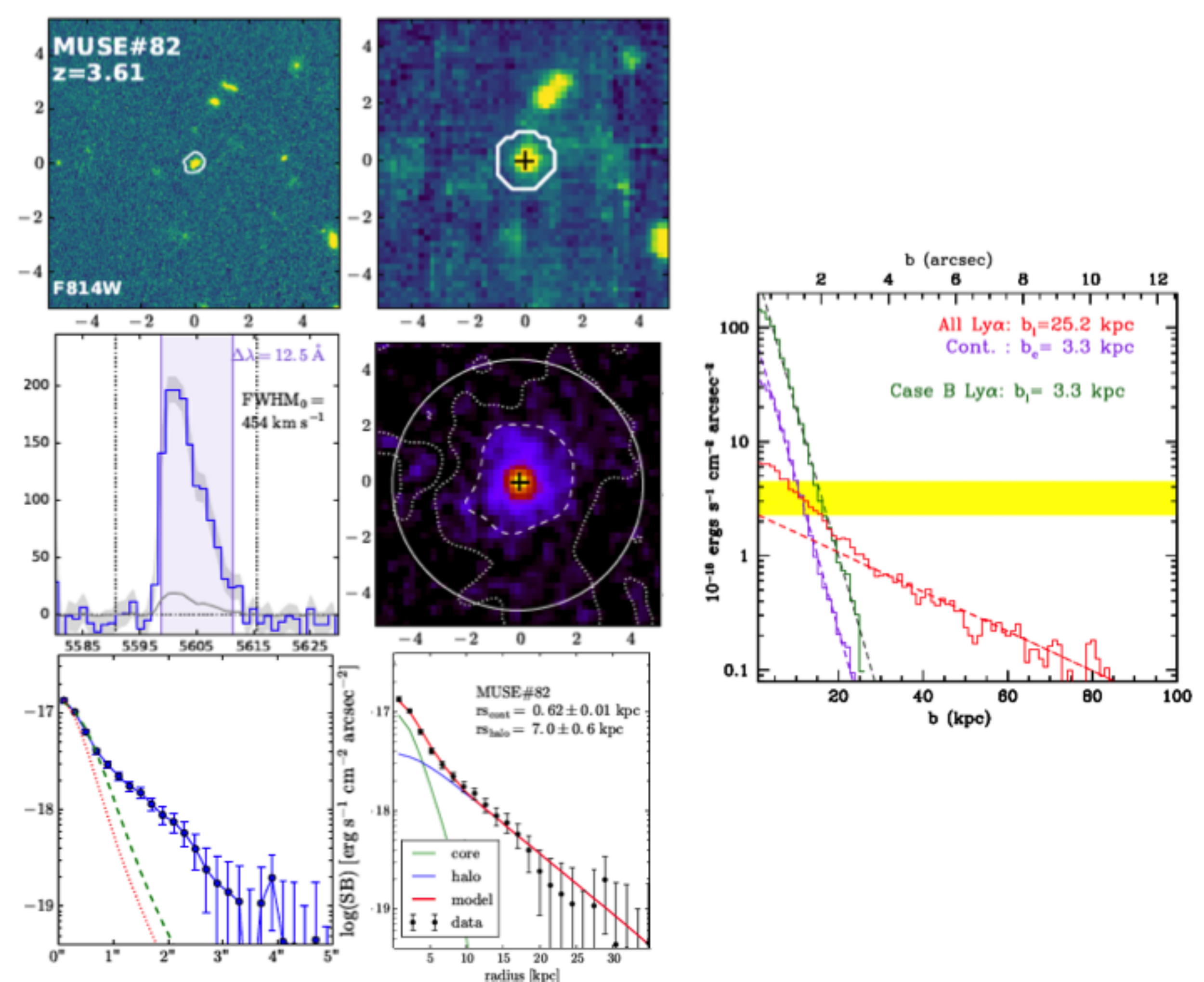}
\caption{
Left six panels: Images, Ly$\alpha$ spectrum, and radial profiles of a galaxy, MUSE\#82, 
at $z=3.61$ \citep{leclercq2017}. The top left and right panels present
continuum images, an HST $F814W$ image and a MUSE white-light image,
respectively. 
The middle left and right panels show a Ly$\alpha$ spectrum of the central part of this galaxy, and a Ly$\alpha$ image,
respectively. The bottom left panel displays a Ly$\alpha$ SB radial profile (blue circles) and 
a UV continuum profile (green line) together with the PSF (red line). The bottom right panel presents the
best-fit model (red line) and its decomposition to core (green line) and LAH (blue line) components,
to compare with the observed profile (black circles). 
Here the core component is modeled with
an exponential profile.
Right panel: Stacked SB radial profiles of galaxies at $z\sim 2.7$ \citep{steidel2011}. 
The red and blue solid lines denote Ly$\alpha$ and UV-continuum SB radial profiles, respectively.
The green solid line indicates an estimated Ly$\alpha$ SB radial profile for no
LAH case that is calculated with the UV-continuum SB radial profile under the assumption
of the Case B recombination. The dashed lines are the best-fit exponential functions
of 
these profiles.
This figure is reproduced by permission of the A\&A and the AAS.
}
\label{fig:leclercq2017_fig3_fig4_steidel_fig5}
\end{figure}

 \citet{hayashino2004} and \citet{steidel2011} have identified
diffuse Ly$\alpha$ halos (LAHs) around star-forming galaxies at $z\sim 2-3$
by deep spectroscopy and narrowband-image stacking analyses
(Figure \ref{fig:leclercq2017_fig3_fig4_steidel_fig5}).
To date, LAHs are found in star-forming galaxies 
including LAEs in a wide-redshift range, $z\sim 2-7$ \citep{matsuda2012,feldmeier2013,momose2014}.
LAHs extend to a scale of 
\noindent
a few $\times 10$ kpc with a Ly$\alpha$ SB of 
$\lesssim 1\times 10^{-18}$ erg s$^{-1}$ cm$^{-2}$ arcsec$^{-2}$
that is about $10-100$ times fainter than that of LABs.
Their Ly$\alpha$ SB profiles roughly follow a power law 
(Figure \ref{fig:leclercq2017_fig3_fig4_steidel_fig5})
with the LAH scale length $r_n$,
\begin{equation}
S(r) = C_n \exp (-r/r_n),
\label{eq:sb_lah}
\end{equation}
where $S(r)$, $r$, and $C_n$ are the Ly$\alpha$ SB, radius, 
and normalization factor, respectively.

\begin{figure}[H]
\centering
\includegraphics[scale=.47]{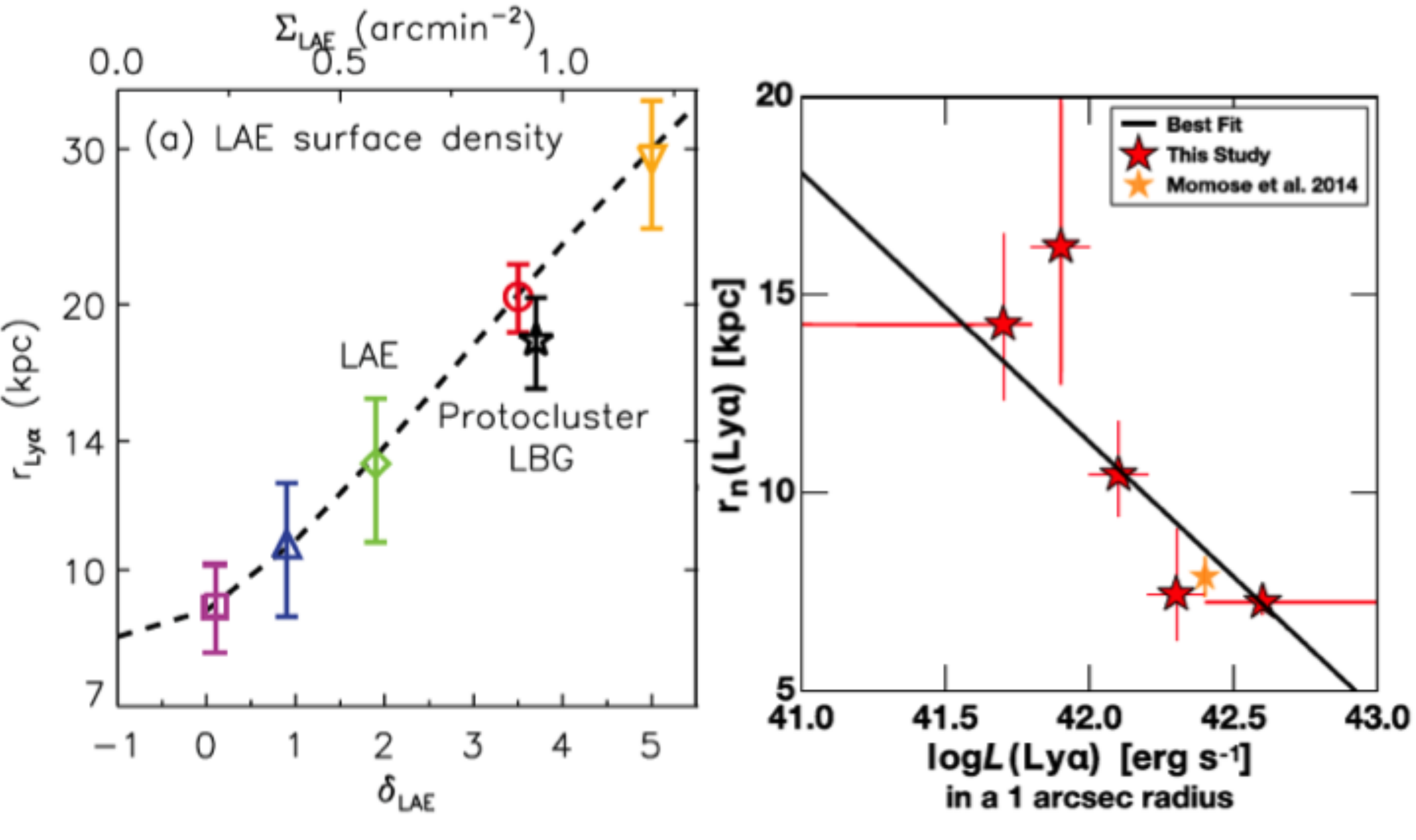}
\caption{
Left: LAH scale length as a function of the LAE sky overdensity
at $z=3$ in the SSA22 region \citep{matsuda2012}.
The star mark denotes the median LAH scale length of LBGs in
the galaxy overdense region, while the other symbols represent 
those of LAEs.
The dashed line is the best-fit quadratic function to all data points.
(Note the non-regular scale on the y-axis.)
Right: LAH scale length as a function of Ly$\alpha$ luminosity 
for LAEs at $z=2$ \citep{momose2016}. 
The star marks are measurements by \citet{momose2014} and \citet{momose2016}, 
and the solid line is the best-fit linear function to them.
This figure is reproduced by permission of MNRAS.
}
\label{fig:matsuda2012_fig4_momose2016_fig9}
\end{figure}

The parameter  
$r_n$ characterizes the size of an LAH.
Figure \ref{fig:matsuda2012_fig4_momose2016_fig9}
presents the relations between $r_n$ and several physical properties
of LAEs.
%
%
\citet{matsuda2012} claim that $r_n$ positively correlates
with the local LAE surface density $\delta_{\rm LAE}$.
The result of \citet{matsuda2012} would imply that galaxies in a dense environment
have a large Ly$\alpha$ halo (cf. \citealt{xue2017}).
\citet{momose2016} find that $r_n$ negatively correlates with
the Ly$\alpha$ luminosity of 
the main body of galaxies,
$L_{\rm Ly\alpha}^{\rm cent}$, at $r<8$ kpc ($r<1"$).
Because $L_{\rm Ly\alpha}^{\rm cent}$ depends on \hi\ column density through resonant scattering, 
galaxies with the ISM rich in \hi\ 
would have a faint $L_{\rm Ly\alpha}^{\rm cent}$ and a large extended Ly$\alpha$ halo.
It is suggestive that LAEs (galaxies with a bright $L_{\rm Ly\alpha}^{\rm cent}$) have the ISM whose  
\hi\ column density is lower than that of LBGs (galaxies with a faint $L_{\rm Ly\alpha}^{\rm cent}$).

\begin{figure}[H]
\centering
\includegraphics[scale=.48]{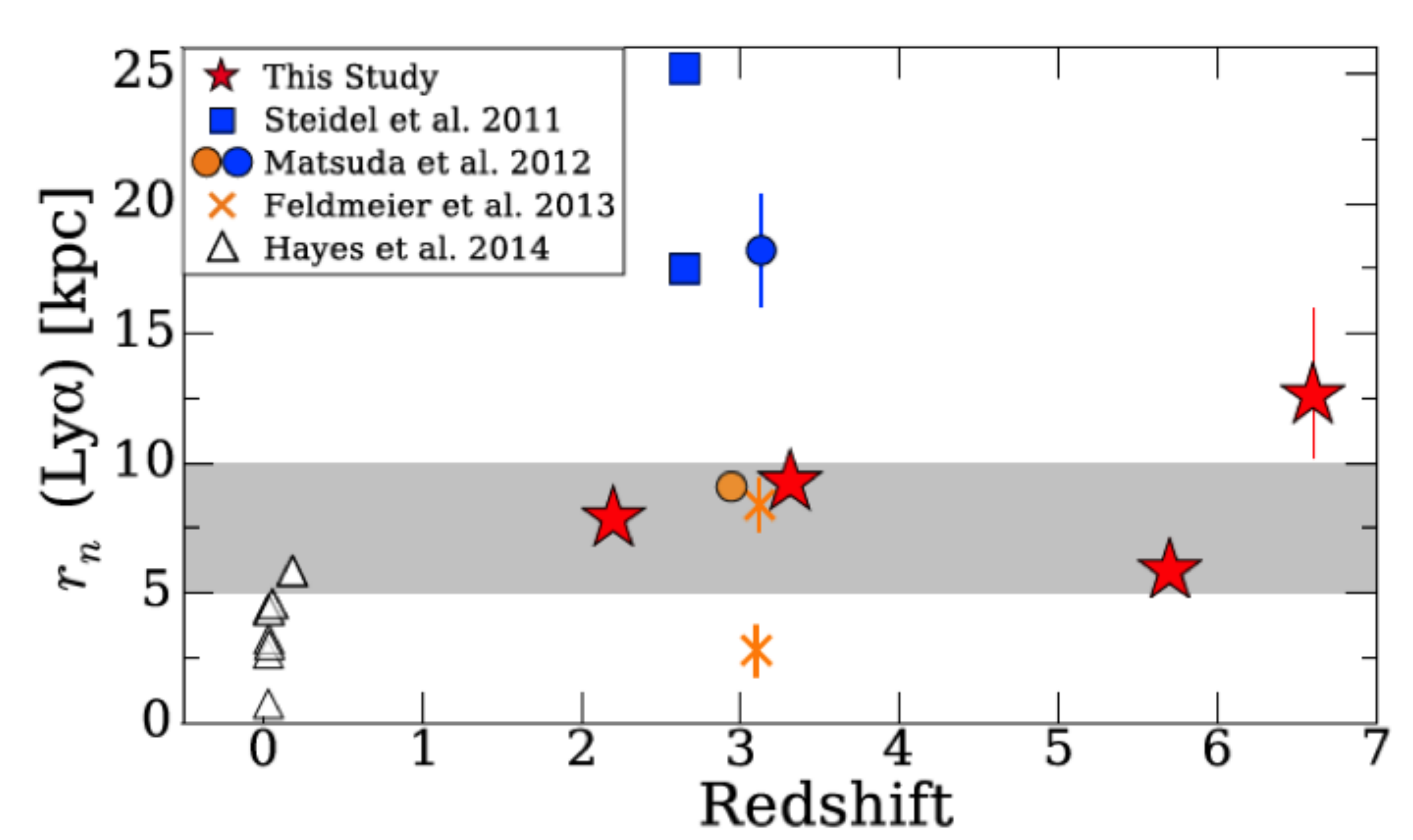}
\caption{
Redshift evolution of the LAH scale length \citep{momose2014}.
The red star marks represent the median LAH scale lengths of LAEs
with $L_{\rm Ly \alpha} \gtrsim 2 \times 10^{42}$ erg s$^{-1}$
at $z=2.2-6.6$. The gray region indicates the range of $5-10$ kpc
where the LAH scale lengths of LAEs at the post reionization epoch ($z=2.2-5.7$) fall.
The blue squares and the orange and blue circles denote the LAH scale lengths of LAEs
in the overdense region of SSA 22. The orange crosses are the measurements of bright LAEs.
The open triangles show Ly$\alpha$ Petrosian radii of local LAEs.
This figure is reproduced by permission of MNRAS.
}
\label{fig:momose2014_fig9}
\end{figure}

Because $r_n$ depends on 
some 
galaxy properties, 
the evolution of $r_n$ should be investigated carefully
with uniformly selected samples at different redshifts.
The red star marks in Figure \ref{fig:momose2014_fig9} indicate
the $r_n$ values of field ($\delta_{\rm LAE}\lesssim 1$) 
LAEs with $L_{\rm Ly\alpha}^{\rm cent}\gtrsim 2\times 10^{42}$ erg s$^{-1}$
at $z=2.2-6.6$. 
It is found that $r_n$ is nearly constant over $z=2.2-5.7$, falling in the range 
$r_n=5-10$ kpc \citep{momose2014}.
There is a hint of 
an increase in 
$r_n$ from $z=5.7$ to $6.6$
that could be relevant to cosmic reionization, but the error bars 
are too large to conclude whether this is a real signature.

\begin{figure}[H]
\centering
\includegraphics[scale=.47]{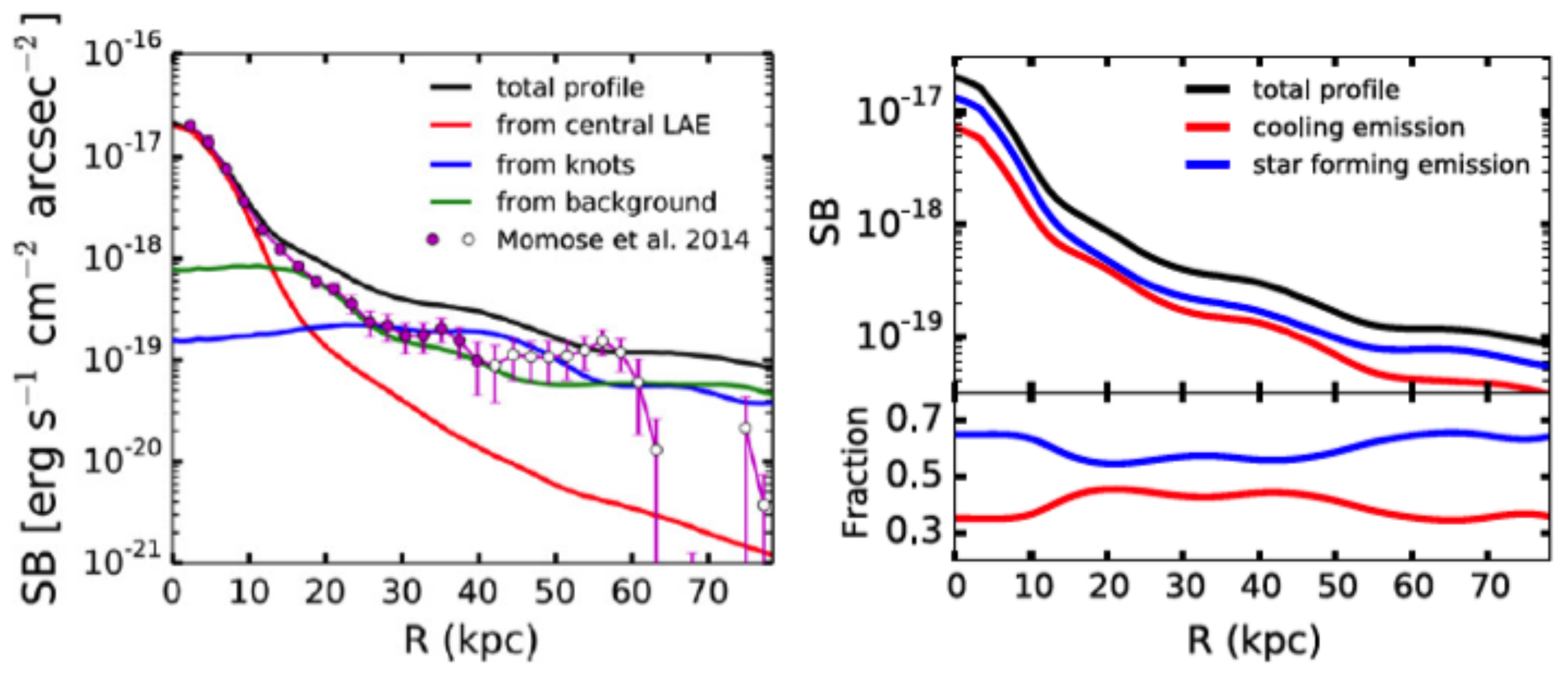}
\caption{
Average Ly$\alpha$ SB profiles of nine model LAEs at $z=3$ obtained by numerical simulations \citep{lake2015}.
Left: The black curve represents the total Ly$\alpha$ SB profile,
while the red, blue, and green curves denote 
a decomposition into
Ly$\alpha$ photons 
originating 
from the central star-forming regions, surrounding knotty star-forming regions,
and background regions, respectively.
The purple filled circles show the observational data of \citet{momose2014}. The open circles are the same
as the filled circles, but for data points potentially with large systematic errors. 
The purple line simply connects all data points.
Right: Same as the left panel, but for 
a decomposition into 
cooling radiation (red line) and 
star-formation radiation (blue line) in the top panel.
The fractional contributions of these two radiation components 
(to the total Ly$\alpha$ SB profile) are shown in the bottom panel.
This figure is reproduced by permission of the AAS.
}
\label{fig:lake2015_fig4}
\end{figure}

The physical origin of LAHs is not well understood yet.
There are four possible scenarios,
i) CGM's \hi\ gas scattering Ly$\alpha$ photons that originate
from star-forming regions,
ii) cooling radiation, 
iii) unresolved dwarf satellite galaxies, and
iv) fluorescence.
%
%
\citet{lake2015} perform radiative transfer calculations in hydrodynamical simulations (Figure \ref{fig:lake2015_fig4}),
and find that the scenario i) cannot explain 
high Ly$\alpha$ SB at large radii 
found by observations (red line in the left panel of Figure \ref{fig:lake2015_fig4}).
The high Ly$\alpha$ SB at 
large radii
requires
either the mechanism ii) or iii)
(right panel of Figure \ref{fig:lake2015_fig4}).
To distinguish the contributions of ii) and iii), 
UV-continuum SB profiles in stacked broadband images are useful. 
This is because the mechanism iii) produces stellar UV-continuum emission,
while the mechanism ii) creates 
only negligible UV-continuum emission.
It is, however, difficult to investigate UV-continuum SB profiles
in the stacked images due to 
systematic errors in sky subtraction
\citep{momose2016}.
Recent deep spectroscopic observations with VLT/MUSE
find a diffuse LAH on an individual basis with 
no use of stacking data
(Figure \ref{fig:leclercq2017_fig3_fig4_steidel_fig5}; \citealt{wisotzki2016,leclercq2017}). This
discovery rules out the possibility that
moderately large unresolved dwarf 
satellite galaxies mimic a diffuse LAH
in the scenario iii). 
Studies of LAHs 
are proceeding 
rapidly,
and much progress can be expected in the coming few years.
%

%
%
\begin{figure}[h]
\centering
\includegraphics[scale=.47]{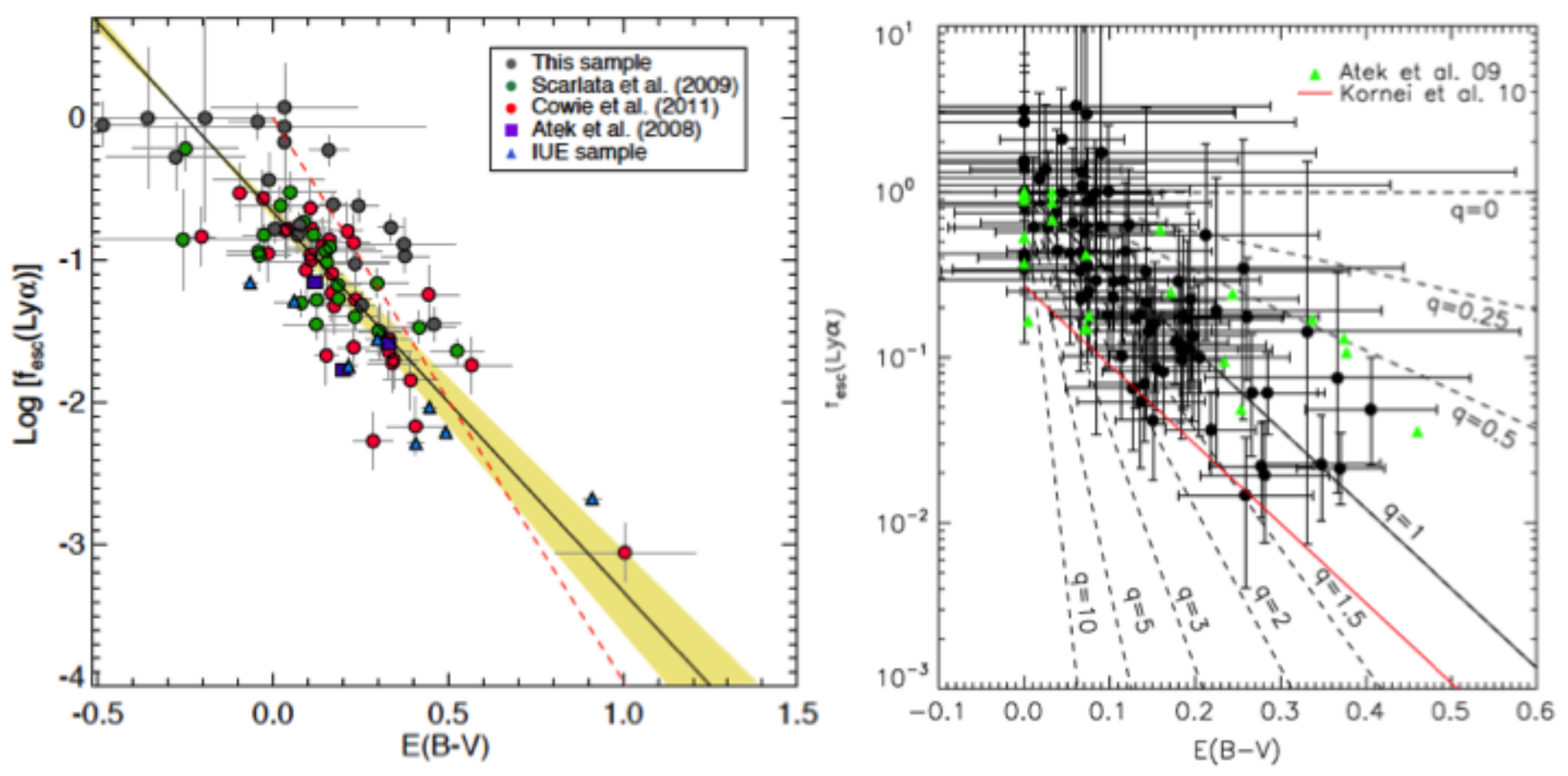}
\caption{
Lya escape fraction $f_{\rm esc}^{\rm Ly\alpha}$ as a function of color excess E(B-V). 
Left: The data points are all measured at $z=0-0.3$ \citep{atek2014a}.
The black line with a yellow shade indicates the best-fit linear function with
the $1\sigma$ fitting error. The red dashed line represents a simple attenuation
with the \citet{cardelli1989} extinction law.
Right: The black circles and the green triangles denote LAEs at $z\sim 2-4$ and $0.3$, respectively \citep{blanc2011}.
The black lines present correlations suggested from eq. \ref{eq:q_value} with various $q$ values, with a solid one corresponding to $q=1$. 
The red line is the best-fit
linear function to LBGs at $z\sim 2$.
This figure is reproduced by permission of the A\&A and the AAS.
}
\label{fig:atek2014_fig10_blanc2011_fig11}
\end{figure}

\subsection{Ly$\alpha$ Escape Fraction}
\label{sec:lya_escape_fraction}

\footnote{In the Saas Fee lecture, 
this subsection dealt with 
outflows and Ly$\alpha$ profiles as well.
For the readers' convenience, I have moved these
topics to Section \ref{sec:outflow_lya_profile}.
} 
One of the most important questions about LAEs is
%
how they emit strong Ly$\alpha$ light.
%
To discuss this question, let us introduce 
the Ly$\alpha$ escape fraction, $f_{\rm esc}^{\rm Ly\alpha}$,
defined by
\begin{equation}
f_{\rm esc}^{\rm Ly\alpha} = \frac{L_{\rm Ly\alpha}^{\rm obs}}{L_{\rm Ly\alpha}^{\rm int}},
\label{eq:fesc_lya_indiv}
\end{equation}
where $L_{\rm Ly\alpha}^{\rm obs}$ and $L_{\rm Ly\alpha}^{\rm int}$
are observed and intrinsic Ly$\alpha$ luminosities, respectively. 
Intrinsic Ly$\alpha$ luminosities can be estimated from a UV continuum
or an \ha\ line luminosity.
Note that $f_{\rm esc}^{\rm Ly\alpha}$ is similar to the number/luminosity average Ly$\alpha$ escape fraction,
$\left < f_{\rm esc}^{\rm Ly\alpha} \right >$ (Equation \ref{eq:fesc_lya_avg}), but that
$f_{\rm esc}^{\rm Ly\alpha}$ is defined for one individual galaxy.
The left panel of Figure \ref{fig:atek2014_fig10_blanc2011_fig11} presents
the Ly$\alpha$ escape fraction as a function of color excess
for LAEs at $z\sim 0$ and $2-4$.
There is a clear anti-correlation between 
these two quantities, 
suggesting that 
a certain fraction of 
Ly$\alpha$ photons are 
absorbed by 
dust in the ISM.
In the left panel of Figure \ref{fig:atek2014_fig10_blanc2011_fig11}, the $f_{\rm esc}^{\rm Ly\alpha}$ values of LAEs
are compared with 
the amount of dust extinction predicted by a simple screen model 
that includes no Ly$\alpha$ resonance scattering effects,
\begin{equation}
f_{\rm esc}^{\rm Ly\alpha, screen} = 10^{-0.4 k_{1216} E(B-V)_{\rm neb}},
\label{eq:fesc_lya_indiv_screen}
\end{equation}
where $k_{1216}$ is the extinction coefficient at 1216\AA.
Calzetti's law provides $k_{1216}=12.0$ \citep{konno2016}.
There is a hint of an excess of $f_{\rm esc}^{\rm Ly\alpha}$ beyond the dust screen model
for some LAEs, albeit with large measurement errors.
The right panel of Figure \ref{fig:atek2014_fig10_blanc2011_fig11}
shows various model lines with different $q$ values \citep{finkelstein2008}
defined as
\begin{equation}
q=\frac{\tau({\rm Ly \alpha})}{\tau_{1216}},
\label{eq:q_value}
\end{equation}
where $\tau({\rm Ly \alpha})$ and $\tau_{1216}$ are the optical depths
for the Ly$\alpha$ line and 1216\AA\ UV-continuum emission.
In the right panel of Figure \ref{fig:atek2014_fig10_blanc2011_fig11},
the $q=1$ model line corresponds to Equation (\ref{eq:fesc_lya_indiv_screen}).

LAEs with a $f_{\rm esc}^{\rm Ly\alpha}$ excess
have $q<1$. 
A selectively 
large Ly$\alpha$ extinction ($q>1$) 
is simply explained, if Ly$\alpha$ photons cross a long effective distance 
in a dusty ISM due to a number of Ly$\alpha$ resonance scatterings,
which enhances the probability of absorption by dust.

\begin{figure}[H]
\centering
\includegraphics[scale=.47]{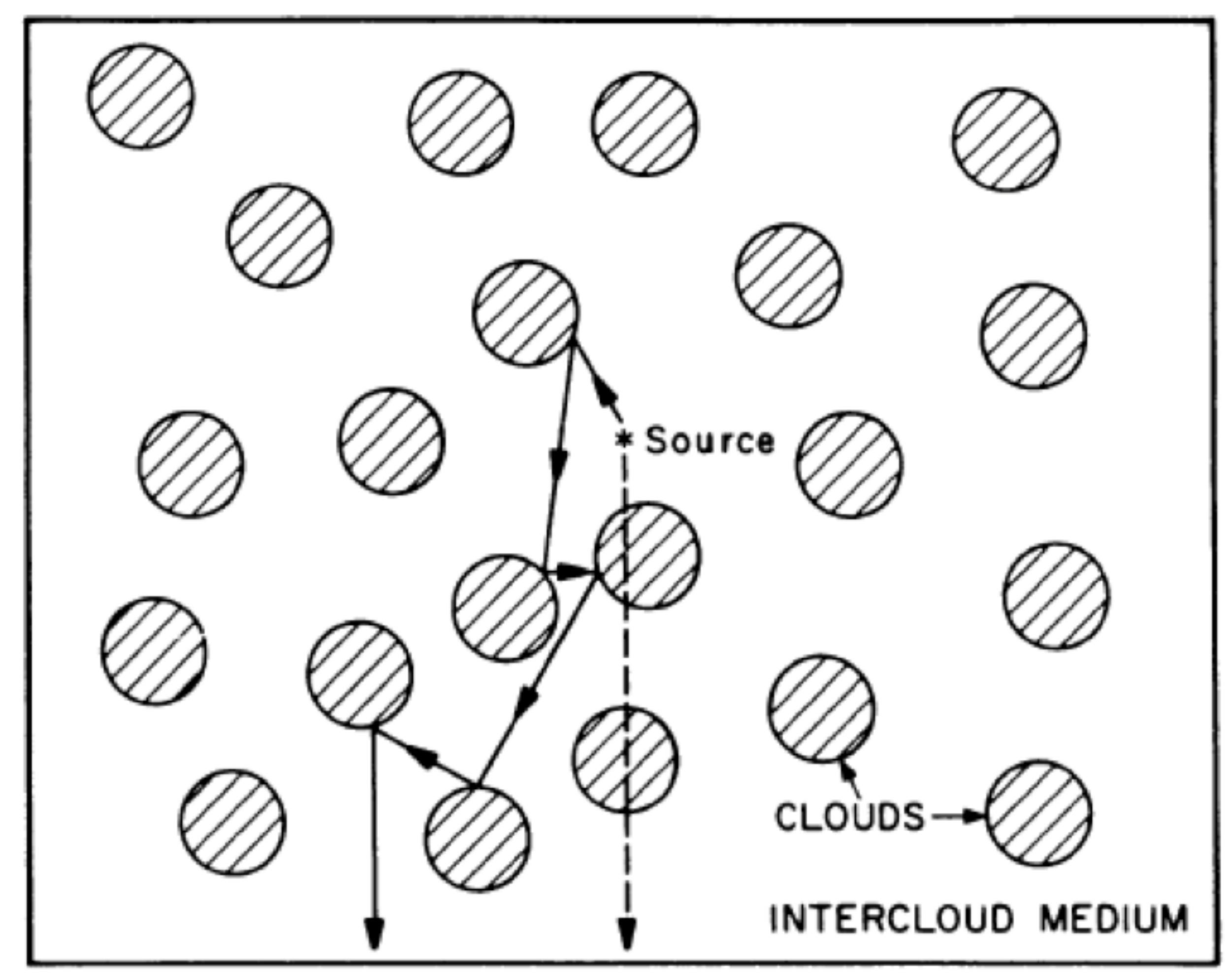}
\caption{
Illustration of \citeauthor{neufeld1991}'s clumpy cloud model \citep{neufeld1991}.
In this model, the ISM is made of gas clumps 
whose central regions contain dust (in cold molecular gas)
that can absorb UV-continuum photons.
The solid-line arrows indicate the light path of a Ly$\alpha$ (resonance) photon 
from the source. The Ly$\alpha$ photon is scattered on the surfaces of gas clumps with 
negligible dust absorptions. The dashed-line arrow represents the light path of 
an UV-continuum (non-resonance) photon
that penetrates clumps.
This figure is reproduced by permission of the AAS.
}
\label{fig:neufeld1991_fig1}
\end{figure}

However, a selectively small Ly$\alpha$ extinction ($q<1$) is difficult to understand.
%
%
Clumpy gas clouds in the ISM may explain $q<1$, 
as has been 
originally suggested by \citet{neufeld1991}.
Figure \ref{fig:neufeld1991_fig1} illustrates 
this idea.
If the ISM is made of clumpy gas clouds,
Ly$\alpha$ (resonance) photons that encounter clumpy clouds are scattered on their surface with a negligible dust absorption. 
On the other hand, UV-continuum (non-resonance) photons
can go into the clouds and are eventually absorbed by dust inside them.
Through these scattering and absorption processes, 
Ly$\alpha$ photons are absorbed less than UV-continuum photons,
resulting in 
a very high Ly$\alpha$ $EW_0$. The clumpy cloud model is sometime 
referred to as \citeauthor{neufeld1991}'s effect.
This model predicts narrow Ly$\alpha$ line widths because Ly$\alpha$ 
photons experience only a small number of resonant scattering 
before escaping from galaxies 
\citep{neufeld1991,hansen2006}.

\begin{figure}[H]
\centering
\includegraphics[scale=.46]{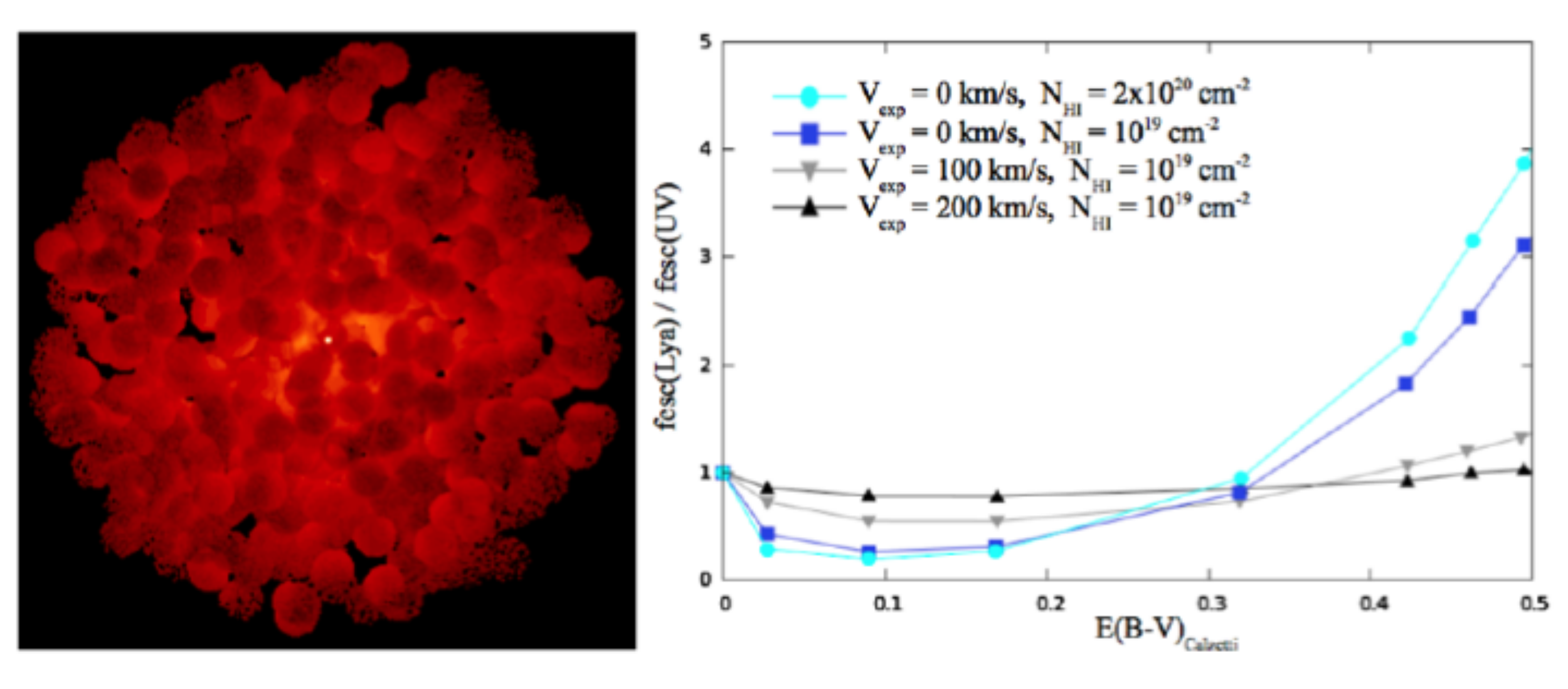}
\caption{
Left: ISM gas geometry of 500 clumpy clouds with a radius of $350$ pc 
produced in the simulations by \citet{laursen2013}.
Right: Ratio of Ly$\alpha$ to UV-continuum photon escape fractions, $f_{\rm esc}^{\rm Ly\alpha}/f_{\rm esc}^{\rm UV}$,
as a function of color excess E(B-V) for models of the clumpy cloud ISM \citep{duval2014}.
The cyan circles, blue squares, gray inverse-triangles, and black triangles represent
models with outflow velocities of $V_{\rm exp} = 0-200$ km s$^{-1}$ and {\sc Hi} column densities of $N_{\rm HI}=10^{19}-2\times 10^{20}$ cm$^{-2}$ indicated in the legend. An enhancement of Ly$\alpha$ photon escape, 
$f_{\rm esc}^{\rm Ly\alpha}/f_{\rm esc}^{\rm UV}>1$, 
is achieved only with $V_{\rm exp} = 0$--$100$ km s$^{-1}$ 
and $E(B-V) \gtrsim 0.3$.
This figure is reproduced by permission of the A\&A.
}
\label{fig:laursen2013_fig1_duval2014_fig18}
\end{figure}

%
Although \citet{neufeld1991} investigated this model 
only in a simple case of static and very clumpy/dusty media,
recent studies have used radiative transfer simulations 
to test this model
in realistic ISM conditions 
(\citealt{laursen2013,duval2014}; cf. \citealt{hansen2006}).
These simulations have found that 
\citeauthor{neufeld1991}'s effect is seen 
 (Figure \ref{fig:laursen2013_fig1_duval2014_fig18}) 
only under special conditions: 
a low outflow velocity ($<200$ km s$^{-1}$), 
very high extinction ($E(B-V)>0.3$), and an extremely clumpy gas distribution
with a density contrast larger than $10^7$ (i.e. most gas is locked up in clumps),
many of which do not meet the observed properties of LAEs (Table \ref{tab:stellar_population}).
Moreover, under these special conditions, observed Ly$\alpha$ lines 
can have neither a velocity shift nor an asymmetric profile.
\citet{laursen2013} and \citet{duval2014} have concluded that 
while it is true that \citeauthor{neufeld1991}'s effect is working to some degree,
this effect cannot explain the fact that a large fraction of LAEs have
high $f_{\rm esc}^{\rm Ly\alpha}$ values
(i.e. $q<1$).
In summary, the physical origin of 
the high $f_{\rm esc}^{\rm Ly\alpha}$ values found 
for LAEs is still under debate (see Section \ref{sec:popIII_in_lae} for more discussion).

\begin{figure}[H]
\centering
\includegraphics[scale=.40]{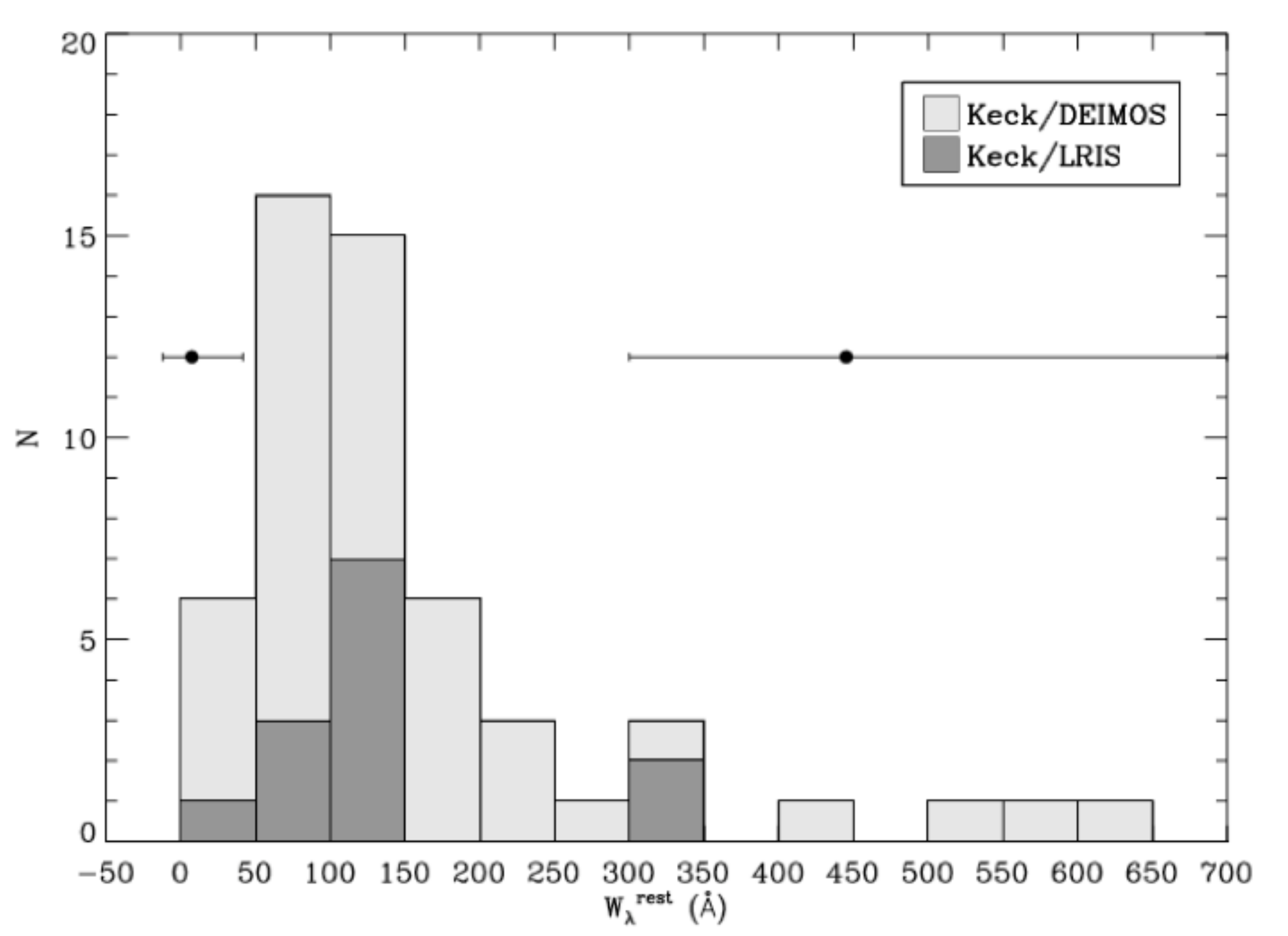}
\caption{
Ly$\alpha$ $EW_0$ distribution for LAEs at $z=4.5$ \citep{dawson2007}.
The light and dark gray histograms are obtained by Keck/DEIMOS and LRIS
spectroscopy, respectively.
The two filled circles with horizontal error bars show 
the typical errors at two Ly$\alpha$ $EW_0$ values.
This figure is reproduced by permission of the AAS.
}
\label{fig:dawson2007_fig7}
\end{figure}

\subsection{Large Ly$\alpha$ and He{\sc ii} Equivalent Widths: Pop III in LAEs?}
\label{sec:popIII_in_lae}

\begin{figure}[H]
\centering
\includegraphics[scale=.50]{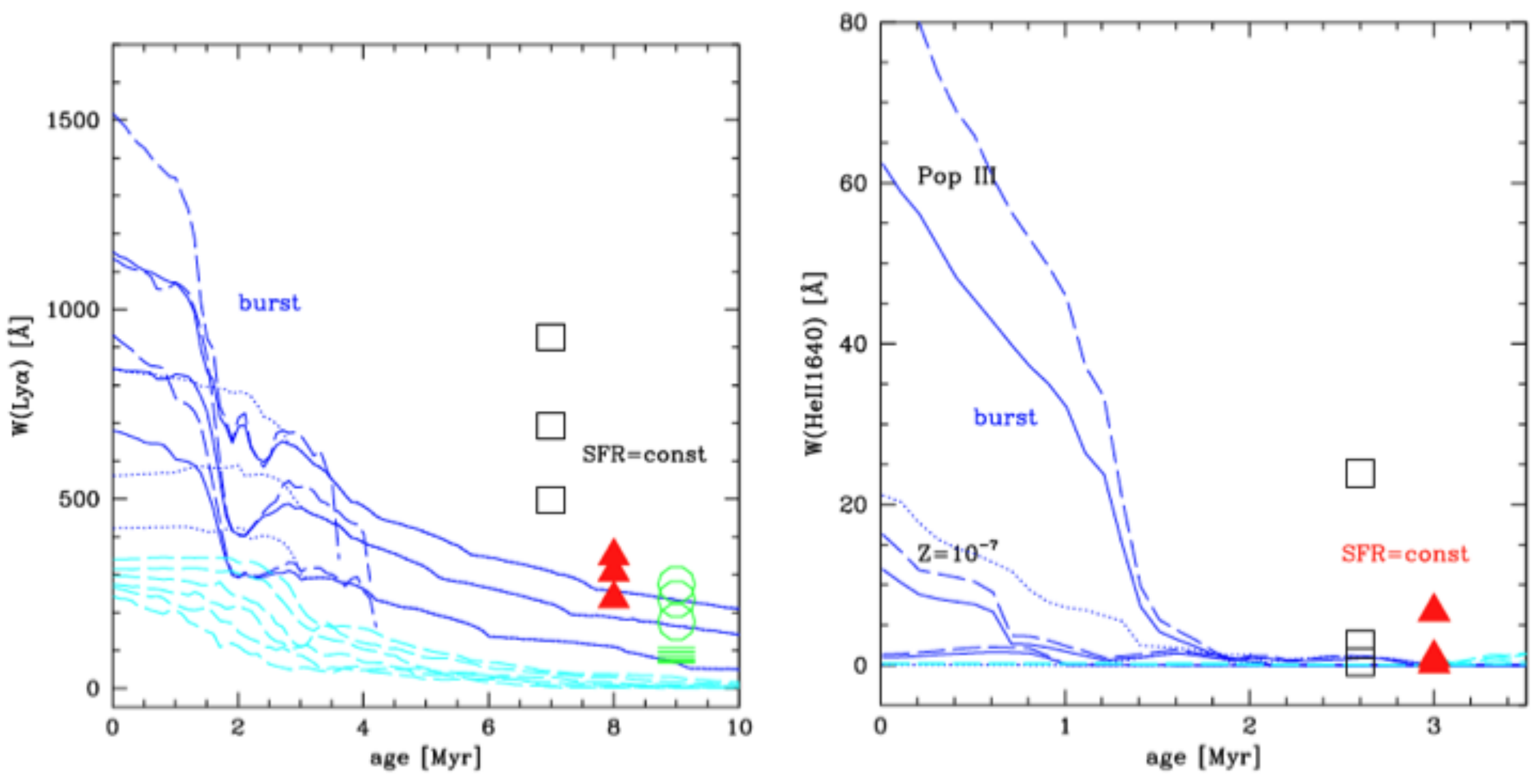}
\caption{
Ly$\alpha$ (left) and {\heii} (right) $EW_0$ 
as a function of stellar age for an instantaneous burst of star-formation
predicted by the stellar evolution and photoionization models 
of \citet{schaerer2003}.
The three blue dashed lines represent metallicity $Z=0$ models
with a Saltpeter IMF and mass ranges of $50-500$, $1-500$, and $1-100 M_\odot$
from top to bottom. The blue solid and dotted lines are the same as the blue dashed lines,
but for $Z=10^{-7}$ and $10^{-5}$. The cyan dashed lines
denote models with $Z=0.0004$, $0.001$, $0.004$, $0.008$, $0.020$, and $0.040$
from top to bottom. 
The three squares show constant SFR models with a $50-500 M_\odot$ 
Salpeter IMF for $Z=0$, $10^{-7}$, and $10^{-5}$
from top to bottom.
The red triangles and green circles are the same as the squares, 
but for the mass ranges $1-500$ and $1-100 M_\odot$, respectively.
The green short lines are the same as the green circles but with metallicities of 
$Z=0.0004$, $0.001$, $0.004$, $0.008$, $0.020$, and $0.040$
from top to bottom. Note that, in the right panel, {\heii} $EW_0\gtrsim 5$\AA\ is
reached only for models with very low metallicities of Z $\lesssim 10^{-7}$, 
except for the mass range $1-100 M_\odot$.
This figure is reproduced by permission of the A\&A.
}
\label{fig:schaerer2003_fig7}
\end{figure}

LAEs with large Ly$\alpha$ $EW_0$ values are potentially important objects
that have excessive Ly$\alpha$ emission at a given stellar continuum.
\citet{malhotra2002} claim the existence of
LAEs with Ly$\alpha$ $EW_0$ $\gtrsim 240$\AA\ 
that cannot be explained by
young star formation with the solar metallicity and a Salpeter IMF \citep{salpeter1955}.
Figure \ref{fig:dawson2007_fig7} presents 
a Ly$\alpha$ $EW_0$ histogram of $z=4.5$ LAEs. 
Although observational 
EW
estimates include 
large uncertainties and systematics due to weak or undetected continua 
(see, e.g., Figure 14 of \citealt{shimasaku2006}), LAE studies have shown
that $\sim 10-30$\% of LAEs in a narrowband-selected sample have
large ($\gtrsim 200-300$\AA) Ly$\alpha$ $EW_0$ 
at $z\sim 2-7$ 
\citep{dawson2007,shimasaku2006,ouchi2008,kashikawa2012}.
The physical origins of the large Ly$\alpha$ $EW_0$ objects
are not well understood. These LAE studies discuss the possibilities
of the Neufeld effect (Section \ref{sec:lya_escape_fraction}), 
cooling radiation (Section \ref{sec:lya_blob}), and pop III star formation.
The relation between large Ly$\alpha$ $EW_0$ and pop III star formation
is presented in the left panel of Figure \ref{fig:schaerer2003_fig7}.
This panel shows theoretically calculated Ly$\alpha$ $EW_0$ 
as a function of stellar age for various stellar populations. 
%
%
For a star-formation history of an instantaneous burst 
with solar metallicity, Salpeter IMF, and a mass range of $1-100$ $M_\odot$, 
the Ly$\alpha$ $EW_0$ does not exceed
$\sim 200-300$\AA\ even at 
the birth time.
A stellar population with 
a larger number of 
massive young stars that produce more 
ionizing photons has a higher Ly$\alpha$ $EW_0$.
%
Ly$\alpha$ $EW_0$ is thus sensitive to
the shape of the IMF and metallicity as well as stellar age.
%
It is predicted that a top heavy IMF 
is realized in metal poor gas clouds 
because they contain only 
a small amount of coolants 
that are needed for low-mass gas clumps to collapse.
Moreover, metal poor stars efficiently produce ionizing photons,
because ionizing photons are not absorbed by metals in the stellar atmosphere.
%
The left panel of Figure \ref{fig:schaerer2003_fig7} indicates that Ly$\alpha$ $EW_0$
can reach $\sim 500-1500$\AA\ for galaxies having
metal poor instantaneous star-formation with top heavy IMFs.
%
%
As demonstrated in this panel,
LAEs with large Ly$\alpha$ $EW_0$ values can be
candidates of pop III galaxies, 
although not definitive ones (e.g. \citealt{yang2006}). 
Moreover, the production rate of ionizing photons 
is sensitive not only to IMF, metallicity, and stellar age, but also 
to the binary fraction of massive stars and many other physical conditions 
such found in the BPASS model \citep{eldridge2017}.

Thus, another test is necessary to isolate pop III star formation
from the candidates. 
{\heii}1640 is an ideal emission line for such a test
\footnote{
The {\heii}1640 line corresponds to \heii\ \ha.
Note that the {\heii}304 line corresponding to {\heii} Ly$\alpha$ cannot be easily observed.
}.
Because He$^+$ has a 
high 
ionization potential of $54.4$ eV, 
He$^+$ can be ionized by hard spectra of very massive young stars 
that can be found in {\sc Hii} regions of pop III star formation.
The right panel of Figure \ref{fig:schaerer2003_fig7} presents 
{\heii}1640 $EW_0$ as a function of stellar age for instantaneous star-formation,
and suggests that a large {\heii}1640 $EW_0$ ($\gtrsim 10$\AA) is indicative of pop III.
Although the hot outflowing gas from a WR star and 
the broad-line region of an AGN can also produce {\heii}1640 emission with 
$EW_0 \gtrsim 10$\AA, both lines are predicted to be much broader, 
with a line-width velocity of $\sim 1000$ km s$^{-1}$, than those from 
{\sc Hii} regions of pop III star formation (a few hundred km s$^{-1}$).
However, the {\heii}1640 emission of narrow-line (type II) AGNs
has similarly small line widths.
To isolate pop III stars from such AGNs, one needs to
investigate high ionization lines such as \nv, \ovi, and strong X-ray emission
that cannot be produced by the photoionization by very massive stars (cf. fast radiative shocks; \citealt{thuan2005}).

 \begin{figure}[H]
\centering
\includegraphics[scale=.52]{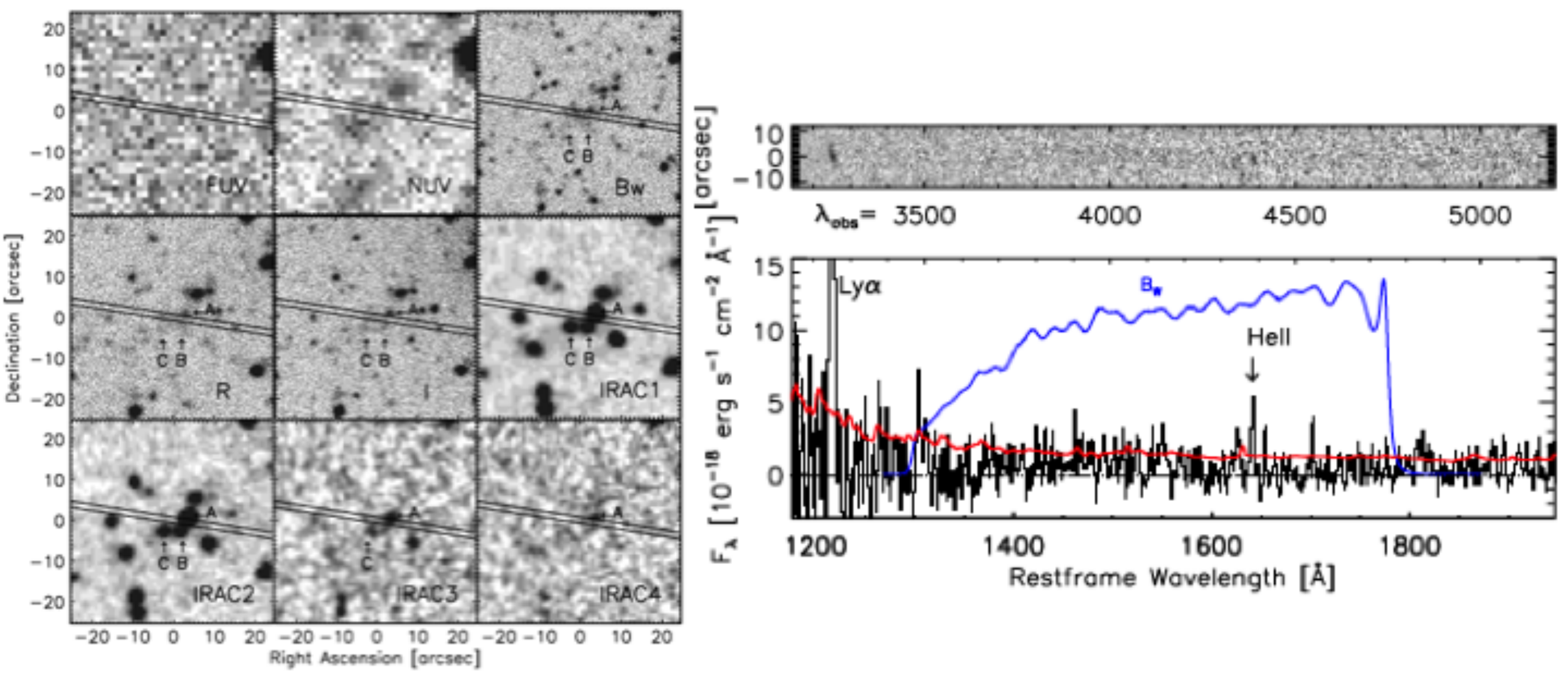}
\caption{
Left: Thumbnail images of PRG1  
taken from
GALEX (FUV and NUV), Kitt-Peak 4m ($B_{\rm w}$, $R$, and $I$), and
Spitzer ($3.6$, $4.5$, $5.8$, and $8.0 \mu$m; \citealt{prescott2009}).
Spectroscopy slit positions are shown with lines. The labels A, B, and C
denote the positions of three IRAC counterparts.
Right: Two dimensional (top) and one-dimensional (bottom) spectra of PRG1 \citep{prescott2009}. 
The black line presents the spectrum of PRG1, while the red line denotes $1 \sigma$ errors.
The Ly$\alpha$ and \heii\ lines are detected. The blue line indicates 
the $B_{\rm w}$ band transmission curve.
This figure is reproduced by permission of the AAS.
}
\label{fig:prescott2009_fig1_fig2}
\end{figure}

A strong narrow \heii\ line is found in a $z=2$ LAE 
with an extended Ly$\alpha$ halo, named PRG1 
(Figure \ref{fig:prescott2009_fig1_fig2}; \citealt{prescott2009}).
The \heii\ line of PRG1 is strong, \heii\ $EW_0=37\pm 10$\AA, and
the ratio of \heii\ to Ly$\alpha$ fluxes is ${\rm HeII}/{\rm Ly\alpha}=0.12$.
Metal lines 
are not detected
with upper limits of
\civ$1548/$Ly$\alpha$ and \ciii]1909$/$Ly$\alpha \lesssim 0.03$.
These properties of a strong narrow \heii\ and very weak (or no)
metal lines are suggestive of 
pop III star formation.
However, the subsequent deep spectroscopy of \citet{prescott2015}
has clearly detected \civ\ and \ciii] lines 
with a line ratio of
\civ$/$\heii, \ciii]$/$\heii$\sim 0.5$, 
indicating that PRG1 is photoionized by an AGN, not by pop III stars.

\begin{figure}[H]
\centering
\includegraphics[scale=.47]{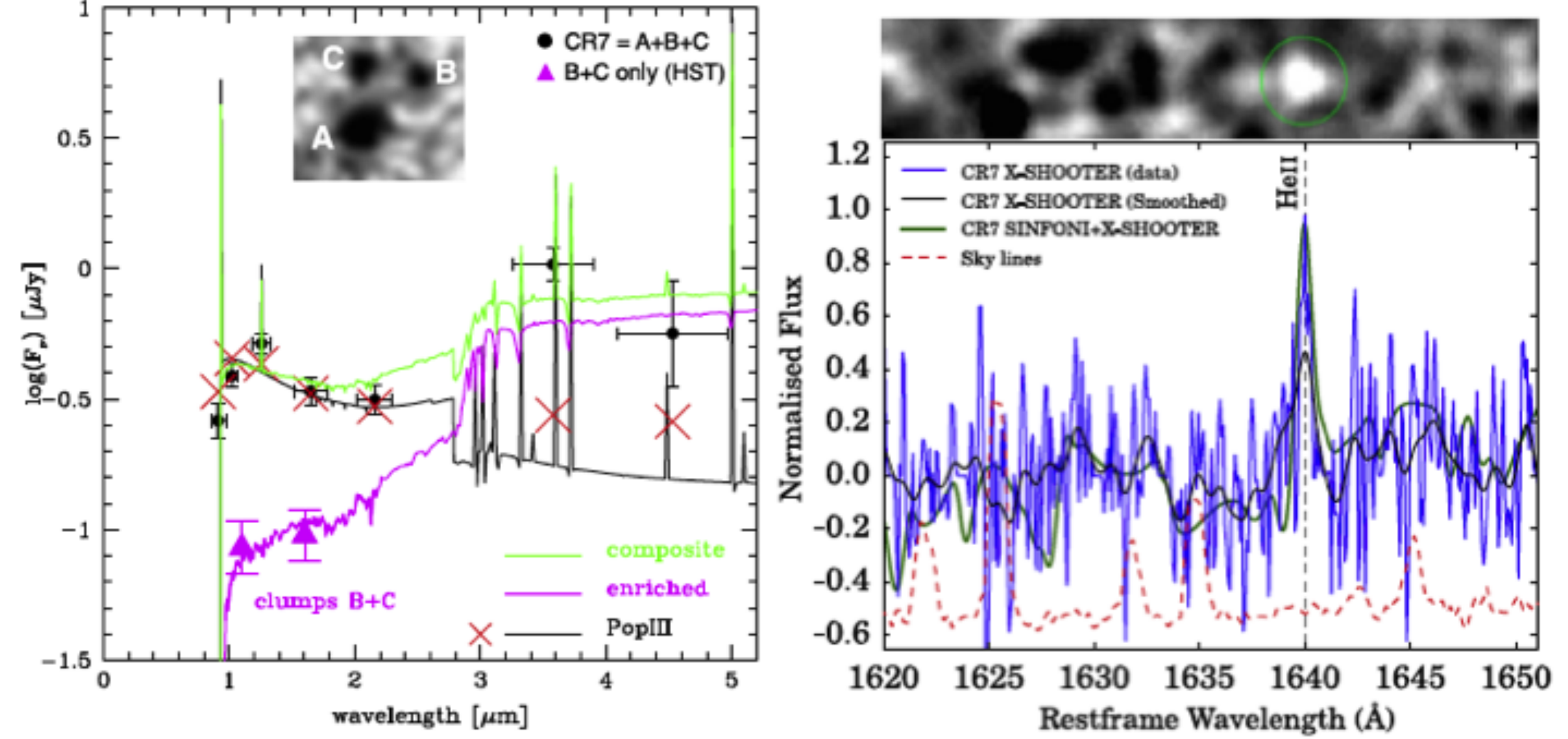}
\caption{
Left: The SED (main panel) and a thumbnail image (inset panel) of CR7 \citep{sobral2015}.
The thumbnail image is an HST NIR image that resolves the three continuum components of CR7 
named clumps A, B, and C, where clump A is the candidate of pop III star formation reported
by \citet{sobral2015}. In the main panel, the black circles represent total flux densities, while 
the magenta triangles denote HST NIR photometry of clumps B+C. The black and magenta
lines are the pop III and $Z=0.2Z_\odot$ SED models best-fit to 
the $1-2\mu$m data points of the black circles and the magenta triangles, respectively, 
and the green line indicates the sum of these two model SEDs.
The red crosses are the photometry 
predicted by 
the best-fit pop III model.
Right: Two-dimensional (top) and one-dimensional (bottom) spectra of CR7 reported by \citet{sobral2015}
(cf. \citealt{shibuya2018b,sobral2018}).
The black and blue lines represent X-Shooter spectra with and without a spectrum smoothing process, respectively.
The green line denotes a stack of the X-Shooter and VLT/SINFONI spectra. The red line indicates
the sky background spectrum.
This figure is reproduced by permission of the AAS.
}
\label{fig:sobral2015_fig5_fig4}
\end{figure}

More recently, 
%
\citet{sobral2015} have reported a detection of a strong narrow \heii\ emission 
in an LAE at $z=6.6$ that is dubbed CR7.
The left and right panels of Figure \ref{fig:sobral2015_fig5_fig4} present
the SED and the \heii\ spectrum of CR7. 
This object
is made of three stellar components
whose total SED exhibits a mature stellar population with a Balmer break.
Although the SEDs of the three stellar components are not clearly distinguished,
there is a possibility that one 
component, A, (Figure \ref{fig:sobral2015_fig5_fig4}) 
would have a very young population with a blue SED.
\citet{sobral2015} report that CR7 has a very large \heii\ equivalent width of $EW_0=80\pm 20$\AA\
as well as a large Ly$\alpha$ equivalent width of $EW_0=211\pm 20$\AA.
The reported line ratio of \heii$/$Ly$\alpha$$=0.22$ 
is about twice as large as 
the one of PRG1. No metal lines are detected in VLT/X-Shooter spectra
covering the entire NIR wavelength range accessible from the ground.
%
Some theoretical studies suggest that CR7 is a candidate of 
a direct collapse black hole because of a strong \heii\ line without
detection of metal lines from moderately-massive stellar components 
\citep{pallottini2015,dijkstra2016}. 
%
Recently, \citet{shibuya2018b} present reanalysis results of the CR7 X-Shooter spectra,
and find no \heii\ line,
placing only an upper limit.
A similar upper limit is also reported by \citet{sobral2018}.
Moreover, ALMA observations reveal the metal {\sc [Cii]} line \citep{matthee2017b}.
To summarize, 
although CR7 was a promising candidate of pop III star formation or a direct collapse black hole,  
subsequent studies find no such evidence.



\begin{figure}[H]
\centering
\includegraphics[scale=.45]{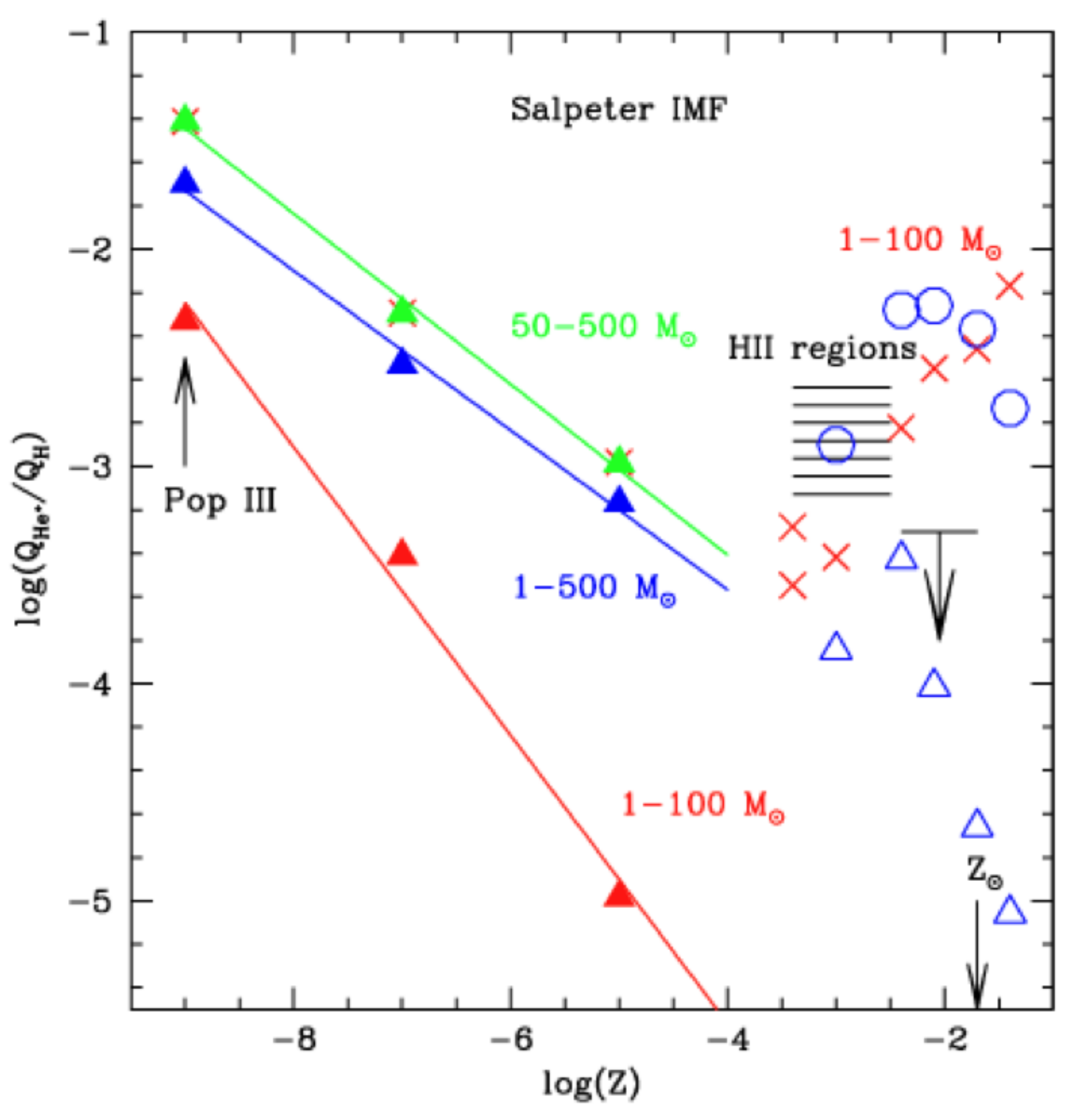}
\caption{
Spectral hardness $Q_{\rm He^+}/Q_{\rm H}$ as a function of metallicity \citep{schaerer2003}. 
The red, blue, and green
filled triangles are the spectral hardness values for three very metal poor cases with mass cuts
of $1-100$, $1-500$, and $50-500 M_\odot$, respectively, predicted with
the stellar evolution and photoionization models of \citet{schaerer2003}.
The red, blue, and green lines indicate the linear functions in the log-log plot
that are best fit to the red, blue, and green filled triangles, respectively.
The red crosses are the same as the red filled triangles, but for metal rich cases.
The blue open circles and open triangles are other model predictions (see \citealt{schaerer2003}).
The shaded region and the upper limit denote the spectral hardness of {\sc Hii} regions
estimated with observational data.
This figure is reproduced by permission of the A\&A.
}
\label{fig:schaerer2003_fig5}
\end{figure}

Spectral hardness measurements are useful to diagnose the presence of pop III star formation
in a galaxy. Figure \ref{fig:schaerer2003_fig5} shows theoretical predictions of 
the spectral hardness $Q_{\rm He^+}/Q_{\rm H}$
as a function of metallicity \citep{schaerer2003},
where $Q_{\rm He^+}$ and $Q_{\rm H}$ are the fluxes of ionizing photons for He$^+$ ($>54.4$ eV) and H ($>13.6$ eV), respectively.
This spectral hardness 
can be estimated from observed \heii\ and Ly$\alpha$ fluxes, 
$f_{\rm HeII 1640}$ and $f_{\rm Ly \alpha}$: 
\begin{equation}
\frac{f_{\rm HeII 1640}}{f_{\rm Ly \alpha}}\sim 0.55 \frac{Q_{\rm He^+}}{Q_{\rm H}}.
\label{eq:spectral_hardness}
\end{equation}
Reported $f_{\rm HeII 1640}/f_{\rm Ly \alpha}$ measurements 
give $Q_{\rm He^+}/Q_{\rm H}\sim 0.22$ for PRG1 \citep{prescott2009} and $\sim 0.42$ for CR7 \citep{sobral2015}. 
In Figure \ref{fig:schaerer2003_fig5}, the estimated $Q_{\rm He^+}/Q_{\rm H}$ values 
are larger than the predictions for pop III star formation by an order of magnitude 
even for a top heavy IMF with a mass range of $50-500 M_\odot$.
The large $Q_{\rm He^+}/Q_{\rm H}$ value of PRG1 is explained by the existence of an 
AGN, while that of CR7 is probably explained by the
recent reanalysis results of no \heii\ line detection \citep{shibuya2018b,sobral2018}.
%
%
In addition to analyses on an individual basis,
one can also measure an average $f_{\rm HeII 1640}/f_{\rm Ly \alpha}$ 
from stacked LAE spectra.
Composite spectra using large LAE samples
show no clear detection of \heii\ emission, placing upper limits
of $f_{\rm HeII 1640}/f_{\rm Ly \alpha}<2$\% and $20$\% 
at $z=3$ and $5$, respectively \citep{dawson2004,ouchi2008}. Although the upper limit for 
$z=5$ LAEs is not strong enough to give a meaningful constraint ($Q_{\rm He^+}/Q_{\rm H}\lesssim 0.1$),
the one for $z=3$ LAEs ($Q_{\rm He^+}/Q_{\rm H}\lesssim 0.01$) indicates that 
on average $z=3$ LAEs do not have star-formation dominated by pop III
with a top heavy IMF with a mass cut of $50-500$ or $1-500$ $M_\odot$.



\subsection{Summary of Galaxy Formation III}
\label{sec:summary_galaxy_formationIII}

Section \ref{sec:galaxy_formationIII} has presented three important questions about LAEs:
extended Ly$\alpha$ halos, Ly$\alpha$ escape mechanisms, and the connection between
LAEs and pop III star formation.  Deep observations have revealed
largely extended Ly$\alpha$ nebulae, dubbed LABs, Ly$\alpha$ filaments, and LAHs,
with a size of about a few $10$ kpc to $500$ kpc
around high-$z$ star-forming galaxies and AGNs.
Because tangential polarization signals are detected in the Ly$\alpha$ blob LAB1,
\hi\ gas scattering of Ly$\alpha$ photons should exist in LABs.
However, it is not clear what is the major source(s) of the Ly$\alpha$ photons. 
Proposed candidate sources are 
\hii\ regions in the ISM, cooling radiation, and unresolved dwarf satellite galaxies.
There exist LAEs with a Ly$\alpha$ $EW_0$ as large as a few $100$\AA.  
Recent theoretical calculations 
suggest that a clumpy ISM made of discrete clouds would
%
boost Ly$\alpha$ $EW_0$, but that
the 
boosting is only found in physical conditions (high extinction and low outflow velocities)
that are clearly different from those seen in typical LAEs. The mechanism producing large $EW_0$ values is still unknown.
Several observational studies have reported LAEs with narrow and strong \heii\ emission lines. 
These LAEs may be candidates of galaxies with pop III star formation whose young massive stars 
emit moderately high energy photons ionizing He$^+$. 
However, these pop III star-formation candidates can also be
narrow-line AGNs or may include erroneous \heii\ emission measurements. 
The spectral hardness $Q_{\rm He^+}/Q_{\rm H}$ is useful to diagnose pop III star formation and AGNs, 
although, to date, $Q_{\rm H}$ is often estimated from a Ly$\alpha$ flux that includes a large uncertainty in 
the Ly$\alpha$ escape fraction.

%
%

\section{Cosmic Reionization I: Reionization History}
\label{sec:cosmic_reionizationI}


There are two major questions about cosmic reionization: reionization history and reionization sources.
This section addresses the first question with an emphasis on LAE studies,
starting with a brief introduction to cosmic reionization. 
The second question, reionization sources,
is discussed in Section \ref{sec:cosmic_reionizationII}.

\subsection{What is Cosmic Reionization?}
\label{sec:cosmic_reionization_introduction}

\begin{figure}[H]
\centering
\includegraphics[scale=.45]{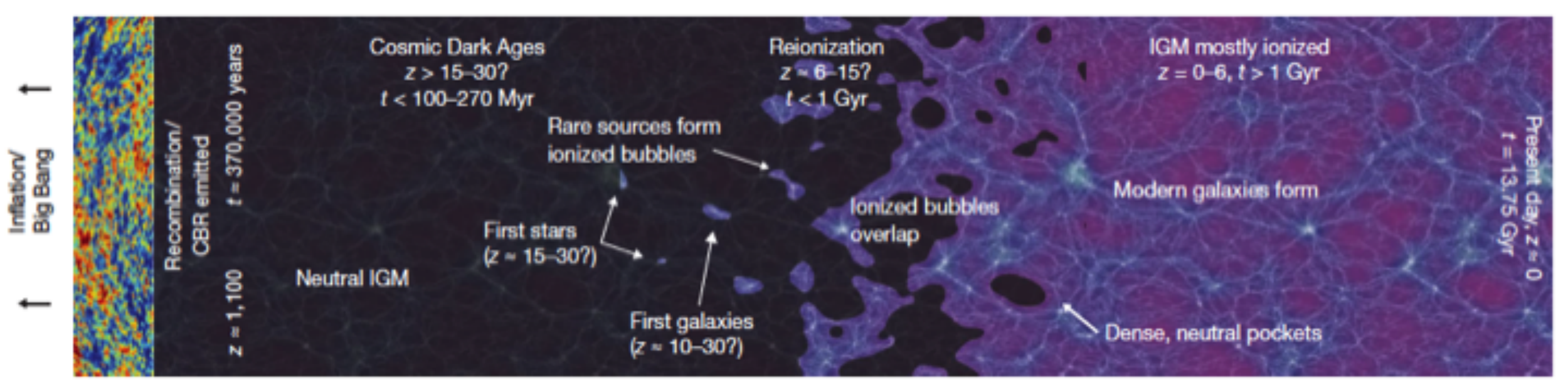}
\caption{
Picture of cosmic reionization in the cosmic history \citep{robertson2010}.
This figure is reproduced by permission of the Nature Publishing Group.
}
\label{fig:robertson2010_fig1}
\end{figure}

Cosmic reionization is a cosmic event that took place at a high redshift (Figure \ref{fig:robertson2010_fig1}).
By
the recombination of hydrogen at $z\sim 10^3$, 
the early universe with hot plasma gas 
evolved into one filled with neutral gas (i.e. atomic hydrogen gas);
the last photon scattering surface made by this transition
is observed as the cosmic microwave background (CMB).
On the other hand, today's universe does not contain abundant neutral gas, 
but harbors fully ionized gas in the inter-galactic space
that makes no Ly$\alpha$ absorption lines
at $z\sim 0$ in UV spectra of QSOs \citep{bahcall1991}.
These two pieces of evidence suggest that 
hydrogen atoms that became neutral at $z\sim 10^3$ were ionized again 
by today.
This event is known as cosmic reionization
(see the review of \citealt{fan2006b}). 
In this lecture, I focus on hydrogen reionization that is 
deeply related to LAEs and galaxy formation. See, e.g., \citet{worseck2014} for 
observational progresses in helium reionization studies. Hereafter, 'reionization' indicates
hydrogen reionization, if not otherwise specified.

Cosmic reionization is driven by ionizing photon radiation, $\gamma$,
the origin of which being stellar or non-stellar or both:

\begin{equation}
{\rm H} + \gamma \rightarrow {\rm H}^+ + {\rm e}^-,
\label{eq:reionization}
\end{equation}
where ${\rm H}^+$ is an ionized hydrogen atom (proton) 
and e$^-$ is an electron.

\begin{figure}[H]
\centering
\includegraphics[scale=.50]{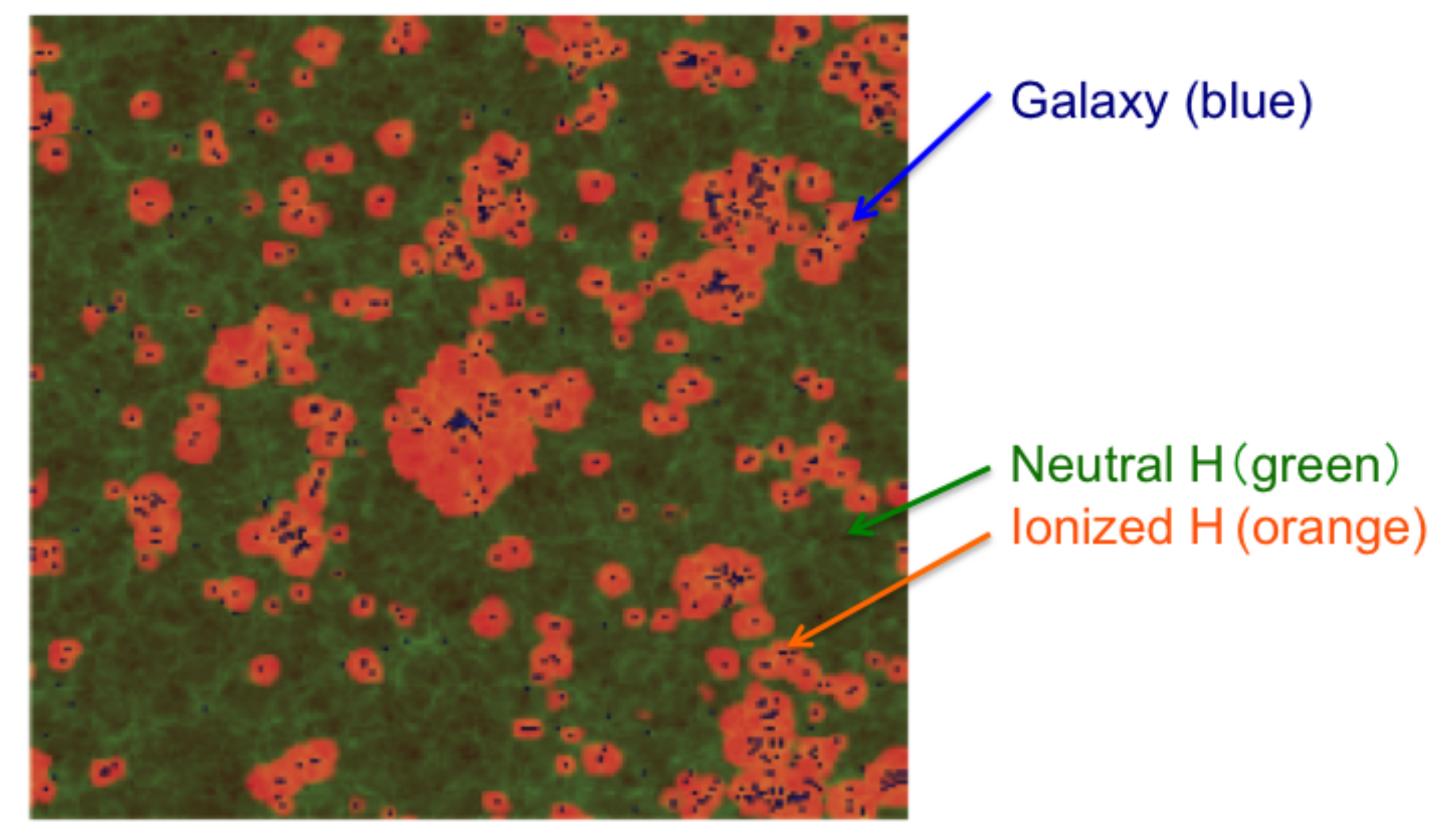}
\caption{
A snapshot of the simulations of cosmic reionization by \citet{iliev2006}.
The blue dots indicate galaxies emitting ionizing photons. 
The green and orange regions represent, respectively, regions with 
a neutral and ionized IGM.
The orange regions are called ionized bubbles.
This figure is reproduced by permission of the MNRAS.
}
\label{fig:bubbles_galaxies}
\end{figure}

Although the cosmic reionization process and reionization sources are not
well understood, many theoretical studies suggest a picture where
massive stars in galaxies provide the majority of ionizing photons and 
first ionize the IGM in the vicinity of galaxies (Figure \ref{fig:bubbles_galaxies}). 
Ionized regions made around galaxies 
are called ionized bubbles or cosmic \hii\ regions. Ionized bubbles
grow by time and merge, 
eventually 
making the universe fully ionized.
%
%
In this process, star-formation in low-mass galaxies is suppressed 
by heating of their cold gas 
by background ionizing photons (UV background radiation) 
if they are located in ionized bubbles 
(e.g. \citealt{susa2004,wyithe2006}). 
This physical picture indicates that galaxies
drive cosmic reionization by supplying ionizing photons, 
while star-formation in galaxies is strongly influenced 
by the UV background radiation.
Thus, cosmic reionization and galaxy formation have a tight
physical relation. Because the UV background radiation in ionized bubbles is 
originally produced by galaxies, one can also find 
that cosmic reionization is a cosmological-scale feedback process
for galaxies. In reionization studies,
one of the observational goals is to test this physical picture.



The key quantity for describing 
cosmic reionization is the neutral
hydrogen fraction, 
\begin{equation}
x_{\rm HI} = \frac{n_{\rm HI}}{n_{\rm H}},
\label{eq:x_HI}
\end{equation}
where $n_{\rm HI}$ is the neutral hydrogen density 
and $n_{\rm H}$ the neutral+ionized hydrogen density.
\footnote{
This is the volume-averaged 
fraction.
There is another definition of the neutral hydrogen fraction,
the mass-averaged neutral hydrogen fraction, that is sometime used. 
Because the mass-averaged neutral hydrogen fraction
is difficult to evaluate, 
the volume-averaged 
fraction is
referred to as the neutral hydrogen fraction in most observational studies.
}
The evolution of the IGM ionization, i.e., the history of reionization, 
is described by $x_{\rm HI}$ as a function of redshift.
Note that the ionized hydrogen fraction
\begin{equation}
Q_{\rm HII} =  1-x_{\rm HI} 
\label{eq:q_HII}
\end{equation}
is often used in place of $x_{\rm HI}$.
%

It is not easy to estimate $x_{\rm HI}$ (or $Q_{\rm HII}$).
Emission from ionized gas (e.g. Ly$\alpha$)
and neutral gas (e.g. 21 cm line) in the IGM is too diffuse 
to be directly detected even with today's technology.
Instead, 
one needs 
to detect an absorption or scattering signal by neutral gas 
imprint in spectra of bright background sources 
such as 
QSOs, CMB, LAEs, and gamma ray bursts (GRBs). 
The following subsections
summarize $x_{\rm HI}$ estimates 
obtained with this method.






\subsection{Probing Reionization History I: Gunn Peterson Effect}
\label{sec:gunn_peterson}

\begin{figure}[H]
\centering
\includegraphics[scale=.45]{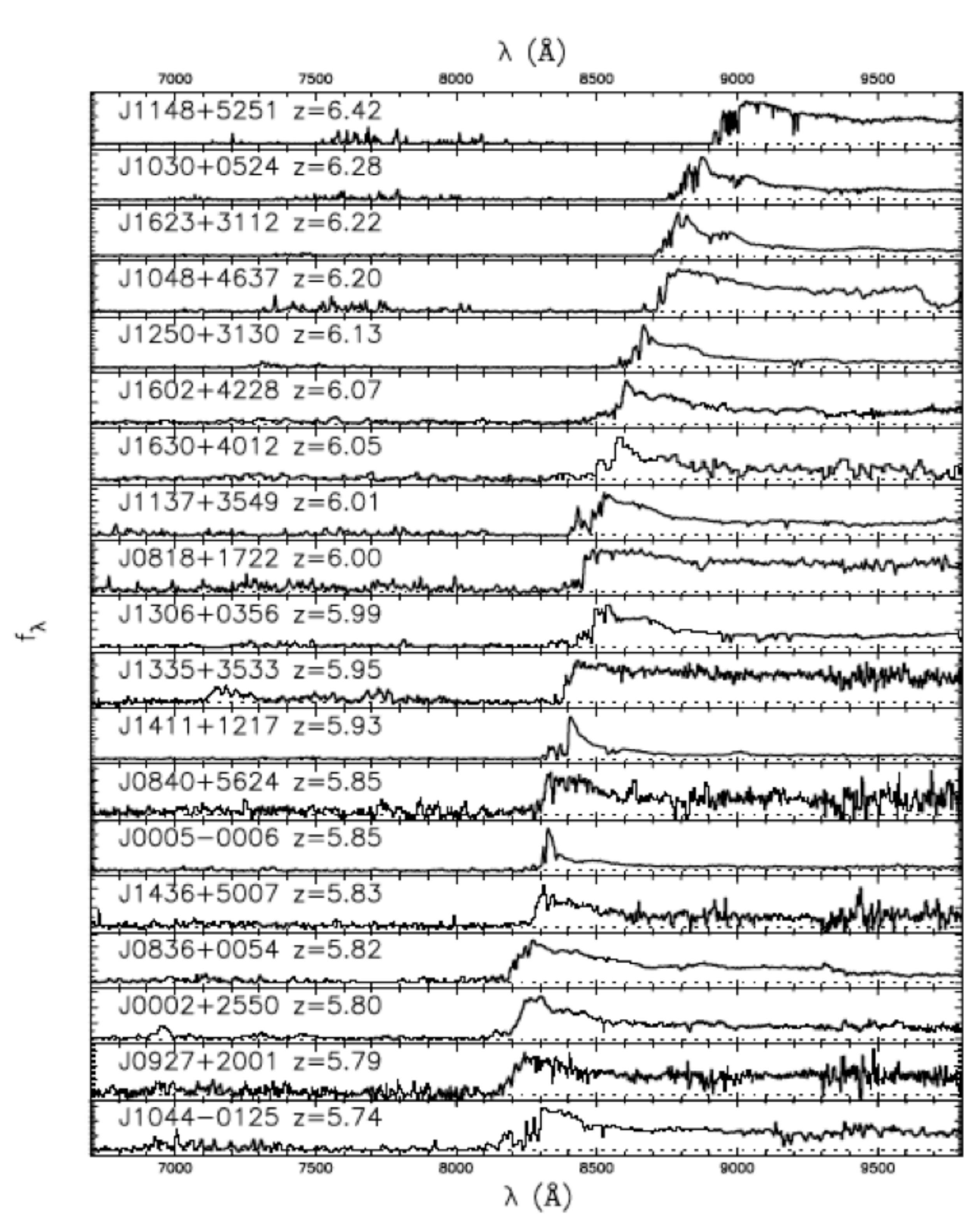}
\caption{
Optical spectra of QSOs at $z=5.7-6.4$ \citep{fan2006a}.
This figure is reproduced by permission of the AAS.
}
\label{fig:fan2006a_fig1}
\end{figure}

\begin{figure}[H]
\centering
\includegraphics[scale=.40]{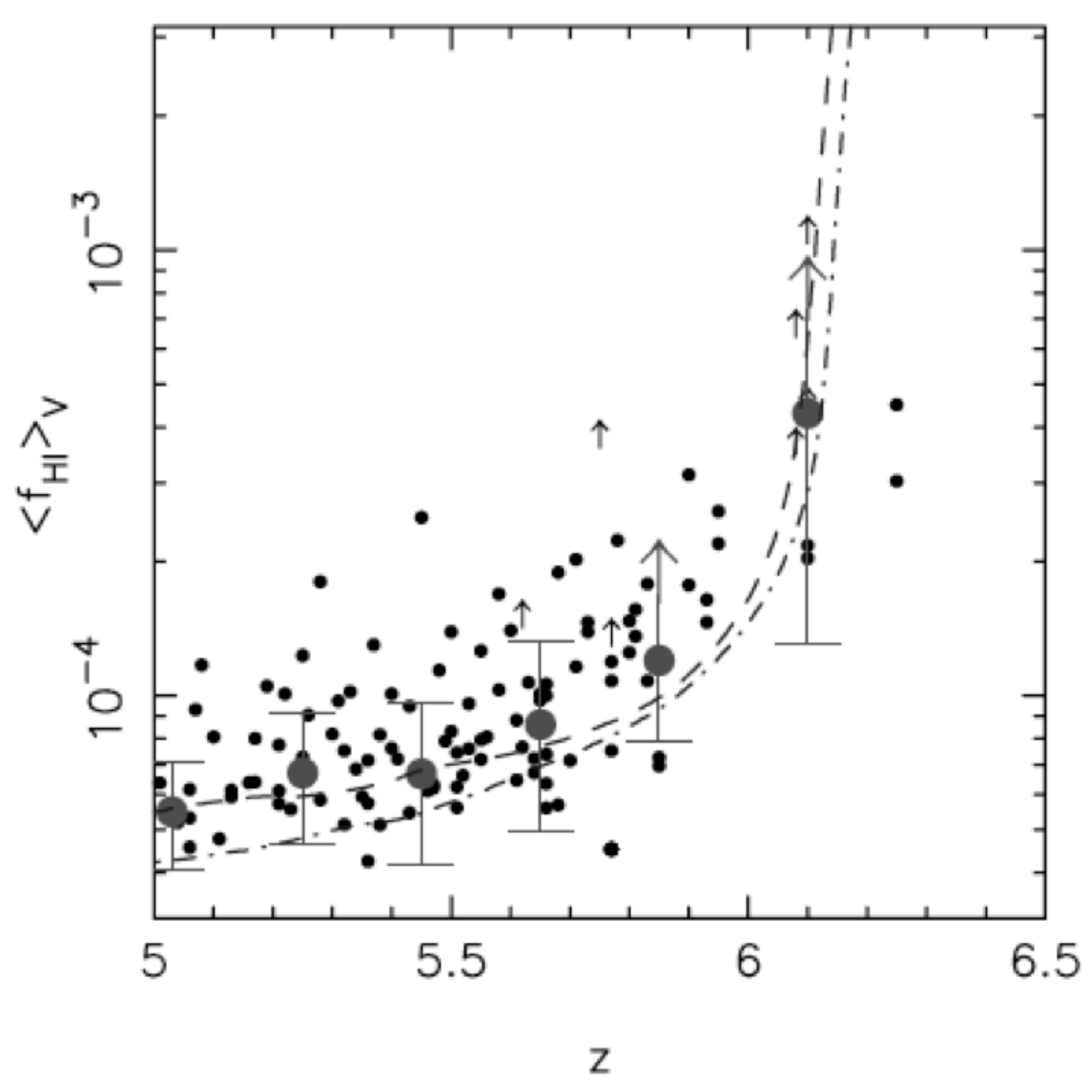}
\caption{
Volume-averaged neutral hydrogen fraction as a function of redshift \citep{fan2006a}.
The small circles (arrows) indicate neutral hydrogen fraction estimates (lower limits) 
from QSO $\tau_{\rm GP}$ measurements, while
the large circles (large circles with an arrow) represent average values (lower limits) for individual redshift bins.
The dashed and dot-dashed lines denote the simulation results of \citet{gnedin2004b}.
This figure is reproduced by permission of the AAS.
}
\label{fig:fan2006a_fig7}
\end{figure}

The classic method for estimating $x_{\rm HI}$ is to use \hi\ Ly$\alpha$ absorption
lines in high-$z$ QSO spectra, 
where QSOs play the role of bright background light.
Before the end of reionization, neutral hydrogen of the IGM makes
a complete absorption trough in QSO spectra at wavelengths shorter than Ly$\alpha$,
which is called the Gunn Peterson effect (\citealt{gunn1965}; see also \citealt{field1959,shklovskii1964,bahcall1965}). 
The strength of this effect for a given QSO spectrum
is evaluated with the Gunn-Peterson optical depth $\tau_{\rm GP}$
\begin{equation}
I/I_0 = e^{-\tau_{\rm GP}},
\label{eq:tau_gp}
\end{equation}
where $I$ and $I_0$ are the observed and intrinsic QSO continuum 
flux densities 
at the wavelength of the Ly$\alpha$ absorption of the redshifted 
IGM neutral hydrogen. Here, $I_0$ is estimated by 
a power-law extrapolation of the observed QSO continuum at $>1216$\AA.
The relation between $\tau_{\rm GP}$ and $x_{\rm HI}$ is written as
\begin{equation}
\tau_{\rm GP} (z) = 4.9\times 10^5 \left( \frac{{\rm \Omega_m} h^2}{0.13} \right)^{-1/2} \left( \frac{{\rm \Omega_b} h^2}{0.02} \right) \left( \frac{1+z}{7} \right)^{3/2} x_{\rm HI}(z)
\label{eq:tau_xHI}
\end{equation}
\citep{fan2006b}.
One can use this equation to estimate $x_{\rm HI}$ from $\tau_{\rm GP}$.
Figure \ref{fig:fan2006a_fig1} presents optical spectra of QSOs at $z=5.7-6.4$. 
In this figure, 
QSO residual fluxes escaping from the IGM absorption 
are found at $\lesssim 8000$\AA, while no significant continuum fluxes
remain at $\gtrsim 8000$\AA\ up to rest-frame 1216\AA\ (i.e. Ly$\alpha$)
\footnote{
Except at wavelengths very close to the QSO Ly$\alpha$. These wavelengths
correspond to the proximity region where hydrogen is completely 
ionized by strong UV radiation from the QSO.
}. 
These QSO spectra indicate large $\tau_{\rm GP}$ values at $\gtrsim 8000$\AA, 
or at $z\gtrsim 6$.
Figure \ref{fig:fan2006a_fig7} shows 
$\tau_{\rm GP}$-based $x_{\rm HI}$ measurements over $z \sim 5$--6.5,
indicating that $x_{\rm HI}$ 
rapidly increases at $z\gtrsim 5.7$. This rapid increase in $x_{\rm HI}$ 
suggests that cosmic reionization is completing 
at $z\sim 6$.


In this figure, 
only lower limits of $x_{\rm HI}$ are obtained at $z\gtrsim 5.7$.
This is because this method can estimate $x_{\rm HI}$ only when
the IGM is highly ionized with $x_{\rm HI}\lesssim 10^{-4}$.
Since Ly$\alpha$ is a resonance line, $\tau_{\rm GP}$ cannot be 
accurately measured for the IGM even with moderately small 
neutral fractions of $x_{\rm HI}\sim 10^{-4}-1$,
owing to the saturation of Ly$\alpha$ absorption.
%
%
In other words, this method is useful only 
at the final stage of cosmic reionization, i.e., $z\lesssim 6$.
To probe $x_{\rm HI}$ higher than $\sim 10^{-4}$,
one can use Ly$\beta$ and Ly$\gamma$ lines whose absorption
is weaker than that of Ly$\alpha$ by factors of 3 and 5, respectively.
With Ly$\alpha$ absorption lines in QSO spectra, 
there is another technique to evaluate $x_{\rm HI}$, which measures
the wavelength range over which the spectrum is completely absorbed.
This measurement is called the dark gap \citep{fan2006b}. 
However, this additional technique
extends the redshift range of $x_{\rm HI}$ measurements only up to $z\sim 6.5$.

\subsection{Probing Reionization History II: Thomson Scattering of the Cosmic Microwave Background}
\label{sec:cmb}

The CMB is another background light useful for probing 
the reionization history.
%
%
Because CMB photons are Thomson-scattered by 
free electrons existing between $z=0$ and $z=1100$ (the redshift when
CMB photons are created), 
one can identify signatures of the scattering in 
E-mode polarization and temperature fluctuation smearing seen in the CMB.
These signatures allow us to estimate the column density of free electrons
that is quantitatively expressed 
with the optical depth of Thomson scattering $\tau_{\rm e}$.

\begin{figure}[H]
\centering
\includegraphics[scale=.40]{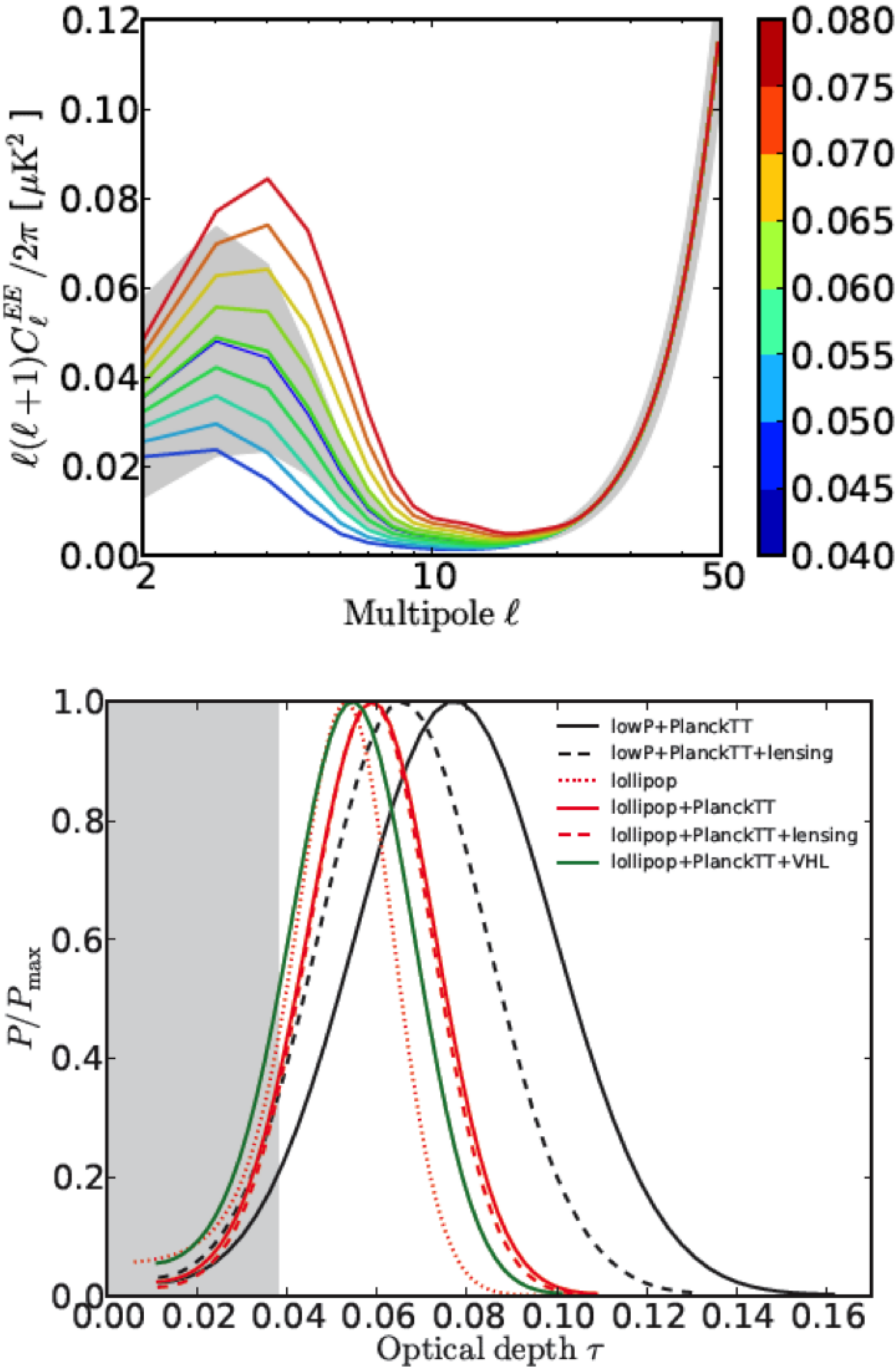}
\caption{
Top: E-mode polarization auto-power spectrum \citep{planck2016}. 
The colored curves represent power spectra for $\tau_{\rm e}$ values
indicated with the color code on the right hand side, while the gray region
denotes the cosmic variance for full sky observations
in the case of $\tau_{\rm e}=0.06$.
Bottom: Posterior distributions of $\tau_{\rm e}$ for various combinations of 
Planck data \citep{planck2016}. The gray region indicates the range of $\tau_{\rm e}$ 
that is ruled out by observational constraints from the QSO Gunn-Peterson effect.
This figure is reproduced by permission of the A\&A.
}
\label{fig:planck2016_fig3_fig5}
\end{figure}

Here, $\tau_{\rm e}$ is particularly sensitive to
large-scale (low multipole $\ell<10$) anisotropies of CMB polarization.
The top panel of Figure \ref{fig:planck2016_fig3_fig5} presents auto-power spectra 
of CMB E-mode polarization anisotropies with various $\tau_{\rm e}$ values,
demonstrating that measurements of CMB polarization can constrain $\tau_{\rm e}$.
The auto-power spectra of CMB polarization depend 
not only on $\tau_{\rm e}$ but also on 
the cosmic reionization history, i.e., redshift evolution of $x_{\rm HI}$.
However, the dependence on the latter 
is much smaller than the uncertainties in $\tau_{\rm e}$ measurements to date \citep{planck2016}.
Thus, when deriving $\tau_{\rm e}$, one can safely assume that 
the universe is instantaneously ionized 
at a redshift that is referred to as $z_{\rm re}$.
The bottom panel of Figure \ref{fig:planck2016_fig3_fig5} shows posterior probability
distributions of $\tau_{\rm e}$ given by Planck 2016 observations.
The best-estimate $\tau_{\rm e}$ from the Planck 2016 study
is $\tau_{\rm e}=0.058\pm 0.012$.

The Thomson scattering optical depth up to a given redshift 
is expressed as: 
\begin{equation}
\tau_{\rm e} (z) = \sigma_{\rm T} \int^{z}_{0} n_e(z') \frac{dl(z')}{dz'} dz',
\label{eq:tau_e}
\end{equation}
where $\sigma_{\rm T}$ is the cross section of Thomson scattering
and $n$ the number density of free electrons.
Setting $z$ to $1100$ gives the total optical depth between today and 
the time when CMB photons are created.
In the standard picture, there is a negligible contribution to $\tau_{\rm e}$ 
before the formation of the first stars ($z\gtrsim 20$).

In the case of instantaneous reionization, Equation (\ref{eq:tau_e}) is simplified to
\begin{equation}
\tau_{\rm e} (z_{\rm re}) \simeq 0.07 \left( \frac{h}{0.7} \right) \left( \frac{ {\rm \Omega_b} }{ 0.04 } \right)  \left( \frac{ {\rm \Omega_m} }{ 0.3 } \right)^{-1/2} 
\left( \frac{1+z_{\rm re}}{ 10 } \right)^{3/2}.
%
\label{eq:tau_e_simp}
\end{equation}
Based on the $\tau_{\rm e}$ estimate above,
\citet{planck2016} obtain the instantaneous reionization redshift 
to be $z_{\rm re} \simeq 7.8-8.8$  
that is a moderately late epoch.
%
%
However, note again that the cosmic reionization history cannot be
constrained well by this method because the power spectra of CMB 
polarization are not sensitively dependent on it.
%
\footnote{
There is another probe for the cosmic reionization history that uses
Kinetic Sunyaev-Zeldovich effects of the CMB temperature anisotropies 
made by the bulk motion of free electrons at the EoR.
However, the constraints on the cosmic history, so far obtained, 
are not strong (see the summary of \citealt{planck2016}).
}



%

\subsection{Probing Reionization History III: Ly$\alpha$ Damping Wing}
\label{sec:lya_damping_wing}

Sections \ref{sec:gunn_peterson} and \ref{sec:cmb} have reviewed
two methods to probe the cosmic reionization history.
The method using the Gunn Peterson effect can pinpoint the completion 
epoch of cosmic reionization at $z\sim 6$, but it cannot probe 
$z\gtrsim 6$ due to the saturation of Ly$\alpha$ absorptions.
The method using CMB Thomson scattering, on the other hand, can probe 
the entire cosmic history with free electrons between $z=0$ and 
the CMB epoch, but it has not been able to clearly distinguish 
different cosmic reionization histories. 
%

\begin{figure}[H]
\centering
\includegraphics[scale=.40]{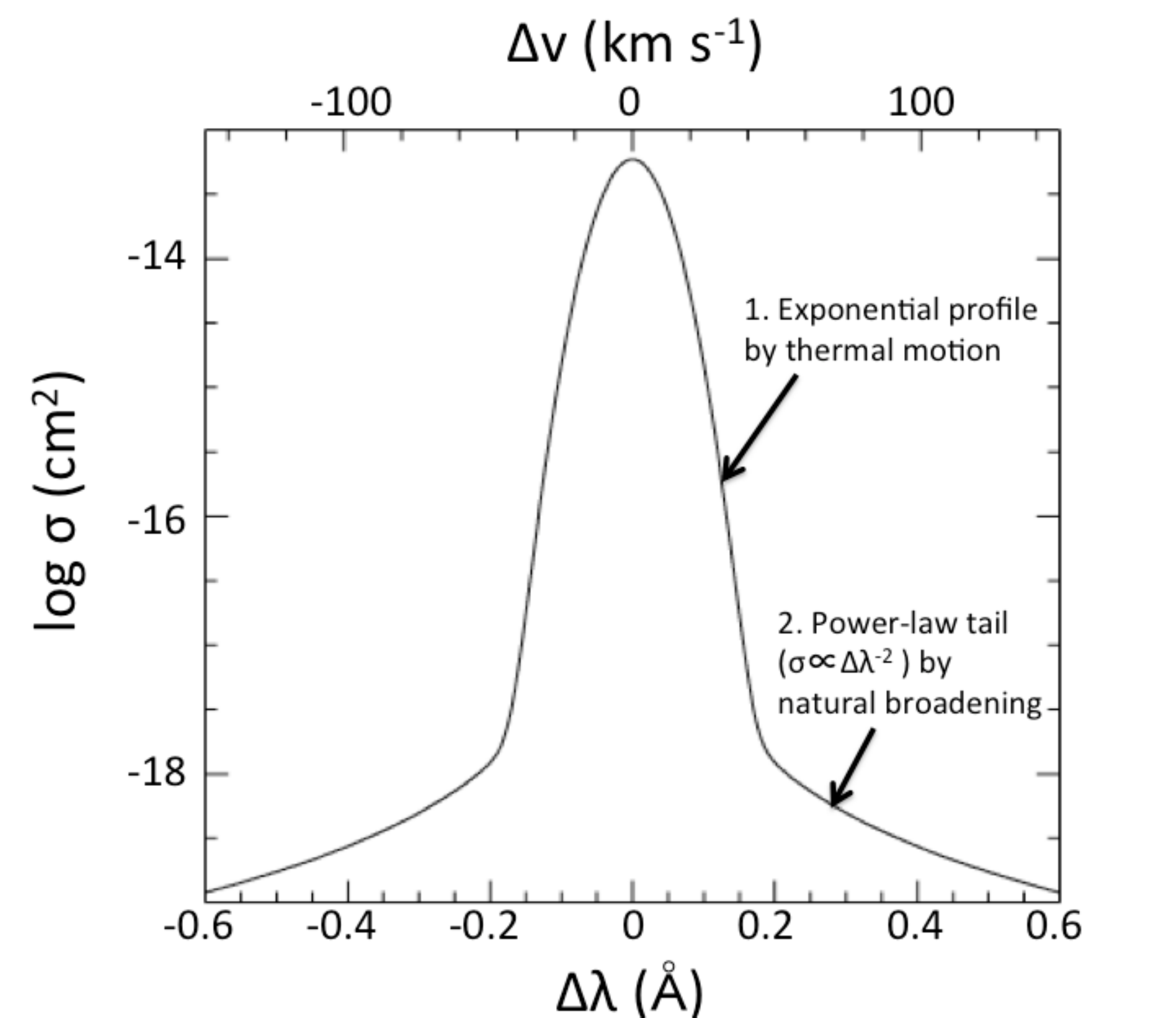}
\caption{
Ly$\alpha$ scattering cross section as a function of wavelength 
(or velocity).
The cross section profile consists of two components: 
an exponential profile made by thermal motions and 
a power-law tail by natural broadening.
}
\label{fig:lya_scattering_cross_section}
\end{figure}

To probe $x_{\rm HI}$ at the missing epoch of $z\gtrsim 6$, one can use 
HI Ly$\alpha$ damping wing (DW) absorptions seen
in LAE, GRB, and QSO spectra.
Briefly, Ly$\alpha$ DW is the tail of Ly$\alpha$ absorption.
Figure \ref{fig:lya_scattering_cross_section} presents the profile of Ly$\alpha$ cross section.
The Ly$\alpha$ absorption profile consists of two components. 
One is the main component with an exponential profile produced by thermal 
motions, and the other a weak, power-law component due to natural 
broadening 
(i.e., quantum mechanics's uncertainty principle in energy and time). 
Ly$\alpha$ DW corresponds to the latter.
The Ly$\alpha$ DW absorption is more than 5 orders of magnitude
weaker than the peak of the main absorption component. Moreover,
since the DW absorption has a power-law shape ($\sigma \propto \Delta \lambda^{-2}$),
it can extend to much redder wavelengths beyond $1216$\AA\ 
than the main component.
Because the Ly$\alpha$ DW absorption is significantly weaker than 
that by 
the Gunn-Peterson effect,
the DW absorption allows us to investigate the IGM with a moderately high neutral hydrogen fraction, $x_{\rm HI} \sim 0.1-1.0$.
Moreover, the extended profile of the DW absorption is useful to study 
continua at $>1216$\AA\ free from 
the Gunn-Peterson effect.
Below, I detail $x_{\rm HI}$ constraints from the Ly$\alpha$ DW absorption
so far obtained with GRB, QSO, and LAE spectra.



\begin{figure}[H]
\centering
\includegraphics[scale=.40]{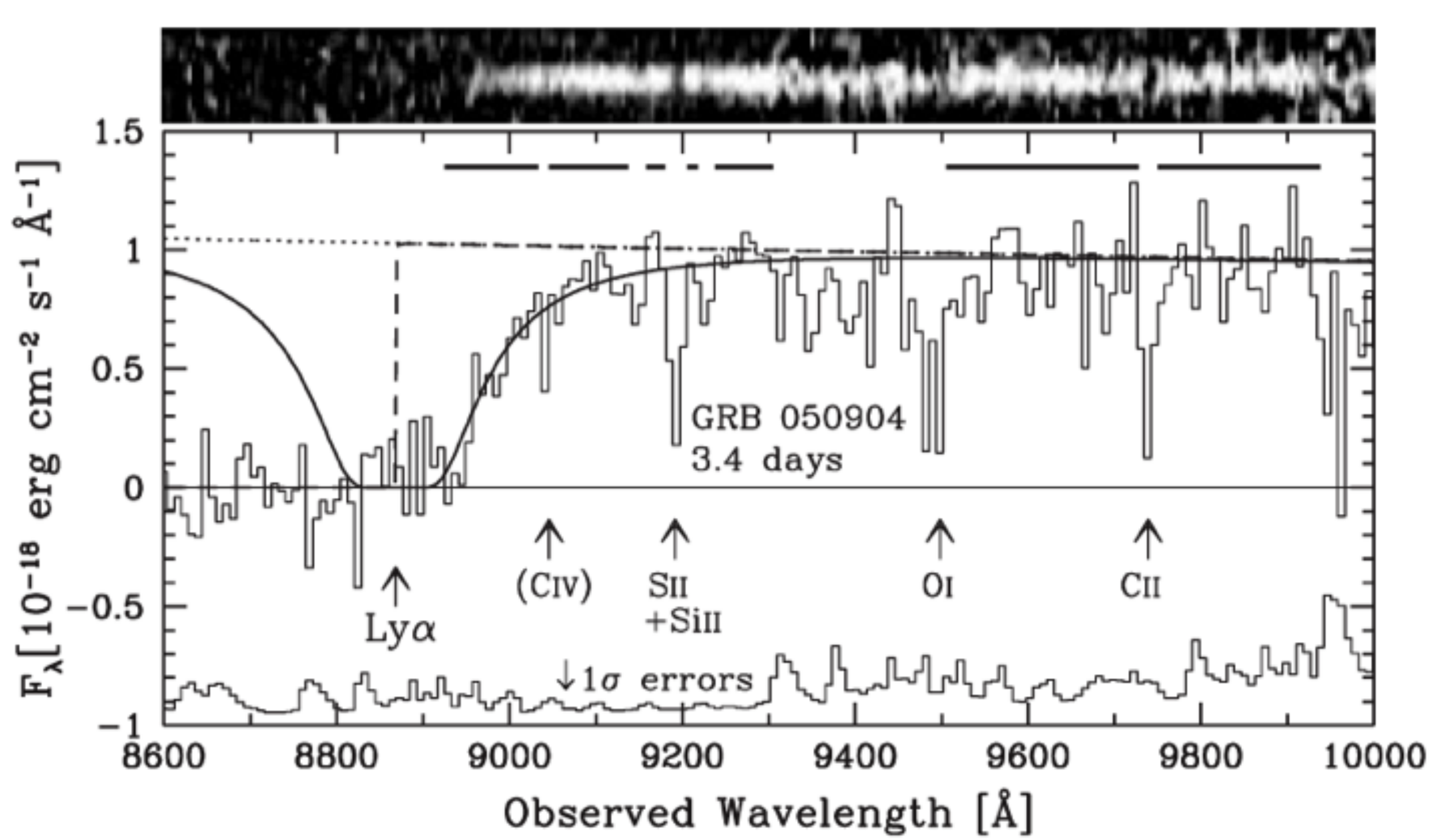}
\caption{
GRB 050904 afterglow spectrum \citep{totani2006}. 
Top: Two dimensional spectrum.
Middle: One dimensional spectrum. 
The solid curve represents the best-fit model of the absorption by
the neutral hydrogen of the host galaxy in the case of no IGM absorption.
The dotted line is an intrinsic power-law spectrum determined 
from the observed continuum in the wavelength ranges indicated 
by the horizontal lines at 
$F_{\rm \lambda}\simeq 1.35 \times 10^{-18}$ erg cm$^{-2}$ s$^{-1}$ 
\AA$^{-1}$.
The dashed line denotes a model spectrum when only the IGM absorption
shortward of Ly$\alpha$ is taken into account; 
a very low neutral fraction of $x_{\rm HI}=10^{-3}$ is assumed here.
Bottom: One sigma errors in the observed spectrum.
This figure is reproduced by permission of the PASJ.
}
\label{fig:totani2006_fig1}
\end{figure}

\subsubsection{GRBs}
\label{sec:grb}

To date, four GRBs at $z\sim 6-7$ have been used 
to constrain $x_{\rm HI}$: 
GRB 050904 at $z=6.3$ \citep{totani2006},
GRB 080913 at $z=6.7$ \citep{patel2010},
GRB 130606A at $z =5.9$ \citep{chornock2013,totani2014,totani2016}, and
GRB 140515A at $z=6.3$ \citep{chornock2014}.
Figure \ref{fig:totani2006_fig1} presents an observed spectrum of 
GRB 050904
and the best-fit continuum model with the Ly$\alpha$ DW absorption.
Here, the intrinsic spectrum shortward of Ly$\alpha$ is an extrapolation 
of a power-law function fitted to the observed continuum
longward of Ly$\alpha$.
Note that Ly$\alpha$ DW absorption modeling should consider
not only IGM neutral hydrogen but also that in the GRB host galaxy. 
The Ly$\beta$ line is also used to resolve the
degeneracy between the IGM and host-galaxy components 
\citep{totani2006}.
Among the four GRBs, one gives an estimate of $x_{\rm HI}$ of 
a few percent, while the others only place an upper or lower limit of 
$x_{\rm HI}\lesssim 0.1-0.7$ at $z\sim 6-7$.
%
Although there are many GRBs found at the EoR, $z\gtrsim 6$, that include
GRB 090423 at $z=8.2$, the most distant GRB confirmed to date \citep{tanvir2009}, 
all of them 
are too faint to identify the Ly$\alpha$ DW absorption.


\begin{figure}[H]
\centering
\includegraphics[scale=.40]{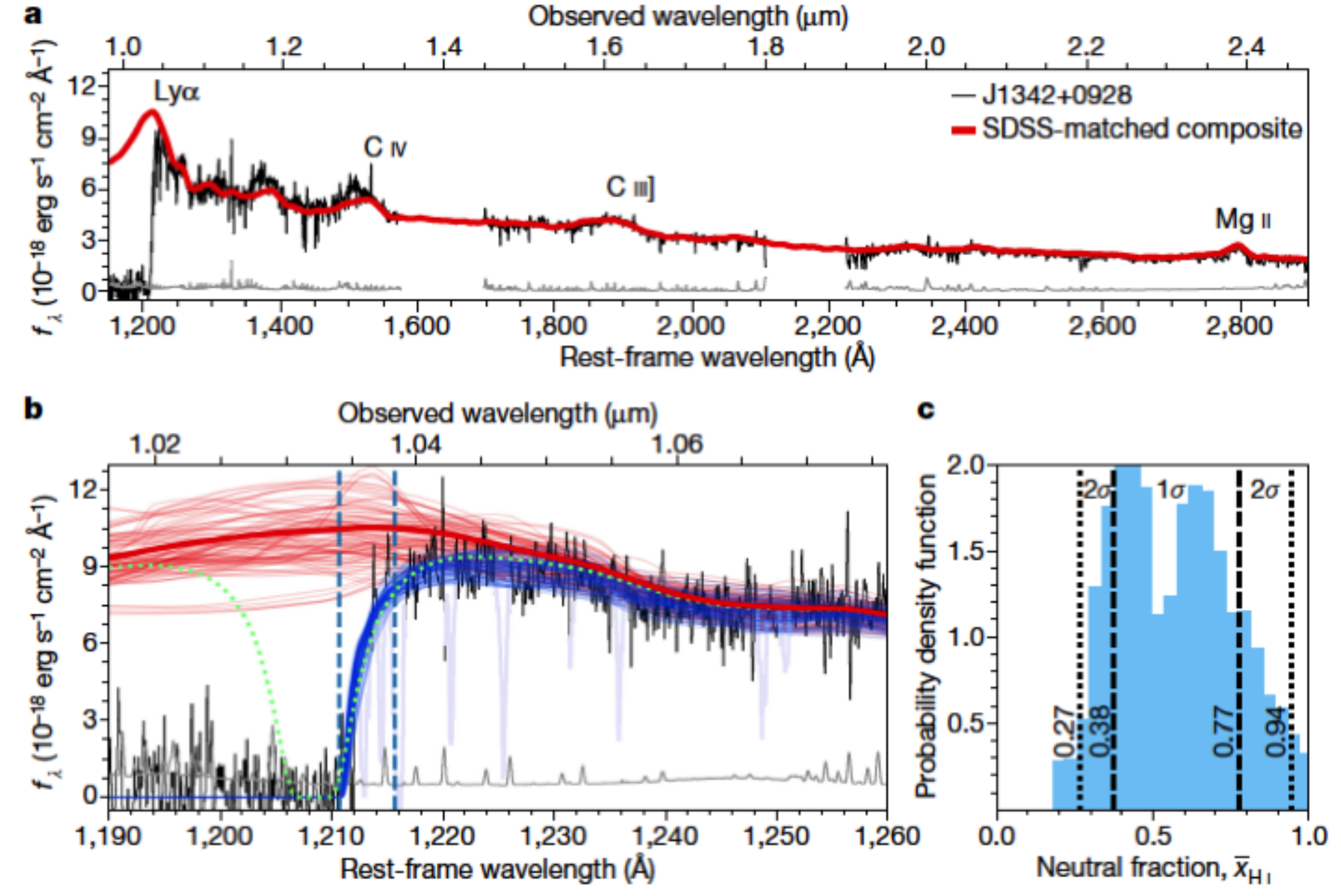}
\caption{
Spectra of QSO ULAS J1342+0928 at $z=7.5$ and $x_{\rm HI}$ estimation \citep{banados2018}.
Top: Observed spectrum of ULAS J1342+0928 (black line) and the best-matched
SDSS QSO composite spectrum (red line). The gray line represents the
$1\sigma$ error. 
Bottom left: Same as the top panel, but for a close-up view around 
the Ly$\alpha$ wavelength.
The thick blue line represents the best-matched composite spectrum with the DW absorption
by the IGM whose neutral fraction is set to the average value 
over the redshift range between $z=7$ and the end of 
the QSO proximity zone (blue dashed lines).
The green dotted line denotes a spectrum with a single absorber, such as a foreground galaxy, 
that appears different from the observed spectrum.
Bottom right: Probability density distribution (blue histogram) of $x_{\rm HI}$ values that are obtained by a fitting
%
%
of the composite spectrum with DW absorptions. 
The $1$ and $2 \sigma$ ranges of $x_{\rm HI}$ are indicated
with vertical dashed and dotted lines, respectively.
This figure is reproduced by permission of the Nature Publishing Group.
%
}
\label{fig:banados2018_fig3}
\end{figure}

\subsubsection{QSOs}
\label{sec:qso}

DW absorptions 
in QSO spectra also
provide constraints on $x_{\rm HI}$.
As in the GRB DW absorption analyses,
one needs to assume an intrinsic spectrum for the QSO in question.
While the major uncertainty in GRB analyses
is the DW absorption by \hi\ gas in the host galaxy,
QSO analyses include potential systematic uncertainties in 
the Ly$\alpha$ emission profile and the QSO near-zone size
that impact on the shape of the intrinsic QSO spectrum 
before IGM absorption.
To mitigate these systematic uncertainties, 
one can use a low-$z$ QSO spectrum template to estimate the intrinsic spectrum.
Figure \ref{fig:banados2018_fig3} presents the spectrum of 
QSO ULAS J1342+0928 at $z=7.5$. 
The bottom left panel of Figure \ref{fig:banados2018_fig3}
is a close-up around the Ly$\alpha$ wavelength of this object,
overplotted with the best-estimate intrinsic spectrum 
that is a composite of SDSS QSO spectra (thick red line).
The spectrum with the DW absorption by neutral hydrogen is presented
with a thick blue line. The bottom right panel of Figure \ref{fig:banados2018_fig3}
indicates the obtained probability density function of the IGM 
neutral hydrogen fraction,
from which the best-estimate IGM neutral hydrogen fraction is 
found to be 
$x_{\rm HI}=0.56^{+0.21}_{-0.18}$ at $z=7.5$ \citep{banados2018}.
With a lower-redshift QSO than this object,
ULAS J112010641 at $z=7.1$,
\citet{mortlock2011} obtain $x_{\rm HI}>0.1$. 
This neutral hydrogen constraint at $z=7.1$ is a lower limit  
because of an uncertain contribution by a damped Ly$\alpha$ (DLA) system 
associated with this object.
Subsequently, \citet{greig2016} carefully
reconstruct the intrinsic spectrum of 
this QSO 
from SDSS BOSS data,
and obtain $x_{\rm HI}=0.40^{+0.21}_{-0.19}$ at $z=7.1$ in conjunction with 
patchy reionization modeling.
%
%
%
%


\subsubsection{LAEs}
\label{sec:lae}

LAEs play a unique role in estimating $x_{\rm HI}$. In contrast with
bright continuum sources, i.e., GRBs and QSOs, DW absorptions
in the Ly$\alpha$ emission lines of LAEs are used to quantify $x_{\rm HI}$
(e.g. \citealt{malhotra2004,kashikawa2006,kashikawa2011,ouchi2010}).
Although spectra of LAEs are too faint to be modeled 
with a comparable accuracy as those of GRBs and QSOs, 
the abundance of LAEs is orders of magnitude higher than those of GRBs and QSOs.
Thus, LAEs can probe the \hi\ of the IGM with a large number of sightlines,
which reduces the field variance systematics.
Moreover, one can evaluate the IGM absorption amount of Ly$\alpha$ DW
with simple comparisons of Ly$\alpha$ statistics,
exploiting the large statistics given by abundant LAEs.

\begin{figure}[H]
\centering
\includegraphics[scale=.50]{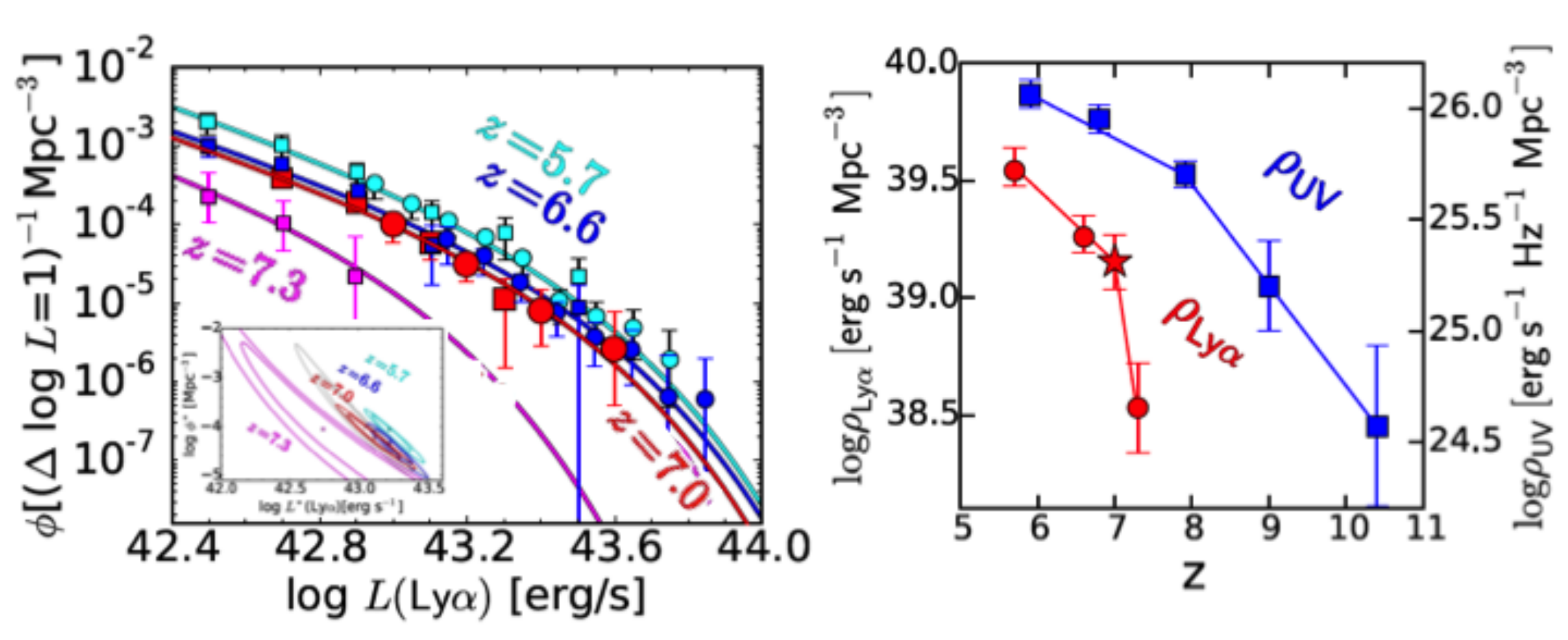}
\caption{
Left: Evolution of the Ly$\alpha$ luminosity function over $z=5.7-7.3$ \citep{itoh2018}.
The cyan, blue, red, and magenta data points (curves) present Ly$\alpha$ luminosity functions
(the best-fit Schechter functions) at $z=5.7$, $6.6$, $7.0$, and $7.3$, respectively.
The inset panel presents the error contours of 
the Schechter parameters $\phi^*$ and $L^*_{\rm Ly\alpha}$ 
at the 68 and 90\% confidence levels.
Right: Redshift evolution of the Ly$\alpha$ luminosity density 
(red symbols and lines) and the UV luminosity density 
(blue symbols and lines) \citep{itoh2018}. 
The Ly$\alpha$ luminosity density drops at $z\gtrsim 7$
faster than the UV continuum density,
suggestive of strong Ly$\alpha$ absorption by the IGM with a moderately high $x_{\rm HI}$ at $z\gtrsim 7$.
The left and right ordinate axes indicate the Ly$\alpha$ and $UV$-continuum
luminosity densities, respectively.
This figure is reproduced by permission of the AAS.
}
\label{fig:itoh2018_fig9_fig10_fig11}
\end{figure}

Figure \ref{fig:itoh2018_fig9_fig10_fig11}
shows the evolution of the Ly$\alpha$ luminosity function and $\rho_{\rm Ly\alpha}$ over $z=5.7-7.3$
obtained by deep narrowband imaging and spectroscopic surveys.
Both the Ly$\alpha$ luminosity function and $\rho_{\rm Ly\alpha}$ decrease from $z=5.7$
towards higher $z$. Moreover, it is also suggested that $\rho_{\rm Ly\alpha}$ evolution is accelerated at $z\gtrsim 7$,
and that the decrease in $\rho_{\rm Ly\alpha}$ at $z>7$ is clearly faster than that in 
$\rho_{\rm UV}$ that represents the star-formation rate density evolution.
This accelerated evolution of $\rho_{\rm Ly\alpha}$, 
which cannot be explained by any observed evolutionary trends in 
the star-formation properties of LAEs, 
suggests high neutral hydrogen fractions of
$x_{\rm HI}=0.2\pm 0.2$ ($z=6.6$), 
$x_{\rm HI}=0.25\pm 0.25$ ($z=7.0$), 
and 
$x_{\rm HI}=0.55\pm 0.25$ ($z=7.3$) 
from comparisons of various reionization models, where the errors include model variances \citep{ouchi2010,konno2014,itoh2018}.
To estimate $x_{\rm HI}$, the Ly$\alpha$ emitting galaxy fraction $X_{\rm Ly\alpha}$ 
is also measured by deep follow-up spectroscopy of dropout galaxies.
Here, $X_{\rm Ly\alpha}$ is defined by the ratio of galaxies with Ly$\alpha$ emission 
to all galaxies down to a given UV-continuum magnitude limit.
In contrast to the LAE selection, 
the UV-continuum selection does not depend on cosmic reionization (i.e., the value of $x_{\rm HI}$).
%
%
\begin{figure}[H]
\centering
\includegraphics[scale=.36]{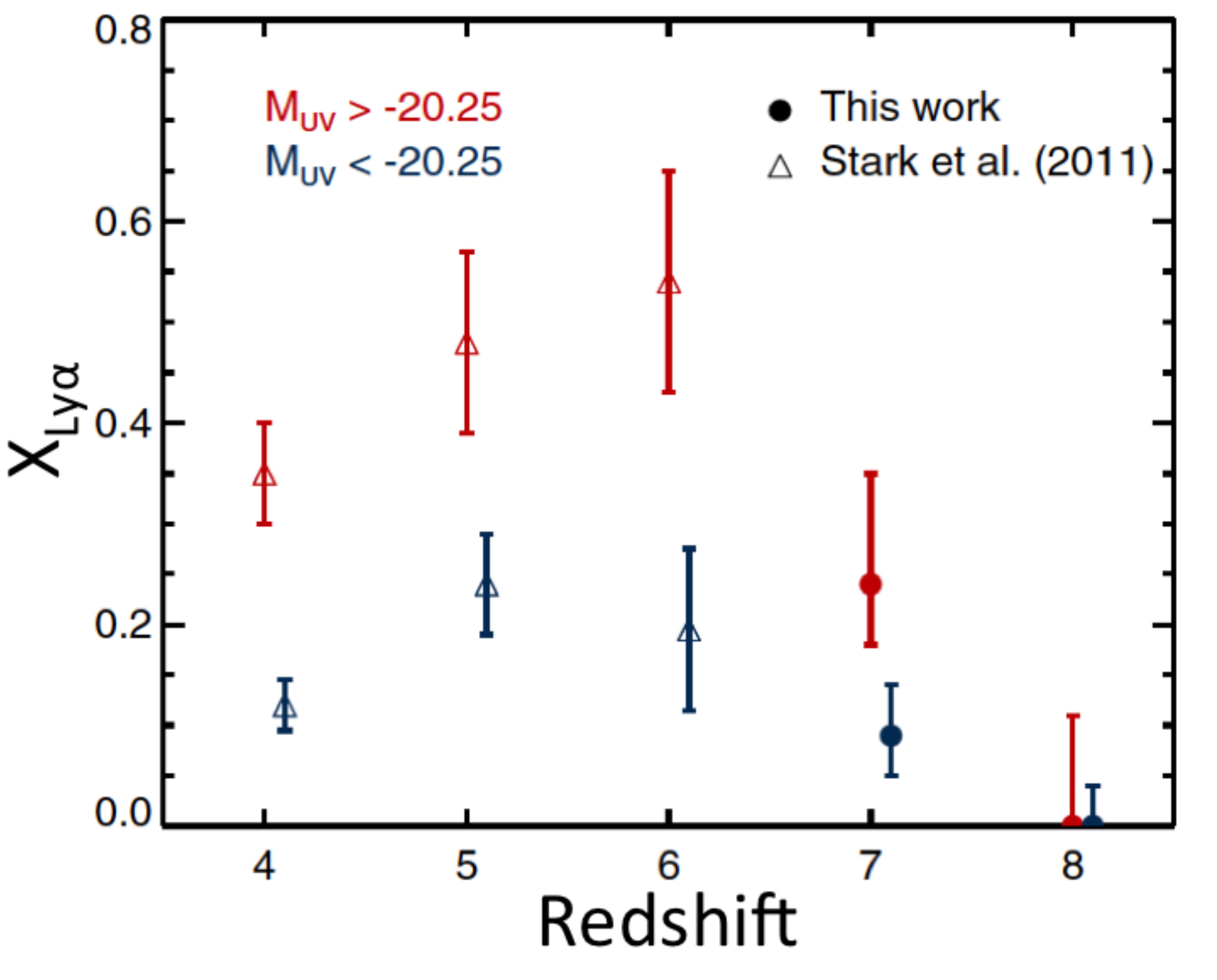}
\caption{
Ly$\alpha$ emitting galaxy fraction $X_{\rm Ly\alpha}$ as a function of
redshift \citep{schenker2014}. 
Ly$\alpha$ emitting galaxies are defined as those 
with Ly$\alpha$ $EW\ge 25$\AA. 
The red and blue data points represent $X_{\rm Ly\alpha}$ for UV-continuum faint ($M_{\rm UV}>-20.25$) 
and bright ($M_{\rm UV}<-20.25$) dropout galaxies, respectively.
This figure is reproduced by permission of the AAS.
}
\label{fig:schenker2014_fig9}
\end{figure}
Figure \ref{fig:schenker2014_fig9} indicates that
$X_{\rm Ly\alpha}$ peaks at $z\sim 6$ and decreases
towards higher $z$ \citep{stark2011,pentericci2011,pentericci2014,
ono2012,schenker2012,schenker2014,treu2013}.
The $x_{\rm HI}$ value is estimated to be 
$x_{\rm HI}=0.39^{+0.08}_{-0.09}$ ($z\sim 7$)
and 
$x_{\rm HI}>0.64$ ($z\sim 8$; \citealt{schenker2014}).
All of these Ly$\alpha$ emission observations suggest that 
the Ly$\alpha$ emissivity of galaxies decreases from $z\sim 6$ 
towards higher $z$,
and that the neutral hydrogen fraction is moderately high at $z\sim 7-8$.

%

\subsection{Reionization History}
\label{sec:reionization_history}

\begin{figure}[H]
\centering
\includegraphics[scale=.48]{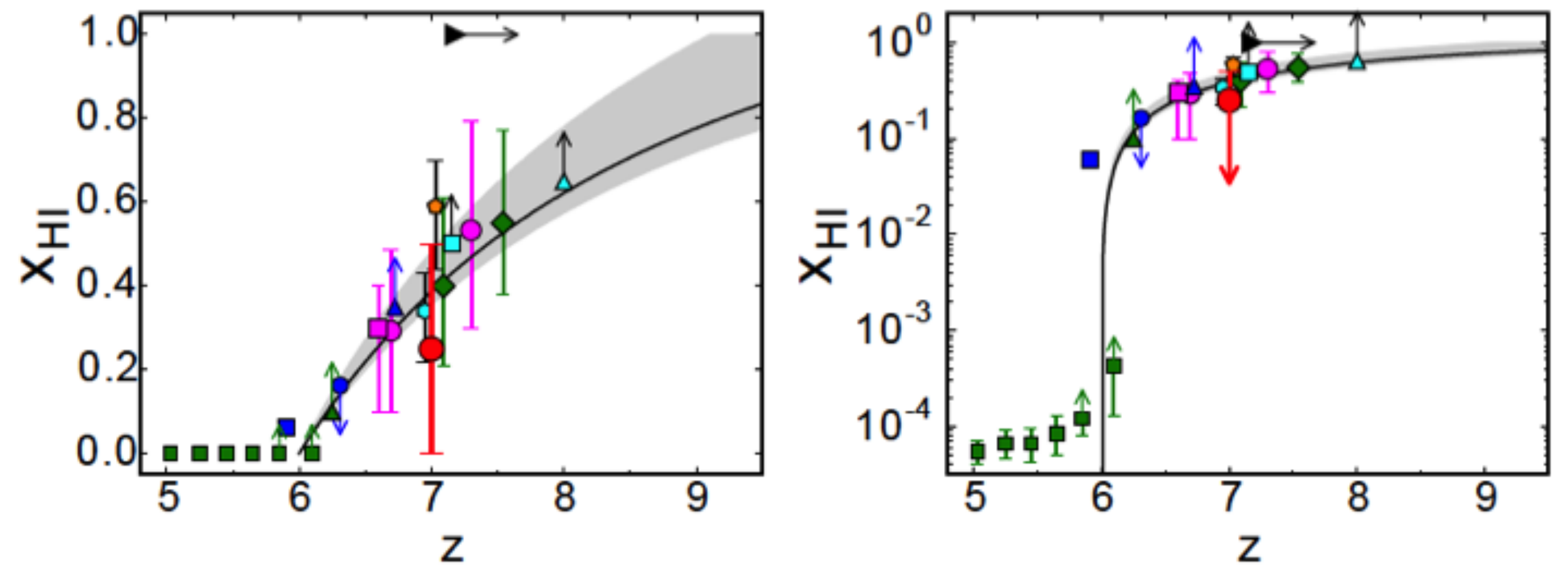}
\caption{
Redshift evolution of $x_{\rm HI}$ \citep{itoh2018}. The left and right panels
are the same, but with linear and log-scale ordinate axes, respectively.
The green squares represent estimates from QSO Gunn Peterson optical depths,
while the other green symbols indicate results of QSO DW measurements.
The magenta and red data points denote $x_{\rm HI}$ 
estimated from LAE DW absorptions.
The cyan symbols are given by Ly$\alpha$ emitting galaxy fractions.
The blue symbols show the results of GRB DW measurements.
The orange pentagon presents an estimate obtained with the Ly$\alpha$ EW 
distribution of dropout galaxies. The black triangle with an arrow indicates the
$1\sigma$ lower limit of $z_{\rm re}$ obtained by \citet{planck2016}.
The black curve and gray shade show $x_{\rm HI}$ and its uncertainty 
suggested from the evolution of the UV luminosity function \citep{ishigaki2018}.
This figure is reproduced by permission of the AAS.
}
\label{fig:itoh2018_fig12}
\end{figure}

Figure \ref{fig:itoh2018_fig12} 
summarizes $x_{\rm HI}$ estimates
given by the Gunn Peterson effect, Thomson scattering of the CMB, and Ly$\alpha$ DW absorptions
of GRBs, QSOs, and LAEs. 
This figure clarifies that 
the measurements of the Gunn Peterson effect 
reveal the completion epoch of cosmic reionization at $z\sim 6$ (Section \ref{sec:gunn_peterson}).
The Ly$\alpha$ DW absorption measurements suggest a moderately high
$x_{\rm HI}$ at $z\sim 6-8$, indicative of late reionization, 
albeit with large uncertainties (Section \ref{sec:lya_damping_wing}).
Although the CMB Thomson scattering results have no time resolution, 
they also imply a moderately
late reionization epoch of 
$z_{\rm re} \simeq 7.8-8.8$
for the case of instantaneous reionization (Section \ref{sec:cmb}). 
However, because all the $x_{\rm HI}(z)$ data plotted here 
have large uncertainties, the duration of cosmic reionization, 
by which the reionization process is characterized, e.g., 
as being sharp or extended,
remains to be determined \citep{ishigaki2018}.


\subsection{\hi\ 21 cm Observations: Direct Emission from the IGM}
\label{sec:hi_21cm}

Sections \ref{sec:gunn_peterson}-\ref{sec:lya_damping_wing}
introduce studies of cosmic reionization probed with bright background radiation.
Although it is difficult to directly detect emission from the diffuse IGM at the EoR 
with the technology today, there are many efforts to find such a signal.
The most important direct signal is 
the \hi\ hyperfine structure line of 21 cm wavelength 
that is produced when the spins of the proton and 
electron in a neutral hydrogen atom flip from antiparallel to parallel.
Catching this emission produced at the EoR ($z\sim 10$) 
needs a low frequency radio observation 
because it is redshifted to $\sim 100$ MHz.
%
Future 21 cm emission data will allow us
to study the cosmic reionization history $x_{\rm HI} (z)$ and 
the topology of \hi\ distribution that depends on 
ionizing sources (i.e., galaxies vs. AGNs; ionizing photons from 
these two populations have different mean-free paths against 
neutral hydrogen gas because of different spectral shapes).

\subsubsection{Basic Picture of EoR \hi\ 21 cm Emission}
\label{sec:21cm_signal_from_EoR}

\begin{figure}[H]
\centering
\includegraphics[scale=.45]{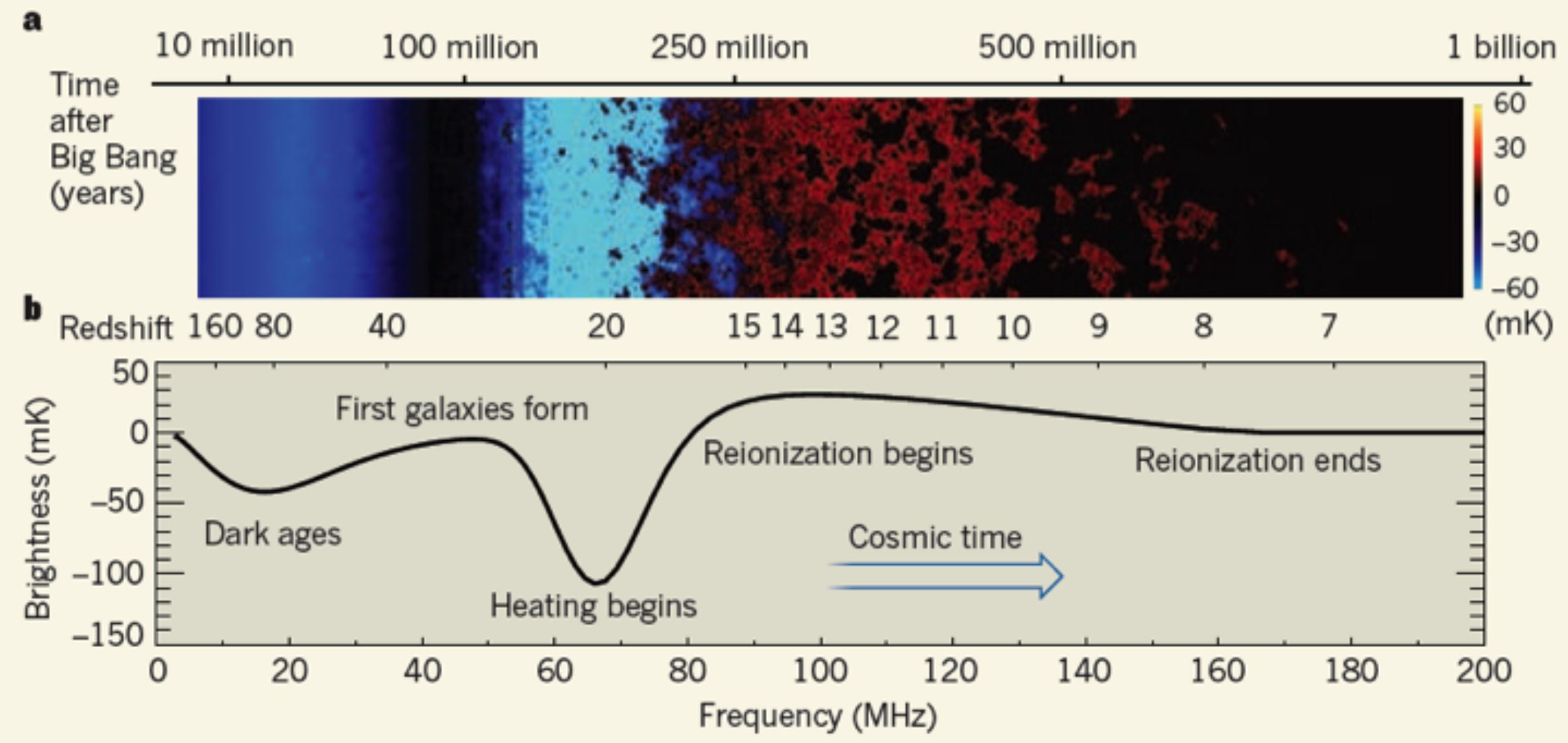}
\caption{
Evolution of brightness temperature from $z=200$ to $6$ 
suggested by a theoretical model \citep{pritchard2010b}.
Top: Brightness temperature with spatial fluctuations as a function of redshift (cosmic age).
The brightness temperature is color-coded following 
the color bar on the right hand side.
Bottom: Sky-averaged brightness temperature as a function of redshift (observed frequency).
Also indicated are the epochs of several major cosmic events
that change the brightness temperature of the 21cm line.
This figure is reproduced by permission of the Nature Publishing Group.
}
\label{fig:pritchard2010b_fig1}
\end{figure}

The strength of the 21 cm emission depends on 
the spin temperature $T_{\rm s}$ 
(from Maxwell-Boltzmann equation) that is 
defined by 
\begin{equation}
\frac{n_{\rm \uparrow \uparrow}}{n_{\rm \uparrow \downarrow}}=3 \exp \left( - \frac{h \nu_{\rm 21cm}}{k T_{\rm s}} \right),
\label{eq:spin_temperature}
\end{equation}
where $n_{\rm \uparrow \uparrow}$ and $n_{\rm \uparrow \downarrow}$ are the number densities of
parallel and antiparallel hydrogen atoms, 
$h$ the Planck constant, $k$ the Boltzmann constant, 
and $\nu_{\rm 21cm}$ the frequency of 21 cm wavelength.
%
When $T_{\rm s}$ is lower (higher) than the CMB temperature $T_{\rm CMB}$,
the 21 cm line is observed as absorption (emission) in the CMB spectrum.
The observable is thus an increment of brightness temperature relative to the CMB, $\delta T_{\rm B}$, that is described as
\begin{eqnarray}
\delta T_{\rm B} 
 & \simeq & \frac{T_{\rm s} - T_{\rm CMB}}{1+z} \tau_{\rm 21cm}\\
 & \simeq & 7(1+\delta) x_{\rm HI} \left( 1- \frac{T_{\rm CMB}}{T_{\rm s}} \right) (1+z)^{1/2} \ \ \ \ \ {\rm mK},
\label{eq:brightness_temperature}
\end{eqnarray}
where $\delta$ is the baryon overdensity and 
$\tau_{\rm 21cm}$ the \hi\ optical depth at 21 cm 
\citep{fan2006b,pritchard2010a}.
Figure \ref{fig:pritchard2010b_fig1} presents a theoretical prediction of the brightness temperature increment
evolution, together with an \hi\ map illustration.
%
At $z\sim 150$, baryons and CMB decouple because collisions and cooling dominate
in baryon gas. 
After $z\sim 80$, the cosmic baryon density is sufficiently low
that the collisional cooling is inefficient.
After the first stars and QSOs, i.e. galaxies, form at $z\sim 20-30$,
Ly$\alpha$ photons from galaxies are scattered by \hi\ gas in the IGM.
This process redistributes the two spin states of \hi, and enlarges 
the difference between the spin and CMB temperatures
(Ly$\alpha$ cooling aka Wouthuysen-Field effect). Then, the \hi\ of the IGM is heated by X-ray emission from objects.
At $z\sim 15$, the reionization begins, and the brightness temperature increment becomes small, due to an 
increase in ionized regions in the IGM. In this way, the evolution of $\delta T_{\rm B}$ is predicted.
The spatial fluctuations of $\delta T_{\rm B}$ 
also vary with evolutionary phase 
due to differences in heating and cooling sources.

\subsubsection{Early \hi\ 21cm Observation Results and Expectations}
\label{sec:early_21cm_observation_results}

\begin{figure}[H]
\centering
\includegraphics[scale=.45]{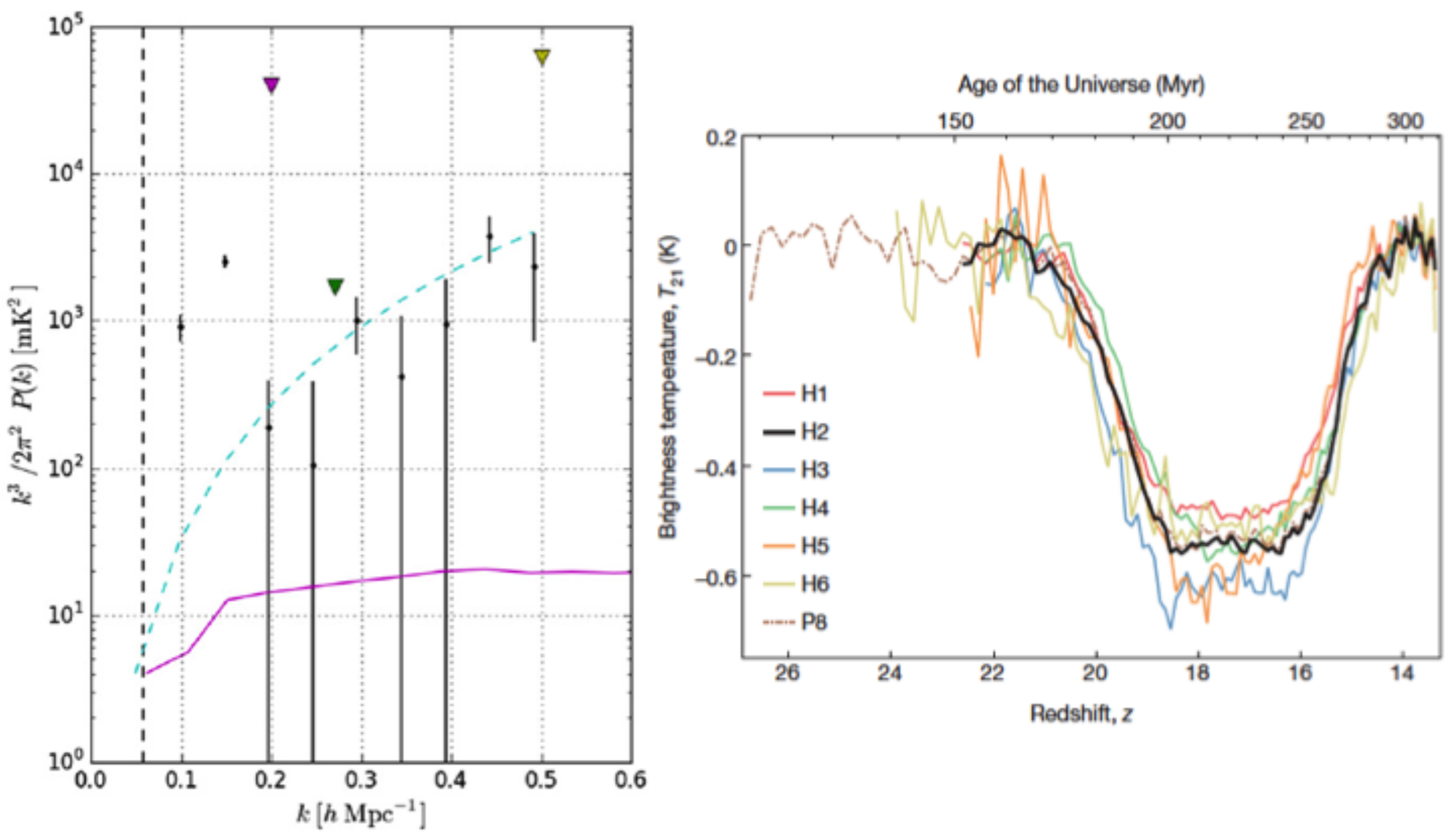}
\caption{
Left: Observed and predicted power spectra of 21 cm emission \citep{ali2015}. 
All observational data points are upper limits. 
The yellow, magenta, and green triangles show $2\sigma$ upper limits 
at $z=8.6$, $9.5$, and $7.7$ that are given by GMRT, MWA, and PAPER experiments, respectively.
The magenta curve represents the model 21 cm power spectrum for the 50\% reionization case
predicted by \citet{lidz2008}. 
The black dots, the cyan dashed line, and the black dashed vertical line indicate
the results of the PAPER experiment shown in \citet{ali2015}, while these results
are negated by \citet{ali2018} due to the underestimations of the signal uncertainties.
%
%
This figure is reproduced by permission of the AAS.
Right: 21 cm absorption profiles at $z=17$ best-fitted to the EDGES data \citep{bowman2018}.
The eight lines with different colors indicate absorption profiles
for different observational set-up data, 
among which the thick black line corresponding to
the highest signal-to-noise ratio case.
This figure is reproduced by permission of the Nature Publishing Group.
}
\label{fig:ali2015_fig18_bowman2018_fig2} 
\end{figure}

Measuring the brightness temperature of the IGM at the EOR 
requires radio observations in low frequencies ($\sim 100$ MHz)
(see Figure \ref{fig:pritchard2010b_fig1}), 
%
%
and several programs have conducted such observations: 
Giant Metrewave Radio Telescope (GMRT; \citealt{paciga2011}),
Precision Array for Probing the Epoch of Reionization (PAPER; \citealt{parsons2014,ali2015}),
LOw Frequency ARray (LOFAR; \citealt{yatawatta2013,jelic2014}), and
Murchison Widefield Array (MWA; \citealt{dillon2014}).
The left panel of Figure \ref{fig:ali2015_fig18_bowman2018_fig2} 
summarizes 21-cm power spectrum ($z\sim 8-9$) results from these programs.
So far, no 
programs have identified a signal of the 21-cm emission from the EoR,
and only upper limits have been obtained. 
The left panel of Figure \ref{fig:ali2015_fig18_bowman2018_fig2} indicates that
the $2\sigma$ upper limits are about 2 orders of magnitude higher than 
predicted signals.
%
Although some programs have expected sensitivities high enough to detect 
the EoR 21 cm emission, in practice the expected sensitivities 
cannot be reached 
due to difficulties in subtraction of bright foreground emission.
There are many Galactic and telluric foreground sources. 
One of the most challenging foregrounds 
is
ionospheric radio emission that varies with sky position and time.
Thus, the most important challenge in detecting 21 cm signals from 
the EOR is to properly model the foreground emission that is 
a few orders of magnitude brighter.
%
Because such foreground emission dominates
in specific Fourier spaces and wavelengths,
it would be possible to isolate the EoR 21 cm emission
in the parameter space that is referred to as the "EoR window" \citep{deboer2016},
free from the foreground emission.
%
%
%
%

Although EoR 21-cm emission signals have not been detected yet,
there is a report of EoR 21-cm absorption detection. 
\citet{bowman2018} have conducted low-frequency radio observations
with the Experiment to Detect the Global Epoch of
Reionization Signature (EDGES) low-band instruments, and 
found an absorption at 78 MHz corresponding to $z=17$ 
(right panel of Figure \ref{fig:ali2015_fig18_bowman2018_fig2}).
This absorption may be a signature of first stars and QSOs whose 
Ly$\alpha$ photons 
lower 
the brightness temperature by the
Wouthuysen-Field effect. 
However, the observed absorption is significantly
stronger than model predictions. 
In other words, 
the hydrogen gas at $z\sim 17$ is suggested to be 
colder than the gas kinetic temperature as well as the CMB temperature.
\citet{barkana2018} claims that cold hydrogen gas can be produced by
the interaction of hydrogen gas with dark matter whose temperature is low enough
to explain the low brightness temperature suggested by the EDGES observations.
Because the detection of an absorption by the EDGES, if true, 
will have a great impact on
our understanding of the thermal history of the universe,
it should be confirmed by independent projects.

\begin{figure}[H]
\centering
\includegraphics[scale=.38]{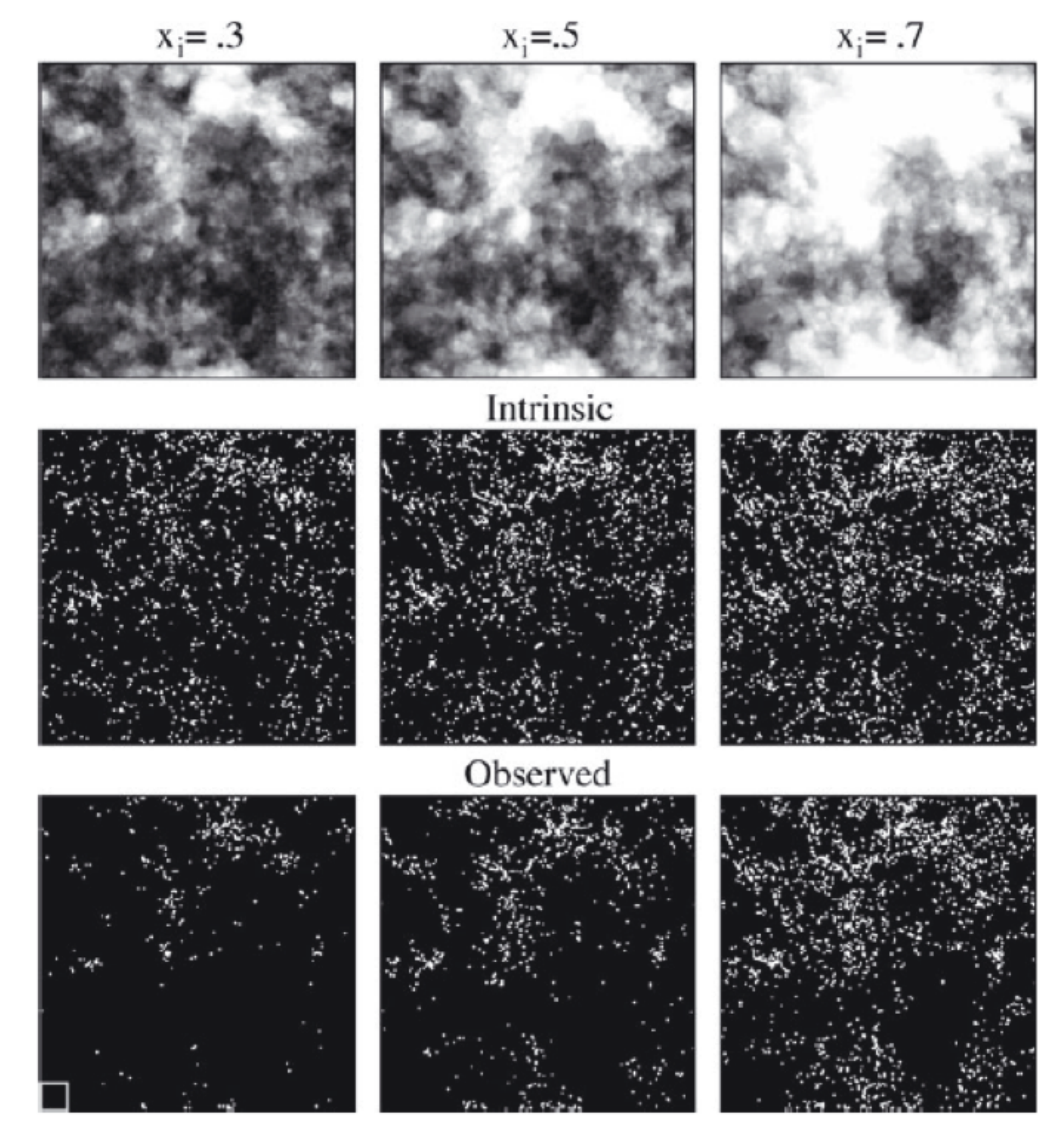}
\caption{
Spatial distributions of ionized gas (top panels), all LAEs (middle panels), and observed LAEs (bottom panels) 
over a cosmological volume ($94\times 94\times 35$ Mpc$^{3}$) 
with average ionized fractions of 0.3 (or $x_{\rm HI}=0.7$; left column), 
$0.5$ (middle column), and $0.7$ ($x_{\rm HI}=0.3$; right column)
predicted by numerical simulations \citep{mcquinn2007}.
In the top panels, ionized regions are
indicated in white. In the middle and bottom panels, LAEs are shown with white dots.
This figure is reproduced by permission of MNRAS.
}
\label{fig:mcquinn2007_fig3} 
\end{figure}

\begin{figure}[H]
\centering
\includegraphics[scale=.40]{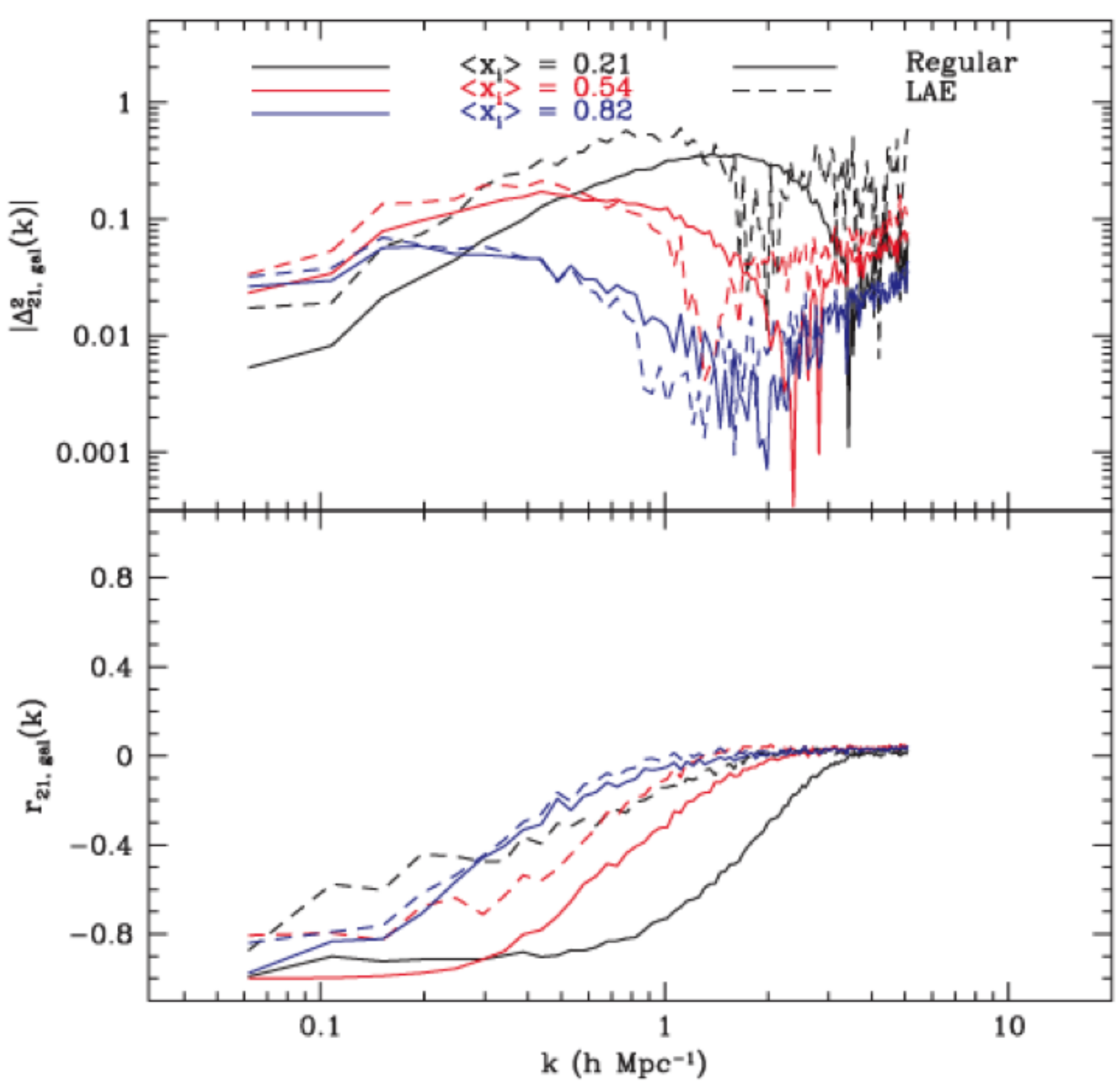}
\caption{
Top: Galaxy-21 cm cross-power spectra predicted by \citet{lidz2009}.
The solid and dashed lines denote the 21 cm cross-power spectra 
for all galaxies and LAEs, respectively.
The black, red, and blue colors mean the ionized fraction of 0.21,
0.54, and 0.82 (corresponding to $x_{\rm HI}=0.79$, $0.46$, and $0.18$)
at $z=8.3$, $7.3$, and $6.9$, respectively.
Bottom: Same as the top panel, but for the cross-correlation coefficients 
between 21 cm radiation and galaxies.
This figure is reproduced by permission of the AAS.
}
\label{fig:lidz2009_fig10} 
\end{figure}

%
Theoretical models predict that the cross-correlation function between \hi\ 21 cm emission
and LAEs is key for understanding the reionization process.
If cosmic reionization proceeds from high to low density regions, so called in the inside-out manner,
star-forming galaxies including LAEs exist 
preferentially 
in cosmic ionized bubbles. 
Moreover, cosmic ionized bubbles allow Ly$\alpha$ photons escaping 
from LAEs to survive, 
thus enhancing the observed overdensity of LAEs in ionized bubbles
(Figure \ref{fig:mcquinn2007_fig3}; \citealt{mcquinn2007}). 
It is thus expected that \hi\ 21 cm emission and LAEs anti-correlate strongly. 
Figure \ref{fig:lidz2009_fig10}
presents theoretical predictions of the cross-power spectrum 
between \hi\ 21 cm emission and LAEs \citep{lidz2009}.
The distance scale where the sign of the correlation changes
can be used to constrain the typical size of ionized bubbles. 
The distance scale
becomes large towards the end of reionization (i.e. small $x_{\rm HI}$;
Figure \ref{fig:lidz2009_fig10}).

%

%

\subsection{Summary of Cosmic Reionization I}
\label{sec:summary_cosmic_reionizationI}

This section has introduced the basic physical picture of cosmic 
reionization,
and showcased various techniques to probe the cosmic reionization history,
discussing constraints obtained by those techniques 
on the neutral hydrogen fraction $x_{\rm HI}$ 
(or ionized hydrogen fraction $Q_{\rm HII}$) as a function of redshift. 
%
There are three major techniques to estimate $x_{\rm HI}$ 
that require bright background light. 
One uses the Gunn-Peterson effect of background QSO spectra, while
the other two measure, respectively, 
the Thomson optical depth of the CMB and
Ly$\alpha$ DW absorptions seen in background GRB, QSO, and LAE spectra.
%
The measurements of the Gunn Peterson effect suggest that 
cosmic reionization ended at $z\sim 6$ (Section \ref{sec:gunn_peterson}).
The Thomson scattering optical depth of the CMB obtained from 
the recent Planck2016 data is small ($\tau_{\rm e}\sim 0.06$) 
(Section \ref{sec:cmb}),
%
and the Ly$\alpha$ DW absorption strengths indicate a moderately high
$x_{\rm HI}$ at $z\sim 6-8$ (Section \ref{sec:lya_damping_wing}). These two pieces of information support
late reionization.
%
Since the CMB Thomson scattering results available to date have no 
time resolution and the Ly$\alpha$ DW constraints on $x_{\rm HI}$ are 
not very strong, the duration of cosmic reionization (i.e. sharp or 
extended reionization) has been constrained only weakly.

%

\section{Cosmic Reionization II: Sources of Reionization}
\label{sec:cosmic_reionizationII}

This section presents progresses in observations
for understanding sources of reionization 
that is one of the two major questions of cosmic reionization
(see the first paragraph of Section \ref{sec:cosmic_reionizationI}).

\subsection{What are the Major Sources Responsible for Reionization?}
\label{sec:what_are_reionization_sources}

There are several candidates for sources of cosmic reionization
that supply ionizing photons at the EoR. These candidates
include galaxies, AGNs, high-mass X-ray binaries (HMXBs),
primordial blackholes (PBHs), and dark-matter annihilation.

Although it is obvious that galaxies and AGNs should contribute to 
cosmic reionization because they are bright in UV, the question is 
the relative contributions of individual candidate populations.
%
AGNs produce not only UV ionizing photons, but also X-ray photons
whose mean-free paths in the \hi\ IGM are as large as the sizes of LSSs.
If the X-ray emission of AGNs dominates in reionizing the universe, 
the structures of ionized regions should be smooth.
%
%
HMXBs can also contribute via X-ray radiation, but
it is not yet clear whether they play a major role 
%
because the observed X-ray background 
is mostly explained by known AGNs
at redshifts up to $z\sim 6$ \citep{hickox2007}. 
%
PBHs would emit Hawking radiation that would heat
the IGM, but observational studies place moderately tight upper limits
on the fraction of the total mass of PBHs to dark matter that is less than $\sim 10$\%
over the PBH masses of $\sim 10^{20}$ to $1 M_\odot$ \citep{niikura2017}.
It is predicted that dark-matter particles annihilate into high energy 
particles including neutrinos and gamma rays that produce X-ray radiation.
However, this happens only if dark matter 
is made of supersymmetric
particles such like axions.

In this lecture, I only consider reionization by UV ionizing photons 
and discuss two promising reionization sources, galaxies and AGNs. 



\subsection{Ionization Equation for Cosmic Reionization}
\label{sec:ionization_equation}

The key quantity for sources of reionization
is the production rate of ionizing photons $\dot{n}_{\rm ion}$
that is defined by the number of ionizing photons 
per volume and time. The $\dot{n}_{\rm ion}$ values 
should be estimated by observations for galaxies and AGNs.
The $\dot{n}_{\rm ion}$ value is related with 
%
%
the ionized hydrogen fraction of the IGM $Q_{\rm HII}$ (eq. \ref{eq:q_HII}) 
via the simple one zone model of the ionization equation \citep{madau1999,robertson2013,ishigaki2015,robertson2015},
%
%
%
\begin{equation}
\dot{Q}_{\rm HII} = \frac{ \dot{n}_{\rm ion} }{ \left< n_{\rm H} \right> } - \frac{ Q_{\rm HII} }{ t_{\rm rec} },
\label{eq:ionization_equation}
\end{equation}
where $\left< n_{\rm H} \right>$ and $t_{\rm rec}$ are
the average hydrogen number density and the recombination time, respectively, 
given by
\begin{eqnarray}
\label{eq:hydrogen_number_density}
\left< n_{\rm H} \right>  & = & \frac{X_{\rm p} \Omega_{\rm b} \rho_{\rm c}}{m_{\rm H}}\\
\label{eq:recombination_time1}
t_{\rm rec} & = & \frac{1}{C_{\rm HII} \alpha_{\rm B}(T) (1+Y_{\rm p}/4X_{\rm p}) \left< n_{\rm H} \right> (1+z)^3}.
\end{eqnarray}
Here, $X_{\rm p}$ ($Y_{\rm p}$), $\rho_{\rm c}$, and $m_{\rm H}$
are the primordial mass fraction of hydrogen (helium),
the critical density, and 
the mass of the hydrogen atom, respectively.
%
%
In eq. (\ref{eq:recombination_time1}), 
$\alpha_{\rm B} (T)$ is the case B hydrogen recombination coefficient
for the IGM temperature $T$ at a mean density.
%
%
The value of $C_{\rm HII}$ is the clumping factor,
\begin{equation}
C_{\rm HII} = \frac{ \left< n^2_{\rm HII} \right> }{ \left< n_{\rm HII} \right>^2},
\label{eq:clumping_factor}
\end{equation}
where $n_{\rm HII}$ is the density of ionized hydrogen gas in the IGM.
With the brackets, $\left< n^2_{\rm HII} \right>$ and $\left< n_{\rm HII} \right>^2$
are the spatially averaged values.
One can derive the evolution of $Q_{\rm HII}$ with 
the observational estimates of $\dot{n}_{\rm ion}$ via eq. (\ref{eq:ionization_equation}),
where most of the parameters
are determined by physics and cosmology.
By this technique, the budget of ionizing photons is evaluated (see Section \ref{sec:galaxy_contribution}).

For the specific case of ionization equilibrium,
one can substitute $\dot{Q}_{\rm HII}=0$ and $Q_{\rm HII}=1$
in eq. (\ref{eq:ionization_equation}), and obtain
\begin{eqnarray}
\dot{n}_{\rm ion}  & = & \frac{ \left< n_{\rm H} \right> }{ t_{\rm rec} }\\
 & = & 10^{50.0} C_{\rm HII} \left( \frac{1+z}{7} \right)^3 \ \ \ \ \ {\rm s^{-1} Mpc^{-3}}.
\label{eq:recombination_time2}
\end{eqnarray}
This condition of $\dot{n}_{\rm ion}$ gives the lower limit of the ionizing photon
production rate that can keep the ionized universe (e.g. \citealt{bolton2007,ouchi2009b}).

\begin{figure}[H]
\centering
\includegraphics[scale=.48]{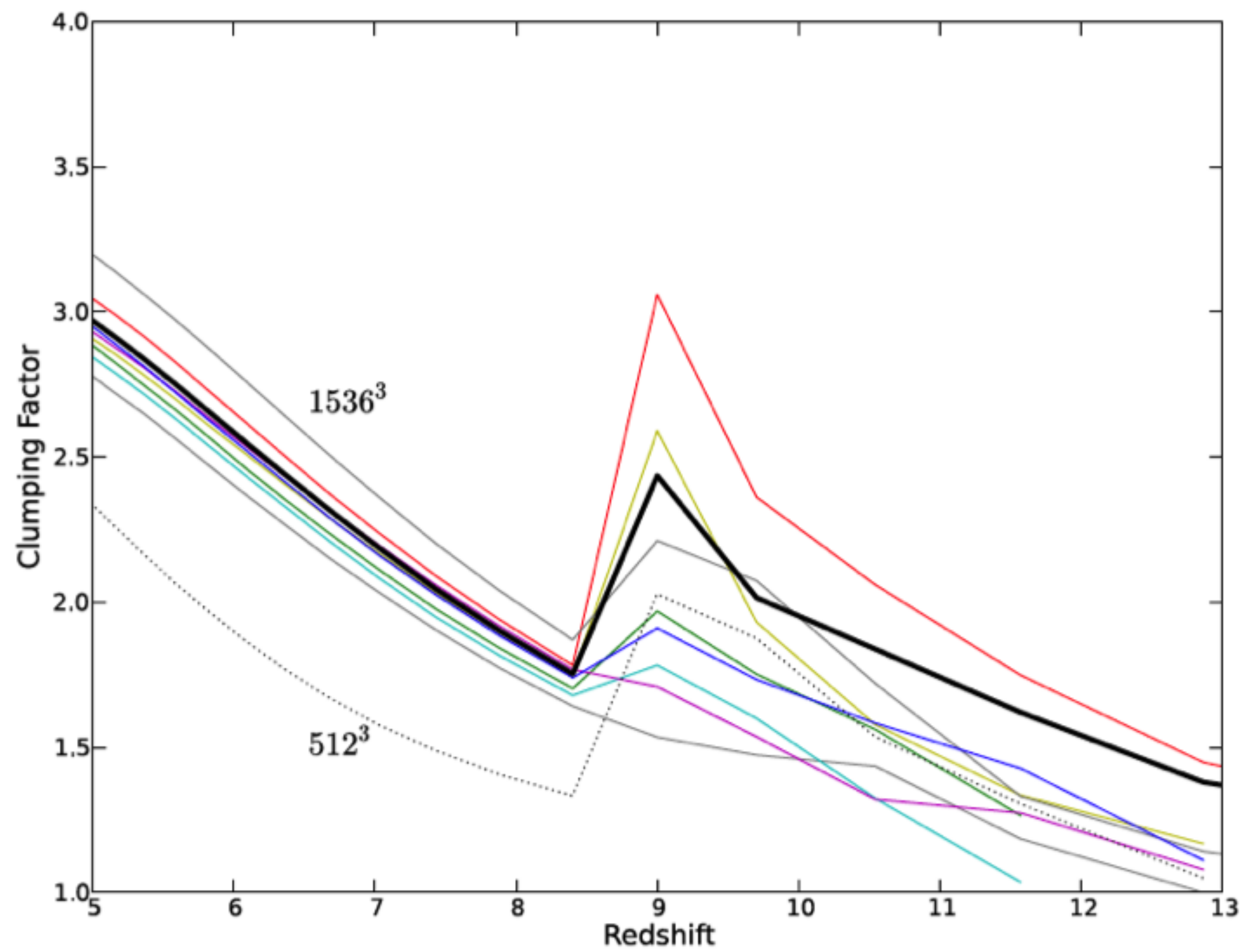}
\caption{
Evolution of clumping factor predicted by the numerical simulations \citep{shull2012}.
The black solid line indicates the results of the simulations with $1536^3$ cells
in the $50 h^{-1}$ Mpc box, which are approximated with the function of 
equation (\ref{eq:clumping_factor}) at $z=5-9$. The rest of the solid lines
are the same as the black solid lines, but for $768^3$-cell sub-volumes.
The dotted line represents the $512^3$-cell simulations.
This figure is reproduced by permission of the AAS.
}
\label{fig:shull2012_fig3top}
\end{figure}

One of the free parameters in the ionization equation (eq. \ref{eq:ionization_equation})
is the clumping factor $C_{\rm HII}$ (eq. \ref{eq:clumping_factor}) that determines
the recombination rate of ionized hydrogen in the IGM. Based on the ionizing photon
emissivity measurements from QSO absorption line data, 
the clumping factor is estimated to be as low as $C_{\rm HII}\sim 3$ at $z\sim 6$ \citep{bolton2007}.
%
Because the universe becomes homogeneous $C_{\rm HII}=1$ with negligibly small fluctuations
at the Big Bang epoch, the clumping factor is low, $C_{\rm HII}\sim 1-3$, over the EoR ($z>6$).
In fact, cosmological numerical simulations with the QSO UV background radiation
predict monotonically decreasing values of $C_{\rm HII}$ towards high-$z$ with $C_{\rm HII}\sim 1-3$
(Figure \ref{fig:shull2012_fig3top}; \citealt{shull2012}; see also \citealt{pawlik2009}).
The numerical simulation results of Figure \ref{fig:shull2012_fig3top}
are approximated by the power law,
\begin{equation}
C_{\rm HII} (z) = 2.9 \left(\frac{1+z}{6}\right)^{-1.1}
\label{eq:clumping_factor}
\end{equation}
at $z=5-9$ \citep{shull2012}.
Although there remain systematic uncertainties 
related with
the mass resolution and
the radiative transfer implementation, 
the majority of
theoretical models
agree with these small clumping factors ($C_{\rm HII}\sim 1-3$) at the EoR. If it is true,
the uncertainties of clumping factors are not as large as those
of the other free parameters (see Section \ref{sec:galaxy_contribution}).

\subsection{Galaxy Contribution}
\label{sec:galaxy_contribution}


To evaluate the ionizing photon contribution of galaxies 
to the cosmic reionization with eq. (\ref{eq:ionization_equation}), 
one needs to estimate $\dot{n}_{\rm ion}$ of galaxies.
The value of $\dot{n}_{\rm ion}$ is calculated by
\begin{eqnarray}
\dot{n}_{\rm ion} & = & \int^{M_{\rm trunc}}_{-\infty} 
f_{\rm esc}^{\rm ion} (M_{\rm UV}) \xi_{\rm ion} (M_{\rm UV})
\phi(M_{\rm UV}) L(M_{\rm UV}) dM_{\rm UV}\\
 & = & f_{\rm esc}^{\rm ion}\ \xi_{\rm ion}\ \rho_{\rm UV} (M_{\rm trunc}) \ \ \ \ \ \ \ \ \ \ \ \ \ {\rm [for \ no\ M_{\rm UV}\ dependences]},
\label{eq:n_ion_galaxy}
\end{eqnarray}
where 
$f_{\rm esc}^{\rm ion}$ and $\xi_{\rm ion}$ are the ionizing photon escape fraction 
(eq. \ref{eq:fesc_ion}; Section \ref{sec:escape_fraction_ionizing_photon})
and the ionizing photon production efficiency
(eq. \ref{eq:xi_ion}; Section \ref{sec:xi_ion}), respectively.
%
%
Here, for simplicity, it is assumed that
$f_{\rm esc}^{\rm ion}$ and $\xi_{\rm ion}$ do not depend on $M_{\rm UV}$.
It should be noted that $f_{\rm esc}^{\rm ion}$ is the escape fraction of ionizing photons
that is different from the escape fraction of Ly$\alpha$ photons (eq. \ref{eq:fesc_lya_indiv}).
The value of $\rho_{\rm UV} (M_{\rm trunc})$ is the UV luminosity density 
defined with eq. (\ref{eq:uv_ld}), where $M_{\rm trunc}$
is the limiting magnitude for the integration, a.k.a. the truncation magnitude \citep{ishigaki2015}.
The truncation magnitude indicates how faint galaxies can exist, which depends on 
the gas cooling and feedback efficiencies in a faint (i.e. low-mass) galaxy.

There are three major parameters for $\dot{n}_{\rm ion}$, i.e.
$\rho_{\rm UV} (M_{\rm trunc})$, $f_{\rm esc}^{\rm ion}$, and $\xi_{\rm ion}$.
\footnote{
To understand the sources of reionization, one needs to solve the equation (\ref{eq:ionization_equation}).
In this case, there are four major parameters, the three parameters 
($\rho_{\rm UV}$, $f_{\rm esc}^{\rm ion}$, $\xi_{\rm ion}$) for $\dot{n}_{\rm ion}$
and one parameter ($C_{\rm HII}$) for $t_{\rm rec}$.
}
%
These three parameters are constrained by observations. 
In the following sections (Sections \ref{sec:uv_luminosity_density}-\ref{sec:xi_ion}),
I introduce constraints on the parameters obtained by observations, to date.

\subsubsection{UV Luminosity Density}
\label{sec:uv_luminosity_density}

A number of deep optical and NIR imaging surveys have derived luminosity functions of 
UV continuum at $\sim 1500$\AA\ at the EoR, and estimated $\rho_{\rm UV}$
(e.g. \citealt{mclure2013,schenker2013,oesch2015,bouwens2015}).
These surveys provide good measurements of UV luminosity functions at $z=6-10$,
and reveal that the faint-end slopes of the UV luminosity functions are as steep as
$\alpha \simeq 2$ 
\citep{bouwens2015}.
The steep faint-end slopes imply that the $\rho_{\rm UV}$ ($\propto \dot{n}_{\rm ion}$) value
is significantly contributed by faint galaxies ($M_{\rm UV}\gtrsim -15$) that are not luminous 
but abundant (see, e.g., \citealt{robertson2010}). 
Because these conventional deep surveys only reach the moderately bright 
magnitude limit of $M_{\rm UV}\sim -17$ at $z\sim 7$
even in the Hubble Ultra Deep Field (HUDF) program, a $\rho_{\rm UV}$ value estimate
requires an extrapolation of the UV luminosity function 
from $M_{\rm UV}\sim -17$ to $M_{\rm UV}> -15$
to obtain $\rho_{\rm UV}$ via eq. (\ref{eq:uv_ld}).
Moreover, the limiting magnitude of $M_{\rm trunc}$ is unknown,
requiring an assumption such as $M_{\rm trunc}=-13$ \citep{robertson2013}.
The major uncertainty in the $\rho_{\rm UV}$ determination 
is the extrapolation of the UV luminosity function at the faint end 
below the detection limit.

\begin{figure}[H]
\centering
\includegraphics[scale=.50]{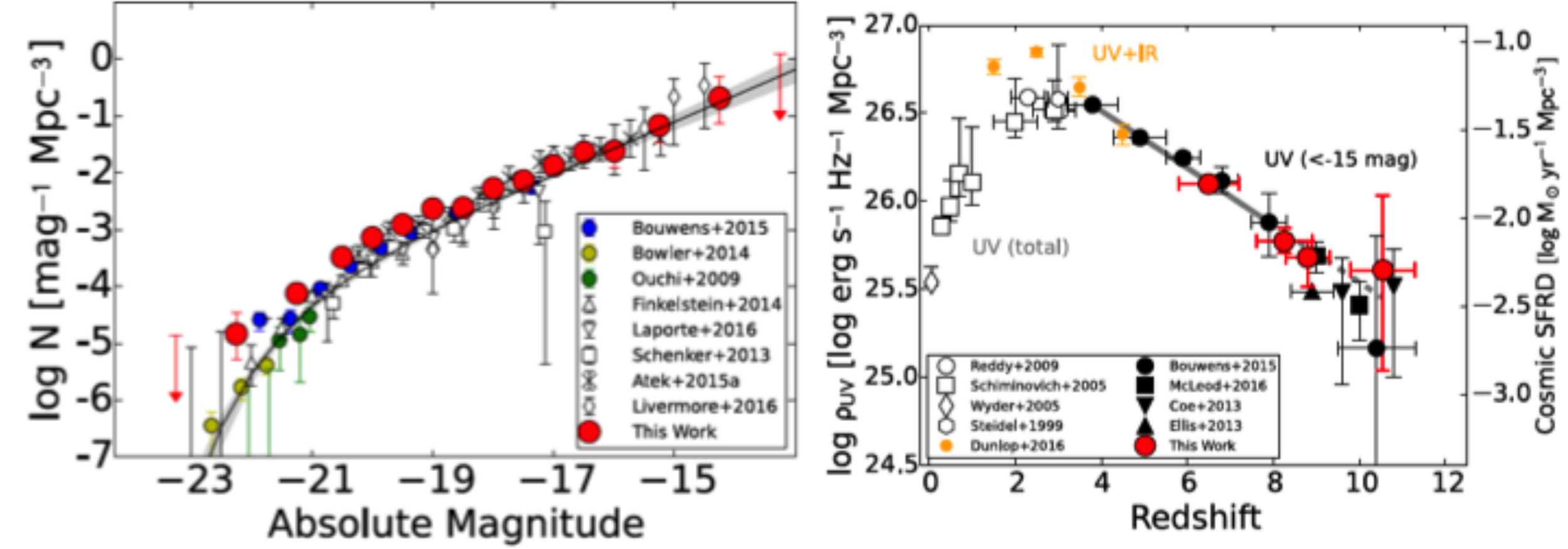}
\caption{
Left: UV luminosity function up to $M_{\rm UV}\sim -14$ \citep{ishigaki2018}.
The red circles (and the black open diamonds and crosses) are the luminosity functions
derived from the HFF data. The other symbols including blue circles are the luminosity functions
obtained without HFF data, which reach up to $M_{\rm UV}\sim -17$.
The black line and the gray shade denote the best-fit Schechter function and the fitting error, respectively.
Right: Evolution of UV luminosity density that is derived under the assumption of $M_{\rm trunc}=-15$ \citep{ishigaki2018}.
The ordinate axis on the right-hand side indicates the cosmic SFR density.
The red and black symbols represent the observational data points calculated with
UV luminosity functions. The orange circles show the cosmic SFR density
estimated from the sum of the UV and FIR luminosity densities.
This figure is reproduced by permission of the AAS.
}
\label{fig:hff_lf_rhoUV}
\end{figure}

To determine $\rho_{\rm UV}$ at the EoR 
with the measurements of the faint-end UV luminosity function,
the Hubble Frontier Fields (HFF) project is conducted \citep{lotz2017}.
The HFF project has performed ultra-deep optical and NIR imaging
with HST/ACS and WFC3-IR, respectively, in six massive galaxy clusters at $z\sim 0.3-0.5$,
and targeted intrinsically very faint background galaxies lensed by the clusters. Exploiting the lensing
magnifications, one can probe the UV luminosity functions down to the detection limit 
deeper than the one of the HUDF program by a few magnitudes (e.g. \citealt{atek2014b,atek2015,ishigaki2015,ishigaki2018,coe2015,oesch2015,oesch2018,mcleod2015,mcleod2016,livermore2017}). 
%
The left panel of Figure \ref{fig:hff_lf_rhoUV}
presents the UV luminosity function 
at $z\sim 7$ thus obtained. 
Although it reaches $M_{\rm UV}\sim -14$ mag,
no signature of the truncation of the luminosity function is found.
The truncation magnitudes would exist at even fainter magnitudes.
Nevertheless, the HFF project has revealed the UV luminosity function up to $M_{\rm UV}\sim -14$ mag
with no extrapolations, thereby imposing the constraint on the truncation magnitude, that it must be fainter than $M_{UV}\sim -14$
mag at $z\sim 7$. 
The results of the HFF project significantly reduce the uncertainty on $\rho_{UV}$ estimates 
that is given by the extrapolation and the assumed $M_{\rm trunc}$ value.
The right panel of Figure \ref{fig:hff_lf_rhoUV} shows
the redshift evolution of $\rho_{\rm UV}$ calculated from these UV luminosity functions
under the assumption of $M_{\rm trunc}=-15$.  In this panel, $\rho_{\rm UV}$ monotonically
decreases from $z\sim 2$ towards high-$z$. The values of $\rho_{\rm UV}$ at $z\sim 9-10$
still include moderately large statistical uncertainties due to the small number of galaxies
identified at these redshifts.



%



\subsubsection{Escape Fraction of Ionizing Photon}
\label{sec:escape_fraction_ionizing_photon}

The ionizing photon escape fraction is measured
at $\simeq 900$\AA\ with the ratio of the observed flux $f_{900}^{\rm obs}$ to intrinsic 
Lyman-continuum (LyC) flux ($f_{900}^{\rm int}$);
\begin{equation}
f_{\rm esc}^{\rm ion}=\left ( f_{900}^{\rm obs}/f_{900}^{\rm int} \right ) e^{\tau_{900}},
\label{eq:fesc_ion}
\end{equation}
where $e^{\tau_{900}}$ is the line-of-sight average IGM opacity to the LyC photons  
that is determined by QSO absorption line observations \citep{steidel2001}.
Because the estimate of $f_{900}^{\rm int}$ includes large uncertainties with assumptions,
observers introduce the relative escape fraction 
\begin{equation}
f_{\rm esc,rel}^{\rm ion}= \left ( f_{900}^{\rm obs}/f_{1500}^{\rm obs} \right ) e^{\tau_{900}},
\label{eq:fesc_ion_rel}
\end{equation}
where $f_{1500}^{\rm obs}$ is the observed 1500\AA\ UV continuum flux
\citep{steidel2001,inoue2006,shapley2006}.
Because the observation study papers discuss 
$f_{\rm esc}^{\rm ion}$ (i.e. the absolute escape fraction)
and $f_{\rm esc,rel}^{\rm ion}$ (i.e. the relative escape fraction),
one needs to carefully check the definition of the ionizing photon escape fraction.
Hereafter, the absolute escape fraction $f_{\rm esc}^{\rm ion}$ is discussed,
unless otherwise specified.

\begin{figure}[H]
\centering
\includegraphics[scale=.45]{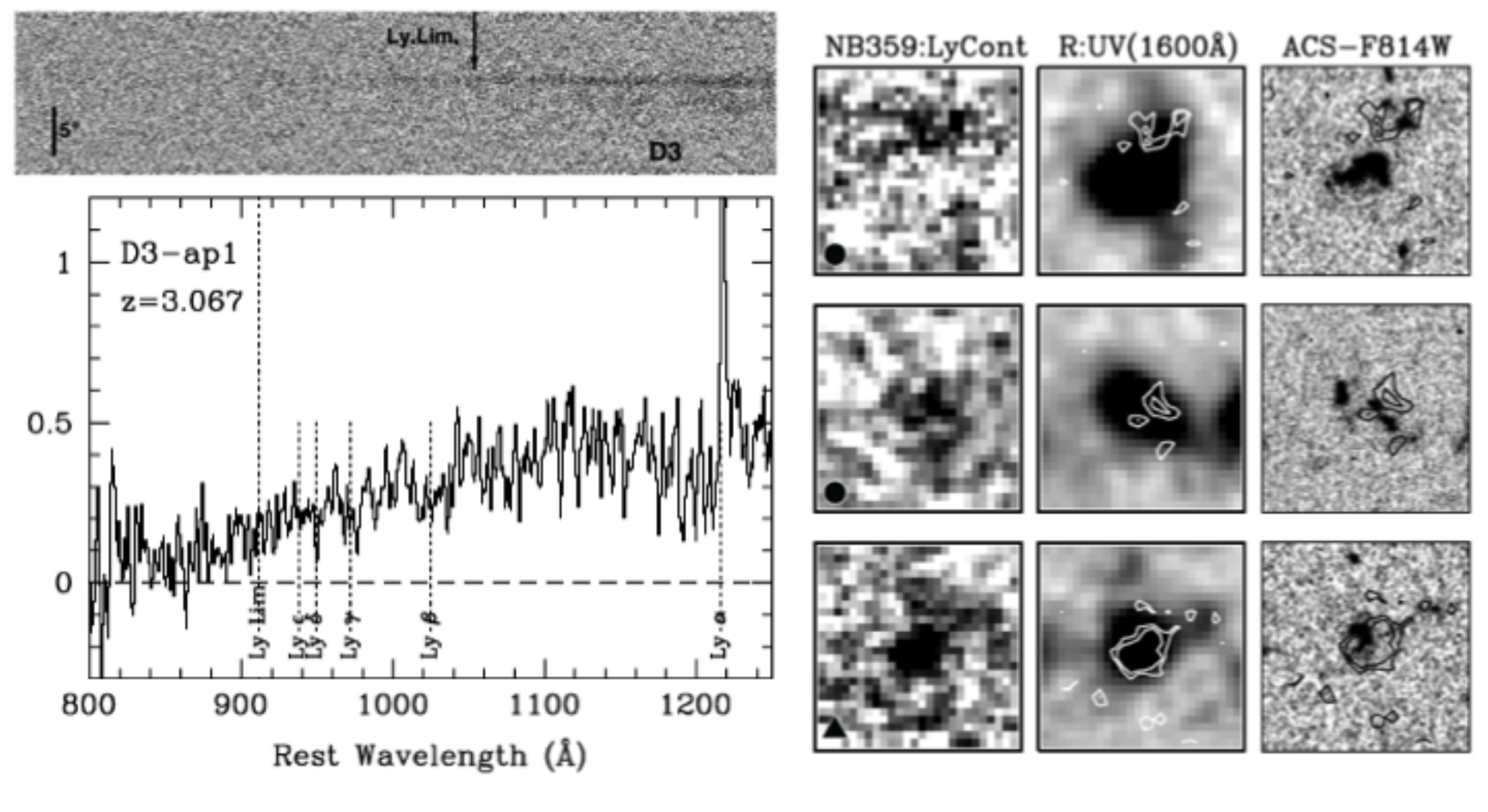}
\caption{
Left: LyC spectra of a galaxy at $z=3$ \citep{shapley2006}. The two- and one-dimensional spectra
are shown in the top and bottom panels, respectively. LyC emission 
is found at wavelengths shorter than the Lyman limit corresponding to the label "Ly Lim" on the left-hand side.
Right: Ground-based imaging data of three galaxies at $z=3$ taken with a narrowband
covering the LyC wavelengths (left panels; \citealt{iwata2009}). The central panels are the same
as the left panels, but with the broadband ($R$ band) images. The right panels
display the HST broadband $I_{814}$ images. The contours indicate the LyC emission detected 
in the narrowband data.
This figure is reproduced by permission of the AAS and PASJ.
}
\label{fig:shapley2006_fig3-4_iwata2009_fig2}
\end{figure}

\begin{figure}[H]
\centering
\includegraphics[scale=.45]{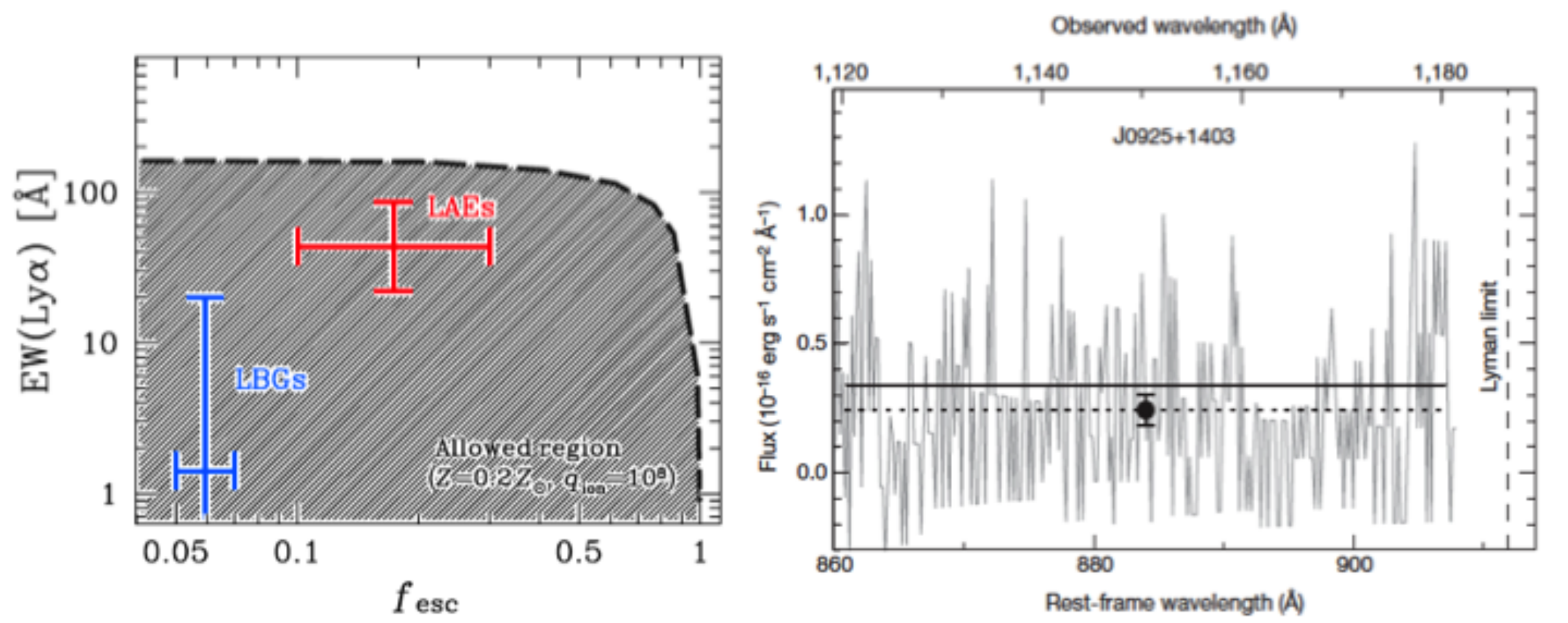}
\caption{
Left: Rest-frame Ly$\alpha$ EW and $f_{\rm esc}^{\rm ion}$ for LAEs (red) and LBGs (blue) at $z=3$ \citep{nakajima2014}.
The dashed curve is the results of the CLOUDY model calculations 
with a metallicity 0.2 times solar and an ionization parameter $q_{\rm ion}=8.0$, 
indicating the upper limits of Ly$\alpha$ EW.
The shaded region represents the parameter space allowed for the model.
Right: LyC spectrum 
of a local star-forming galaxy 
(gray line; \citealt{izotov2016a}).
Both the filled circle and the dotted line represent the LyC flux averaged 
over the range $860-913$\AA.
The error bar associated with the filled circle shows the $3\sigma$ 
uncertainty
for the average LyC flux measurement.
The horizontal solid line is the same as the dotted one, but corrected for the Milky Way extinction.
This figure is reproduced by permission of the MNRAS and Nature Publishing Group.
}
\label{fig:nakajima2014_fig13_izotov2016_fig4}
\end{figure}

A determination of $f_{\rm esc}^{\rm ion}$ requires very deep
observations of galaxies for LyC detections.
There are two observational approaches to detect LyC of galaxies, extremely deep
spectroscopy and narrowband imaging for star-forming galaxies.
The left and right panels of Figure \ref{fig:shapley2006_fig3-4_iwata2009_fig2} present
LyC emission of $z\sim 3$ galaxies 
found in the spectrum and the narrowband image, respectively.
With the spectra and narrowband images,
the average $f_{\rm esc}^{\rm ion}$ value is estimated to be $\sim 5$\% for LBGs
and $\sim 20$\% for LAEs at $z\sim 3$ (\citealt{shapley2006,iwata2009,nestor2013};
see the other cases in \citealt{vanzella2016,debarros2016}).
Interestingly, the average $f_{\rm esc}^{\rm ion}$ of LAEs is higher than the one of LBGs,
suggesting a positive correlation between $f_{\rm esc}^{\rm ion}$ and Ly$\alpha$ EW
in the moderately low Ly$\alpha$ EW regime
(left panel of Figure \ref{fig:nakajima2014_fig13_izotov2016_fig4}).
Moreover, the average $f_{\rm esc}^{\rm ion}$ of $z\sim 3$ galaxies is significantly
higher than that of most well known star-forming galaxies
($f_{\rm esc}^{\rm ion}<3$\%) 
such as Haro11 and Tol 1247-232 \citep{leitet2011,leitet2013}.
%
%
In addition to 
the high estimated value of 
$f_{\rm esc}^{\rm ion}$ at $z\sim 3$,
the LyC emission in narrowband images show a spatial offset from the intensity peak of the UV ($\sim 1500$\AA) continuum
(right panel of Figure \ref{fig:shapley2006_fig3-4_iwata2009_fig2}).
Although the spatial offset of the LyC emission may indicate that 
the major LyC emitting region is different from the UV-continuum emitting region,
there is a possibility that the spatial offset would be a signature of the chance alignment
of a foreground
(low-$z$) object
whose rest-frame UV continuum can mimic the LyC. However, the probability
for such chance alignments
is estimated to be only $2-3$\% \citep{iwata2009}.
With this small probability, one cannot explain all of the LyC emitting galaxy candidates
by chance alignments of foreground objects.

Although foreground contamination objects may not be the reason for the high observational value of
$f_{\rm esc}^{\rm ion}$ at $z\sim 3$, 
there remains the question why
observations do not find star-forming galaxies at $z\sim 0$ with
$f_{\rm esc}^{\rm ion}\simeq 5-20$\% that is as high as that of 
$z\sim 3$ LBGs and LAEs. Recent HST/COS observations
identify a total of 5 star-forming galaxies at $z\sim 0$ with a high escape fraction,
$f_{\rm esc}^{\rm ion}\simeq 6-13$\% \citep{izotov2016a,izotov2016b}, which is the definitive evidence that
there exist local galaxies with a high $f_{\rm esc}^{\rm ion}$ value comparable with
those of high-$z$ galaxies (right panel of Figure \ref{fig:nakajima2014_fig13_izotov2016_fig4}).
\footnote{
It should be noted that \citet{borthakur2014} have identified 
a local star-forming galaxy 
with an absolute escape fraction 
of $f_{\rm esc}^{\rm ion}\simeq 1$\%. This absolute escape fraction
corresponds to 21\%, if one does not include dust extinction effects. 
}
These 5 star-forming galaxies are selected with the criterion of $O_{32}\gtrsim 5$ \citep{izotov2016b}.
Because there is a possibility that such a high $O_{32}$ value indicates a high ionization parameter
and perhaps a density-bounded nebula, there would exist
a positive correlation between $f_{\rm esc}^{\rm ion}$ and $O_{32}$ 
\citep{nakajima2014}. In this case, one can easily understand that
high-$z$ galaxies, especially LAEs, have the high escape fraction of $\simeq 5-20$\%,
because high-$z$ galaxies have a large value of $O_{32}\sim 10$ that is significantly larger
than the average $O_{32}$ value of local galaxies (Figure \ref{fig:nakajima2014_fig2_fig12}; Section \ref{sec:ionization_state}).
The question of the high $f_{\rm esc}^{\rm ion}$ value for high-$z$ galaxies
is being answered by recent studies.



%

%




\subsubsection{Ionizing Photon Production Efficiency}
\label{sec:xi_ion}

\begin{figure}[H]
\centering
\includegraphics[scale=.48]{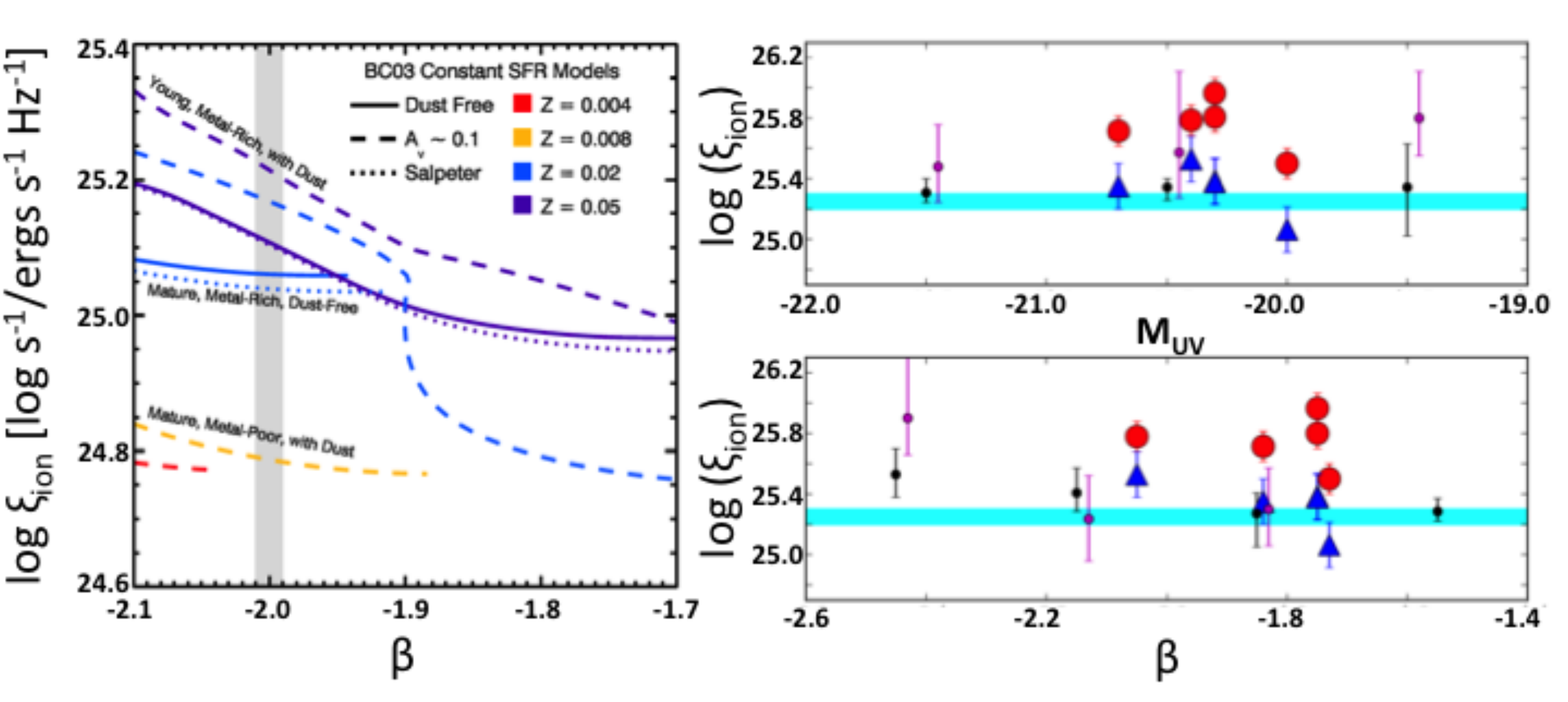}
\caption{
Left: Model predictions for $\xi_{\rm ion}$ as a function of $\beta$ \citep{robertson2013}.
The solid and dashed curves represent dust-free and dusty ($A_{\rm V}\sim 0.1$)
galaxies, respectively, that are calculated with the stellar population synthesis model of \citet{bruzual2003}
under the assumptions of the \citet{chabrier2003} IMF and constant star-formation history.
The red, orange, blue, and purple colors of the curves indicate the metallicities of $Z=0.004$, $0.008$, $0.02$, and $0.05$,
respectively. The dotted curves are the same as the solid curves, but for the \citet{salpeter1955} IMF.
The calculations stop at $7.8\times 10^8$ yr corresponding to the cosmic age at $z\sim 7$.
The gray shade denotes the $\beta$ range for the average $z\sim 7$ galaxy that is obtained by observations. 
Right: $\xi_{\rm ion}$ as a function of $M_{\rm UV}$ (top) and $\beta$ (bottom; \citealt{schaerer2016}).
The red circles (blue triangles) show $\xi_{\rm ion}$ with (without) a dust attenuation correction of the UV continuum
for five star-forming galaxies at $z\sim 0.3$. Note that two blue triangles are indistinguishable 
in this plot. The black and magenta circles represent LBGs at $z=3.8-5$ and $5.1-5.4$, respectively.
The cyan region indicates the canonical value for the dust-extinction corrected $\xi_{\rm ion}$.
%
This figure is reproduced by permission of the AAS and A\&A.
}
\label{fig:robertson2013_fig1_schaerer2016_fig1}
\end{figure}

The ionizing photon production efficiency is defined as
\begin{equation}
\xi_{\rm ion}=\dot{n}_{\rm ion}^0/L_{\rm UV},
\label{eq:xi_ion}
\end{equation}
where $\dot{n}_{\rm ion}^0$ and $L_{\rm UV}$ are
the intrinsic ionizing photon production rate (before the escape from the ISM) 
and the UV ($\sim 1500$\AA) continuum luminosity,
respectively. The left pane of Figure \ref{fig:robertson2013_fig1_schaerer2016_fig1} 
presents $\xi_{\rm ion}$ as a function of UV spectral slope predicted by 
stellar synthesis models with various metallicities and IMFs \citep{robertson2013}. 
In this way, $\xi_{\rm ion}$ depends on stellar populations.
Although many parameters of stellar populations
are constrained with galaxy SEDs in UV ($>1216$\AA), optical, 
and NIR bands including the UV spectral slope,
there remain large differences of $\xi_{\rm ion}$ for a given SED shape
(see Figure \ref{fig:robertson2013_fig1_schaerer2016_fig1} for a given UV slope).
This is because $\xi_{\rm ion}$ is very sensitive 
to the metallicity, IMF, and star-formation history
that are key parameters for ionizing photon production, while the observable galaxy SED 
at $>1216$\AA\ does not change.
Moreover, there is another large uncertainty in the choice of stellar synthesis models 
for ionizing production rates that depend on the physical properties of massive binary stars \citep{eldridge2017}.
Nevertheless, the galaxy SED approach can suggest a value of $\log \xi_{\rm ion} [{\rm erg^{-1} Hz}]\sim 25.2$ 
that should include systematic uncertainties by a factor of a few (\citealt{robertson2013}; 
Figure \ref{fig:robertson2013_fig1_schaerer2016_fig1}).
To determine $\xi_{\rm ion}$ with no such systematics, one needs other approaches.
A promising method to estimate $\xi_{\rm ion}$ is to use hydrogen Balmer lines such as
H$\alpha$ and H$\beta$. Because Balmer lines are produced in \hii\ regions via photoionization,
one can estimate intrinsic ionizing photon production rates with Balmer line fluxes
with the simple analytical relation,
%
\begin{eqnarray}
\label{eq:intrinsic_n_ion}
\dot{n}_{\rm ion}^0 & = & 2.1\times10^{12} (1-f_{\rm esc}^{\rm ion})^{-1} L_{\rm H\beta},\\
\label{eq:xi_ion}
\log (\xi_{\rm ion}) & = &  \log (\dot{n}_{\rm ion}^0) + 0.4\ M_{\rm UV} - 20.64,
\end{eqnarray}
where $L_{\rm H\beta}$ and $M_{\rm UV}$ are the extinction-corrected H$\beta$ luminosity
(erg s$^{-1}$) 
and the extinction-corrected UV continuum magnitude, respectively
\citep{schaerer2016,storey1995}. In eq. (\ref{eq:intrinsic_n_ion}), the term of $(1-f_{\rm esc}^{\rm ion})$
subtracts the ionizing photons escaping from the galaxy to the IGM.
The right panel of Figure \ref{fig:robertson2013_fig1_schaerer2016_fig1} shows
$\xi_{\rm ion}$ as a function of UV magnitude or UV slope that is obtained by
the Balmer line method. The right panel of Figure \ref{fig:robertson2013_fig1_schaerer2016_fig1}
indicates $\log \xi_{\rm ion} [{\rm erg^{-1} Hz}]\simeq 25.2-25.4$ for $z=4-5$ LBGs \citep{bouwens2015}
and $z\sim 0$ Lyman-continuum leaking galaxies of \citet{izotov2016a,izotov2016b} 
that are similar to (or slightly higher than) the value determined by the galaxy SED study \citep{schaerer2016}. 
There would be a trend of increasing $\xi_{\rm ion}$ towards small UV slopes (i.e. blue UV continuum; \citealt{bouwens2015}).
Hereafter, the value of $\log \xi_{\rm ion} [{\rm erg^{-1} Hz}]\simeq 25.2-25.4$ is referred to as the fiducial value.
%
%


\subsubsection{Galaxy Contribution to Reionization: Comparisons of $Q_{\rm HII}(z)$ and $\tau_{\rm e}$}
\label{sec:comparisons_tau_q}
 
Because the observational studies have constrained 
the three major parameters of galaxies, i.e.
$\rho_{\rm UV} (M_{\rm trunc})$, $f_{\rm esc}^{\rm ion}$, and $\xi_{\rm ion}$
(Sections \ref{sec:uv_luminosity_density}-\ref{sec:xi_ion}),
one can determine $\dot{n}_{\rm ion}$ via eq. (\ref{eq:n_ion_galaxy})
that is the amount of the ionizing photon contribution of galaxies.

The $\dot{n}_{\rm ion}$ value is used in the ionization equation, eq. (\ref{eq:ionization_equation}).
In eq. (\ref{eq:ionization_equation}), $\dot{n}_{\rm ion}$ should include not only the ionizing photons from galaxies
but also those from the other ionizing sources such as AGNs. However, $\dot{n}_{\rm ion}$ values of AGNs are
poorly determined as discussed in Section \ref{sec:AGN_contribution}. Here, 
one can first assume that the contribution to $\dot{n}_{\rm ion}$ from the other ionizing sources 
is negligible (i.e. galaxies' $\dot{n}_{\rm ion}$ is dominant), 
and obtain the evolution of the ionized fraction $Q_{\rm HII} (z)$ and the inferred $\tau_{\rm e}$
with the equations (\ref{eq:ionization_equation}) and (\ref{eq:tau_e}) via (\ref{eq:q_HII}).
Comparing these $Q_{\rm HII} (z)$ and $\tau_{\rm e}$ values with those from 
the direct measurements shown in Section \ref{sec:cosmic_reionizationI},
one can test whether galaxies can reionize the universe or the other sources of reionization
are necessary. Below, I discuss the galaxy contribution to reionization based on these arguments.

\begin{figure}[H]
\centering
\includegraphics[scale=.47]{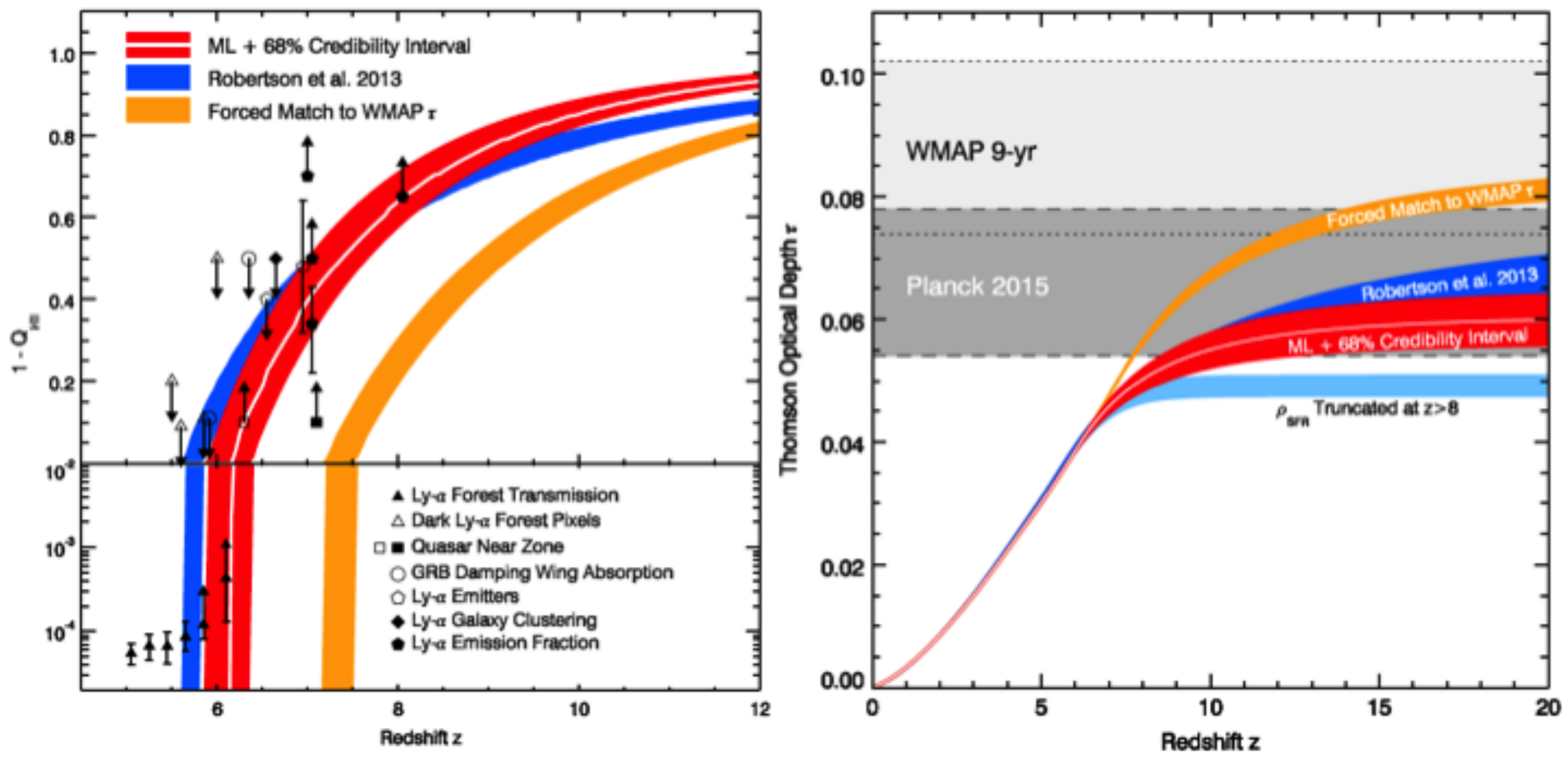}
\caption{
Neutral hydrogen fraction ($1-Q_{\rm HII}$) as a function of redshift (left) and 
Thomson scattering optical depth integrated over redshift (right; \citealt{robertson2015}).
The red region indicates the best estimates obtained from the cosmic star-formation history
determined by observations 
in the case of $f_{\rm esc}^{\rm ion}=0.2$ and $\log \xi_{\rm ion}/[{\rm s^{-1} M_\odot^{-1} yr}]=53.14$. 
The blue, orange, and cyan regions are the results of the previous study \citep{robertson2013}, 
the forced-match to the WMAP data \citep{hinshaw2013}, and a different cosmic star-formation history with a rapid
decrease at $z\sim 8$, respectively.
In the left panel, the data points and the arrows represent the neutral hydrogen fraction estimates
obtained from the Gunn-Peterson effect and Ly$\alpha$ damping wing measurements.
In the right panel, the dark and light-gray shades denote the Thomson scattering optical depth measurements
given by Planck 2015 \citep{planck2015} and the nine-year WMAP \citep{hinshaw2013}.
This figure is reproduced by permission of the AAS.
}
\label{fig:robertson2015_fig3_fig2}
\end{figure}

The $\dot{n}_{\rm ion}$ value of galaxies is estimated with the three parameters,
$\rho_{\rm UV}$ 
shown in Figure \ref{fig:hff_lf_rhoUV}
(Section \ref{sec:uv_luminosity_density}),
$f_{\rm esc}^{\rm ion}\simeq 20$\% (Section \ref{sec:escape_fraction_ionizing_photon}), and
$\log \xi_{\rm ion} [{\rm erg^{-1} Hz}]\sim 25.2$ (Section \ref{sec:xi_ion}),
where no redshift evolutions of $f_{\rm esc}^{\rm ion}$ and $\xi_{\rm ion}$ are included.
Assuming the value of $C_{\rm HII}=3$ (cf. eq. \ref{eq:clumping_factor}),
one can derive $Q_{\rm HII} (z)$ with eq. (\ref{eq:ionization_equation}).
The left panel of Figure \ref{fig:robertson2015_fig3_fig2} presents $Q_{\rm HII} (z)$ calculated
with the galaxies' $\dot{n}_{\rm ion}$ thus obtained.
Although the direct measurements of $Q_{\rm HII} (z)$ have large uncertainties,
the inferred $Q_{\rm HII} (z)$ is consistent with the existing direct measurements.
The right panel of Figure \ref{fig:robertson2015_fig3_fig2} shows $\tau_{\rm e}$
as a function of redshift. Again, the inferred $\tau_{\rm e}$ value agrees
with the direct CMB measurement of $\tau_{\rm e}$ given by \citet{planck2015}
\footnote{
The up-to-date best measurement of $\tau_{\rm e}$ 
%
%
is systematically smaller than the value of \citet{planck2015} 
beyond the 1 sigma error level, $\tau_{\rm e}=0.058\pm 0.012$ 
\citep{planck2016}.
The value is even smaller in 
the latest Planck result,
$\tau_{\rm e}=0.0561\pm 0.0071$ \citep{planck2018}.
}.
Although the direct measurements
have large statistical errors in $Q_{\rm HII} (z)$ and
potentially significant systematic errors in $\tau_{\rm e}$,
the $\dot{n}_{\rm ion}$ value of galaxies alone 
explain both the direct measurements of $Q_{\rm HII} (z)$ and $\tau_{\rm e}$.
These results may suggest that galaxies are major sources of cosmic reionization.

However, it should be noted that there are a factor of $\gtrsim 2$ uncertainties
in the relatively poor determinations of the three parameters 
as well as the direct measurements. There remain possibilities that
these large errors would not allow us to identify an inconsistency
between the inferred value and the direct measurement of 
$Q_{\rm HII} (z)$ or $\tau_{\rm e}$.
One can conclude whether galaxies
are major sources of cosmic reionization or not, after 
the errors on these parameters and the direct measurements 
become considerably small.






%
%

\subsection{AGN Contribution}
\label{sec:AGN_contribution}

Recent galaxy studies suggest that 
a majority of ionizing photons for reionization would be supplied by
galaxies (Section \ref{sec:galaxy_contribution}).
However, these study results still include large systematic
uncertainties. One needs to test whether the other 
sources can contribute to cosmic reionization.
AGNs are prominent sources supplying ionizing photons,
because a large amount of ionizing photons 
are efficiently produced in AGNs and escape from them.
%

\begin{figure}[H]
\centering
\includegraphics[scale=.45]{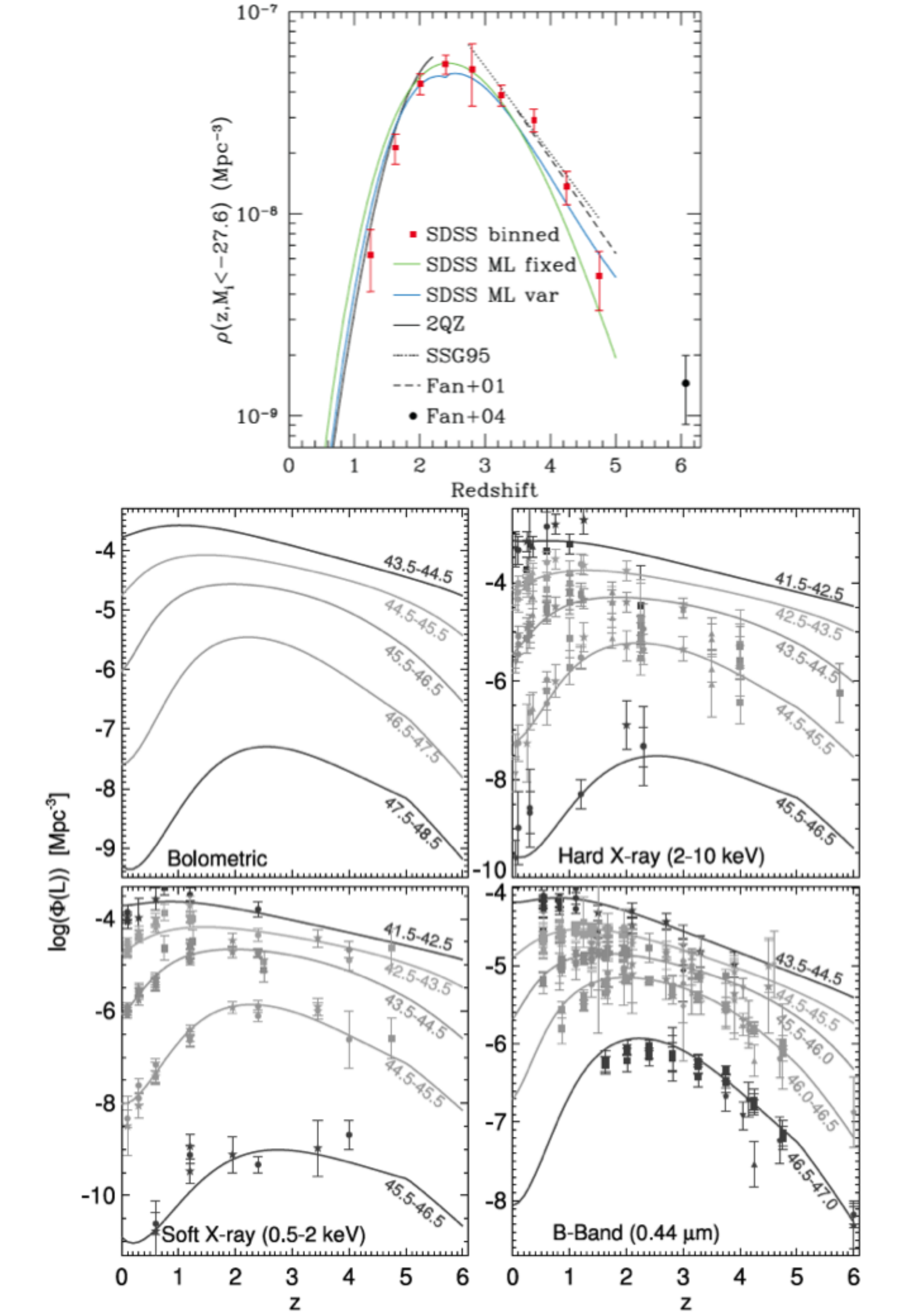}
\caption{
Top panel: Evolution of the QSO number density \citep{richards2006}.
The red squares and the black circle are the number densities integrated up to 
$i=-27.6$ mag in QSO luminosity functions obtained from the SDSS data.
The lines indicate the fitting results of these number densities and previous studies.
Bottom four panels: Evolution of number densities for AGNs with different luminosities \citep{hopkins2007}.
The top left, top right, bottom left, and bottom right panels 
present the redshift evolution of AGNs for a given luminosity interval of bolometric luminosity, hard X-ray ($2-10$ keV), 
soft X-ray ($0.5-2$ keV), and $B$ band ($0.44\mu$m), respectively, where the luminosity 
intervals are indicated with the labels. The lines represent
the results of the best-fit evolving double power-law models.
This figure is reproduced by permission of the AAS.
}
\label{fig:richards2006_fig20_hopkins2007_fig9}
\end{figure}

The top panel of Figure \ref{fig:richards2006_fig20_hopkins2007_fig9} presents the number density evolution of QSOs,
where QSOs are defined as AGNs brighter than $-27.6$ magnitude. The number density of QSOs peaks at $z\sim 2-3$,
and decreases towards high-$z$. 
At $z\sim 6$, the QSO number density is only $10^{-9}$ Mpc$^{-3}$ \citep{fan2004,richards2006}.
Similarly, UV luminosity functions of QSOs decrease very rapidly towards $z\sim 6$. Based on the UV luminosity functions,
the production rate of ionizing photons of QSOs are estimated in the same manner as those of galaxies (Section \ref{sec:galaxy_contribution}).
Although it is assumed that QSO spectra have an ionizing photon escape fraction of unity, and that QSOs have a power law spectrum,
these assumptions are plausible for QSOs whose LyC escape and production are well understood in contrast with 
galaxies (Section \ref{sec:galaxy_contribution}). The QSO contribution is estimated 
to be only $\sim 1-10$\% of ionizing photons at $z\sim 6$ \citep{srbinovsky2007}. 
The ionizing photon production rate of QSOs is negligibly small,
due to the very small number density of QSOs at $z\sim 6$.
A further decrease of QSO
number density is suggested from $z\sim 6$ to $z\sim 7$ \citep{venemans2013}. These results indicate that
the ionizing photon production rate of QSOs becomes smaller towards high-$z$.

Although QSOs (i.e. bright AGNs) do not significantly contribute to cosmic reionization,
there remains the possibility that 
faint AGNs could be major contributors of cosmic reionization.
As shown in the bottom four panels of Figure \ref{fig:richards2006_fig20_hopkins2007_fig9},
the number density of faint AGNs is larger than QSOs by orders of magnitudes.
Given the efficient ionizing photon production with the power-law continuum of AGNs,
faint AGNs would be important in cosmic reionization. Moreover, the number density of 
faint AGNs does not drop as steeply as those of bright AGNs towards the epoch of reionization.

Properties of faint AGNs at the EoR are not well understood, due to the difficulties
in identifying faint AGNs at such a high redshift.
\citet{treister2011} report the $>5 \sigma$ detections in the stacked X-ray spectra
of dropout galaxies at $z\sim 6$, suggesting the existence of faint AGNs in
dropout galaxies
\footnote{
Because these faint AGNs are not identified in UV but X-ray,
\citet{treister2011} claim that these faint AGNs do not contribute to
cosmic reionization due to the obscuration of UV photons.
}.
However, subsequent studies identify no X-ray emission in
a similar stacked X-ray data of $z\sim 6$ dropout galaxies \citep{willott2011,fiore2012}.
\citet{cowie2012} argue that the background subtraction of the X-ray data would
produce wrong detections, and that there are no signatures of faint AGNs
in dropout galaxies at $z\sim 6$ on average.
Nevertheless, the contribution of faint AGNs is still under debate.

\begin{figure}[H]
\centering
\includegraphics[scale=.48]{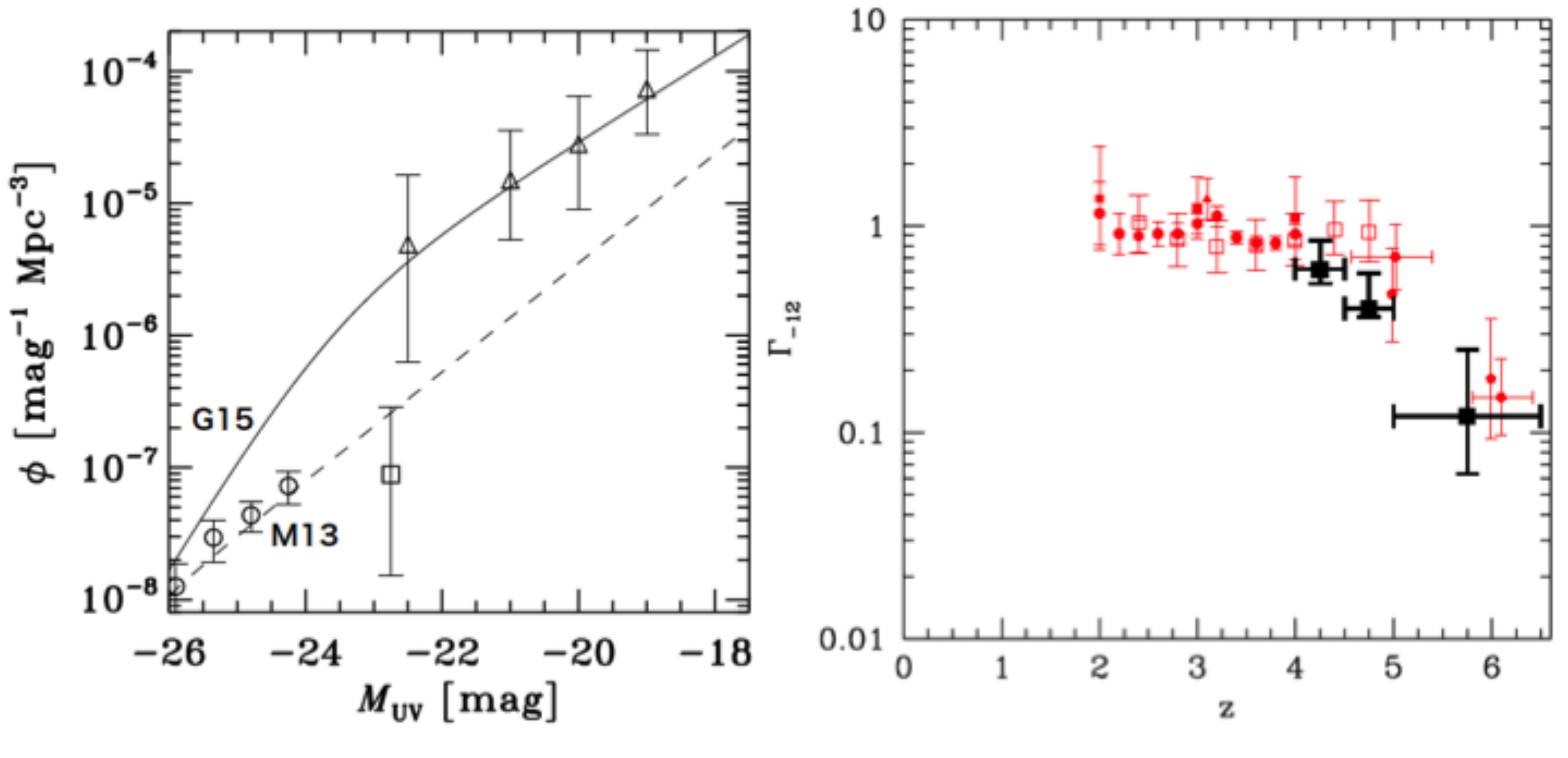}
\caption{
Left: UV luminosity function of AGN at $z\sim 5$. The triangles, circles, and square are the AGN luminosity
functions obtained by \citet{giallongo2015}, \citet{mcgreer2013}, \citet{ikeda2012}, respectively.
The solid and dashed lines are the double-power laws best-fit to
the data of \citet{giallongo2015} and \citet{mcgreer2013}, respectively.
Right: Redshift evolution of the cosmic photoionization rate $\Gamma_{-12}$
in units of $10^{-12}$ s$^{-1}$ \citep{giallongo2015}.
The black squares are the AGN contribution to $\Gamma_{-12}$ (connected to
$\dot{n}_{\rm ion}$ of AGNs) , which are obtained by \citet{giallongo2015}, 
while the red data points denote the total $\Gamma_{-12}$
estimated from QSO Ly$\alpha$ absorption line systems.
This figure is reproduced by permission of the A\&A.
}
\label{fig:ono2016_chorus_giallongo2015_fig6}
\end{figure}

\citet{giallongo2015} show a number of dropouts at $z\sim 4-6$ with X-ray detections
on the individual basis, and claim a steep faint-end slope of AGN UV luminosity functions
(left panel of Figure \ref{fig:ono2016_chorus_giallongo2015_fig6}). If this steep faint-end slope
is true, ionizing photons are mostly originated from faint AGNs at $z\sim 6$ 
(\citealt{giallongo2015}; right panel of \ref{fig:ono2016_chorus_giallongo2015_fig6}).
However, 
the faint-end slope value of the AGN UV luminosity function at $z \sim 4-6$ remains an open question.
Although the UV magnitude range of the \citeauthor{giallongo2015}'s luminosity function
has only a small overlap  with the one of the previous study at $M_{\rm UV}\sim -22$ \citep{mcgreer2013},
the AGN number density of \citet{giallongo2015} is about an order of magnitude higher than
that of \citet{mcgreer2013} (left panel of Figure \ref{fig:ono2016_chorus_giallongo2015_fig6}). 
Moreover, the ionizing photon escape fraction of faint AGNs is not well understood \citep{grazian2018},
while QSOs have an escape fraction as high as unity that is indicated by QSO proximity effects.
In this way, the contribution of faint AGNs is not clearly understood yet.

%


%


%
%
%

\subsection{Summary of Cosmic Reionization II}
\label{sec:summary_cosmic_reionizationII}

This section has discussed what are the major sources that reionize the universe
with the latest observational progresses,
explaining the procedures to quantify the budget of ionizing photons.
Starting with the galaxy contribution to cosmic reionization,
I clarify three major parameters, $\rho_{\rm UV} (M_{\rm trunc})$, 
$f_{\rm esc}^{\rm ion}$, and $\xi_{\rm ion}$, to evaluate 
the production rate of ionizing photons, i.e. $\dot{n}_{\rm ion}$.
There is another parameter, the clumping factor of the ionized hydrogen IGM ($C_{\rm HII}$).
With the latest observational results for these parameters, 
the ionizing photon production rate of galaxies alone agrees with
the direct observational measurements of $Q_{\rm HII} (z)$ and $\tau_{\rm e}$
obtained with high-$z$ objects and CMB (Section \ref{sec:cosmic_reionizationI}), respectively.
These results may indicate that galaxies are major sources of cosmic reionization,
although the agreements 
are not strong, due to the large errors on these parameters.
Because the number density of bright AGNs, i.e. QSOs, is very small at the EoR,
the ionizing photon contribution of QSOs is negligibly small. However, there is
a possibility that ionizing photons of faint AGNs may significantly contribute to cosmic reionization.
The number density as well as the escape fraction of the faint AGNs is not clearly understood by observations, to date,
and various observational studies are trying to determine these AGN properties.

%
%
%

\section{On-Going and Future Projects}
\label{sec:ongoing_and_future_projects}

This section summarizes the open questions about LAEs discussed 
in the previous sections, and introduces on-going and future projects 
potentially answering them.


%

\subsection{Open Questions}
\label{sec:open_questions}

Major open questions about LAEs discussed in this lecture are listed below.\\

\noindent
{\bf Galaxy Formation : }
\begin{itemize}
\item[$-$]  
What are the physical reasons for the LAE's star-formation and ISM characteristics 
that distinguish LAEs from the other galaxy populations?
For example, are the high-ionization parameters of LAEs due to high ionization production rate, 
to high electron density, or to density-bounded ISM?\\
%
%
\item[$-$]  
What are the Ly$\alpha$ blobs and diffuse halos? Are they related to cold accretion?
Where is the cold accretion that produces Ly$\alpha$ emission? \\
\item[$-$] What makes the large Ly$\alpha$ EWs ($\gtrsim 200$\AA) of LAEs? What is the major physical reason for the 
$f_{\rm esc}^{\rm Ly\alpha}$ increase from $z\sim 0$ to $6$?\\
\item[$-$] Have we already identified real popIII star-formation in $z\lesssim 7$ LAEs with strong \heii\ emission and no detectable metal lines? \\
\end{itemize}

\noindent
{\bf Cosmic Reionization : }
\begin{itemize}
\item[$-$]  How did cosmic reionization proceed? 
Is it true that the reionization occurred late, as suggested by the LAEs and recent CMB studies?
Is the evolution of $Q_{\rm HII}$ extended or sharp? How can some Ly$\alpha$ photons
escape from $z>8$ where the IGM is highly neutral (see Section \ref{sec:future_projects})?\\
\item[$-$] Are star-forming galaxies major sources of reionization, especially
for low-mass galaxies most of which 
have intrinsically strong Ly$\alpha$ emission?
Do faint AGNs play an important role in supplying ionizing photons?
Can we conclude what are the major sources of cosmic reionization, given the large uncertainties on the parameters,
$\rho_{\rm UV} (M_{\rm trunc})$,  $f_{\rm esc}^{\rm ion}$, and $\xi_{\rm ion}$ and $C_{\rm HII}$?
\end{itemize}

%
%
%
%
%

\subsection{New Projects Addressing the Open Questions}
\label{sec:new_projects}



\subsubsection{On-Going Projects}
\label{sec:new_projects}

\begin{figure}[H]
\centering
\includegraphics[scale=.40]{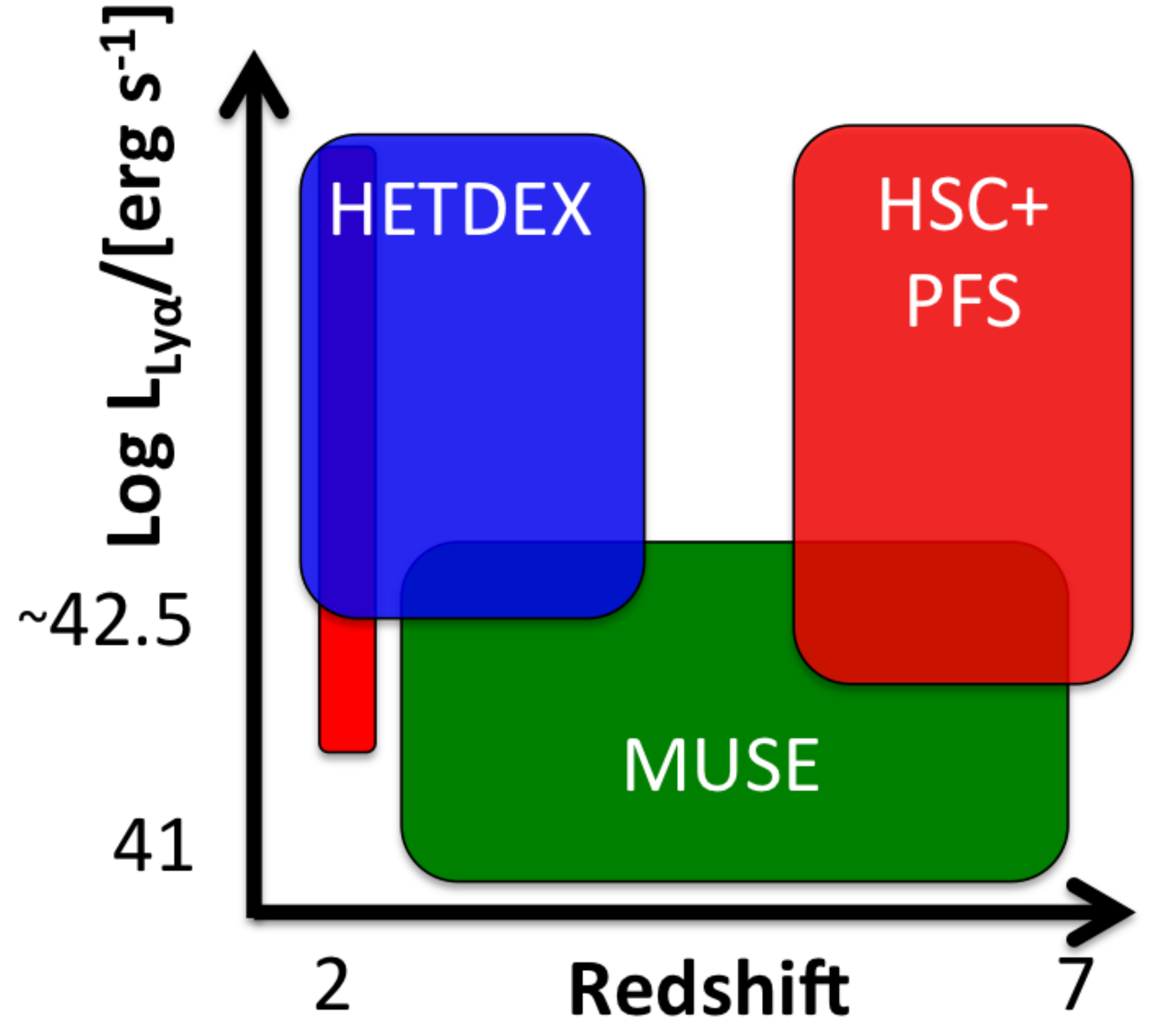}
\caption{
Parameter space of Ly$\alpha$ luminosity vs. redshift that is covered by
HSC+PFS (red), HETDEX (blue), and MUSE (green) surveys.
The HSC+PFS observations will cover $z\sim 2$ and $z\sim 5-7$.
}
\label{fig:hsc_hetdex_muse}
\end{figure}

There are three major on-going projects that can study
LAEs up to $z\lesssim 7$, Subaru/Hyper Suprime-Cam (HSC; \citealt{miyazaki2018})
+ Prime-Focus Spectrograph (PFS; \citealt{tamura2016}),
Hobby-Eberly Telescope Dark Energy Experiment (HETDEX; \citealt{hill2012a}),
and 
VLT/Multi-Unit Spectroscopic Explorer (MUSE; \citealt{bacon2010}).
These three projects can target LAEs, covering the complementary parameter 
space of LAEs in depth and redshift (Figure \ref{fig:hsc_hetdex_muse}).
Moreover, these three projects are complementary
in the covering areas.
Subaru/HSC+PFS and HETDEX
cover large areas of a few $10-100$ deg$^2$, while
VLT/MUSE will observe small fields with an area up to a few $10$ arcmin$^2$.





\noindent
{\bf Subaru/HSC+PFS : }

\begin{figure}[H]
\centering
\includegraphics[scale=.48]{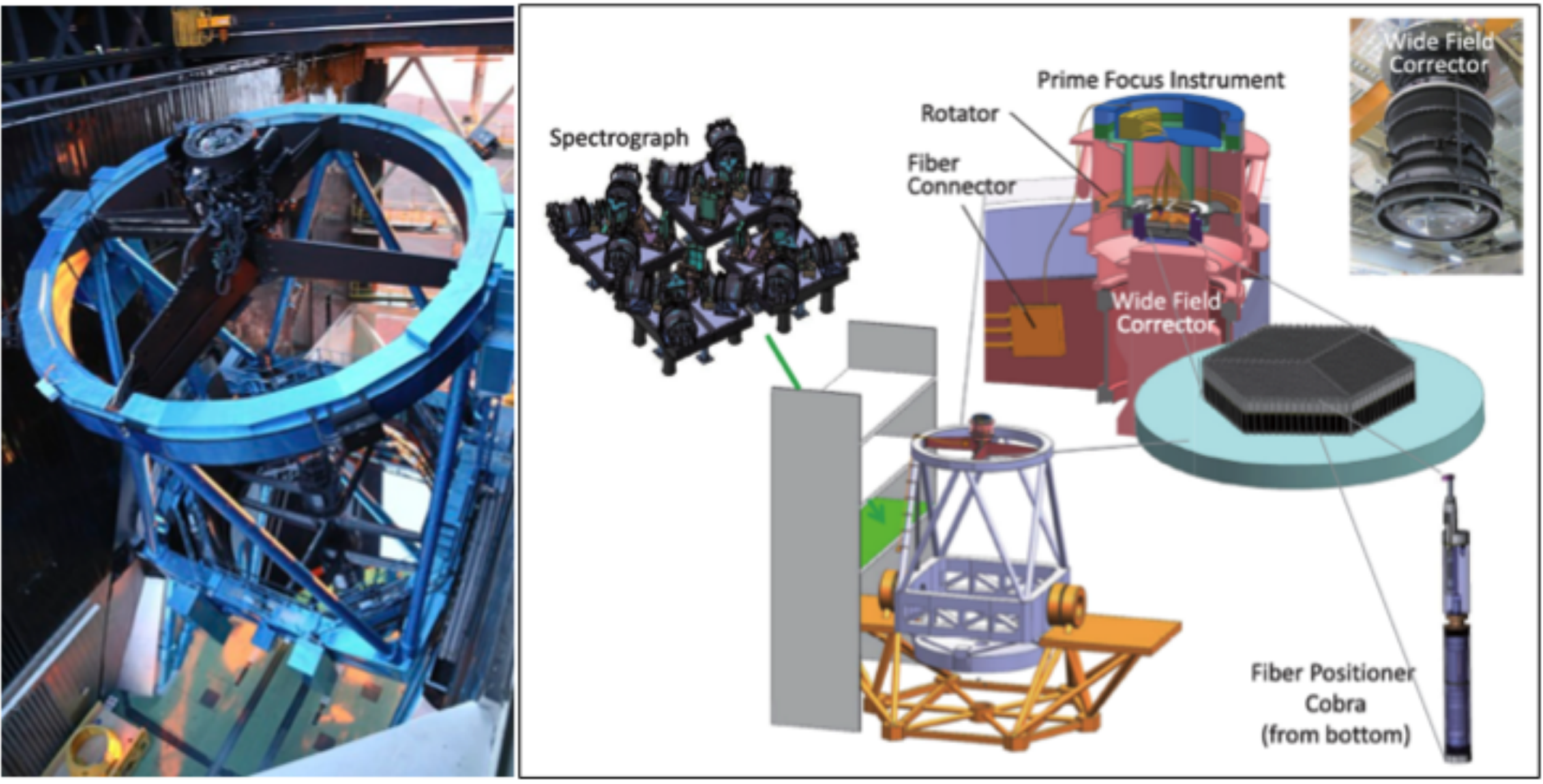}
\caption{
Left: Subaru/HSC mounted on the Subaru telescope (Courtesy: Y. Utsumi).
Right: Subaru/PFS conceptual design \citep{takada2014}.
The PFS instrument with its 2400 fibers will be installed at the Subaru prime focus, just like the HSC instrument.
The massive spectrographs fed by the fibers
are placed on the floor of the Subaru dome.
This figure is reproduced by permission of the PASJ.
}
\label{fig:hscblog_takada2014_fig1}
\end{figure}

Subaru/HSC is a wide-field optical imager with a field-of-view (FoV) of 1.5 deg-diameter circle (Figure \ref{fig:hscblog_takada2014_fig1}).
The FoV of HSC is seven times larger than that of its wide-field imager predecessor at Subaru, Suprime-Cam.
The HSC large program, which is the Subaru Strategic Program (SSP) survey, has started in March 2014; 
it is a collaborative project involving Japanese institutes, Princeton, and Taiwanese institutes.
A total of 300 nights are allocated to the HSC SSP survey 
that will be completed around 2019.
%
The HSC SSP survey has the wedding cake type survey 
design: it is planned to obtain 
$g$, $r$, $i$, $z$, and $y$
band imaging data down to the $5$ sigma limiting magnitudes 
$i\sim 26$ in the wide layer (1400 deg$^2$),
$i\sim 27$ in the deep layer (30 deg$^2$), and
$i\sim 28$ in the ultra-deep layer (3 deg$^2$).
The deep and ultra-deep layers are covered by narrowband images
with 5 sigma limiting magnitudes of $25-26$.
With these imaging data, the HSC SSP survey has a wide variety
of scientific goals from cosmology to solar-system objects.
For LAE studies, the HSC SSP survey 
will uncover
about 20,000 LAEs and 1,000 Ly$\alpha$
blobs down to $L_{\rm Ly\alpha}\sim 3\times 10^{42}$ erg s$^{-1}$
at $z=5.7$ and $6.6$. Additional samples of $z=2.2$ and $7.3$ LAEs will be 
gathered
in the deep and ultra-deep layers, respectively.
It should be noted that the $z=5.7$ and $6.6$ LAEs are found in
a total area nearing 1 comoving Gpc$^2$, allowing 
studies of $z\sim 6-7$ LAEs
in a cosmological large scale. These LAE samples 
will allow the researchers to address
a number of issues
of galaxy formation and cosmic reionization by analyses with Ly$\alpha$ luminosity functions and LAE clustering measurements.
Some of the early HSC SSP survey results on LAEs have been published.
For examples, the evolution of  Ly$\alpha$ blobs 
is determined 
on the basis of an unprecedentedly large sample
(e.g. \citealt{shibuya2018a}), and
the constraints on $x_{\rm HI}$ are being obtained with the goal of
an $x_{\rm HI}$ determination accuracy comparable to the model uncertainty of 10\%
for the reionization history \citep{ouchi2018,konno2018}.


%


%

The wide-field spectrograph Subaru/PFS complements
the wide-field imaging capability of Subaru/HSC (Figure \ref{fig:hscblog_takada2014_fig1}).
PFS is a multi-object fiber spectrograph that is being developed 
by a consortium including Japan, 
Princeton, JHU, Caltech/JPL,
LAM, Brazil, ASIAA, and many other contributors. 
PFS accommodates 2400 fibers with a diameter of 
$1.\!\!^{\prime\prime}0-1.\!\!^{\prime\prime}1$
that 
cover a FoV of $1.3$~deg$^2$, 
sharing the Subaru wide-field optical corrector with HSC.
There are three arms for blue, red, and NIR bands that take spectra at $0.38-0.65$, $0.63-0.97$,
and $0.94-1.26 \mu$m, respectively, with spectral resolutions
of $R=2400-4200$. The first light of PFS is planned around 2020.
There is a plan of large galaxy survey with PFS, and the
strategy of the galaxy survey is being built. 
Follow-up spectroscopy for the LAEs detected by the Subaru/HSC SSP survey
is envisaged; it will provide unique data sets that will significantly enhance 
our understanding of 
galaxy formation and cosmic reionization.\\



\begin{figure}[H]
\centering
\includegraphics[scale=.48]{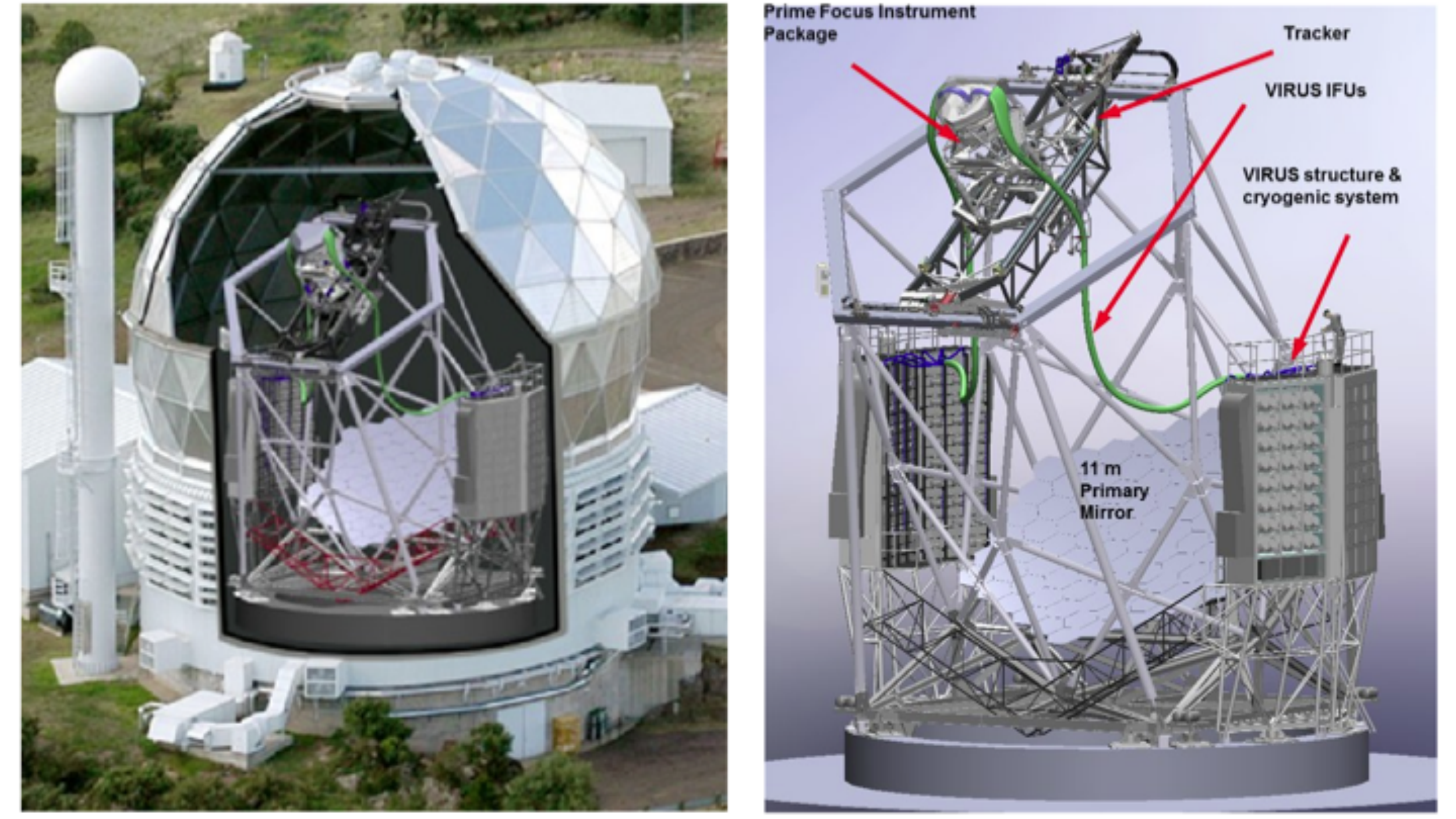}
\caption{
Schematic view of HETDEX \citep{kelz2014}.
Left: Hobby Eberly Telescope seen though the dome.
Right: Schematics of the prime-focus instrument and spectrographs
of VIRUS mounted on the prime focus and the dome, respectively. 
The IFU fibers are connected from the prime-focus instrument to 
the spectrographs.
This figure is reproduced by permission of the SPIE.
}
\label{fig:kelz2014_fig1}
\end{figure}

\noindent
{\bf HETDEX : }

HETDEX is a survey with the Hobby Eberly Telescope (HET) and its visible integral-field replicable unit 
spectrograph (VIRUS \citealt{hill2012b}). The schematic view of HET and VIRUS is shown in Figure \ref{fig:kelz2014_fig1}.
The VIRUS instrument on HET accommodates 78 integral-field units (IFUs) each of which
holds 448 fibers. The total number of the VIRUS fibers is over 30 thousands.
Because HET/VIRUS is optimized to target redshifted Ly$\alpha$ emission
at $z=1.9-3.5$ very efficiently to detect BAO and to constrain the equation of state of dark energy,
HET/VIRUS has the narrow wavelength coverage of
$3500-5500$ \AA\ and the low spectral resolution of $R\sim 700$.
The VIRUS instrument in part saw the first light in the middle of 2016. 
The VIRUS instrument is being built, while the HETDEX survey is conducted.
The HETDEX survey will identify 0.8 million LAEs at 
$z=1.9-3.5$ in a total area of 434 deg$^2$ ($\sim 20$\% filling area) down to 
$\sim 4\times 10^{-17}$ erg s$^{-1}$ cm$^{-2}$. Although the major scientific
goal of HETDEX is understanding the state of equation of dark energy by BAO analysis
with the LAEs, a number of statistical studies are being planned with the HETDEX LAEs;
for examples, the LAE luminosity function evolution and environment effect
in conjunction with clustering analysis. Because the HETDEX LAE data are large and unique,
the HETDEX survey 
will impact many 
areas of LAE studies.

It should be noted that galaxy-LSS connections
can be investigated by such large-volume spectroscopic surveys
with HETDEX as well as Subaru/PFS (e.g. \citealt{adelberger2003,rudie2012,
rakic2012,lee2014,mawatari2017,mukae2017}). 
Deep spectroscopy with Subaru/PFS
would reveal the IGM \hi\ and metal gas distribution
of LSS with Ly$\alpha$ and metal absorption lines found in spectra of
background bright AGNs and galaxies. The dense IFU spectroscopy of HETDEX
will give the spatial distribution of LAEs. The combination of these surveys may provide the
three-dimensional maps of the IGM gas and LAEs
addressing the question 
where LAEs form in the LSSs.




\noindent
{\bf VLT/MUSE : }

\begin{figure}[H]
\centering
\includegraphics[scale=.40]{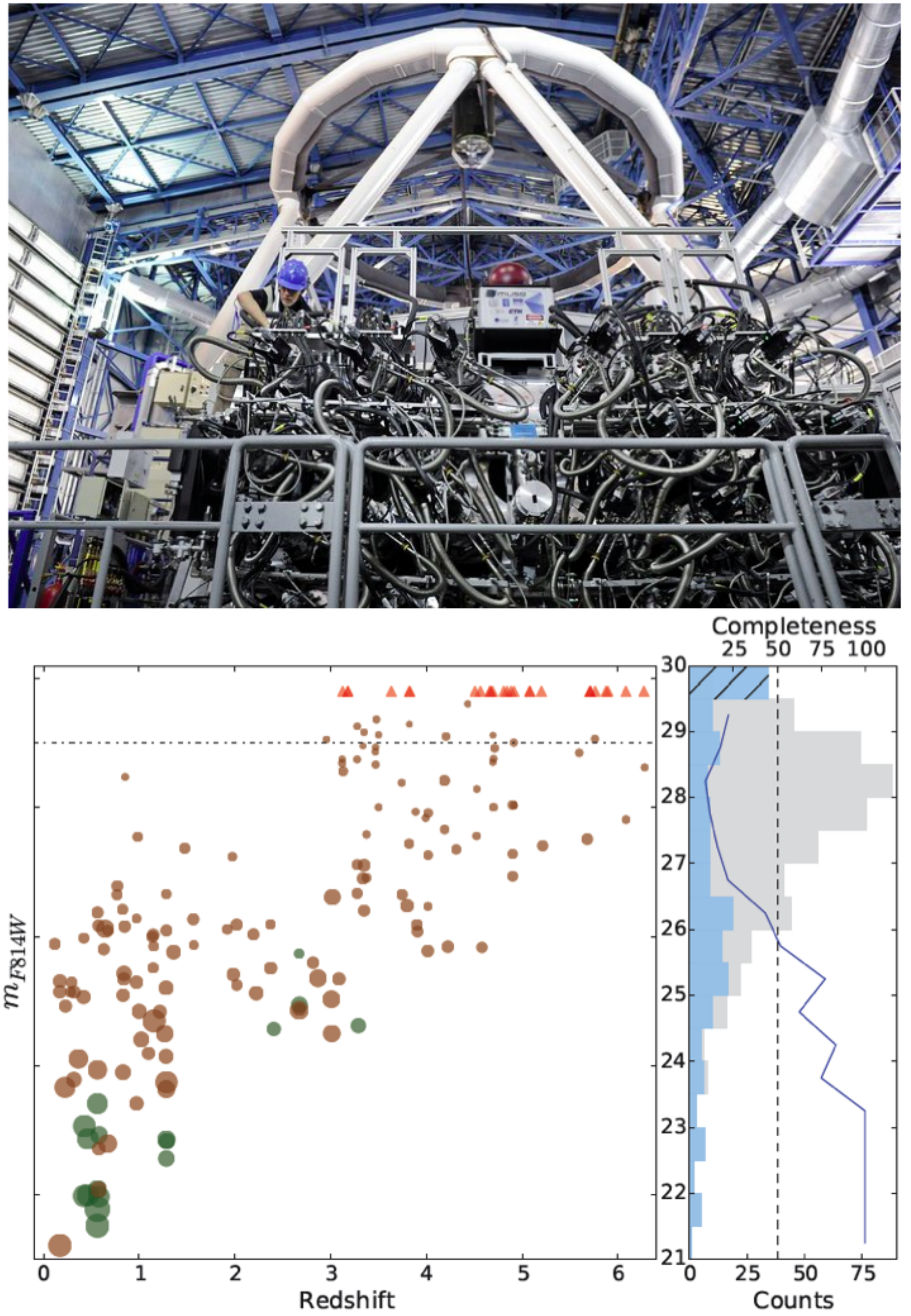}
\caption{
Top: Picture of the MUSE instrument that is installed at the Nasmyth platform (Courtesy: ESO).
Bottom: HST $I_{814}$-band continuum magnitude as a function of redshift (left)
and source number/completeness (right)
for LAEs identified by the deep MUSE observations in the HDF-S \citep{bacon2015}.
Left: The brown circles indicate the spectroscopically confirmed objects
newly identified by the MUSE observations. The green circles are the same as the
brown circles, but  for objects confirmed by previous spectroscopy.
The sizes of the circles represent the continuum size obtained with the HST $I_{814}$-band image.
The red triangles show the MUSE spectroscopically-confirmed objects with
no HST counterparts. The dashed horizontal line presents the $3\sigma$ detection limit
in the HST $I_{814}$-band image.
Right: The gray histogram denotes the magnitude distribution for all of the HST $I_{814}$-band
detected objects. The light-blue histogram (with the shade at $I_{814}=29.5-30.0$ mag) 
indicates the magnitude distribution of the MUSE spectroscopically-confirmed objects with (no) HST counterparts. 
The blue curve represents the completeness of the MUSE spectroscopic confirmation. The dashed vertical line
marks 50\% of completeness.
This figure is reproduced by permission of the A\&A.
}
\label{fig:muse_bacon2015_fig16}
\end{figure}

VLT/MUSE is an IFS 
with a contiguous field coverage of 1 arcmin$^2$ that is 
orders of magnitude larger than those of the other existing IFSs (top panel of Figure \ref{fig:muse_bacon2015_fig16}).
MUSE has a reasonably wide range of wavelength coverage ($4650-9300$\AA)
and a medium high spectral resolution ($R\sim 3000$).
The image slicers of MUSE keep the total throughputs high,
allowing high sensitivity observations. 
A number of exciting MUSE observation results have been
reported since 2015. One of the early MUSE observations 
consists in a
27-hour integration in the central 1 arcmin$^2$ area of the Hubble Deep Field South (HDF-S). The MUSE HDF-S
observations reach the $5\sigma$ detection limit of $\simeq 5\times 10^{-19}$ erg s$^{-1}$ cm$^{-2}$,
and increase the number of the spectroscopically identified galaxies by an order of magnitude 
(bottom panel of Figure \ref{fig:muse_bacon2015_fig16}; \citealt{bacon2015}).
Moreover, there are 26 MUSE-identified LAEs whose continuum is not detected
in the HST HDF-S images with the detection limit of $I\simeq 29.5$.
Further deep and wide surveys are being conducted in HUDF \citep{bacon2017} and Chandra Deep Field South \citep{herenz2017}.
As stated in Sections \ref{sec:lya_blob} and \ref{sec:diffuse_lya_halo}, MUSE observations are playing important roles 
in various Ly$\alpha$ studies, e.g. diffuse Ly$\alpha$ emitters in blank
fields and Ly$\alpha$ emission around QSOs.

It should be noted that there is a counterpart IFS that is developed 
for Keck telescopes, Keck Cosmic Web Imager (KCWI; \citealt{martin2010})
that is moved from the Palomar 5m telescope \citep{martin2010}.
The uniqueness of KCWI is
the blue-band coverage below 4650\AA\ that is not covered by 
VLT/MUSE. The blue-band coverage will allow the studies of
the LAEs-IGM relation (see above) at $z\sim 2$ that is an optimal redshift
for IGM \hi\ detections with a high S/N from the ground.


\subsubsection{Future Projects}
\label{sec:future_projects}

Although the three major LAE surveys of Subaru/HSC+PFS, HETDEX, and VLT/MUSE
cover a wide range of parameter space, these LAE surveys 
only target redshifts up to $z\sim 7$ (Figure \ref{fig:hsc_hetdex_muse}), 
due to the limited wavelength
coverages of optical bands (Section \ref{sec:new_projects}).
There remains the unexplored redshift of $z\gtrsim 8$
for LAE studies.

\begin{figure}[H]
\centering
\includegraphics[scale=.48]{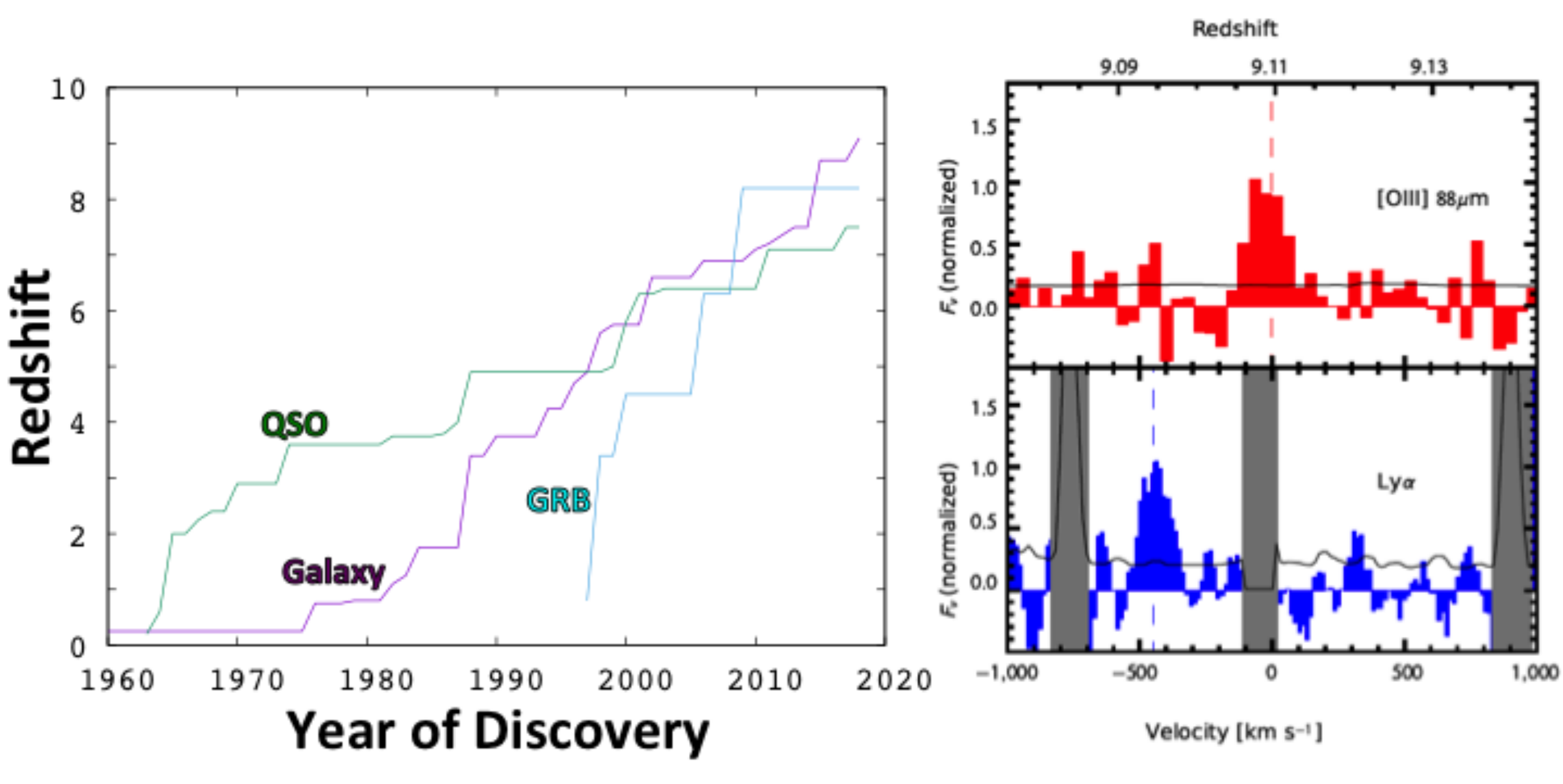}
\caption{
Left: Observational redshift limit vs. year of discovery for the redshift limit object.
The purple, green, and cyan lines represent a galaxy, QSO, and GRB, respectively.
Right: {\sc [Oiii]}$88\mu$m emission of a galaxy (top panel) confirmed at $z=9.1$ \citep{hashimoto2018}.
The bottom panel presents the reported Ly$\alpha$ emission significantly blueshifted by $\simeq 450$ km s$^{-1}$ 
from the systemic ({\sc [Oiii]}$88\mu$m emission) velocity.
The black solid lines indicate the $1\sigma$ noise.
The grey regions show the wavelength ranges strongly contaminated by the night sky emission.
This figure is reproduced by permission of the Nature Publishing Group.
}
\label{fig:year_redshift_hashimoto2018_edfig3}
\end{figure}

Recent deep HST imaging and Keck/MOSFIRE NIR spectroscopy have 
pushed the redshift frontier of spectroscopically-identified galaxies
from $z\sim 7$ to $8-9$ with Ly$\alpha$ emission 
(Figure \ref{fig:oesch2015_fig3_zitrin2015_fig1};
\citealt{finkelstein2013,schenker2014,oesch2015,zitrin2015}).
These observations find galaxies in the range of the highest redshift
confirmed by spectroscopy
with a strong Ly$\alpha$ line as a signpost (left panel of Figure \ref{fig:year_redshift_hashimoto2018_edfig3}).
Moreover, there is an interesting and puzzling report of the significantly blueshifted Ly$\alpha$ emission associated with
a galaxy at a spectroscopic redshift of $z=9.1$ confirmed with an {\sc [Oiii]}$88\mu$m line (right panel of Figure \ref{fig:year_redshift_hashimoto2018_edfig3};
\citealt{hashimoto2018}).

To date, only a few LAEs are spectroscopically identified up to $z\sim 8-9$.
The small number of the $z\sim 8-9$ LAE identifications would be partly explained by
the limits of the sensitivities of the existing observation facilities.
Moreover, there is an important effect
that Ly$\alpha$ emission of LAEs at $z>6$ is weakened by the damping wing absorption of 
the IGM \hi\ at the EoR. 
In fact, the Ly$\alpha$ luminosity function evolution suggests
strong dimming of Ly$\alpha$ luminosity from $z\sim 6$ to $7.3$ (Figure \ref{fig:itoh2018_fig9_fig10_fig11}).

\begin{figure}[H]
\centering
\includegraphics[scale=.35]{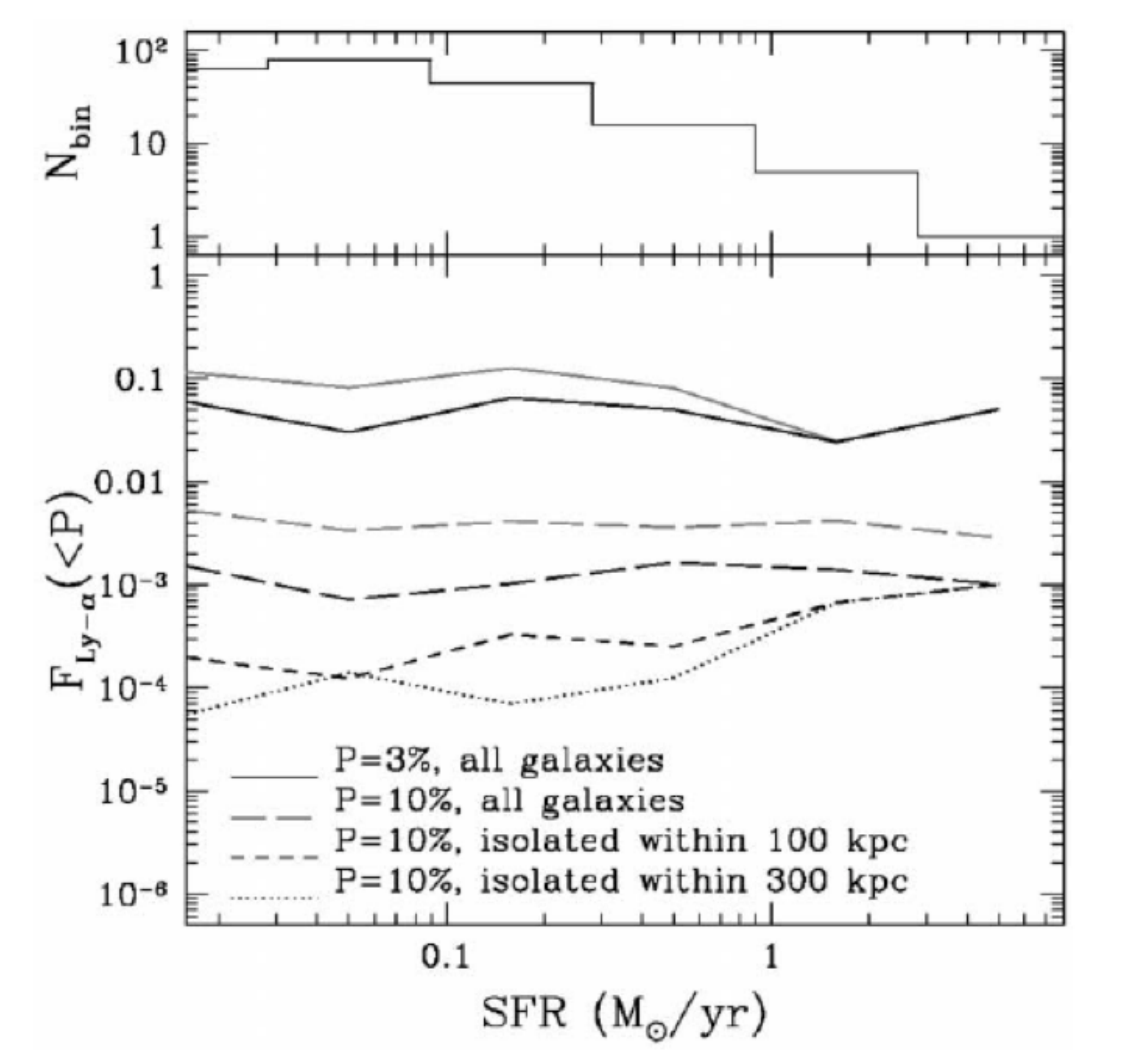}
\caption{
Bottom: 
Fraction of observed to intrinsic Ly$\alpha$ fluxes
as a function of central galaxy SFR that is
predicted by numerical simulations \citep{gnedin2004a}.
The black (gray) solid and dashed lines indicate the top 3 and 10\% Ly$\alpha$ flux survival, respectively, 
in all galaxies at $z=9$, for the case of the Ly$\alpha$ offset velocity of 150 (200) km s$^{-1}$.
The short-dashed and dotted lines are the same as the dashed lines,
but for galaxies with no companion galaxy within a distance of 100 and 300 physical kpc, respectively.
Top: Number of galaxies in each SFR bin that are used for the predictions shown in the bottom panel.
This figure is reproduced by permission of the AAS.
}
\label{fig:gnedin2004a_fig2}
\end{figure}

If one assumes that both the IGM and the LAEs are uniformly distributed and static,
Ly$\alpha$ emission of LAEs cannot escape from the universe as early as the epoch of first galaxies. However, 
clustering and peculiar motions of LAEs would help Ly$\alpha$ photons escape
from the highly neutral hydrogen IGM. Theoretical models suggest that up to 10\% of Ly$\alpha$ fluxes
can escape from an LAE in galaxy-clustered regions at the highly neutral epoch 
(Figure \ref{fig:gnedin2004a_fig2}; \citealt{gnedin2004a}).
To understand LAEs' physical properties as well as the evolution of $x_{\rm HI}$, 
it is important to know how much fraction of high-redshift galaxies show Ly$\alpha$ emission
escaping from the highly neutral hydrogen IGM. In other words, at what redshift LAEs disappear in the observations,
while continuum-selected star-forming galaxies 
are still seen at the same redshift in sufficiently deep observations.
%


%

\begin{figure}[H]
\centering
\includegraphics[scale=.47]{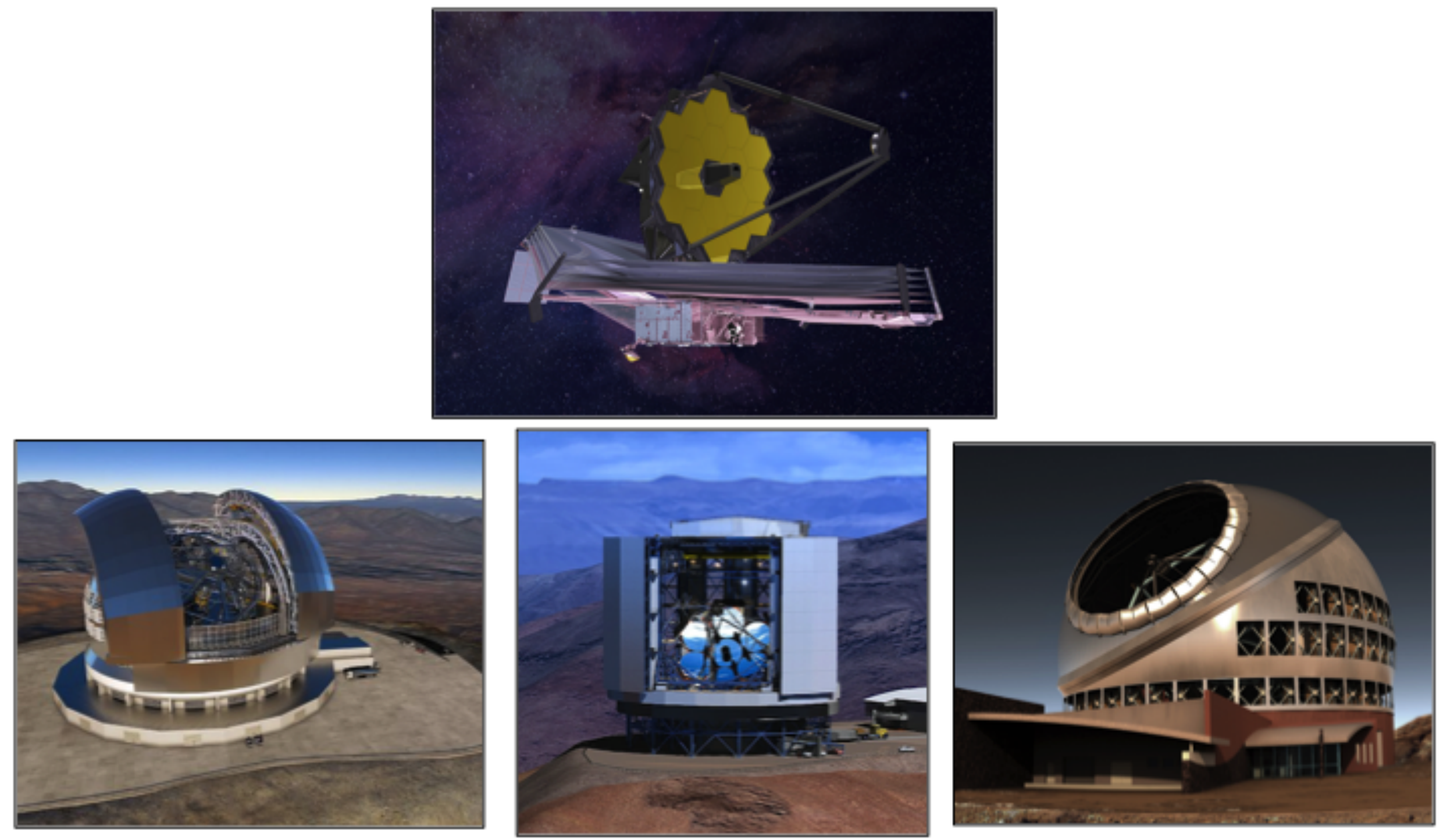}
\caption{
Artist rendering images of JWST (top), E-ELT (bottom left), GMT (bottom center), and TMT (bottom right).
Figure courtesy: NASA (JWST), ESO (E-ELT), GMTO Corporation (GMT), and TMT International Observatory (TMT).
}
\label{fig:jwst_eelts}
\end{figure}

Next generation large telescopes useful for observations of LAEs at $z\gtrsim 8-9$ 
are being built (Figure \ref{fig:jwst_eelts}). 
In space, James Webb Space Telescope (JWST) is planned
to be launched in or after 2021. JWST is a successor of HST 
that covers optical to IR bands ($0.6-28\mu$m)
with a large 6.5m segmented primary mirror.
On the ground, there are three projects of extremely large telescopes
for optical-IR observations;
the European Extremely Large Telescope (E-ELT),
the Giant Magellan Telescope (GMT), and
the Thirty Meter Telescope (TMT) that have
segmented primary mirrors with 39, 24, and 30m effective diameters, respectively.
The E-ELT and the GMT are being constructed in Cerro Armazones and Las Campanas Chile, respectively, in the southern hemisphere,
while the TMT is planned to be placed in Mauna Kea, Hawaii in the northern hemisphere.
These three projects will cover both the northern and southern hemispheres.
There exists a possibility that the TMT may move out from Mauna Kea to 
La Palma on the Canary Islands, Spain, due to difficult issues of the construction site permission 
in the Hawaii Island. All three projects plan first light in the middle/late 2020s.

It is expected that the first-light of JWST comes 
earlier than these three ground-based next generation telescopes.
Here I introduce two JWST science cases for LAEs.

The first JWST science case is to probe rest-frame optical nebular lines of LAEs at $z\gtrsim 5$
that are redshifted beyond the $K$ band, which require a space telescope
for deep observations. For the test of popIII star-formation
(Figure \ref{fig:schaerer2003_fig5}; Section \ref{sec:popIII_in_lae}), 
one can measure the spectral hardness of $Q_{\rm He^+}/Q_{\rm H}$
for popIII LAE candidates with flux measurements of an \heii\ line
and a Balmer line such as H$\alpha$ and/or H$\beta$. The Balmer lines 
fall in the JWST's Near-Infrared Spectrograph (NIRSpec) wavelength window of 
$1-5\mu$m for high sensitivity spectroscopy.
As discussed in Section \ref{sec:popIII_in_lae}, 
$Q_{\rm H}$ estimated from Ly$\alpha$-line measurements have large systematic uncertainties
raised by the Ly$\alpha$ escape fraction. If one can constrain 
the nebular extinction with a Balmer decrement of H$\alpha$/H$\beta$
(or H$\beta$/H$\gamma$ for high-$z$ sources),
the Balmer-line method provides a reliable
$Q_{\rm He^+}/Q_{\rm H}$ value that is critical for popIII star-formation tests.
It should be noted that the other JWST LAE studies will 
be also conducted with rest-frame optical nebular lines including [\oii]3727 and [\oiii]5007,
indicators of metallicity and ionization parameter (Sections \ref{sec:gas_phase_metallicity}-\ref{sec:ionization_state}),
and that systemic velocities from the nebular lines are key for understanding
the relation between Ly$\alpha$ photon escape and reionization \citep{hashimoto2013}.

\begin{figure}[H]
\centering
\includegraphics[scale=.30]{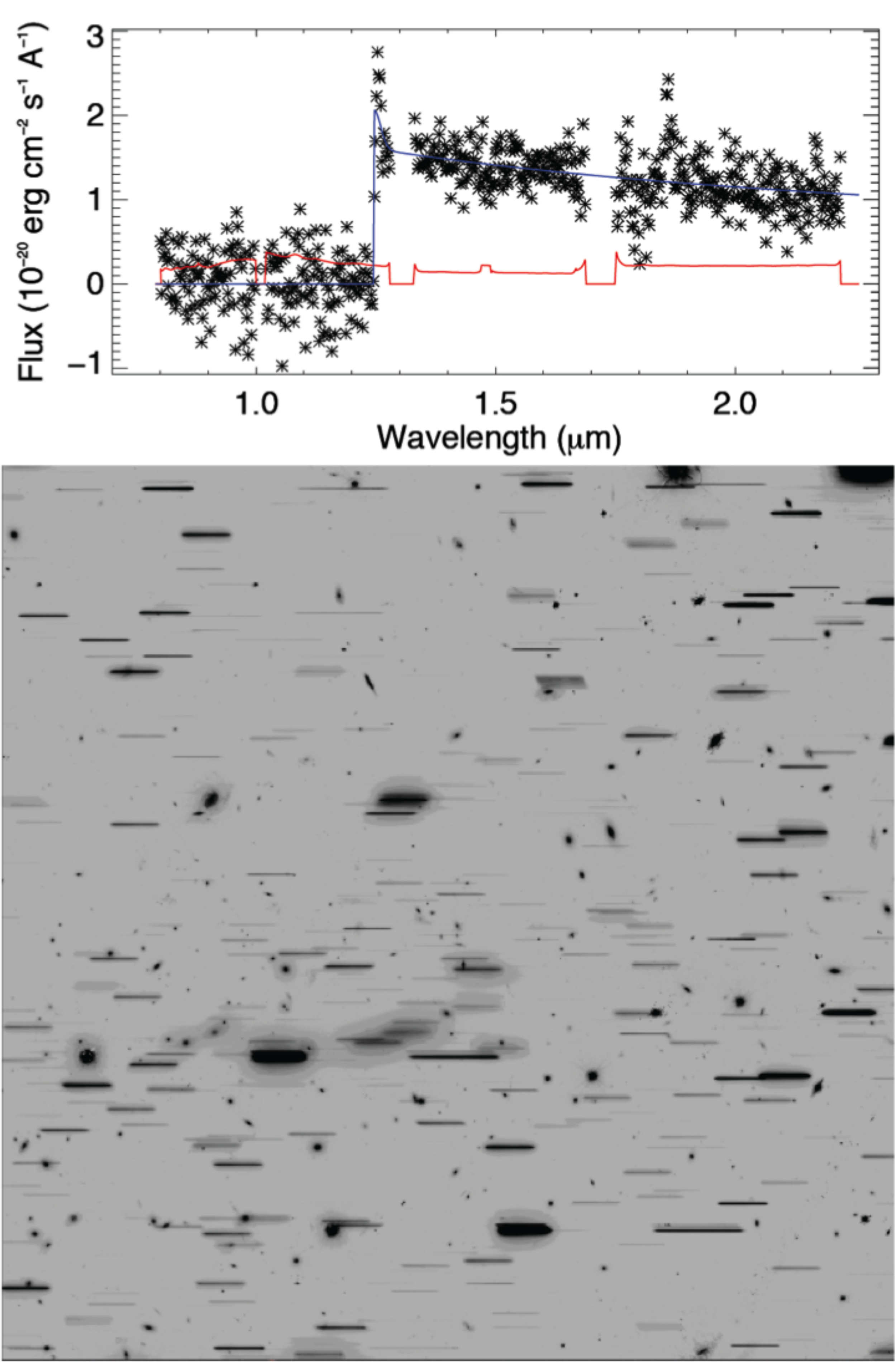}
\caption{
Simulation results of JWST/NIRISS observations \citep{dixon2015}.
Bottom: Simulated grism GR150R image with the F200W filter for MACS J0647+7015.
The assumed integration time is 10 hours.
Top: One-dimensional spectrum of a galaxy at $z=9.3$ retrieved from the simulated grism image. 
The black symbols and the red line represent the simulated data points and the error, respectively.
The blue curve denotes the best-fit model with a Ly$\alpha$ emission line.
}
\label{fig:dixon2014_poster_fig1}
\end{figure}

The second JWST science case is to conduct an unbiased search for LAEs at a very high redshift.
Such an LAE search can be performed with JWST/Near Infrared Imager and Slitless Spectrograph (NIRISS) 
that is packaged with the guide camera. NIRISS offers the wide-field slitless spectroscopy (WFSS)
with grisms whose FoV and spectral resolution are 4 arcmin$^2$ and $R\sim 150$, respectively,
at $0.8-2.2\mu$m. The deep slitless spectroscopy can target Ly$\alpha$ emission redshifted to $z\sim 6-17$.
Figure \ref{fig:dixon2014_poster_fig1} presents simulations of deep (10-hour integration) 
NIRISS/WFSS grism spectroscopy in the lensing galaxy cluster MACS J0647+7015 \citep{dixon2015}.
This simulation predicts the detections of 180 LBGs and LAEs with $F200W=26-28$ mag 
at $z\sim 6-15$ with the help of gravitational lensing magnification, assuming a uniform
random redshift distribution of LAEs. Investigating the number of detected LAEs as a function of redshift,
one can address the problem of Ly$\alpha$ emission escaping from the highly neutral hydrogen IGM
(Figure \ref{fig:gnedin2004a_fig2}).



%

In addition to these next generation projects for optical-IR observations, 
there are many multiwavelength projects useful for LAE studies. 
Programs of 21cm observations are especially important for understanding 
a reionization-LAE connection.
There are two major projects for next generation 21 cm observations,
Hydrogen Epoch of Reionization Array (HERA; \citealt{pober2014}) and
Square Kilometer Array (SKA
\footnote{
https://www.skatelescope.org
}).

HERA will be a 350-element interferometer 
that is composed of 14-m parabolic dishes covering $50-250$ MHz. 
HERA is under construction. In 2016, a few percent of the dishes are already 
built on the site of South Africa \citep{deboer2016}.
HERA is not an interferometer that is significantly larger than the existing 
21 cm instruments.
However, HERA aims to accomplish detections of the weak EoR signals, 
improving calibration and foreground isolation accuracies
with redundant baselines for the "EoR window" (Section \ref{sec:early_21cm_observation_results})
that are designed with the experiences of PAPER and MWA. 
HERA expects a 21 cm power spectrum detection with an S/N of $\sim 20$ at $z\sim 9$.

\begin{figure}[H]
\centering
\includegraphics[scale=.48]{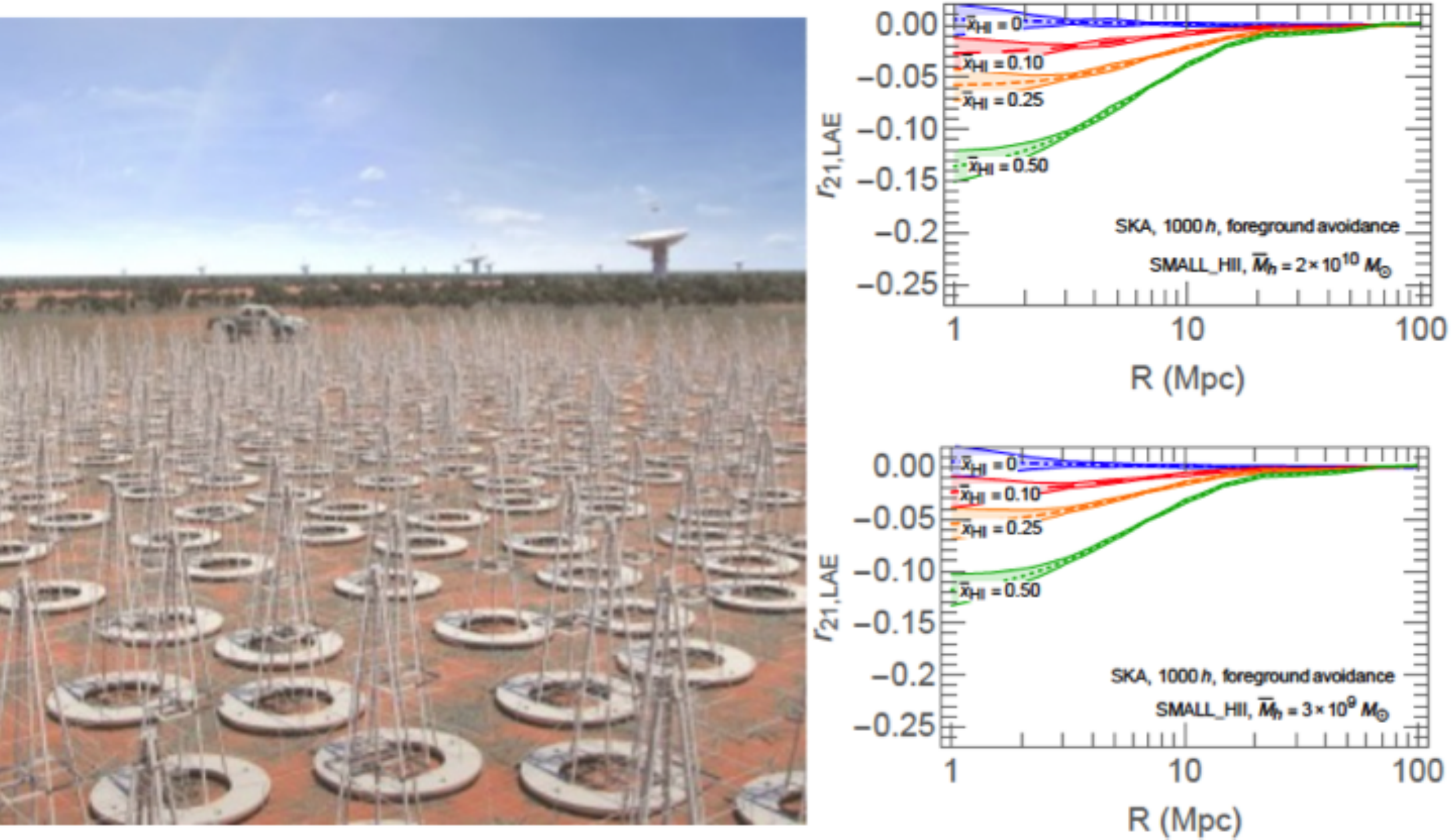}
\caption{
Left: Artist impression image of SKA low frequency array (courtesy: SKA 2018).
Right: LAE-21cm cross correlation functions for the average neutral hydrogen fractions of
$x_{\rm HI}=0$ (blue dot-dashed lines), $0.10$ (red long-dashed lines), $0.25$ (orange dashed lines), and $0.50$ (green short-dashed lines) 
that are predicted by numerical simulations \citep{sobacchi2016}.
The shades indicate the total uncertainties of theoretical models and observations including foreground effects.
These calculations assume SKA1-Low 1000 hour observations for the 21cm data and the HSC survey for the LAE data \citep{ouchi2018}.
The top and bottom panels present the results of LAEs' host halo masses of $2\times 10^{10}$ and $3\times 10^9$ $M_\odot$,
respectively.
This figure is reproduced by permission of the MNRAS.
}
\label{fig:skalow_sobacchi2016_fig7}
\end{figure}

SKA 
will consist in
radio interferometers whose total photon collecting area 
covers
one square kilometer. Towards the completion of the interferometers,
SKA has the two-phase approach: Phase 1 for the initial deployment
starting in 2018 and Phase 2 for the full operation in the mid 2020s.
In the Phase 1, SKA integrates two precursor telescopes of MeerKAT
\footnote{
http://www.ska.ac.za/gallery/meerkat/
}
in Karoo, South Africa
and Australian Square Kilometre Array Pathfinder (ASKAP
\footnote{
http://www.csiro.au/en/Research/Facilities/ATNF/ASKAP
})
in Murchison, Australia. The SKA Phase 1 instrument consists of three elements 
of SKA1-Mid (Karoo), SKA1-Survey, and SKA1-Low (Murchison; Figure \ref{fig:skalow_sobacchi2016_fig7}).
Among the three elements, SKA1-Low is a low frequency array consisting of 
more than 100 thousand antennas covering 50-350 MHz 
corresponding to the wavelength of the redshifted EoR 21 cm emission.
%
%
%
%
Before the full SKA construction, it is expected that SKA1-Low would provide 
21 cm data useful for the cross-correlation analysis with LAEs \citep{sobacchi2016,hutter2016,kubota2018}. 
Figure \ref{fig:skalow_sobacchi2016_fig7} presents the LAE-21cm brightness temperature
cross correlation functions predicted with reionization models \citep{sobacchi2016}.
The models assume LAEs from the Subaru/HSC survey and $100-1000$ hour SKA1-Low observations,
and suggest that the two different cases with $x_{\rm HI}=0$ and $0.5$ will be clearly distinguished
over the uncertainties. Although the large systematic uncertainties (such as foreground subtraction errors) 
are serious, as discussed in Section \ref{sec:early_21cm_observation_results},
there is a possibility that a new probe of LAE-21 cm cross correlation 
(Section \ref{sec:early_21cm_observation_results}) may be powerful,
due to the capability for removing the systematics in the cross correlation analysis.

%
%
%

\subsubsection{Trying Other Approaches}
\label{sec:other_approach}

In Sections \ref{sec:new_projects} and \ref{sec:future_projects}, I have introduced
the next-generation powerful instruments in the on-going and future programs. 
These next-generation instruments cover the unexplored parameter space, and 
are useful for resolving the issues of the open questions. 
However, even with no next-generation instruments,
one can make a breakthrough with the present-generation
instruments by a new approach. 
There are two observation examples in such breakthrough studies. 
One is the polarization observations for LABs conducted by
\citet{hayes2011b} and \citet{prescott2011}. 
\citet{hayes2011b} have performed the polarization observations
with VLT/FORS for a bright LAB, and detected the tangential polarization
indicating the Ly$\alpha$ photon scattering in the LAB (Section \ref{sec:lya_blob}).
Another example is the identifications of luminous SNe (SNe IIn) hosted by 
high-$z$ galaxies, $z\sim 2-4$ LBGs and LAEs
\citep{cooke2012}. \citet{cooke2012} have investigated variables in 
these high-$z$ galaxies selected in the CFHT Legacy Survey data.
Although it was well known that the popular bright SNe, SNe Ia, at $z\sim 2-4$ were too faint to be detected 
with the present data, \citet{cooke2012} successfully found luminous SNe in the $z\sim 2-4$ galaxies.
These are two good examples of making a breakthrough discovery
with the existing facilities and data, if one takes new approaches.
I would suggest young astronomers not to simply wait for next-generation instruments,
but to try new observational approaches with existing telescopes and data.


\subsection{Summary of On-Going and Future Projects}
\label{sec:summary_ongoing_and_future_projects}

I have showcased the five major open questions
related to
LAEs about galaxy formation and cosmic reionization.
For galaxy formation, there are four questions
about 
i) the LAE's distinguishing characteristics of star-formation and ISM,
ii) physical origins of LABs and diffuse halos,
iii) physical reasons for the large Ly$\alpha$ EW + Ly$\alpha$ escape fraction,
and 
iv) the LAE-popIII connection.
For cosmic reionization, two questions are presented:
a) reionization history (late or early / sharp or extended reionization?) and 
b) major sources of reionization (star-forming galaxies and/or faint AGN?).
These open questions are being addressed by
on-going observations with the three new instruments,
Subaru/HSC(+PFS), HETDEX, and VLT/MUSE.
Although the observing programs of these three new instruments
cover the complementary survey parameter space in redshift and depth, 
these programs can investigate LAEs only at $z\sim 2-7$. Beyond $z\sim 7-8$,
one needs next generation large telescopes with great sensitivities,
JWST, E-ELT, GMT, and TMT in the optical-IR wavelength range.
Among these next generation projects, forthcoming JWST projects
will probe rest-frame optical nebular lines of LAEs at $z\gtrsim 5$
for testing popIII, and conduct the unbiased search for LAEs at $z\sim 6-17$.
Beyond the optical-IR wavelength range, 21cm observations with
HERA and SKA will provide important results. Theoretical studies
suggest that the early-phase low-frequency array of SKA1-Low and Subaru/HSC
will identify a signal of spatial anti-correlation between 21 cm emission and LAEs.
Although these new facilities are key for exciting discoveries, there exist 
examples that new observational approaches with existing facilities can also make a breakthrough.
With these examples in hand,
I encourage young astronomers 
not to simply wait
for next-generation instruments, but try new observational approaches for 
LAE studies with the existing facilities.

\section{Grand Summary for this Series of Lectures}
\label{sec:grand_summary}

Observations of high-$z$ LAEs and the related scientific subjects 
are detailed in this series of lecturers. After I explain the background of LAE
discoveries, I clarify that LAEs are
important objects for studying galaxy formation and cosmic reionization (Section \ref{sec:introduction}).
This is because LAEs are unique probes for high-$z$/low-mass (popIII-like) galaxies
and the IGM \hi\ gas via resonance scattering and damping wing absorptions.
I then overview the basic theoretical ideas of galaxy formation including structure formation,
gas cooling/feedback, and the five Ly$\alpha$ emission mechanisms
(Section \ref{sec:galaxy_formationI}).
In Section \ref{sec:galaxy_formationII}, I summarize the physical properties of LAEs, known to date, 
stellar populations, luminosity functions, morphologies, ISM state, LAE-AGN connection,
and hosting halos.
Section \ref{sec:galaxy_formationIII} discusses the observational challenges of LAE properties;
extended Ly$\alpha$ halos, Ly$\alpha$ escape fraction, and LAE-popIII connection.
This series of lectures moves to cosmic reionization science 
in Section \ref{sec:cosmic_reionizationI}. Evolution of neutral hydrogen fraction (i.e.
cosmic reionization history) is constrained by the estimates of the IGM Ly$\alpha$ absorption 
including the Ly$\alpha$ damping wing absorption. The combination of the Ly$\alpha$
absorption, CMB, and \hi\ 21 cm data provides the rough picture of reionization history.
In Section  \ref{sec:cosmic_reionizationII}, I explain simple analytic formulae to evaluate
sources of reionization with three parameters for the ionizing photon production rate
and one parameter for the IGM spatial distribution, and discuss whether star-forming galaxies
can be major sources of reionization. The observational results suggest
that star-forming galaxies alone can explain cosmic reionization, although the contribution of faint AGNs
and the other sources are poorly understood, to date. Section \ref{sec:ongoing_and_future_projects}
lists up the five open questions about LAEs, and explains
three on-going projects for LAEs at $z\sim 2-7$ (Subaru/HSC+PFS, HETDEX, and VLT/MUSE) and 
various future optical-IR telescope projects for LAEs at $z\gtrsim 7-8$ (JWST and ground-based extremely large telescopes),
together with low-frequency 21-cm observations (HERA and SKA), including some examples of science cases.

In the past two decades, I find that high-$z$ LAE studies are one of the most important driving forces
for the state-of-the-art observational facilities. The importance of Ly$\alpha$ emission is undoubted, due to the fact
that Ly$\alpha$ is the strongest emission for 
the most abundant element in the universe, namely hydrogen.
Moreover, the resonance
nature of Ly$\alpha$ photons allows us to probe the distribution and kinematics of neutral hydrogen in any cosmic structures
of galaxies and LSSs, from the ISM to the CGM and the IGM. In the coming decade, Ly$\alpha$ observations will continue
leading the field of high-$z$ observations with the on-going and future facilities. I hope that 
some of the readers will conduct new Ly$\alpha$ studies inspired by this series of lectures,
and produce exciting results very soon.

%

%

%
\begin{acknowledgement}
I am grateful to Anne Verhamme, Pierre North, 
Sebastiano Cantalupo, Hakim Atek, and Myriam Burgener Frick 
for inviting me to give these lectures 
in the historic and prestigious Saas Fee Advanced Course. 
I thank my fellow lectures, Mark Dijkstra, J. Xavier Prochaska, and
Matthew Hayes for allowing to accommodate my lectures in the early
days of the week for resolving my scheduling issue.
I also thank the participants of the course who are 
enthusiastic about discussing scientific problems and 
presenting their ideas in the discussion time.
I acknowledge Kazuhiro Shimasaku and Pierre North
for editing these lecture notes and 
giving me 
a lot of valuable inputs.
Finally, I thank my close colleagues, 
Yuichi Harikane, Yoshiaki Ono, and Takatoshi Shibuya 
for improving their figures and 
giving permissions to show them in this lecture note.

%
%
\end{acknowledgement}
%
%



\begin{thebibliography}{}

\bibitem[Abel et al.(2005)]{abel2005} Abel, N.~P., Ferland, G.~J., Shaw, G., \& van Hoof, P.~A.~M.\ 2005, \apjs, 161, 65 

\bibitem[Adams et al.(2011)]{adams2011} Adams, J.~J., Blanc, G.~A., Hill, G.~J., et al.\ 2011, \apjs, 192, 5 

\bibitem[Adelberger et al.(2003)]{adelberger2003} Adelberger, K.~L., Steidel, C.~C., Shapley, A.~E., \& Pettini, M.\ 2003, \apj, 584, 45 

\bibitem[Ahn(2004)]{ahn2004} Ahn, S.-H.\ 2004, \apjl, 601, L25 

\bibitem[Ali et al.(2015)]{ali2015} Ali, Z.~S., Parsons, A.~R., Zheng, H., et al.\ 2015, \apj, 809, 61

\bibitem[Ali et al.(2018)]{ali2018} Ali, Z.~S., Parsons, A.~R., Zheng, H., et al.\ 2018, \apj, 863, 201 
 
\bibitem[Ando et al.(2006)]{ando2006} Ando, M., Ohta, K., Iwata, I., et al.\ 2006, \apjl, 645, L9 
\bibitem[Asplund et al.(2009)]{asplund2009} Asplund, M., Grevesse, N., Sauval, A.~J., \& Scott, P.\ 2009, \araa, 47, 481 

\bibitem[Atek et al.(2014a)]{atek2014a} Atek, H., Kunth, D., Schaerer, D., et al.\ 2014a, \aap, 561, A89 

\bibitem[Atek et al.(2014b)]{atek2014b} Atek, H., Richard, J., Kneib, J.-P., et al.\ 2014b, \apj, 786, 60 

\bibitem[Atek et al.(2015)]{atek2015} Atek, H., Richard, J., Kneib, J.-P., et al.\ 2015, \apj, 800, 18 

\bibitem[Bacon et al.(2010)]{bacon2010} Bacon, R., Accardo, M., Adjali, L., et al.\ 2010, \procspie, 7735, 773508 

\bibitem[Bacon et al.(2015)]{bacon2015} Bacon, R., Brinchmann, J., Richard, J., et al.\ 2015, \aap, 575, A75 

\bibitem[Bacon et al.(2017)]{bacon2017} Bacon, R., Conseil, S., Mary, D., et al.\ 2017, \aap, 608, A1 

\bibitem[Bahcall \& Salpeter(1965)]{bahcall1965} Bahcall, J.~N., \& Salpeter, E.~E.\ 1965, \apj, 142, 1677 

\bibitem[Bahcall et al.(1991)]{bahcall1991} Bahcall, J.~N., Jannuzi, B.~T., Schneider, D.~P., et al.\ 1991, \apjl, 377, L5 

\bibitem[Behroozi et al.(2013)]{behroozi2013} Behroozi, P.~S., Wechsler, R.~H., \& Conroy, C.\ 2013, \apj, 770, 57 



\bibitem[Ba{\~n}ados et al.(2018)]{banados2018} Ba{\~n}ados, E., Venemans, B.~P., Mazzucchelli, C., et al.\ 2018, \nat, 553, 473 

\bibitem[Barger et al.(2012)]{barger2012} Barger, A.~J., Cowie, L.~L., \& Wold, I.~G.~B.\ 2012, \apj, 749, 106 

\bibitem[Barkana(2018)]{barkana2018} Barkana, R.\ 2018, \nat, 555, 71 

\bibitem[Bell et al.(2003)]{bell2003} Bell, E.~F., McIntosh, D.~H., Katz, N., \& Weinberg, M.~D.\ 2003, \apjl, 585, L117 

\bibitem[Baldwin et al.(1981)]{baldwin1981} Baldwin, J.~A., Phillips, M.~M., \& Terlevich, R.\ 1981, \pasp, 93, 5 

\bibitem[Basu-Zych \& Scharf(2004)]{basu-zych2004} Basu-Zych, A., \& Scharf, C.\ 2004, \apjl, 615, L85 

\bibitem[Blanc et al.(2011)]{blanc2011} Blanc, G.~A., Adams, J.~J., Gebhardt, K., et al.\ 2011, \apj, 736, 31 

\bibitem[Blanton et al.(2001)]{blanton2001} Blanton, M.~R., Dalcanton, J., Eisenstein, D., et al.\ 2001, \aj, 121, 2358 

\bibitem[Bolton \& Haehnelt(2007)]{bolton2007} Bolton, J.~S., \& Haehnelt, M.~G.\ 2007, \mnras, 382, 325 

\bibitem[Bond et al.(2010)]{bond2010} Bond, N.~A., Feldmeier, J.~J., Matkovi{\'c}, A., et al.\ 2010, \apjl, 716, L200 

\bibitem[Borisova et al.(2016)]{borisova2016} Borisova, E., Cantalupo, S., Lilly, S.~J., et al.\ 2016, \apj, 831, 39 

\bibitem[Borthakur et al.(2014)]{borthakur2014} Borthakur, S., Heckman, T.~M., Leitherer, C., \& Overzier, R.~A.\ 2014, Science, 346, 216 


\bibitem[Bouwens et al.(2010)]{bouwens2010} Bouwens, R.~J., Illingworth, G.~D., Oesch, P.~A., et al.\ 2010, \apjl, 708, L69 
\bibitem[Bouwens et al.(2015)]{bouwens2015} Bouwens, R.~J., Illingworth, G.~D., Oesch, P.~A., et al.\ 2015, \apj, 803, 34 


\bibitem[Bower et al.(2006)]{bower2006} Bower, R.~G., Benson, A.~J., Malbon, R., et al.\ 2006, \mnras, 370, 645 

\bibitem[Bowman et al.(2018)]{bowman2018} Bowman, J.~D., Rogers, A.~E.~E., Monsalve, R.~A., Mozdzen, T.~J., \& Mahesh, N.\ 2018, \nat, 555, 67 

\bibitem[Brocklehurst(1971)]{brocklehurst1971} Brocklehurst, M.\ 1971, \mnras, 153, 471 
\bibitem[Bruzual \& Charlot(2003)]{bruzual2003} Bruzual, G., \& Charlot, S.\ 2003, \mnras, 344, 1000 

\bibitem[Cardamone et al.(2009)]{cardamone2009} Cardamone, C., Schawinski, K., Sarzi, M., et al.\ 2009, \mnras, 399, 1191 

\bibitem[Calzetti et al.(2000)]{calzetti2000} Calzetti, D., Armus, L., Bohlin, R.~C., et al.\ 2000, \apj, 533, 682 
\bibitem[Calzetti(2001)]{calzetti2001} Calzetti, D.\ 2001, \pasp, 113, 1449 

\bibitem[Campbell et al.(1986)]{campbell1986} Campbell, A., Terlevich, R., \& Melnick, J.\ 1986, \mnras, 223, 811 

\bibitem[Cantalupo et al.(2014)]{cantalupo2014} Cantalupo, S., Arrigoni-Battaia, F., Prochaska, J.~X., Hennawi, J.~F., \& Madau, P.\ 2014, \nat, 506, 63 

\bibitem[Capak et al.(2011)]{capak2011} Capak, P.~L., Riechers, D., Scoville, N.~Z., et al.\ 2011, \nat, 470, 233 
\bibitem[Capak et al.(2015)]{capak2015} Capak, P.~L., Carilli, C., Jones, G., et al.\ 2015, \nat, 522, 455 

\bibitem[Cardelli et al.(1989)]{cardelli1989} Cardelli, J.~A., Clayton, G.~C., \& Mathis, J.~S.\ 1989, \apj, 345, 245 

\bibitem[Cassata et al.(2011)]{cassata2011} Cassata, P., Le F{\`e}vre, O., Garilli, B., et al.\ 2011, \aap, 525, A143 

\bibitem[Chabrier(2003)]{chabrier2003} Chabrier, G.\ 2003, \pasp, 115, 763 

\bibitem[Chapman et al.(2005)]{chapman2005} Chapman, S.~C., Blain, A.~W., Smail, I., \& Ivison, R.~J.\ 2005, \apj, 622, 772 

\bibitem[Chiang et al.(2013)]{chiang2013} Chiang, Y.-K., Overzier, R., \& Gebhardt, K.\ 2013, \apj, 779, 127 

\bibitem[Chornock et al.(2013)]{chornock2013} Chornock, R., Berger, E., Fox, D.~B., et al.\ 2013, \apj, 774, 26 
\bibitem[Chornock et al.(2014)]{chornock2014} Chornock, R., Berger, E., Fox, D.~B., et al.\ 2014, arXiv:1405.7400 

\bibitem[Ciardullo et al.(2012)]{ciardullo2012} Ciardullo, R., Gronwall, C., Wolf, C., et al.\ 2012, \apj, 744, 110 

\bibitem[Coe et al.(2015)]{coe2015} Coe, D., Bradley, L., \& Zitrin, A.\ 2015, \apj, 800, 84 

\bibitem[Cooke et al.(2012)]{cooke2012} Cooke, J., Sullivan, M., Gal-Yam, A., et al.\ 2012, \nat, 491, 228 

\bibitem[Cowie \& Hu(1998)]{cowie1998} Cowie, L.~L., \& Hu, E.~M.\ 1998, \aj, 115, 1319 

\bibitem[Cowie et al.(2010)]{cowie2010} Cowie, L.~L., Barger, A.~J., \& Hu, E.~M.\ 2010, \apj, 711, 928 

\bibitem[Cowie et al.(2011)]{cowie2011} Cowie, L.~L., Barger, A.~J., \& Hu, E.~M.\ 2011, \apj, 738, 136 

\bibitem[Cowie et al.(2012)]{cowie2012} Cowie, L.~L., Barger, A.~J., \& Hasinger, G.\ 2012, \apj, 748, 50 

\bibitem[Croft et al.(2016)]{croft2016} Croft, R.~A.~C., Miralda-Escud{\'e}, J., Zheng, Z., et al.\ 2016, \mnras, 457, 3541

\bibitem[Croft et al.(2018)]{croft2018} Croft, R.~A.~C., Miralda-Escud{\'e}, J., Zheng, Z., Blomqvist, M., \& Pieri, M.\ 2018, \mnras, 481, 1320 

\bibitem[Daddi et al.(2007)]{daddi2007} Daddi, E., Dickinson, M., Morrison, G., et al.\ 2007, \apj, 670, 156 

\bibitem[Dawson et al.(2004)]{dawson2004} Dawson, S., Rhoads, J.~E., Malhotra, S., et al.\ 2004, \apj, 617, 707 

\bibitem[Dawson et al.(2007)]{dawson2007} Dawson, S., Rhoads, J.~E., Malhotra, S., et al.\ 2007, \apj, 671, 1227 

\bibitem[de Barros et al.(2016)]{debarros2016} de Barros, S., Vanzella, E., Amor{\'{\i}}n, R., et al.\ 2016, \aap, 585, A51 

\bibitem[DeBoer et al.(2016)]{deboer2016} DeBoer, D.~R., Parsons, A.~R., Aguirre, J.~E., et al.\ 2016, arXiv:1606.07473 

\bibitem[Deharveng et al.(2008)]{deharveng2008} Deharveng, J.-M., Small, T., Barlow, T.~A., et al.\ 2008, \apj, 680, 1072-1082 

\bibitem[Dekel et al.(2009)]{dekel2009} Dekel, A., Birnboim, Y., Engel, G., et al.\ 2009, \nat, 457, 451 

\bibitem[Dijkstra et al.(2007)]{dijkstra2007} Dijkstra, M., Lidz, A., \& Wyithe, J.~S.~B.\ 2007, \mnras, 377, 1175 

\bibitem[Dijkstra \& Loeb(2008)]{dijkstra2008} Dijkstra, M., \& Loeb, A.\ 2008, \mnras, 386, 492 
\bibitem[Dijkstra \& Loeb(2009)]{dijkstra2009} Dijkstra, M., \& Loeb, A.\ 2009, \mnras, 400, 1109 
\bibitem[Dijkstra et al.(2016)]{dijkstra2016} Dijkstra, M., Gronke, M., \& Sobral, D.\ 2016, \apj, 823, 74 

\bibitem[Dillon et al.(2014)]{dillon2014} Dillon, J.~S., Liu, A., Williams, C.~L., et al.\ 2014, \prd, 89, 023002 

\bibitem[Dixon et al.(2015)]{dixon2015} Dixon, W.~V., Ravindranath, S., \& Willott, C.\ 2015, IAU General Assembly, 22, 2250490 

\bibitem[Djorgovski \& Thompson(1992)]{djorgovski1992} Djorgovski, S., \& Thompson, D.~J.\ 1992, The Stellar Populations of Galaxies, 149, 337 

\bibitem[Duval et al.(2014)]{duval2014} Duval, F., Schaerer, D., {\"O}stlin, G., \& Laursen, P.\ 2014, \aap, 562, A52 

\bibitem[Elbaz et al.(2007)]{elbaz2007} Elbaz, D., Daddi, E., Le Borgne, D., et al.\ 2007, \aap, 468, 33 

\bibitem[Eldridge et al.(2017)]{eldridge2017} Eldridge, J.~J., Stanway, E.~R., Xiao, L., et al.\ 2017, PASA, 34, e058 

\bibitem[Erb et al.(2014)]{erb2014} Erb, D.~K., Steidel, C.~C., Trainor, R.~F., et al.\ 2014, \apj, 795, 33 

\bibitem[Erb et al.(2016)]{erb2016} Erb, D.~K., Pettini, M., Steidel, C.~C., et al.\ 2016, arXiv:1605.04919

\bibitem[Fall \& Efstathiou(1980)]{fall1980} Fall, S.~M., \& Efstathiou, G.\ 1980, \mnras, 193, 189 

\bibitem[Fan et al.(2004)]{fan2004} Fan, X., Hennawi, J.~F., Richards, G.~T., et al.\ 2004, \aj, 128, 515 

\bibitem[Fan et al.(2006a)]{fan2006a} Fan, X., Strauss, M.~A., Becker, R.~H., et al.\ 2006a, \aj, 132, 117 
\bibitem[Fan et al.(2006b)]{fan2006b} Fan, X., Carilli, C.~L., \& Keating, B.\ 2006b, \araa, 44, 415 

\bibitem[Fardal et al.(2001)]{fardal2001} Fardal, M.~A., Katz, N., Gardner, J.~P., et al.\ 2001, \apj, 562, 605 

\bibitem[Faucher-Gigu{\`e}re \& Kere{\v s}(2011)]{faucher-giguere2011} Faucher-Gigu{\`e}re, C.-A., \& Kere{\v s}, D.\ 2011, \mnras, 412, L118 

\bibitem[Feldmeier et al.(2013)]{feldmeier2013} Feldmeier, J.~J., Hagen, A., Ciardullo, R., et al.\ 2013, \apj, 776, 75 

\bibitem[Field(1959)]{field1959} Field, G.~B.\ 1959, \apj, 129, 551 

\bibitem[Finkelstein et al.(2007)]{finkelstein2007} Finkelstein, S.~L., Rhoads, J.~E., Malhotra, S., Pirzkal, N., \& Wang, J.\ 2007, \apj, 660, 1023
\bibitem[Finkelstein et al.(2008)]{finkelstein2008} Finkelstein, S.~L., Rhoads, J.~E., Malhotra, S., Grogin, N., \& Wang, J.\ 2008, \apj, 678, 655-668 
\bibitem[Finkelstein et al.(2011a)]{finkelstein2011a} Finkelstein, S.~L., Hill, G.~J., Gebhardt, K., et al.\ 2011a, \apj, 729, 140 
\bibitem[Finkelstein et al.(2011b)]{finkelstein2011b} Finkelstein, S.~L., Cohen, S.~H., Windhorst, R.~A., et al.\ 2011b, \apj, 735, 5 

\bibitem[Finkelstein et al.(2013)]{finkelstein2013} Finkelstein, S.~L., Papovich, C., Dickinson, M., et al.\ 2013, \nat, 502, 524 

\bibitem[Fiore et al.(2012)]{fiore2012} Fiore, F., Puccetti, S., Grazian, A., et al.\ 2012, \aap, 537, A16 


\bibitem[Garel et al.(2012)]{garel2012} Garel, T., Blaizot, J., Guiderdoni, B., et al.\ 2012, \mnras, 422, 310 
\bibitem[Garnett(1992)]{garnett1992} Garnett, D.~R.\ 1992, \aj, 103, 1330 

\bibitem[Gawiser et al.(2007)]{gawiser2007} Gawiser, E., Francke, H., Lai, K., et al.\ 2007, \apj, 671, 278 

\bibitem[Geach et al.(2009)]{geach2009} Geach, J.~E., Alexander, D.~M., Lehmer, B.~D., et al.\ 2009, \apj, 700, 1 

\bibitem[Giallongo et al.(2015)]{giallongo2015} Giallongo, E., Grazian, A., Fiore, F., et al.\ 2015, \aap, 578, A83 


\bibitem[Gnedin \& Prada(2004)]{gnedin2004a} Gnedin, N.~Y., \& Prada, F.\ 2004, \apjl, 608, L77 

\bibitem[Gnedin(2004)]{gnedin2004b} Gnedin, N.~Y.\ 2004, \apj, 610, 9 

\bibitem[Goerdt et al.(2010)]{goerdt2010} Goerdt, T., Dekel, A., Sternberg, A., et al.\ 2010, \mnras, 407, 613 

\bibitem[Grazian et al.(2018)]{grazian2018} Grazian, A., Giallongo, E., Boutsia, K., et al.\ 2018, \aap, 613, A44 

\bibitem[Greig et al.(2016)]{greig2016} Greig, B., Mesinger, A., Haiman, Z., \& Simcoe, R.~A.\ 2016, arXiv:1606.00441 

\bibitem[Gronwall et al.(2007)]{gronwall2007} Gronwall, C., Ciardullo, R., Hickey, T., et al.\ 2007, \apj, 667, 79 

\bibitem[Groth \& Peebles(1977)]{groth1977} Groth, E.~J., \& Peebles, P.~J.~E.\ 1977, \apj, 217, 385 

\bibitem[Guaita et al.(2011)]{guaita2011} Guaita, L., Acquaviva, V., Padilla, N., et al.\ 2011, \apj, 733, 114 
\bibitem[Guaita et al.(2013)]{guaita2013} Guaita, L., Francke, H., Gawiser, E., et al.\ 2013, \aap, 551, A93 

\bibitem[Gunn \& Peterson(1965)]{gunn1965} Gunn, J.~E., \& Peterson, B.~A.\ 1965, \apj, 142, 1633 

\bibitem[Hagen et al.(2014)]{hagen2014} Hagen, A., Ciardullo, R., Gronwall, C., et al.\ 2014, \apj, 786, 59 
\bibitem[Hagen et al.(2016)]{hagen2016} Hagen, A., Zeimann, G.~R., Behrens, C., et al.\ 2016, \apj, 817, 79 

\bibitem[Hansen \& Oh(2006)]{hansen2006} Hansen, M., \& Oh, S.~P.\ 2006, \mnras, 367, 979 

\bibitem[Harikane (2016a)]{harikane2016a} Harikane, 2016a, master thesis, the University of Tokyo
\bibitem[Harikane et al.(2016b)]{harikane2016b} Harikane, Y., Ouchi, M., Ono, Y., et al.\ 2016b, \apj, 821, 123 

\bibitem[Harikane et al.(2018a)]{harikane2018a} Harikane, Y., Ouchi, M., Ono, Y., et al.\ 2018a, \pasj, 70, S11 
\bibitem[Harikane et al.(2018b)]{harikane2018b} Harikane, Y., Ouchi, M., Shibuya, T., et al.\ 2018b, \apj, 859, 84 


\bibitem[Hashimoto et al.(2013)]{hashimoto2013} Hashimoto, T., Ouchi, M., Shimasaku, K., et al.\ 2013, \apj, 765, 70 
\bibitem[Hashimoto et al.(2015)]{hashimoto2015} Hashimoto, T., Verhamme, A., Ouchi, M., et al.\ 2015, \apj, 812, 157 

\bibitem[Hashimoto et al.(2018)]{hashimoto2018} Hashimoto, T., Laporte, N., Mawatari, K., et al.\ 2018, \nat, 557, 392 


\bibitem[Hayashino et al.(2004)]{hayashino2004} Hayashino, T., Matsuda, Y., Tamura, H., et al.\ 2004, \aj, 128, 2073 

\bibitem[Hayes et al.(2011a)]{hayes2011a} Hayes, M., Schaerer, D., {\"O}stlin, G., et al.\ 2011a, \apj, 730, 8 
\bibitem[Hayes et al.(2011b)]{hayes2011b} Hayes, M., Scarlata, C., \& Siana, B.\ 2011b, \nat, 476, 304 
\bibitem[Hayes et al.(2013)]{hayes2013} Hayes, M., {\"O}stlin, G., Schaerer, D., et al.\ 2013, \apjl, 765, L27 

\bibitem[Heckman et al.(2015)]{heckman2015} Heckman, T.~M., Alexandroff, R.~M., Borthakur, S., Overzier, R., \& Leitherer, C.\ 2015, \apj, 809, 147 

\bibitem[Herenz et al.(2017)]{herenz2017} Herenz, E.~C., Urrutia, T., Wisotzki, L., et al.\ 2017, \aap, 606, A12 


\bibitem[Hickox \& Markevitch(2007)]{hickox2007} Hickox, R.~C., \& Markevitch, M.\ 2007, \apjl, 661, L117 


\bibitem[Hill et al.(2012a)]{hill2012a} Hill, G.~J., Gebhardt, K., Drory, N., et al.\ 2012a, American Astronomical Society Meeting Abstracts \#219, 219, 424.01 
\bibitem[Hill et al.(2012b)]{hill2012b} Hill, G.~J., Tuttle, S.~E., Lee, H., et al.\ 2012b, \procspie, 8446, 84460N 

\bibitem[Hinshaw et al.(2013)]{hinshaw2013} Hinshaw, G., Larson, D., Komatsu, E., et al.\ 2013, \apjs, 208, 19 

\bibitem[Hopkins et al.(2007)]{hopkins2007} Hopkins, P.~F., Richards, G.~T., \& Hernquist, L.\ 2007, \apj, 654, 731 


\bibitem[Hu \& McMahon(1996)]{hu1996} Hu, E.~M., \& McMahon, R.~G.\ 1996, \nat, 382, 231 
\bibitem[Hu et al.(1998)]{hu1998} Hu, E.~M., Cowie, L.~L., \& McMahon, R.~G.\ 1998, \apjl, 502, L99 
\bibitem[Hu et al.(1999)]{hu1999} Hu, E.~M., McMahon, R.~G., \& Cowie, L.~L.\ 1999, \apjl, 522, L9 
\bibitem[Hu et al.(2002)]{hu2002} Hu, E.~M., Cowie, L.~L., McMahon, R.~G., et al.\ 2002, \apjl, 568, L75 


\bibitem[Hu et al.(2010)]{hu2010} Hu, E.~M., Cowie, L.~L., Barger, A.~J., et al.\ 2010, \apj, 725, 394 

\bibitem[Hutter et al.(2016)]{hutter2016} Hutter, A., Dayal, P., M{\"u}ller, V., \& Trott, C.\ 2016, arXiv:1605.01734 

\bibitem[Ikeda et al.(2012)]{ikeda2012} Ikeda, H., Nagao, T., Matsuoka, K., et al.\ 2012, \apj, 756, 160 

\bibitem[Iliev et al.(2006)]{iliev2006} Iliev, I.~T., Mellema, G., Pen, U.-L., et al.\ 2006, \mnras, 369, 1625 

\bibitem[Inoue et al.(2006)]{inoue2006} Inoue, A.~K., Iwata, I., \& Deharveng, J.-M.\ 2006, \mnras, 371, L1 

\bibitem[Inoue et al.(2016)]{inoue2016} Inoue, A.~K., Tamura, Y., Matsuo, H., et al.\ 2016, Science, 352, 1559 

\bibitem[Ishigaki et al.(2015)]{ishigaki2015} Ishigaki, M., Kawamata, R., Ouchi, M., et al.\ 2015, \apj, 799, 12 

\bibitem[Ishigaki et al.(2018)]{ishigaki2018} Ishigaki, M., Kawamata, R., Ouchi, M., et al.\ 2018, \apj, 854, 73 

\bibitem[Ishiyama et al.(2013)]{ishiyama2013} Ishiyama, T., Rieder, S., Makino, J., et al.\ 2013, \apj, 767, 146 
\bibitem[Ishiyama et al.(2015)]{ishiyama2015} Ishiyama, T., Enoki, M., Kobayashi, M.~A.~R., et al.\ 2015, \pasj, 67, 61 

\bibitem[Itoh et al.(2018)]{itoh2018} Itoh, R., Ouchi, M., Zhang, H., et al.\ 2018, arXiv:1805.05944 

\bibitem[Iwata et al.(2009)]{iwata2009} Iwata, I., Inoue, A.~K., Matsuda, Y., et al.\ 2009, \apj, 692, 1287 

\bibitem[Iye et al.(2006)]{iye2006} Iye, M., Ota, K., Kashikawa, N., et al.\ 2006, \nat, 443, 186 

\bibitem[Izotov et al.(2006)]{izotov2006} Izotov, Y.~I., Stasi{\'n}ska, G., Meynet, G., Guseva, N.~G., \& Thuan, T.~X.\ 2006, \aap, 448, 955 

\bibitem[Izotov et al.(2016a)]{izotov2016a} Izotov, Y.~I., Orlitov{\'a}, I., Schaerer, D., et al.\ 2016a, \nat, 529, 178 

\bibitem[Izotov et al.(2016b)]{izotov2016b} Izotov, Y.~I., Schaerer, D., Thuan, T.~X., et al.\ 2016b, \mnras, 461, 3683 




\bibitem[Jaskot \& Oey(2014)]{jaskot2014} Jaskot, A.~E., \& Oey, M.~S.\ 2014, \apjl, 791, L19 

\bibitem[Jenkins et al.(2001)]{jenkins2001} Jenkins, A., Frenk, C.~S., White, S.~D.~M., et al.\ 2001, \mnras, 321, 372 

\bibitem[Jeli{\'c} et al.(2014)]{jelic2014} Jeli{\'c}, V., de Bruyn, A.~G., Mevius, M., et al.\ 2014, \aap, 568, A101 

\bibitem[Kashikawa et al.(2006)]{kashikawa2006} Kashikawa, N., Shimasaku, K., Malkan, M.~A., et al.\ 2006, \apj, 648, 7 
\bibitem[Kashikawa et al.(2011)]{kashikawa2011} Kashikawa, N., Shimasaku, K., Matsuda, Y., et al.\ 2011, \apj, 734, 119 
\bibitem[Kashikawa et al.(2012)]{kashikawa2012} Kashikawa, N., Nagao, T., Toshikawa, J., et al.\ 2012, \apj, 761, 85 

\bibitem[Kashino et al.(2013)]{kashino2013} Kashino, D., Silverman, J.~D., Rodighiero, G., et al.\ 2013, \apjl, 777, L8 

\bibitem[Katz et al.(2003)]{katz2003} Katz, N., Keres, D., Dave, R., \& Weinberg, D.~H.\ 2003, The IGM/Galaxy Connection.~The Distribution of Baryons at z=0, 281, 185 

\bibitem[Kelz et al.(2014)]{kelz2014} Kelz, A., Jahn, T., Haynes, D., et al.\ 2014, \procspie, 9147, 914775 


\bibitem[Kennicutt(1998a)]{kennicutt1998a} Kennicutt, R.~C., Jr.\ 1998a, \araa, 36, 189 
\bibitem[Kennicutt(1998b)]{kennicutt1998b} Kennicutt, R.~C., Jr.\ 1998b, \apj, 498, 541 


\bibitem[Kere{\v s} et al.(2005)]{keres2005} Kere{\v s}, D., Katz, N., Weinberg, D.~H., \& Dav{\'e}, R.\ 2005, \mnras, 363, 2 
\bibitem[Kere{\v s} et al.(2009)]{keres2009} Kere{\v s}, D., Katz, N., Fardal, M., Dav{\'e}, R., \& Weinberg, D.~H.\ 2009, \mnras, 395, 160 


\bibitem[Knudsen et al.(2016)]{knudsen2016} Knudsen, K.~K., Richard, J., Kneib, J.-P., et al.\ 2016, arXiv:1603.02277 

\bibitem[Kodaira et al.(2003)]{kodaira2003} Kodaira, K., Taniguchi, Y., Kashikawa, N., et al.\ 2003, \pasj, 55, L17 

\bibitem[Kojima et al.(2017)]{kojima2017} Kojima, T., Ouchi, M., Nakajima, K., et al.\ 2017, \pasj, 69, 44 


\bibitem[Kollmeier et al.(2010)]{kollmeier2010} Kollmeier, J.~A., Zheng, Z., Dav{\'e}, R., et al.\ 2010, \apj, 708, 1048 


\bibitem[Konno et al.(2014)]{konno2014} Konno, A., Ouchi, M., Ono, Y., et al.\ 2014, \apj, 797, 16 
\bibitem[Konno et al.(2016)]{konno2016} Konno, A., Ouchi, M., Nakajima, K., et al.\ 2016, \apj, 823, 20 

\bibitem[Konno et al.(2018)]{konno2018} Konno, A., Ouchi, M., Shibuya, T., et al.\ 2018, \pasj, 70, S16 


\bibitem[Koo \& Kron(1980)]{koo1980} Koo, D.~C., \& Kron, R.~T.\ 1980, \pasp, 92, 537 

\bibitem[Kothes \& Kerton(2002)]{kothes2002} Kothes, R., \& Kerton, C.~R.\ 2002, \aap, 390, 337 

\bibitem[Kova{\v c} et al.(2007)]{kovac2007} Kova{\v c}, K., Somerville, R.~S., Rhoads, J.~E., Malhotra, S., \& Wang, J.\ 2007, \apj, 668, 15 

\bibitem[Kravtsov(2003)]{kravtsov2003} Kravtsov, A.~V.\ 2003, \apjl, 590, L1 

\bibitem[Kubota et al.(2018)]{kubota2018} Kubota, K., Yoshiura, S., Takahashi, K., et al.\ 2018, \mnras, 479, 2754 

\bibitem[Kurk et al.(2004)]{kurk2004} Kurk, J.~D., Cimatti, A., di Serego Alighieri, S., et al.\ 2004, \aap, 422, L13 

\bibitem[Kusakabe et al.(2015)]{kusakabe2015} Kusakabe, H., Shimasaku, K., Nakajima, K., \& Ouchi, M.\ 2015, \apjl, 800, L29 

\bibitem[Lai et al.(2008)]{lai2008} Lai, K., Huang, J.-S., Fazio, G., et al.\ 2008, \apj, 674, 70-74 

\bibitem[Lake et al.(2015)]{lake2015} Lake, E., Zheng, Z., Cen, R., et al.\ 2015, \apj, 806, 46 

\bibitem[Landy \& Szalay(1993)]{landy1993} Landy, S.~D., \& Szalay, A.~S.\ 1993, \apj, 412, 64 

\bibitem[Laporte et al.(2017)]{laporte2017} Laporte, N., Nakajima, K., Ellis, R.~S., et al.\ 2017, arXiv:1708.05173 

\bibitem[Laursen et al.(2013)]{laursen2013} Laursen, P., Duval, F., {\"O}stlin, G.\ 2013, \apj, 766, 124 


\bibitem[Leauthaud et al.(2012)]{leauthaud2012} Leauthaud, A., Tinker, J., Bundy, K., et al.\ 2012, \apj, 744, 159 

\bibitem[Leclercq et al.(2017)]{leclercq2017} Leclercq, F., Bacon, R., Wisotzki, L., et al.\ 2017, \aap, 608, A8 

\bibitem[Lee et al.(2014)]{lee2014} Lee, K.-G., Hennawi, J.~F., Stark, C., et al.\ 2014, \apjl, 795, L12 


\bibitem[Leitet et al.(2011)]{leitet2011} Leitet, E., Bergvall, N., Piskunov, N., \& Andersson, B.-G.\ 2011, \aap, 532, A107 

\bibitem[Leitet et al.(2013)]{leitet2013} Leitet, E., Bergvall, N., Hayes, M., Linn{\'e}, S., \& Zackrisson, E.\ 2013, \aap, 553, A106 

\bibitem[Lidz et al.(2008)]{lidz2008} Lidz, A., Zahn, O., McQuinn, M., Zaldarriaga, M., \& Hernquist, L.\ 2008, \apj, 680, 962 
\bibitem[Lidz et al.(2009)]{lidz2009} Lidz, A., Zahn, O., Furlanetto, S.~R., et al.\ 2009, \apj, 690, 252 

\bibitem[Livermore et al.(2017)]{livermore2017} Livermore, R.~C., Finkelstein, S.~L., \& Lotz, J.~M.\ 2017, \apj, 835, 113 

\bibitem[Lotz et al.(2017)]{lotz2017} Lotz, J.~M., Koekemoer, A., Coe, D., et al.\ 2017, \apj, 837, 97 

\bibitem[Madau et al.(1999)]{madau1999} Madau, P., Haardt, F., \& Rees, M.~J.\ 1999, \apj, 514, 648 

\bibitem[Madau \& Dickinson(2014)]{madau2014} Madau, P., \& Dickinson, M.\ 2014, \araa, 52, 415 
\bibitem[Maiolino et al.(2015)]{maiolino2015} Maiolino, R., Carniani, S., Fontana, A., et al.\ 2015, \mnras, 452, 54 
\bibitem[Malhotra \& Rhoads(2002)]{malhotra2002} Malhotra, S., \& Rhoads, J.~E.\ 2002, \apjl, 565, L71 
\bibitem[Malhotra \& Rhoads(2004)]{malhotra2004} Malhotra, S., \& Rhoads, J.~E.\ 2004, \apjl, 617, L5 
\bibitem[Malhotra et al.(2012)]{malhotra2012} Malhotra, S., Rhoads, J.~E., Finkelstein, S.~L., et al.\ 2012, \apjl, 750, L36 
\bibitem[Mao et al.(2007)]{mao2007} Mao, J., Lapi, A., Granato, G.~L., de Zotti, G., \& Danese, L.\ 2007, \apj, 667, 655 
\bibitem[Marlowe et al.(1995)]{marlowe1995} Marlowe, A.~T., Heckman, T.~M., Wyse, R.~F.~G., \& Schommer, R.\ 1995, \apj, 438, 563 
\bibitem[Martin(1998)]{martin1998} Martin, C.~L.\ 1998, \apj, 506, 222 
\bibitem[Martin et al.(2010)]{martin2010} Martin, C., Moore, A., Morrissey, P., et al.\ 2010, \procspie, 7735, 77350M 
\bibitem[Matthee et al.(2017a)]{matthee2017a} Matthee, J., Sobral, D., Best, P., et al.\ 2017a, \mnras, 465, 3637 
\bibitem[Matthee et al.(2017b)]{matthee2017b} Matthee, J., Sobral, D., Boone, F., et al.\ 2017b, \apj, 851, 145 
\bibitem[Matsuda et al.(2004)]{matsuda2004} Matsuda, Y., Yamada, T., Hayashino, T., et al.\ 2004, \aj, 128, 569 
\bibitem[Matsuda et al.(2011)]{matsuda2011} Matsuda, Y., Yamada, T., Hayashino, T., et al.\ 2011, \mnras, 410, L13 
\bibitem[Matsuda et al.(2012)]{matsuda2012} Matsuda, Y., Yamada, T., Hayashino, T., et al.\ 2012, \mnras, 425, 878 

\bibitem[Mawatari et al.(2017)]{mawatari2017} Mawatari, K., Inoue, A.~K., Yamada, T., et al.\ 2017, \mnras, 467, 3951 


\bibitem[McCarthy et al.(1987)]{mccarthy1987} McCarthy, P.~J., Spinrad, H., Djorgovski, S., et al.\ 1987, \apjl, 319, L39 

\bibitem[McGreer et al.(2013)]{mcgreer2013} McGreer, I.~D., Jiang, L., Fan, X., et al.\ 2013, \apj, 768, 105 

\bibitem[McLeod et al.(2015)]{mcleod2015} McLeod, D.~J., McLure, R.~J., Dunlop, J.~S., et al.\ 2015, \mnras, 450, 3032 
\bibitem[McLeod et al.(2016)]{mcleod2016} McLeod, D.~J., McLure, R.~J., \& Dunlop, J.~S.\ 2016, \mnras, 459, 3812 

\bibitem[McLinden et al.(2011)]{mclinden2011} McLinden, E.~M., Finkelstein, S.~L., Rhoads, J.~E., et al.\ 2011, \apj, 730, 136 
\bibitem[McLure et al.(2013)]{mclure2013} McLure, R.~J., Dunlop, J.~S., Bowler, R.~A.~A., et al.\ 2013, \mnras, 432, 2696 

\bibitem[McQuinn et al.(2007)]{mcquinn2007} McQuinn, M., Hernquist, L., Zaldarriaga, M., \& Dutta, S.\ 2007, \mnras, 381, 75 

\bibitem[Miyazaki et al.(2018)]{miyazaki2018} Miyazaki, S., Komiyama, Y., Kawanomoto, S., et al.\ 2018, \pasj, 70, S1 

\bibitem[Mo et al.(2010)]{mo2010} Mo, H., van den Bosch, F.~C., \& White, S.\ 2010, Galaxy Formation and Evolution, by Houjun Mo , Frank van den Bosch , Simon White, Cambridge, UK: Cambridge University Press, 2010,
\bibitem[Momcheva et al.(2013)]{momcheva2013} Momcheva, I.~G., Lee, J.~C., Ly, C., et al.\ 2013, \aj, 145, 47 
\bibitem[Momose et al.(2014)]{momose2014} Momose, R., Ouchi, M., Nakajima, K., et al.\ 2014, \mnras, 442, 110 
\bibitem[Momose et al.(2016)]{momose2016} Momose, R., Ouchi, M., Nakajima, K., et al.\ 2016, \mnras, 457, 2318 

\bibitem[Mukae et al.(2017)]{mukae2017} Mukae, S., Ouchi, M., Kakiichi, K., et al.\ 2017, \apj, 835, 281 

\bibitem[Mortlock et al.(2011)]{mortlock2011} Mortlock, D.~J., Warren, S.~J., Venemans, B.~P., et al.\ 2011, \nat, 474, 616 
\bibitem[Muratov et al.(2015)]{muratov2015} Muratov, A.~L., Kere{\v s}, D., Faucher-Gigu{\`e}re, C.-A., et al.\ 2015, \mnras, 454, 2691 

\bibitem[Nagamine et al.(2010)]{nagamine2010} Nagamine, K., Ouchi, M., Springel, V., \& Hernquist, L.\ 2010, \pasj, 62, 1455 

\bibitem[Nagao et al.(2006)]{nagao2006} Nagao, T., Maiolino, R., \& Marconi, A.\ 2006, \aap, 459, 85 

\bibitem[Nakajima et al.(2012)]{nakajima2012} Nakajima, K., Ouchi, M., Shimasaku, K., et al.\ 2012, \apj, 745, 12 
\bibitem[Nakajima et al.(2013)]{nakajima2013} Nakajima, K., Ouchi, M., Shimasaku, K., et al.\ 2013, \apj, 769, 3 
\bibitem[Nakajima \& Ouchi(2014)]{nakajima2014} Nakajima, K., \& Ouchi, M.\ 2014, \mnras, 442, 900 

\bibitem[Nestor et al.(2013)]{nestor2013} Nestor, D.~B., Shapley, A.~E., Kornei, K.~A., Steidel, C.~C., \& Siana, B.\ 2013, \apj, 765, 47

\bibitem[Neufeld(1991)]{neufeld1991} Neufeld, D.~A.\ 1991, \apjl, 370, L85 

\bibitem[Niikura et al.(2017)]{niikura2017} Niikura, H., Takada, M., Yasuda, N., et al.\ 2017, arXiv:1701.02151 

\bibitem[Nilsson et al.(2006)]{nilsson2006} Nilsson, K.~K., Fynbo, J.~P.~U., M{\o}ller, P., Sommer-Larsen, J., \& Ledoux, C.\ 2006, \aap, 452, L23

\bibitem[Nilsson et al.(2006)]{nilsson2006} Nilsson, K.~K., Fynbo, J.~P.~U., M{\o}ller, P., Sommer-Larsen, J., \& Ledoux, C.\ 2006, \aap, 452, L23 


\bibitem[Oesch et al.(2015)]{oesch2015} Oesch, P.~A., van Dokkum, P.~G., Illingworth, G.~D., et al.\ 2015, \apjl, 804, L30 
\bibitem[Oesch et al.(2016)]{oesch2016} Oesch, P.~A., Brammer, G., van Dokkum, P.~G., et al.\ 2016, \apj, 819, 129 

\bibitem[Oesch et al.(2018)]{oesch2018} Oesch, P.~A., Bouwens, R.~J., Illingworth, G.~D., Labb{\'e}, I., \& Stefanon, M.\ 2018, \apj, 855, 105 


\bibitem[Ono et al.(2010a)]{ono2010a} Ono, Y., Ouchi, M., Shimasaku, K., et al.\ 2010a, \mnras, 402, 1580 
\bibitem[Ono et al.(2010b)]{ono2010b} Ono, Y., Ouchi, M., Shimasaku, K., et al.\ 2010b, \apj, 724, 1524 
\bibitem[Ono et al.(2012)]{ono2012} Ono, Y., Ouchi, M., Mobasher, B., et al.\ 2012, \apj, 744, 83 

\bibitem[Osterbrock(1989)]{osterbrock1989} Osterbrock, D.~E.\ 1989, Research supported by the University of California, John Simon Guggenheim Memorial Foundation, University of Minnesota, et al.~Mill Valley, CA, University Science Books, 1989, 422 p.,  

\bibitem[Osterbrock \& Ferland(2006)]{osterbrock2006} Osterbrock, D.~E., \& Ferland, G.~J.\ 2006, Astrophysics of gaseous nebulae and active galactic nuclei, 2nd.~ed.~by D.E.~Osterbrock and G.J.~Ferland.~Sausalito, CA: University Science Books, 2006,  

\bibitem[{\"O}stlin et al.(2014)]{ostlin2014} {\"O}stlin, G., Hayes, M., Duval, F., et al.\ 2014, \apj, 797, 11 

\bibitem[Ota et al.(2014)]{ota2014} Ota, K., Walter, F., Ohta, K., et al.\ 2014, \apj, 792, 34 

\bibitem[Ouchi et al.(2003)]{ouchi2003} Ouchi, M., Shimasaku, K., Furusawa, H., et al.\ 2003, \apj, 582, 60 

\bibitem[Ouchi et al.(2004)]{ouchi2004} Ouchi, M., Shimasaku, K., Okamura, S., et al.\ 2004, \apj, 611, 660 


\bibitem[Ouchi et al.(2005a)]{ouchi2005a} Ouchi, M., Shimasaku, K., Akiyama, M., et al.\ 2005a, \apjl, 620, L1 
\bibitem[Ouchi et al.(2005b)]{ouchi2005b} Ouchi, M., Hamana, T., Shimasaku, K., et al.\ 2005b, \apjl, 635, L117 
\bibitem[Ouchi et al.(2008)]{ouchi2008} Ouchi, M., Shimasaku, K., Akiyama, M., et al.\ 2008, \apjs, 176, 301-330 
\bibitem[Ouchi et al.(2009a)]{ouchi2009a} Ouchi, M., Ono, Y., Egami, E., et al.\ 2009a, \apj, 696, 1164 
\bibitem[Ouchi et al.(2009b)]{ouchi2009b} Ouchi, M., Mobasher, B., Shimasaku, K., et al.\ 2009b, \apj, 706, 1136 
\bibitem[Ouchi et al.(2010)]{ouchi2010} Ouchi, M., Shimasaku, K., Furusawa, H., et al.\ 2010, \apj, 723, 869 
\bibitem[Ouchi et al.(2013)]{ouchi2013} Ouchi, M., Ellis, R., Ono, Y., et al.\ 2013, \apj, 778, 102 

\bibitem[Ouchi et al.(2018)]{ouchi2018} Ouchi, M., Harikane, Y., Shibuya, T., et al.\ 2018, \pasj, 70, S13 

\bibitem[Paciga et al.(2011)]{paciga2011} Paciga, G., Chang, T.-C., Gupta, Y., et al.\ 2011, \mnras, 413, 1174 

\bibitem[Pallottini et al.(2015)]{pallottini2015} Pallottini, A., Ferrara, A., Pacucci, F., et al.\ 2015, \mnras, 453, 2465 

\bibitem[Parsons et al.(2014)]{parsons2014} Parsons, A.~R., Liu, A., Aguirre, J.~E., et al.\ 2014, \apj, 788, 106 

\bibitem[Partridge \& Peebles(1967a)]{partridge1967a} Partridge, R.~B., \& Peebles, P.~J.~E.\ 1967a, \apj, 147, 868 
\bibitem[Partridge \& Peebles(1967b)]{partridge1967b} Partridge, R.~B., \& Peebles, P.~J.~E.\ 1967b, \apj, 148, 377 

\bibitem[Pascarelle et al.(1996a)]{pascarelle1996a} Pascarelle, S.~M., Windhorst, R.~A., Keel, W.~C., \& Odewahn, S.~C.\ 1996a, \nat, 383, 45 
\bibitem[Pascarelle et al.(1996b)]{pascarelle1996b} Pascarelle, S.~M., Windhorst, R.~A., Driver, S.~P., Ostrander, E.~J., \& Keel, W.~C.\ 1996b, \apjl, 456, L21 

\bibitem[Paulino-Afonso et al.(2018)]{paulino-afonso2018} Paulino-Afonso, A., Sobral, D., Ribeiro, B., et al.\ 2018, \mnras, 476, 5479 



\bibitem[Patel et al.(2010)]{patel2010} Patel, M., Warren, S.~J., Mortlock, D.~J., \& Fynbo, J.~P.~U.\ 2010, \aap, 512, L3 

\bibitem[Patr{\'{\i}}cio et al.(2016)]{patricio2016} Patr{\'{\i}}cio, V., Richard, J., Verhamme, A., et al.\ 2016, \mnras, 456, 4191 

\bibitem[Pawlik et al.(2009)]{pawlik2009} Pawlik, A.~H., Schaye, J., \& van Scherpenzeel, E.\ 2009, \mnras, 394, 1812 


\bibitem[Peacock \& Dodds(1996)]{peacock1996} Peacock, J.~A., \& Dodds, S.~J.\ 1996, \mnras, 280, L19 

\bibitem[Peebles(1993)]{peebles1993} Peebles, P.~J.~E.\ 1993, Principles of Physical Cosmology by P.J.E.~Peebles.~Princeton University Press, 1993.~ISBN: 978-0-691-01933-8,  
\bibitem[Pentericci et al.(2011)]{pentericci2011} Pentericci, L., Fontana, A., Vanzella, E., et al.\ 2011, \apj, 743, 132 
\bibitem[Pentericci et al.(2014)]{pentericci2014} Pentericci, L., Vanzella, E., Fontana, A., et al.\ 2014, \apj, 793, 113 
\bibitem[Pentericci et al.(2016)]{pentericci2016} Pentericci, L., Carniani, S., Castellano, M., et al.\ 2016, \apjl, 829, L11 


\bibitem[Pettini et al.(2001)]{pettini2001} Pettini, M., Shapley, A.~E., Steidel, C.~C., et al.\ 2001, \apj, 554, 981 


\bibitem[Pirzkal et al.(2004)]{pirzkal2004} Pirzkal, N., Xu, C., Malhotra, S., et al.\ 2004, \apjs, 154, 501 

\bibitem[Planck Collaboration et al.(2015)]{planck2015} Planck Collaboration, Ade, P.~A.~R., Aghanim, N., et al.\ 2015, arXiv:1502.01589 
\bibitem[Planck Collaboration et al.(2016)]{planck2016} Planck Collaboration, Adam, R., Aghanim, N., et al.\ 2016, arXiv:1605.03507 

\bibitem[Planck Collaboration et al.(2018)]{planck2018} Planck Collaboration, Akrami, Y., Arroja, F., et al.\ 2018, arXiv:1807.06205 


\bibitem[Pober et al.(2014)]{pober2014} Pober, J.~C., Liu, A., Dillon, J.~S., et al.\ 2014, \apj, 782, 66 

\bibitem[Prescott et al.(2009)]{prescott2009} Prescott, M.~K.~M., Dey, A., \& Jannuzi, B.~T.\ 2009, \apj, 702, 554 
\bibitem[Prescott et al.(2011)]{prescott2011} Prescott, M.~K.~M., Smith, P.~S., Schmidt, G.~D., \& Dey, A.\ 2011, \apjl, 730, L25 
\bibitem[Prescott et al.(2015)]{prescott2015} Prescott, M.~K.~M., Martin, C.~L., \& Dey, A.\ 2015, \apj, 799, 62 


\bibitem[Press \& Schechter(1974)]{press1974} Press, W.~H., \& Schechter, P.\ 1974, \apj, 187, 425 

\bibitem[Pritchard \& Loeb(2010a)]{pritchard2010a} Pritchard, J.~R., \& Loeb, A.\ 2010a, \prd, 82, 023006 
\bibitem[Pritchard \& Loeb(2010b)]{pritchard2010b} Pritchard, J., \& Loeb, A.\ 2010b, \nat, 468, 772 


\bibitem[Pritchet \& Hartwick(1987)]{pritchet1987} Pritchet, C.~J., \& Hartwick, F.~D.~A.\ 1987, \apj, 320, 464 
\bibitem[Pritchet \& Hartwick(1990)]{pritchet1990} Pritchet, C.~J., \& Hartwick, F.~D.~A.\ 1990, \apjl, 355, L11 

\bibitem[Rakic et al.(2012)]{rakic2012} Rakic, O., Schaye, J., Steidel, C.~C., \& Rudie, G.~C.\ 2012, \apj, 751, 94 

\bibitem[Rauch et al.(2011)]{rauch2011} Rauch, M., Becker, G.~D., Haehnelt, M.~G., et al.\ 2011, \mnras, 418, 1115 

\bibitem[Rees \& Ostriker(1977)]{rees1977} Rees, M.~J., \& Ostriker, J.~P.\ 1977, \mnras, 179, 541 

\bibitem[Rhoads et al.(2000)]{rhoads2000} Rhoads, J.~E., Malhotra, S., Dey, A., et al.\ 2000, \apjl, 545, L85 

\bibitem[Read \& Trentham(2005)]{read2005} Read, J.~I., \& Trentham, N.\ 2005, Philosophical Transactions of the Royal Society of London Series A, 363,  2693

\bibitem[Richards et al.(2006)]{richards2006} Richards, G.~T., Strauss, M.~A., Fan, X., et al.\ 2006, \aj, 131, 2766 

\bibitem[Robertson et al.(2010)]{robertson2010} Robertson, B.~E., Ellis, R.~S., Dunlop, J.~S., McLure, R.~J., \& Stark, D.~P.\ 2010, \nat, 468, 49 

\bibitem[Robertson et al.(2013)]{robertson2013} Robertson, B.~E., Furlanetto, S.~R., Schneider, E., et al.\ 2013, \apj, 768, 71 

\bibitem[Robertson et al.(2015)]{robertson2015} Robertson, B.~E., Ellis, R.~S., Furlanetto, S.~R., \& Dunlop, J.~S.\ 2015, \apjl, 802, L19 

\bibitem[Roberts-Borsani et al.(2016)]{roberts-borsani2016} Roberts-Borsani, G.~W., Bouwens, R.~J., Oesch, P.~A., et al.\ 2016, \apj, 823, 143 

\bibitem[Rodighiero et al.(2011)]{rodighiero2011} Rodighiero, G., Daddi, E., Baronchelli, I., et al.\ 2011, \apjl, 739, L40 

\bibitem[Rudie et al.(2012)]{rudie2012} Rudie, G.~C., Steidel, C.~C., Trainor, R.~F., et al.\ 2012, \apj, 750, 67 


\bibitem[Salpeter(1955)]{salpeter1955} Salpeter, E.~E.\ 1955, \apj, 121, 161 

\bibitem[Samui et al.(2009)]{samui2009} Samui, S., Srianand, R., \& Subramanian, K.\ 2009, \mnras, 398, 2061 


\bibitem[Sanders et al.(2016)]{sanders2016} Sanders, R.~L., Shapley, A.~E., Kriek, M., et al.\ 2016, \apj, 816, 23 
\bibitem[Santos et al.(2004)]{santos2004} Santos, M.~R., Ellis, R.~S., Kneib, J.-P., Richard, J., \& Kuijken, K.\ 2004, \apj, 606, 683 



\bibitem[Scarlata et al.(2009)]{scarlata2009} Scarlata, C., Colbert, J., Teplitz, H.~I., et al.\ 2009, \apj, 706, 1241 

\bibitem[Schaerer(2003)]{schaerer2003} Schaerer, D.\ 2003, \aap, 397, 527 
\bibitem[Schaerer \& de Barros(2009)]{schaerer2009} Schaerer, D., \& de Barros, S.\ 2009, \aap, 502, 423 




\bibitem[Schechter(1976)]{schechter1976} Schechter, P.\ 1976, \apj, 203, 297 
\bibitem[Schenker et al.(2012)]{schenker2012} Schenker, M.~A., Stark, D.~P., Ellis, R.~S., et al.\ 2012, \apj, 744, 179
\bibitem[Schenker et al.(2013)]{schenker2013} Schenker, M.~A., Robertson, B.~E., Ellis, R.~S., et al.\ 2013, \apj, 768, 196 

\bibitem[Schenker et al.(2014)]{schenker2014} Schenker, M.~A., Ellis, R.~S., Konidaris, N.~P., \& Stark, D.~P.\ 2014, \apj, 795, 20 


\bibitem[Sheth \& Tormen(1999)]{sheth1999} Sheth, R.~K., \& Tormen, G.\ 1999, \mnras, 308, 119 

\bibitem[Shklovskii(1964)]{shklovskii1964} Shklovskii, I.~S.\ 1964, \azh, 41, 801 

\bibitem[Schaerer et al.(2016)]{schaerer2016} Schaerer, D., Izotov, Y.~I., Verhamme, A., et al.\ 2016, \aap, 591, L8 

\bibitem[Shapley et al.(2006)]{shapley2006} Shapley, A.~E., Steidel, C.~C., Pettini, M., Adelberger, K.~L., \& Erb, D.~K.\ 2006, \apj, 651, 688 

\bibitem[Shibuya et al.(2012)]{shibuya2012} Shibuya, T., Kashikawa, N., Ota, K., et al.\ 2012, \apj, 752, 114 
\bibitem[Shibuya et al.(2014a)]{shibuya2014a} Shibuya, T., Ouchi, M., Nakajima, K., et al.\ 2014a, \apj, 785, 64 
\bibitem[Shibuya et al.(2014b)]{shibuya2014b} Shibuya, T., Ouchi, M., Nakajima, K., et al.\ 2014b, \apj, 788, 74 
\bibitem[Shibuya et al.(2015)]{shibuya2015} Shibuya, T., Ouchi, M., \& Harikane, Y.\ 2015, \apjs, 219, 15 
\bibitem[Shibuya et al.(2017a)]{shibuya2017a} Shibuya, T., Ouchi, M., Konno, A., et al.\ 2017a, \pasj,  

\bibitem[Shibuya et al.(2018a)]{shibuya2018a} Shibuya, T., Ouchi, M., Konno, A., et al.\ 2018a, \pasj, 70, S14 
\bibitem[Shibuya et al.(2018b)]{shibuya2018b} Shibuya, T., Ouchi, M., Harikane, Y., et al.\ 2018b, \pasj, 70, S15 
\bibitem[Shibuya et al.(2018c)]{shibuya2018c} Shibuya, T., Ouchi, M., Harikane, Y., \& Nakajima, K.\ 2018c, arXiv:1809.00765 


\bibitem[Shimakawa et al.(2015)]{shimakawa2015} Shimakawa, R., Kodama, T., Steidel, C.~C., et al.\ 2015, \mnras, 451, 1284 

\bibitem[Shimasaku et al.(2006)]{shimasaku2006} Shimasaku, K., Kashikawa, N., Doi, M., et al.\ 2006, \pasj, 58, 313 

\bibitem[Shull et al.(2012)]{shull2012} Shull, J.~M., Harness, A., Trenti, M., \& Smith, B.~D.\ 2012, \apj, 747, 100 

\bibitem[Silk \& Wyse(1993)]{silk1993} Silk, J., \& Wyse, R.~F.~G.\ 1993, \physrep, 231, 293 
\bibitem[Silk(2013)]{silk2013} Silk, J.\ 2013, \apj, 772, 112 

\bibitem[Sobacchi et al.(2016)]{sobacchi2016} Sobacchi, E., Mesinger, A., \& Greig, B.\ 2016, \mnras, 459, 2741 

\bibitem[Sobral et al.(2015)]{sobral2015} Sobral, D., Matthee, J., Darvish, B., et al.\ 2015, \apj, 808, 139 

\bibitem[Sobral et al.(2018)]{sobral2018} Sobral, D., Matthee, J., Brammer, G., et al.\ 2018, arXiv:1710.08422 

\bibitem[Speagle et al.(2014)]{speagle2014} Speagle, J.~S., Steinhardt, C.~L., Capak, P.~L., \& Silverman, J.~D.\ 2014, \apjs, 214, 15 


\bibitem[Springel et al.(2005)]{springel2005} Springel, V., White, S.~D.~M., Jenkins, A., et al.\ 2005, \nat, 435, 629 

\bibitem[Srbinovsky \& Wyithe(2007)]{srbinovsky2007} Srbinovsky, J.~A., \& Wyithe, J.~S.~B.\ 2007, \mnras, 374, 627 

\bibitem[Stark et al.(2010)]{stark2010} Stark, D.~P., Ellis, R.~S., Chiu, K., Ouchi, M., \& Bunker, A.\ 2010, \mnras, 408, 1628 
\bibitem[Stark et al.(2011)]{stark2011} Stark, D.~P., Ellis, R.~S., \& Ouchi, M.\ 2011, \apjl, 728, L2 

\bibitem[Stark et al.(2014)]{stark2014} Stark, D.~P., Richard, J., Siana, B., et al.\ 2014, \mnras, 445, 3200 
\bibitem[Stark et al.(2015a)]{stark2015a} Stark, D.~P., Richard, J., Charlot, S., et al.\ 2015a, \mnras, 450, 1846 
\bibitem[Stark et al.(2015b)]{stark2015b} Stark, D.~P., Walth, G., Charlot, S., et al.\ 2015b, \mnras, 454, 1393 

\bibitem[Steidel et al.(2000)]{steidel2000} Steidel, C.~C., Adelberger, K.~L., Shapley, A.~E., et al.\ 2000, \apj, 532, 170 
\bibitem[Steidel et al.(2001)]{steidel2001} Steidel, C.~C., Pettini, M., \& Adelberger, K.~L.\ 2001, \apj, 546, 665 

\bibitem[Steidel et al.(2010)]{steidel2010} Steidel, C.~C., Erb, D.~K., Shapley, A.~E., et al.\ 2010, \apj, 717, 289 
\bibitem[Steidel et al.(2011)]{steidel2011} Steidel, C.~C., Bogosavljevi{\'c}, M., Shapley, A.~E., et al.\ 2011, \apj, 736, 160 
\bibitem[Steidel et al.(2014)]{steidel2014} Steidel, C.~C., Rudie, G.~C., Strom, A.~L., et al.\ 2014, \apj, 795, 165 

\bibitem[Storey \& Hummer(1995)]{storey1995} Storey, P.~J., \& Hummer, D.~G.\ 1995, \mnras, 272, 41 

\bibitem[Susa \& Umemura(2004)]{susa2004} Susa, H., \& Umemura, M.\ 2004, \apj, 600, 1 

\bibitem[Takada et al.(2014)]{takada2014} Takada, M., Ellis, R.~S., Chiba, M., et al.\ 2014, \pasj, 66, R1 

\bibitem[Takeuchi et al.(2012)]{takeuchi2012} Takeuchi, T.~T., Yuan, F.-T., Ikeyama, A., Murata, K.~L., \& Inoue, A.~K.\ 2012, \apj, 755, 144 

\bibitem[Tamura et al.(2016)]{tamura2016} Tamura, N., Takato, N., Shimono, A., et al.\ 2016, \procspie, 9908, 99081M 

\bibitem[Tanvir et al.(2009)]{tanvir2009} Tanvir, N.~R., Fox, D.~B., Levan, A.~J., et al.\ 2009, \nat, 461, 1254 

\bibitem[Tapken et al.(2007)]{tapken2007} Tapken, C., Appenzeller, I., Noll, S., et al.\ 2007, \aap, 467, 63 

\bibitem[Thompson et al.(1995)]{thompson1995} Thompson, D., Djorgovski, S., \& Trauger, J.\ 1995, \aj, 110, 963 

\bibitem[Thuan \& Izotov(2005)]{thuan2005} Thuan, T.~X., \& Izotov, Y.~I.\ 2005, \apjs, 161, 240 

\bibitem[Totani et al.(2006)]{totani2006} Totani, T., Kawai, N., Kosugi, G., et al.\ 2006, \pasj, 58, 485 

\bibitem[Totani et al.(2014)]{totani2014} Totani, T., Aoki, K., Hattori, T., et al.\ 2014, \pasj, 66, 63 

\bibitem[Totani et al.(2016)]{totani2016} Totani, T., Aoki, K., Hattori, T., \& Kawai, N.\ 2016, \pasj, 68, 15 

\bibitem[Trebitsch et al.(2016)]{trebitsch2016} Trebitsch, M., Verhamme, A., Blaizot, J., \& Rosdahl, J.\ 2016, arXiv:1604.02066 

\bibitem[Tresse et al.(2007)]{tresse2007} Tresse, L., Ilbert, O., Zucca, E., et al.\ 2007, \aap, 472, 403 

\bibitem[Treu et al.(2013)]{treu2013} Treu, T., Schmidt, K.~B., Trenti, M., Bradley, L.~D., \& Stiavelli, M.\ 2013, \apjl, 775, L29 

\bibitem[Treister et al.(2011)]{treister2011} Treister, E., Schawinski, K., Volonteri, M., Natarajan, P., \& Gawiser, E.\ 2011, \nat, 474, 356 

\bibitem[van Breukelen et al.(2005)]{vanbreukelen2005} van Breukelen, C., Jarvis, M.~J., \& Venemans, B.~P.\ 2005, \mnras, 359, 895 

\bibitem[Vanzella et al.(2011)]{vanzella2011} Vanzella, E., Pentericci, L., Fontana, A., et al.\ 2011, \apjl, 730, L35 

\bibitem[Venemans et al.(2002)]{venemans2002} Venemans, B.~P., Kurk, J.~D., Miley, G.~K., et al.\ 2002, \apjl, 569, L11 

\bibitem[Verhamme et al.(2006)]{verhamme2006} Verhamme, A., Schaerer, D., \& Maselli, A.\ 2006, \aap, 460, 397 
\bibitem[Verhamme et al.(2008)]{verhamme2008} Verhamme, A., Schaerer, D., Atek, H., \& Tapken, C.\ 2008, \aap, 491, 89 

\bibitem[van Ojik et al.(1997)]{vanojik1997} van Ojik, R., Roettgering, H.~J.~A., Miley, G.~K., \& Hunstead, R.~W.\ 1997, \aap, 317, 358 

\bibitem[Vanzella et al.(2016)]{vanzella2016} Vanzella, E., de Barros, S., Vasei, K., et al.\ 2016, \apj, 825, 41 

\bibitem[Venemans et al.(2013)]{venemans2013} Venemans, B.~P., Findlay, J.~R., Sutherland, W.~J., et al.\ 2013, \apj, 779, 24 

\bibitem[Vitale et al.(2015)]{vitale2015} Vitale, M., Fuhrmann, L., Garc{\'{\i}}a-Mar{\'{\i}}n, M., et al.\ 2015, \aap, 573, A93 

\bibitem[Wardlow et al.(2014)]{wardlow2014} Wardlow, J.~L., Malhotra, S., Zheng, Z., et al.\ 2014, \apj, 787, 9 

\bibitem[Watson et al.(2015)]{watson2015} Watson, D., Christensen, L., Knudsen, K.~K., et al.\ 2015, \nat, 519, 327 

\bibitem[Weller et al.(2005)]{weller2005} Weller, J., Ostriker, J.~P., Bode, P., \& Shaw, L.\ 2005, \mnras, 364, 823 

\bibitem[Whitaker et al.(2014)]{whitaker2014} Whitaker, K.~E., Franx, M., Leja, J., et al.\ 2014, \apj, 795, 104 

\bibitem[White \& Rees(1978)]{white1978} White, S.~D.~M., \& Rees, M.~J.\ 1978, \mnras, 183, 341 

\bibitem[Willott(2011)]{willott2011} Willott, C.~J.\ 2011, \apjl, 742, L8 


\bibitem[Wisotzki et al.(2016)]{wisotzki2016} Wisotzki, L., Bacon, R., Blaizot, J., et al.\ 2016, \aap, 587, A98 

\bibitem[Worseck et al.(2014)]{worseck2014} Worseck, G., Prochaska, J.~X., Hennawi, J.~F., \& McQuinn, M.\ 2014, arXiv:1405.7405 

\bibitem[Wyithe \& Loeb(2006)]{wyithe2006} Wyithe, J.~S.~B., \& Loeb, A.\ 2006, \nat, 441, 322 

\bibitem[Xue et al.(2017)]{xue2017} Xue, R., Lee, K.-S., Dey, A., et al.\ 2017, \apj, 837, 172 


\bibitem[Yamada et al.(2012a)]{yamada2012a} Yamada, T., Nakamura, Y., Matsuda, Y., et al.\ 2012a, \aj, 143, 79 

\bibitem[Yamada et al.(2012b)]{yamada2012b} Yamada, T., Matsuda, Y., Kousai, K., et al.\ 2012b, \apj, 751, 29 

\bibitem[Yang et al.(2006)]{yang2006} Yang, Y., Zabludoff, A.~I., Dav{\'e}, R., et al.\ 2006, \apj, 640, 539 

\bibitem[Yang et al.(2009)]{yang2009} Yang, Y., Zabludoff, A., Tremonti, C., Eisenstein, D., \& Dav{\'e}, R.\ 2009, \apj, 693, 1579

\bibitem[Yatawatta et al.(2013)]{yatawatta2013} Yatawatta, S., de Bruyn, A.~G., Brentjens, M.~A., et al.\ 2013, \aap, 550, A136 


\bibitem[Zheng et al.(2005)]{zheng2005} Zheng, Z., Berlind, A.~A., Weinberg, D.~H., et al.\ 2005, \apj, 633, 791 

\bibitem[Zitrin et al.(2015)]{zitrin2015} Zitrin, A., Labb{\'e}, I., Belli, S., et al.\ 2015, \apjl, 810, L12 


\end{thebibliography}
\end{document}